\documentclass[]{article}
\usepackage{lipsum}
\usepackage{amsfonts}
\usepackage{graphicx}
\usepackage{epstopdf}
\usepackage{algorithmic}
\usepackage[margin=1in]{geometry}
\usepackage{amsmath,amssymb}
\usepackage{mathtools}
\usepackage{diagbox}
\usepackage{bm}
\usepackage{caption}
\usepackage{subcaption}
\usepackage{parskip}
\usepackage{mathrsfs} 
\usepackage{verbatim}
\usepackage{xcolor}
\usepackage{color}
\usepackage{graphicx}
\usepackage{slashbox}
\usepackage{array}
\usepackage[title]{appendix}
\usepackage{hyperref}
\usepackage{titlesec}
\usepackage{float}
\title{Elastohydrodynamics of a soft coating under fluid-mediated loading by a spherical probe}
\author{Pratyaksh Karan $~~~$ Jeevanjyoti Chakraborty $~~~$ Suman Chakraborty  \\ \\
Department of Mechanical Engineering, \\ Indian Institute of Technology Kharagpur, India
}
\date{}
\begin{document}
\maketitle
\begin{abstract}
Motion of an object near a soft wall with intervening fluid is a canonical problem in elastohydrodynamics, finding presence in subjects spanning biology to tribology. Particularly, motion of a sphere towards a soft substrate with intervening fluid is often encountered in the context of scanning probe microscopy. While there have been fundamental theoretical studies on this setup, they have focussed on specific applications and thus enforced suitable simplifications. Here we present a versatile semi-analytical framework for studying the elastohydrodynamics of axisymmetric loading of a rigid sphere near a soft elastic substrate coated on a rigid platform mediated by an aqueous electrolytic solution. Three loading modes are considered - approach, recession and oscillatory. The framework incorporates - large oscillation frequency and amplitude, two-way coupling between pressure and substrate deformation, and presence of DLVO forces. From computations using the framework, we gain insights on the effects of DLVO forces, substrate thickness, substrate material compressibility (quantified by Poisson's ratio) and high oscillation frequency for different physical setups encountered in SPM and the likes. We list some key observations. A substrate that is thicker and more compressible allows for larger deformation, i.e. is effectively softer. Presence of DLVO forces lead to magnification in force of upto two orders of magnitude and in substrate deformation of upto an order of magnitude for oscillatory loading at low frequencies and approach/recession loading at low speed. For oscillatory loading at high frequencies, DLVO forces do not appreciably affect the force and deflection behaviour of the system. Having demonstrated the versatility and utility of our framework, we expect it to evolve into a diverse and useful tool for solving problems of soft-lubrication.  
\end{abstract}
\section{Introduction}\label{sec:Introduction}
The motion of an object near a wall with fluid being squeezed between the two is an important physical problem in hydrodynamics, motivating numerous fundamental studies during the early days of hydrodynamics, specially low Reynolds hydrodynamics \cite{Brenner1961,Brenner1967,Brenner1967a,Goldman1967,Goldman1967a,Stewartson1967,ONeill1968,Cooley1968,Cooley1969,Johnson1985,Vinogradova1995,Hamrock2004,Williams2005,Zhang2005}. This setup finds presence in research areas like motion of colloidal particles in confined flow domains \cite{Asakura1954,Balmforth2010,Cawthorn2010,Glover2007,Michailidou2009,Tabatabaei2006,Tabatabaei2006a,Toporov1975,Trefalt2016,Urzay2010,VandeVen1993,VandeVen1993a,Warszynski1998,Vinogradova2000,Wu1996}, flow of vesicles and cells near blood vessel wall and other biological processes \cite{Beaucourt2004,Secomb1986,Shen2007,Trouilloud2008}, tribological devices and modules \cite{Higginson1962,Jang1995,Chu2006,Chu2006a,Chu2010,Chakraborty2011,Larsson1997,Li2010,Liu2016,Rallabandi2017,Saintyves2016,Salez2015,Scaraggi2014,Shinkarenko2009,Shinkarenko2009a}, scanning probe microscopy for force laws and surface characterization \cite{Butt1991,Butt2005,Jones2005,Carpentier2015,Kaveh2014,Leroy2011,Leroy2012,Matsuoka1997,Murat1996,Restagno2002,Steinberger2008,Villey2013,Wang2015,Wang2017,Wang2017a,Zhang2019a}, and many more. In some cases, certain tribological devices for instance, the deformation of the object as well as the wall is small enough to be considered absent. In other cases, when the object or the wall is made of soft polymer for example, the combination of involved materials and imposed dynamics can be (and frequently is) such that neglecting the deformation of the object and/or the wall can amount to an erroneous analysis of the associated physics. To account for the presence and effects of this deformation in the `motion of object near wall with intervening fluid', the field of soft-lubrication came into existence \cite{Mahadevan2004,Mahadevan2005,Weekley2006,Yang1991,Yin2005,Zhang2019,Zhao2019,Temizer2016,Pandey2016}, constituting a major topic under the expansive discipline of elastohydrodynamics (EHD) \cite{Dowson1995,Dowson1999,Naik2017,Anand2018,Anand2019a,Christov2018,Wang2019,Karan2018,Mukherjee2013}.\\
The deformation of the object/wall/both is the outcome of interplay between softness of the material (of object/wall), which is constitutive material behaviour for part of the system, and force interactions between object and wall, which is part of the imposed dynamics on the system. While not constrained to, consitutive behaviour of the object/wall material for common setups in soft-lubrication typically falls under either elastic or viscoelastic, and often remains restricted to the linear strain limit. On the other hand, the force interactions between object and wall fall under two categories - hydrodynamic and non-hydrodynamic \cite{Davis1986,Israelachvili1982b,Israelachvili1988,Israelachvili2011,Karan2019,Urzay2010}. Hydrodynamic force is the sole force interaction when there is sufficiently large separation between object and wall for entirety of the former's range of motion. However, when this separation reaches below a threshold, even for part of the object's range of motion, non-hydrodynamic forces emerge and can be comparable to or even significantly larger than hydrodynamic force. These non-hydrodynamic forces are divided into two categories - DLVO forces and non-DLVO molecular forces. DLVO forces comprise van der Waals force and EDL disjoining force. van der Waals force is the aggregate outcome of the induced dipole interactions between the object, the wall and the intervening fluid. EDL disjoining force (often expressed as force per unit area, called EDL disjoining pressure) arises when the intervening fluid is an electrolytic solution. This pressure is simply the osmotic pressure of non-uniform distrobution of the ionic species in the intervening fluid because of the quasi-equilibrium interaction of the EDLs (electrical double layers) on the surfaces of the object and the wall (and hence is strong only when the EDLs overlap appreciably, which occurs at separations smaller than the Debye length) \footnote{While EDL disjoining pressure is the osmotic pressure because of quasi-equilibrium interaction of the EDLs, flow of the intervening fluid leads to a streaming potential between the inner and outer regions of the object-wall gap. This streaming potential acts to oppose the flow that creates it, often quantified as an enhancement in the fluid viscosity and termed as electroviscous effect \cite{Tabatabaei2006,Tabatabaei2006a,VandeVen1993,VandeVen1993a,Chakraborty2008,Chakraborty2010,Zhao2020}. In contrast to EDL osmotic pressure which is conservative, force due to the electroviscous effect is dissipative. While electroviscous effect is a cricual component that should be considered in a more comprehensive treatment of the electrokinetics in soft-lubrication studies, it is sufficiently smaller in magnitude than EDL disjoining pressure that we do not take it into consideration for current analysis.}. At even smaller separations, non-DLVO molecular forces, like solvation force, hydration force, steric hindrance, etc., arise as well \cite{Israelachvili2011}. However, for many contemporary soft-lubrication problems, the separation stays sufficiently large that accounting for only DLVO forces is sufficient. \\
We now turn our attention to the object and its motion. In the most general case, the object can be of any shape. It can execute any arbitrary combination of three base motions - translation parallel to the wall, translation perpendicular to the wall, and rotation at an arbitrary axis to the wall . Furthermore, this motion can be any arbitrary combination of imposed or spontaneous. Assuming the intervening fluid and the solid object/wall are isotropic in their material constitution, the flow dynamics in the squeeze gap (between object and wall) and the deformation characteristics of the soft object/wall will inherit any simplying features of the object's shape and motion. Consequently, the complete system behaviour will fall under one of three categories - planar, axisymmetric, or general. An example of planar system behaviour is rolling or sliding (or a combination of the two) of a cylinder near a wall \cite{Salez2015,Pandey2016}. An example of axisymmetric system behaviour is approach of a spherical particle towards a wall or another spherical particle \cite{Brenner1961,Hinch1986}. Lastly, examples of general behaviour are motion of a vesicle or micro-swimmer near a wall \cite{Beaucourt2004,Trouilloud2008} and the common tribological setup of a ball being dragged on a disk \cite{Li2013,Stupkiewicz2016}.\\
A major domain of experimental physics where soft-lubrication often comes into picture is scanning probe microscopy (SPM). SPM comprises a large family of devices including colloidal probe microscopy (AFM) to surface force apparatus (SFA). Applications of SPM are sample surface topological scanning, obtaining force laws, and elasticity characterization to name a few \cite{Butt2005,Butt1991,Crassous1997,Crassous1993,Shubin1993,Restagno2002,Murat1996,Steinberger2008,Leroy2011,Leroy2012,Wang2015,Wang2017,Wang2017a}. In many situations, a SPM setup consists of a spherical probe constrained to move along an axis perpendicular to the planar surface of a substrate - an oscillatory or a constant velocity motion is typically imposed on the probe \cite{Leroy2011,Leroy2012,Villey2013,Wang2017,Wang2015}. Quite often, the probe is significantly more rigid than the substrate. Furthermore, in some cases, for instance when using SPM to characterize some aspect of a delicate substrate that is prone to damage in direct contact, a fluid is inserted between the probe and the substrate \cite{Leroy2011,Leroy2012,Villey2013}. As a result, the physical setup of `a rigid sphere moving over a soft substrate coated on a rigid platform with a fluid filling the gap between sphere and coating, where the sphere executes sinusoidal oscillations or constant velocity motion along an axis perpendicular to the initially planar fluid-substrate interface' (which can succintly be called `fluid-mediated loading of a soft substrate coating by a rigid sphere') acts as a suitable one for the mathematical modelling of a major subclass of SPM studies \cite{Wang2017a,Leroy2011} \footnote{Even for SFA studies, that involve two cross-cylinders with squeeze-flow in between, sphere near a plane wall is often deemed as the appropriate proxy for theoretical treatment \cite{Wang2015,Wang2017,Wang2017a}.}. Lastly, the substrate material and intervening fluid are usually isotropic and homogeneous. Therefore, the system behaviour falls suitably in the axisymmetric category discussed in the last paragraph.\\
Hence, we set out with the general topic of soft lubrication. Then, we have identified the specific physical setup of fluid-mediated loading on a soft substrate coating by a rigid sphere, that the discipline of soft-lubrication caters to. We have also established that this setup bears immense importance for theoretical modelling a multitude of SPM studies. There have been some key studies on this setup that have made significant contributions to advancement of SPM. Leroy and Charlaix (2011) \cite{Leroy2011} performed a theoretical study on the elastohydrodynamics of a sphere oscillating over a deformable substrate layer on a platform. The objective of their work was to assist in developing a methodology for elasticity characterization of substrates using non-contact mode of AFM. The key aspect of their study (and the associated methodology) was the decomposition of the force response into elastic and viscous components, i.e. storage modulus $G^{\prime}$ and loss modulus $G^{\prime\prime}$. To this end, the spherical probe was oscillated with a small amplitude and a high frequency. Kaveh et al (2014) \cite{Kaveh2014} performed an experimental and theoretical study of approach and recession loading of an spherical AFM probe on a soft deformable wall. They used finite element analysis (FEA) to obtain the computational predictions, which matched well with the experimental findings. The substrate deformation in their study assumed values comparable to the gap between the surfaces, a notable contrast with the study by Leroy and Charlaix (2011) \cite{Leroy2011}. Utilizing a methodology similar to Leroy and Charlaix (2011) \cite{Leroy2011}, Wang et al (2017) \cite{Wang2017a} developed a theoretical framework for the study of fluid drainage between a sphere and a soft coating of an incompressible material. They subsequently used their framework to study softness and roughness of contact surface \cite{Wang2018} and characteristics of particle rebound during fluid-mediated collision of a sphere with a soft coating  \cite{Tan2019}. Their studies are similar to Kaveh et al (2014) \cite{Kaveh2014} in geometry and imposed system dynamics and to Leroy and Charlaix (2011) \cite{Leroy2011} in theoretical framework. Although these studies are crucial contributions to EHD literature on sphere undergoing `axisymmetric motion near a wall', each study consisted of simplifications and limitations that were respectively appropriate but rendered the individual frameworks mutually inapplicable. The study by Leroy and Charlaix (2012) \cite{Leroy2012} studied small oscillation amplitude and small substrate deformation compared to sphere-substrate gap. The study by Kaveh et al (2014) \cite{Kaveh2014} relied on a full computational solution obtained using the propriety computational package COMSOL. The study by Wang et al (2017), Wang and Frechette (2018) and Tan et al (2019) \cite{Wang2017a,Wang2018,Tan2019} considered only incompressible substrate materials. \\
Furthermore, we know that for certain instances of the physical setup in consideration (depending separation of sphere from substrate), forces of non-hydrodynamic origin can be important in determining and perhaps exclusively dictating the force interactions between sphere and substrate and in substrate deformation. There have been few studies towards this end. While not explicitly specified as being applicable to SPM, Serayssol and Davis (1986) \cite{Davis1986} studied the elastohydrodynamic approach and collision of two spheres where DLVO forces were taken into consideration. Similarly, a soft-lubrication study was executed by Urzay (2010) \cite{Urzay2010} which incorporated the effect of DLVO forces, focussing on the nature of adhesion between the sphere and soft substrate. In a pre-cursor to current study, we studied the effects of solvation force on fluid-mediated oscillatory loading of sphere on a compliant ultra-thin coating \cite{Karan2019}. Here too, each study enforced simplifications and restrictions that were respectively appropriate. The geometrical setup of Serayssol and Davis (1986) \cite{Davis1986} consisted of two soft spheres both of which behaved similar to infinite half-spaces. Urzay (2010) \cite{Urzay2010} studied the translation of a sphere about an axis parallel to the wall, i.e. it doesn't fall under the family of `fluid-mediated loading on a soft substrate coating by a rigid sphere' setups. To assess the exclusive effects of solvation force using a simple analytical framework, we \cite{Karan2019} imposed the simplifying restrictions of low oscillation frequency, thinness of coating and substrate material being sufficiently compressible. Collating the different aspects from the \\   
The aforementioned studies constitute the essential foundation for studying the dynamics and characteristics of fluid-mediated loading of a soft coating with a rigid sphere. However, their respective simplifying restrictions, made to cater to the specific respective problems, precluded each from qualifying as a general theoretical framework for the family of problems for `fluid-mediated loading on a soft substrate coating by a rigid sphere' setup. Collating the different crucial aspects from the previous paragraphs, we find that a expansive and general theoretical framework for this setup should be capable of incporporating appreciable range of each of these - oscillation frequncy for oscillatory loading, sphere speed for approach/recession loading, oscillation amplitude, substrate coating thickness, substrate material constitutive parameters (i.e. Lam\'e's parameters), and presence of DLVO forces. Furthermore, the framework should be capable of exhaustively incorporating the different modes of two-way coupling between pressure on the substrate and its deformation, as this coupling occurs frequently and has non-trivial implications on the system behaviour. The absence of such a framework constitutes a compelling gap in the literature on soft-lubrication. Motivated to fill this gap, here we present a semi-analytical framework for modelling of the Newtonian fluid mediated sinusoidal oscillatory or constant velocity approach or constant velocity recession loading of a rigid sphere on a homogeneous isotropic linear elastic substrate coated on a rigid platform. For solving the substrate deformation, we employ a Hankel-transform based formulation \cite{Harding1945,Li1997,Leroy2011,Wang2017a}. In building this framework, we ensure that each of the aforementioned aspects is incorporated. \\
With the developed framework, we have obtained solutions for oscillatory, approach and recession loading, for different substrate thicknesses, different substrate material compressibilities (quantified by Poisson's ratio), in presence and absence of DLVO forces, and for different frequencies for oscillatory loading and different loading speeds for recession loading. Some of the key observations are as follows. Higher substrate thickness and lower values of Poisson's ratio (i.e. higher substrate compressibility) render the substrate `effectively' softer. The nature of the different pressure components considered in this study, hydrodynamic, van der Waals and EDL disjoining, is such that - if the total pressure is attractive, the deflection it causes functions to magnify the pressure magnitude and if the total pressure is repulsive, the deflection it causes functions to diminish the pressure magnitude. This is an effect of the two-way coupling between pressure on substrate and its deformation. Furthermore, since an `effectively' softer allows for higher deflection, it leads to higher magnification of attractive pressure and higher diminution of repulsive pressure. The presence of DLVO forces leads to magnification in force characteristics of upto two orders of magnitude and in deflection characteristics of upto an order of magnitude, as compared to presence of only hydrodynamic force. As we consider lower to higher loading speed for approach and recession loading, the loading speed heavily influences force and deflection when DLVO forces are absent. However, in the presence of DLVO forces, both force and deflection lose sensitivity to loading speed for small and medium separations. We emphasize that in order to avoid adhesion-like characteristics (which emerge as a limitation of the framework), we consider only moderate loading speeds and not too high. On the other hand, we are able to consider moderate as well as high frequencies for oscillatory loading. As the oscillating frequency gets higher, the presence of DLVO forces does not affect the system behaviour significantly, because hydrodynamic pressure becomes dominant over pressure corresponding to DLVO forces (i.e. EDL disjoining pressure and van der Waals pressure). Furthermore, we explore the decomposition of force response into  $G^{\prime}$ and $G^{\prime\prime}$, as was done by Leroy and Charlaix (2011) \cite{Leroy2011}. This decomposition is another effect of the two-way coupling between pressure on substrate and its deformation. For high frequency oscillations, increasing oscillation amplitude leads to some degeneration of this decomposition but does not completely eliminate it. Also, extremely high frequency and low frequency leads to insensitivity of $G^{\prime}$ and $G^{\prime\prime}$ to the average separation of sphere from the undeformed interface of fluid and substrate. The stark effects of the different aspects mentioned in previous paragraph and the strong coupling of pressure on substrate and its deformation on the system behaviour, as observed from implementation of our framework, galvanizes its requirement and commensurate utility. \\
The rest of the article is arranged as follows. In section \ref{sec:Math}, the geometry, imposed dynamics, and notation for the mathematical analysis are presented, alongwith the mathematical formulation - scaling and non-dimensionalization, governing equations and boundary conditions, simplified equations that represent the problem, pressure components corresponding to DLVO forces, and Hankel-space expression connecting pressure on substrate and its deformation. In section \ref{sec:Soln}, the methodology for computing deflection in both the regimes of EHD interaction, i.e. one-way coupling and two-way coupling, is presented. Along with it, simplified approximations of the Hankel-space substrate compliance variable for the thin and semi-infinite limits, and limits of substrate incompressibility, are presented. In section \ref{sec:Results}, we present the results, in terms of the force between sphere and substrate and fluid-substrate interface deflection. The article is concluded in section \ref{sec:Conclusion}.
\section{Mathematical Formulation}\label{sec:Math}
\subsection{Model Setup}\label{subsec:model}
\begin{figure}[!htb]
\centering
\centering
\includegraphics[width=0.5\textwidth]{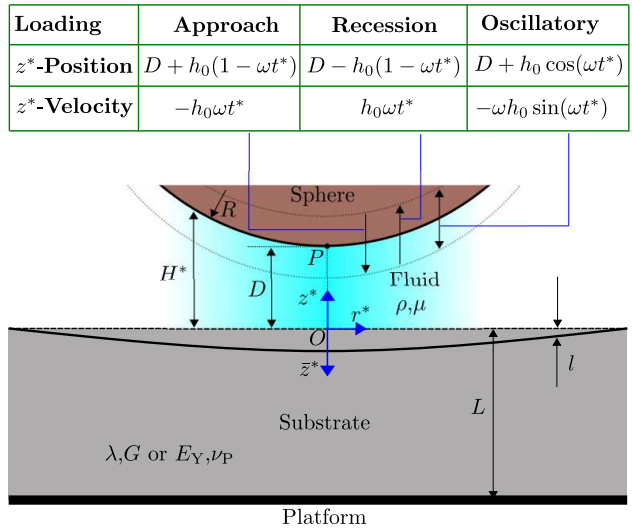}
\caption{Representation of the physical setup; in the table,  `$z^*$-Position' is the $z^*$-position of the fluid-sphere interface at the point closest to the origin, $O$, i.e. the point $P$.}
\label{fig:schematic}
\end{figure}

The physical setup for this study is presented in Figure \ref{fig:schematic}. Time is represented by $t^*$. For the fluid domain, $r^*-z^*$ is used as the co-ordinate system, and, for the substrate domain, $r^*-\bar{z}^*$ is used as the co-ordinate system. The line passing through $O$ and $P$ is henceforth referred to as `Centerline'. Radius of the sphere is $R$. Undeformed substrate thickness is $L$. The ratio of $L$ to $R$ is denoted by $\beta$. The mean separation of sphere from origin is $D$. The ratio of $D$ to $R$ is denoted by $\epsilon$. For oscillatory loading, $\omega$ and $h_0$ are the oscillation amplitude and oscillation frequency. For approach and recession loading, $\omega h_0$ is the constant speed of the sphere. The ratio of $h_0$ to $D$ is denoted by $\alpha$. During approach loading, the sphere moves from $z^* = D+h_0$ to $z^* = D-h_0$ and during recession loading, it moves from $z^* = D-h_0$ to $z^* = D+h_0$, which amount to the same range of motion as the oscillatory loading. Fluid-substrate interface deflection, henceforth referred to as `deflection', is represented by $l^*$. The sphere's profile, i.e. the fluid-sphere interface referenced to the $x^*$ axis is represented by $H^*$. For the mathematical formulation of system domains, $\vec{v}^*$ is the fluid velocity field, $\vec{u}^*$ is the substrate displacement field, $p_{\text{hd}}^*$ is the hydrodynamic pressure, $p_{\text{EDL}}^*$ is the EDL disjoining pressure, $p_{\text{vdW}}^*$ is the van der Waals pressure, $p^*$ is the total pressure (this pressure appears in the fluid-substrate interface traction balance boundary condition), and $F^*$ is the total force between the sphere and the substrate.  \\
The intervening fluid is an incompressible homogeneous isotropic Newtonian fluid with density and viscosity as $\rho$ and $\mu$ respectively. The substrate is a homogeneous isotropic linear-elastic solid, with  Lam\'e's first parameter and shear modulus as $\lambda$ and $G$ respectively. Alternatively, the substrate's compliance is quantified by its Young's modulus and Poisson's ratio, $E_{\text{Y}}$ and $\nu$ respectively.\\
In the lubrication zone, the shape of the sphere can be approximated as parabolic \cite{Leroy2011,Urzay2010}. Thus, the expression for $H^*$ is,
\begin{equation}
\label{eq:H}
H^* = 
\begin{dcases}
D+\frac{r^{*2}}{2R}+h_0\cos(\omega t^*) & \text{\hspace{35pt} oscillatory}, \\
D+\frac{r^{*2}}{2R}+h_0(1-\omega t^*) & \text{\hspace{35pt} approach}, \\
D+\frac{r^{*2}}{2R}-h_0(1-\omega t^*) & \text{\hspace{35pt} recession}.
\end{dcases}
\end{equation}
\subsection{Scaling and Non-Dimensionalization}\label{subsec:scaling}
All the system variables are non-dimensionalized in order to solve the governing equations for system behaviour. The non-dimensionalized variables are represented by the same notation as their dimensional counterparts but with the superscript $^*$ dropped. The pertinent system variables can be categorized into two categories - independent variables ($r^*$, $z^*$, $\bar{z}^*$ and $t^*$) and the dependent variables  ($\vec{v}^*$, $\vec{u}^*$, $p_{\text{hd}}^*$, $p_{\text{EDL}}^*$, $p_{\text{vdW}}^*$ and $p^*$ and  $F^*$). 
\subsubsection{Non-dimensionalization of Independent Variables}\label{subsubsec:independent}
We first consider the non-dimensionalization of independent variables. The scale for $t^*$ is $\displaystyle \frac{1}{\omega}$, giving us $t=\omega t^*$. The scale for the other independent variables are dependent upon what geometry and imposed dynamics are being considered, as discussed ahead. The scale for $\bar{z}^*$ is taken as $\delta R$, giving us $\bar{z}^* = \delta R\bar{z}$; expression for $\delta$ is obtained out of considerations for the deformation characteristics, and is presented ahead in section \ref{subsubsec:domain}. It is emphasized that the scale of $r^*$ is same in the fluid domain and the substrate domain, as the load on the substrate comes from the fluid domain in the form of pressure. \\
We first consider the case of $\alpha \ll 1$, i.e. the sphere's displacement over its range of motion is small compared to its mean separation from the origin. For this case, the scale for $z^*$ is $D = \epsilon R$, giving us $z^*=\epsilon Rz$. Subseqently, following the classical lubrication methodology \cite{Urzay2010,Leal2007}, the scale for $r^*$ is $\sqrt{RD} = \epsilon^{\frac{1}{2}}R$, giving us $r^* = \epsilon^{\frac{1}{2}}Rr$. \\
We now consider the case of $\alpha \sim 1$, i.e. the sphere's displacement over its range of motion is comparable to its mean separation from the origin. The prime consequence of $\alpha \sim 1$ is that the scale for $z^*$ is,
\[ 
z^* \sim d(t^*)=
\left\{
\begin{array}{ll}
D+h_0\cos(\omega t^*) = (1+\alpha\cos(\omega t^*))\epsilon R & \text{\hspace{35pt} oscillatory}, \\
D+h_0(1-\omega t^*) = (1+\alpha (1-\omega t^*))\epsilon R  & \text{\hspace{35pt} approach}, \\
D-h_0(1-\omega t^*) = (1-\alpha (1-\omega t^*))\epsilon R  & \text{\hspace{35pt} recession},
\end{array} 
\right\}
= k(t^*)\epsilon R,
\]
giving us $z^* = k(t^*)\epsilon Rz$. Again, following the classical lubrication methodology, the scale for $r^*$ is $\sqrt{Rd} = k(t^*)^{\frac{1}{2}}\epsilon^{\frac{1}{2}}R$, giving us $r^* = k(t^*)^{\frac{1}{2}}\epsilon^{\frac{1}{2}}Rr$. The expression for $\delta$ as well is prone to being time-dependent for this case, i.e. $\delta = \delta(t^*)$, as seen in section \ref{subsubsec:domain}.
\subsubsection{Conformal Mapping and Transformation of Derivatives}\label{subsubsec:ap_conformal}
Considering time-dependent scale for the independent spatial variables, as done for the case of $\alpha \sim 1$ above, is equivalent to performing a conformal mapping on the system domains. This conformal mapping stems from mapping of $z^*$ to $z$, which is unity for all times. The forward mapping, of the original independent variables ($r^*,z^*,\bar{z}^*,t^*$) to mapped independent variables $(r,z,\bar{z},t)$, as well as backward mapping is presented in table \ref{tab:ap_mapping}.\\
\begin{table}[tbh!]
\def\arraystretch{1.25}
\small
\centering
\caption{Forward and Backward Mapping from $(r^*,z^*,\bar{z}^*,t^*)$ to $(r,z,\bar{z},t)$}
\label{tab:ap_mapping}
\begin{tabular*}{0.80\textwidth}{@{\extracolsep{\fill}}|c|c|c|}
\hline
\textbf{forward} & \textbf{backward} & \textbf{(purely) backward} \\
\hline
$\displaystyle r=\frac{r^*}{k(t^*)^{\frac{1}{2}}\epsilon^{\frac{1}{2}}R}$	&	$\displaystyle r^*=k(t^*)^{\frac{1}{2}}\epsilon^{\frac{1}{2}}Rr$ &	$\displaystyle r^*=k(t)^{\frac{1}{2}}\epsilon^{\frac{1}{2}}Rr$ \\
$\displaystyle z=\frac{z^*}{k(t^*)\epsilon R}$	&	$\displaystyle z^*=k(t^*)\epsilon Rz$ &	$\displaystyle z^*=k(t)\epsilon Rz$ \\
$\displaystyle \bar{z}=\frac{\bar{z}^*}{\delta(t^*)R}$	&	$\displaystyle \bar{z}^*=\delta(t^*) Rz$ &	$\displaystyle \bar{z}^*=\delta(t) Rz$ \\
$\displaystyle t=\omega t^*$	&	$\displaystyle t^*= \frac{t}{\omega}$ &	$\displaystyle t^*= \frac{t}{\omega}$ \\
\hline
\end{tabular*}
\end{table}
Based on the mapping in table \ref{tab:ap_mapping}, the dependent variables (which are functions of various combinations of original independent variables) undergo functional transformations as,
\begin{subequations}
\label{eq:ap_functransforms}
\begin{equation}
\label{eq:ap_ft}
f(t^*) \equiv f(t^*(t)) = f(t)
\end{equation}
\begin{equation}
\label{eq:ap_frt}
f(r^*,t^*) \equiv f(r^*(r,t),t^*(t)) = f(r,t)
\end{equation}
\begin{equation}
\label{eq:ap_frzt}
f(r^*,z^*,t^*) \equiv f(r^*(r,t),z^*(z,t),t^*(t)) = f(r,z,t)
\end{equation}
\begin{equation}
\label{eq:ap_fryt}
f(r^*,\bar{z}^*,t^*) \equiv f(r^*(r,t),\bar{z}^*(\bar{z},t),t^*(t)) = f(r,\bar{z},t)
\end{equation}
\end{subequations}
Using equation \eqref{eq:ap_ft}, we get $k(t^*)\equiv k(t)$ and $\delta(t^*)\equiv\delta(t)$. This gives us the backward mapping of the original independent variables ($r^*,z^*,\bar{z}^*,t^*$) to entirely the mapped independent variables ($r,z,\bar{z},t$), as presented in the third column of table \ref{tab:ap_mapping}. \\
Furthermore, using the chain rule, the derivatives of the independent variables get transformed as,
\begin{equation}
\label{eq:ap_derivatives}
\frac{\partial\mathbb{U}}{\partial\mathbb{X}^*} \equiv \frac{\partial\mathbb{U}}{\partial r}\frac{\partial r}{\partial \mathbb{X^*}} + \frac{\partial\mathbb{U}}{\partial z}\frac{\partial z}{\partial \mathbb{X^*}} + \frac{\partial\mathbb{U}}{\partial \bar{z}}\frac{\partial \bar{z}}{\partial \mathbb{X^*}} + \frac{\partial\mathbb{U}}{\partial t}\frac{\partial t}{\partial \mathbb{X^*}},
\end{equation}
where $\mathbb{U}$ is an arbitrary dependent variable and $\mathbb{X^*}$ is one of the original independent variables. The second order derivatives are obtained recursively by plugging $\displaystyle \frac{\partial\mathbb{U}}{\partial\mathbb{X}^*}$ in place of $\mathbb{U}$ in equation \eqref{eq:ap_derivatives}.\\
As evident from equation \eqref{eq:ap_derivatives}, we require derivatives of mapped independent variables with respect to original independent variables (`mapped w.r.t. original') in order to transform derivatives of dependent variables with respect to original indepdent variables (`dependent w.r.t. original') into derivatives of dependent variables with respect to mapped independent variables (`dependent w.r.t. mapped'). These `mapped w.r.t. original' derivatives are,
\begin{subequations}
\label{eq:ap_derivmap}
\begin{equation}
\label{eq:ap_dtstdt}
\frac{\partial t}{\partial t^*} = \frac{\partial (\omega t^*)}{\partial t^*} = \omega,
\end{equation}
\begin{equation}
\label{eq:ap_drstdt}
\frac{\partial r}{\partial t^*} = \frac{\partial }{\partial t^*}\left(\frac{r^*}{k(t^*)^{\frac{1}{2}}\epsilon^{\frac{1}{2}}R}\right)=-\frac{r^*}{2k^{\frac{3}{2}}\epsilon^{\frac{1}{2}}R}\frac{dk}{dt^*}=-\frac{r}{2k}\frac{dh}{dt^*}\equiv-\frac{\omega r}{2k}\frac{dk}{dt},
\end{equation}
\begin{equation}
\label{eq:ap_dzstdt}
\frac{\partial z}{\partial t^*} \equiv -\frac{\omega z}{k}\frac{dk}{dt},~~~~~
\frac{\partial \bar{z}}{\partial t^*} \equiv -\frac{\omega \bar{z}}{\delta}\frac{d\delta}{dt},
\end{equation}
\begin{equation}
\label{eq:ap_dspdsp}
\frac{\partial r}{\partial r^*} \equiv \frac{1}{k^{\frac{1}{2}}\epsilon^{\frac{1}{2}}R},~~~~~
\frac{\partial z}{\partial z^*} \equiv \frac{1}{k\epsilon R},~~~~~
\frac{\partial \bar{z}}{\partial \bar{z}^*} \equiv \frac{1}{\delta R}.
\end{equation}
\end{subequations}
All the `mapped-original' derivatives that are not presented in equation \eqref{eq:ap_derivmap} are zero. Equation \eqref{eq:ap_dtstdt} is straightforward to derive. In equation \eqref{eq:ap_drstdt}, the second last step involves substituting the expression for $r^*$ in terms of $r$ (from table \ref{tab:ap_mapping}), and the last step (i.e. the `$\equiv$' step) involves transforming $\displaystyle \frac{dk}{dt^*}$ to $\displaystyle \frac{dk}{dt}$ using equation \eqref{eq:ap_derivatives}. Since $k$ is a function of only $t$, the first three terms (i.e. $\displaystyle \frac{\partial k}{\partial r}$, $\displaystyle \frac{\partial k}{\partial z}$, and $\displaystyle \frac{\partial k}{\partial \bar{z}}$) vanish, thus substituting $\displaystyle \frac{\partial t}{\partial t^*}=\omega$ from equation \eqref{eq:ap_dtstdt} gives $\displaystyle \frac{dk}{dt^*}\equiv\omega\frac{dk}{dt}$. Equations \eqref{eq:ap_dzstdt} and \eqref{eq:ap_dspdsp} are obtained using similar procedure as for equation \eqref{eq:ap_drstdt}.\\
Using equations \eqref{eq:ap_derivatives} and \eqref{eq:ap_derivmap}, the pertinent derivatives of dependent variable with respect to mapped independent variables are,
\begin{subequations}
\label{eq:derivatives}
\begin{equation}
\frac{\partial \mathbb{U}}{\partial t^*} \equiv \omega\left[\frac{\partial\mathbb{U}}{\partial t}-\left\{\left(\frac{r}{2}\frac{\partial\mathbb{U}}{\partial r}+z\frac{\partial\mathbb{U}}{\partial z}\right)\frac{1}{k}\frac{dk}{dt}+\frac{\bar{z}}{\delta}\frac{d\delta}{dt}\frac{\partial\mathbb{U}}{\partial\bar{z}}\right\}\right],
\label{eq:deriv_t}
\end{equation}
\begin{equation}
\frac{\partial \mathbb{U}}{\partial r^*} \equiv \frac{1}{k^{\frac{1}{2}}\epsilon^{\frac{1}{2}}R}\frac{\partial\mathbb{U}}{\partial r},~~~~
\frac{\partial^2 \mathbb{U}}{\partial r^{*2}} \equiv \frac{1}{k\epsilon R^2}\frac{\partial^2\mathbb{U}}{\partial r^2},
\label{eq:deriv_r}
\end{equation}
\begin{equation}
\frac{\partial \mathbb{U}}{\partial z^*} \equiv \frac{1}{k\epsilon R}\frac{\partial\mathbb{U}}{\partial z},~~~~\frac{\partial^2 \mathbb{U}}{\partial z^{*2}} \equiv \frac{1}{k^2\epsilon^2 R^2}\frac{\partial^2\mathbb{U}}{\partial z^2},
\label{eq:deriv_z}
\end{equation}
\begin{equation}
\frac{\partial \mathbb{U}}{\partial \bar{z}^*} \equiv \frac{1}{\delta R}\frac{\partial\mathbb{U}}{\partial \bar{z}},~~~~\frac{\partial^2 \mathbb{U}}{\partial \bar{z}^{*2}} \equiv \frac{1}{\delta^2 R^2}\frac{\partial^2\mathbb{U}}{\partial \bar{z}^2},
\label{eq:deriv_y}
\end{equation}
\end{subequations}
\subsubsection{Non-Dimensionalization of Dependent Variables}\label{subsubsec:dependent}
Amongst the dependent variables, $\vec{v}^*$ and $p_{\text{hd}}^*$ are functions of $r^*$, $z^*$ and $t^*$; $\vec{u}^*$ is a function of $r^*$, $\bar{z}^*$ and $t^*$; $p_{\text{EDL}}^*$, $p_{\text{vdW}}^*$, $p_{\text{Sol}}^*$, $p^*$, $H^*$ and $l^*$ are functions of $r^*$ and $t^*$; and $F^*$ is a function of $t^*$. \\
The scale for substrate discplacement, $\vec{u}$ is taken as $\theta R$, giving us $\vec{u}^* = \theta R \vec{u}$. The expression for $\theta$ depends on the applied load, and is obtained ahead in section \ref{subsubsec:domain}. The scale for total pressure, $p^*$ is taken as $\Pi$, whose expression is dependent on the force interactions in the system, and is obtained ahead in section \ref{subsec:nonhd}. The scale for $l^*$ is the same as the scale for $\vec{u}^*$, giving us $l^* = \theta Rl$. Lastly, $H^*$ is scaled similar to $z^*$, giving us $H^*=\epsilon RH$.\\
Turning attention to the flow dynamics, we first consider the case of $\alpha \ll 1$. Following the classical lubrication formalism, the scales for velocity components and hydrodynamic pressure are such that $\displaystyle v_r^* = \epsilon^{\frac{1}{2}}\alpha\omega R v_r$, $\displaystyle v_z^* = \epsilon\alpha\omega R v_z$, and $\displaystyle p_{\text{hd}}^* = \frac{\mu\omega\alpha}{\epsilon}p_{\text{hd}}$. The expression for $H$ for this case is,
\begin{equation}
\label{eq:H_nd_LSM}
H = 
\begin{dcases}
1+\alpha\cos(t)+\frac{r^2}{2} & \text{\hspace{35pt} oscillatory}, \\
1+\alpha (1-t)+\frac{r^2}{2}  & \text{\hspace{35pt} approach}, \\
1-\alpha (1-t)+\frac{r^2}{2}  & \text{\hspace{35pt} recession}.
\end{dcases}
\end{equation}
We now consider the case of $\alpha \sim 1$. For this case, classical lubrication formalism gives $\displaystyle v_r^* = \frac{\epsilon^{\frac{1}{2}}\alpha\omega R}{k^{\frac{1}{2}}} v_r$, $\displaystyle v_z^* = \epsilon\alpha\omega R v_z$, and $\displaystyle p_{\text{hd}}^* = \frac{\mu\omega\alpha}{k^2\epsilon}p_{\text{hd}}$.  The expression for $H$ for this case is,
\begin{equation}
\label{eq:H_nd_HSM}
H = \displaystyle 1+\frac{r^2}{2}.
\end{equation}
The cases of $\alpha \ll 1$ and $\alpha \sim 1$ can be collectively represented as $z^*=k\epsilon Rz$, $r^* = k^{\frac{1}{2}}\epsilon^{\frac{1}{2}}Rr$, $\displaystyle v_r^* = \frac{\epsilon^{\frac{1}{2}}\alpha\omega R}{k^{\frac{1}{2}}} v_r$, $\displaystyle v_z^* = \epsilon\alpha\omega R v_z$, and $\displaystyle p_{\text{hd}}^* = \frac{\mu\omega\alpha}{k^2\epsilon}p_{\text{hd}}$, and $H^*=k\epsilon RH$, where,
\begin{equation}
\label{eq:k}
k =
\left\{ \begin{array}{ll}
1 & \hspace{35pt} \alpha \ll 1 \\
1+\alpha\cos(t) & \hspace{35pt} \alpha \sim 1 \text{\hspace{5pt} oscillatory} \\
1+\alpha (1-t) & \hspace{35pt} \alpha \sim 1 \text{\hspace{5pt} approach} \\
1-\alpha (1-t) & \hspace{35pt} \alpha \sim 1 \text{\hspace{5pt} recession} \\
\end{array} \right.
\end{equation}
Also, the expression for $H$, as obtained by non-dimensionalizing equation \eqref{eq:H} is,
\begin{equation}
\label{eq:H_nd}
H = \frac{r^2}{2} + 
\left\{ \begin{array}{ll}
\displaystyle 1+\alpha\cos(t) & \hspace{35pt} \alpha \ll 1 \text{\hspace{5pt} oscillatory} \\
\displaystyle 1+\alpha (1-t) & \hspace{35pt} \alpha \ll 1 \text{\hspace{5pt} approach} \\
\displaystyle 1-\alpha (1-t) &  \hspace{35pt} \alpha \ll 1 \text{\hspace{5pt} recession} \\
\displaystyle 1 &  \hspace{35pt}  \alpha \sim 1 \\
\end{array} \right.
\end{equation}
We henceforth employ this collective representation for the fluid domain scaling, for both $\alpha \ll 1$ and $\alpha \sim 1$.  \\
It should be noted that since the substrate is deflecting, the deflection $l^*$ can also alter the $z^*$-scale significantly. This would happen in two conditions. First, when is $l^*$ positive (i.e. into the substrate bulk) and much higher than $H^*$, and second, when $l^*$ is negative (i.e. towards the sphere) and has a magnitude close to $H^*$. The second condition is what typically occurs in adhesion. In the interest of not complicating the scaling principles any further, we avoid these cases in the current study and will study such cases in future studies. In mathematical terms, avoiding these restrictions requires,
\begin{equation}
\label{eq:avoiddeflscale}
\Gamma = \frac{\theta}{k\epsilon}
\left\{ \begin{array}{ll}
\displaystyle \not\gg 1, &  l > 0 \\
\displaystyle < 1 \text{ and } \not\approx 1, &  l < 0 \\
\end{array} \right.
\end{equation}
The system dimension ratios and fluid and substrate domain characteristic scales are presented in table \ref{tab:nondim} for quick reference.
\begin{table}[!htb]
\setlength\tabcolsep{1pt}
\centering
\small 
\caption{Assigned notations of length scale ratios and characteristic scales of system variables (fluid domain and substrate domain)}
\label{tab:nondim}
\begin{tabular*}{1.0\textwidth}{@{\extracolsep{\fill}}|cc|cc|cc|cc|}
\hline
\textbf{Ratio}																	&	\textbf{Notation}				&	\textbf{Variable}															&	\textbf{Scale}																	&	\textbf{Variable}				&	\textbf{Scale}															&	\textbf{Variable}																	&	\textbf{Scale}																								\\[10pt]
$\displaystyle \frac{D}{R}$   													& 	$\displaystyle \epsilon$		&	$z^*, H^*$																&									$k \epsilon R$																	&	$\displaystyle \bar{z}^*$			& 	$\displaystyle \delta R$												&	$r^*$																				&	$k^{\frac{1}{2}}\epsilon^{\frac{1}{2}}R$																	\\[10pt]
$\displaystyle \frac{h_0}{D}$   												& 	$\displaystyle \alpha$			&	$v_r^*$																	&								$\displaystyle \frac{\epsilon^{\frac{1}{2}}\alpha\omega R}{k^{\frac{1}{2}}}$	&	$\vec{u}^*, l^*$ 				& 	$\displaystyle \theta R$												&	$t^*$																				&	$\displaystyle \frac{1}{\omega}$																			\\[10pt]
$\displaystyle \frac{L}{R}$   													& 	$\displaystyle \beta$ 			&	$v_z^*$																	&					$\epsilon\alpha\omega R$														&	$\displaystyle p^*$ 				& 	$\displaystyle b\Pi$ 													&								~																				&	~																											\\[10pt]
~									 											& 	~					 			&	$p_{\text{hd}}^*$															&						$\displaystyle \frac{\mu\omega\alpha}{k^2\epsilon}$					&	~								&	~																		&									~																				&	~																											\\[10pt]
\hline
\end{tabular*}
\end{table}
\subsection{Governing Equation and Boundary Conditions}\label{subsec:gdes}
All the governing equations and boundary conditions are first subjected to transformation of the involved differential terms as per equation \ref{eq:derivatives}. Subsequently, the independent variables appearing outside of derivatives and the dependent variables are non-dimensionalized (by substituting each dimensional variable with its scale times its non-dimensional counterpart). This second step applies to the upcoming pressure-component expressions (equations \eqref{eq:ap_piDL_a} and \eqref{eq:pivdW}) as well. For the rest of this subsection, all the equations and expressions have been first subjected to this procedure and then presented.
\subsubsection{Fluid and Substrate Domain}\label{subsubsec:domain}
The flow dynamics of the intervening fluid are governed by the continuity equation,
\begin{equation}
\frac{1}{r} \frac{\partial (rv_{r})}{\partial r} + \frac{\partial v_{z}}{\partial z} = 0,
\label{eq:ap_cont}
\end{equation}
and the momentum-conservation equation (or the Navier-Stokes equation),
\begin{subequations}
\begin{multline}
\frac{k\epsilon^2\rho\omega R^2}{\mu}\left[k\frac{\partial v_{r}}{\partial t} - \frac{v_r}{2}\frac{dk}{dt}- \frac{1}{2}\frac{dk}{dt}\left(r\frac{\partial v_r}{\partial r}+2z\frac{\partial v_r}{\partial z}\right) + \alpha\left(v_{r}\frac{\partial v_{r}}{\partial r}+v_{z}\frac{\partial v_{r}}{\partial z} \right)\right] = \\ -\frac{\partial p_{\text{hd}}}{\partial r} + \frac{\partial^2 v_{r}}{\partial z^{2}} + k\epsilon \left[ \frac{1}{r} \frac{\partial}{\partial r} \left(r \frac{\partial v_{r}}{\partial r}\right) - \frac{v_{r}}{r^{2}} \right],
\label{eq:ap_rmom}
\end{multline}
\begin{multline}
\frac{k^2\epsilon^{3}\rho\omega R^{2}}{\mu}\left[k\frac{\partial v_{z}}{\partial t} -\frac{1}{2}\frac{dk}{dt}\left(r\frac{\partial v_z}{\partial r}+2z\frac{\partial v_z}{\partial z}\right)+ \alpha\left(v_{r}\frac{\partial v_{z}}{\partial r}+v_{z}\frac{\partial v_{z}}{\partial z} \right)\right] = \\ 
-\frac{\partial p_{\text{hd}}}{\partial z} + k\epsilon\frac{\partial^2 v_{r}}{\partial z^{2}} + k^2\epsilon^{2} \left[ \frac{1}{r} \frac{\partial}{\partial r} \left(r \frac{\partial v_{z}}{\partial r}\right) \right].
\label{eq:ap_zmom}
\end{multline}
\label{eq:ap_mom}
\end{subequations}
These governing equations are closed by the no-slip and no-penetration conditions at the fluid-sphere interface, i.e. at $z^*=H^* \text{ or } z=H$,
\begin{subequations} 
\begin{equation}
\label{eq:ap_noslip_fluidsphere_r}
v_r = 0
\end{equation}
\begin{equation}
\label{eq:ap_noslip_fluidsphere_z}
v_z =
\left\{ \begin{array}{ll}
\displaystyle -\sin(t) 	&  	\text{\hspace{35pt} oscillatory}, \\
\displaystyle 1 		&  	\text{\hspace{35pt} approach}, \\
\displaystyle -1 		& 	\text{\hspace{35pt} recession}, \\
\end{array} \right.
\end{equation}
\label{eq:ap_noslip_fluidsphere}
\end{subequations}
no-slip and no-penetration conditions at the fluid-substrate interface, i.e. at $\displaystyle z^* = -l^* \text{ or } z= -\Gamma l$,
\begin{subequations}
\begin{equation}
\label{eq:ap_noslip_fluidsubstrate_r}
v_r = 0
\end{equation}
\begin{equation}
\label{eq:ap_noslip_fluidsubstrate_z}
v_z = -\frac{\theta}{\alpha\epsilon}\left[\frac{\partial l}{\partial t}-\left(\frac{l}{\theta}\frac{d\theta}{dt}+\frac{r}{2k}\frac{dk}{dt}\frac{\partial l}{\partial r}\right)\right]
\end{equation}
\label{eq:ap_noslip_fluidsubstrate}
\end{subequations}
zero-velocity and zero-pressure conditions at far-end of the lubrication zone, i.e. as $r \rightarrow \infty$,
\begin{equation}
v_r\rightarrow 0,~v_z \rightarrow 0,~p_{\text{hd}} \rightarrow 0,
\label{eq:ap_flow_farend}
\end{equation}
and, zero radial velocity and symmetric axial velocity and pressure conditions at the centerline, i.e. at $r = 0$,
\begin{equation}
v_r = \frac{\partial v_z}{\partial r} = \frac{\partial p_{\text{hd}}}{\partial r} = 0.
\label{eq:ap_flow_centerline}
\end{equation}
The substrate displacement is governed by the mechanical equilibrium equation, 
\begin{equation}
\label{eq:mecheq}
\nabla^*\cdot\boldsymbol{\sigma_{S}^*} = 0,
\end{equation}
where $\boldsymbol{\sigma_{S}^*}$ is the substrate domain Cauchy-Green stress tensor. For a linear elastic solid, it is given in terms of the strain tensor, $\boldsymbol{E_{S}^*} = \nabla \vec{u}^*$ as,
\begin{equation}
\label{eq:linelast}
\boldsymbol{\sigma_{S}^*} = \lambda\text{tr}(\boldsymbol{E_{S}^*})\boldsymbol{I}+G(\boldsymbol{E_{S}^*}+\boldsymbol{E_{S}^{*{\text{T}}}}),
\end{equation}
where the superscript T implies matrix transpose. Upon expanding equation \eqref{eq:mecheq} using equation \eqref{eq:linelast}, the two components of the mechanical equilibrium equation are obtained as,
\begin{subequations}
\begin{equation}
\frac{\partial^2 u_{r}}{\partial \bar{z}^2} + \frac{\delta}{k^{\frac{1}{2}}\epsilon^{\frac{1}{2}}}\left(1+\frac{\lambda}{G}\right) \frac{\partial^2 u_{\bar{z}}}{\partial r \partial \bar{z}} + \frac{\delta^2}{k\epsilon} \left(2+\frac{\lambda}{G}\right) \left(\frac{\partial^2 u_{r}}{\partial r^{2}} + \frac{1}{r} \frac{\partial u_{r}}{\partial r} - \frac{u_{r}}{r^2} \right) = 0,
\label{eq:ap_rdisp}
\end{equation}
\begin{equation}
\left(2+\frac{\lambda}{G}\right)\frac{\partial^2 u_{\bar{z}}}{\partial \bar{z}^2} + \frac{\delta}{k^{\frac{1}{2}}\epsilon^{\frac{1}{2}}}\left(1+\frac{\lambda}{G}\right) \left(\frac{\partial^2 u_{r}}{\partial r \partial \bar{z}} + \frac{1}{r} \frac{\partial u_{r}}{\partial \bar{z}}\right) + \frac{\delta^2}{k\epsilon} \left(\frac{\partial^2 u_{\bar{z}}}{\partial r^{2}} + \frac{1}{r} \frac{\partial u_{\bar{z}}}{\partial r} \right) = 0.
\label{eq:ap_ydisp}
\end{equation}
\label{eq:ap_disp}
\end{subequations}
These governing equations are closed by zero-displacement condition at far-end of the lubrication zone, i.e. as $r\rightarrow\infty$, 
\begin{equation}
u_r\rightarrow 0,~u_{\bar{z}} \rightarrow 0,
\label{eq:ap_deform_farend}
\end{equation}
zero-displacement condition at substrate-platform interface, i.e.  at $\displaystyle \bar{z}^* = L$ or $\displaystyle \bar{z} = \frac{\beta}{\delta}$,
\begin{equation}
u_r = u_{\bar{z}} = 0,
\label{eq:ap_deform_platform}
\end{equation}
zero radial displacement and symmetric axial displacement at the centerline, i.e. at $r=0$,
\begin{equation}
u_r = \frac{\partial u_{\bar{z}}}{\partial r} = 0,
\label{eq:ap_deform_centerline}
\end{equation}
and, traction balance condition at fluid-substrate interface, i.e. at $\bar{z} = 0$. The traction balance condition is expressed as,
\begin{equation}
\label{eq:tracbal}
\boldsymbol{\sigma_{S}^*}\cdot\hat{n} = \boldsymbol{\sigma_{F}^*}\cdot\hat{n},
\end{equation}
where $\boldsymbol{\sigma_{F}^*}$ is the fluid domain (Eulerian) stress tensor and is given in terms of the fluid domain strain rate tensor $\boldsymbol{\dot{E}_{F}^*} = \nabla \vec{v}^*$ as,
\begin{equation}
\label{eq:fluidstress}
\boldsymbol{\sigma_{F}^*} = -p^*\boldsymbol{I}+\mu(\boldsymbol{\dot{E}_{F}^*}+\boldsymbol{\dot{E}_{F}^{*{\text{T}}}}),
\end{equation}
and $\hat{n}$ is the normal vector to the fluid-substrate interface in the deformed configuration. Since we restrict our analysis to the infinitesimal strain (i.e. linear elastic) limit, the deformed and the undeformed configurations are sufficiently close that $\hat{n}$ can be approximated to the normal vector in the undeformed configuration. Thus,
\begin{equation}
\label{eq:normal}
\hat{n} = \hat{z}.
\end{equation}
We emphasize that we have exploited this feature of linear elasticity to simplify the following algebra. This is common practice for studies in elastohydrodynamics formulated with linear elastic formulation, where normal stress of solid gets equated to normal stress of fluid and shear stress of solid gets equated to shear stress of fluid. Hence, the expanded form of traction balance condition, equation \eqref{eq:tracbal}, is,
\begin{subequations}
\begin{equation}
\frac{\partial u_{r}}{\partial \bar{z}} + \frac{\delta}{k^{\frac{1}{2}}\epsilon^{\frac{1}{2}}}\frac{\partial u_{\bar{z}}}{\partial r} = \left\{\frac{\delta\Pi}{\theta (
\lambda+2G)}\left(2+\frac{\lambda}{G}\right) \left(\frac{1}{k^2}\frac{\mu\omega\alpha}{\epsilon\Pi}\right)\left(k^{\frac{1}{2}}\epsilon^{\frac{1}{2}} \left(\frac{\partial v_r}{\partial z}+k\epsilon\frac{\partial v_z}{\partial r}\right)  \right) \right\},
\label{eq:ap_rtrac}
\end{equation}
\begin{equation}
\frac{\partial u_{\bar{z}}}{\partial \bar{z}} + \frac{\delta}{k^{\frac{1}{2}}\epsilon^{\frac{1}{2}}}\left(\frac{\lambda}{\lambda+2G}\right)\left(\frac{\partial u_{r}}{\partial r}+\frac{u_{r}}{r}\right) = -\frac{\delta\Pi}{\theta (\lambda+2G)}\left[p-\left\{\left(\frac{1}{k^2}\frac{\mu\omega\alpha}{\epsilon\Pi}\right)\left(2k\epsilon\frac{\partial v_z}{\partial z}\right)\right\}\right],
\label{eq:ap_ytrac}
\end{equation}
\label{eq:ap_trac}
\end{subequations}
We now obtain the expressions for $\delta$ and $\theta$. We focus on $\delta$ first. For a substrate that is sufficiently thin, the substrate deformation extends from the fluid-substrate interface to the substrate-platform interface, and therefore, the substrate thickness $L$ is the scale for $\bar{z}^*$, giving us $\delta = \beta$. One the other hand, for a substrate that is thick enough to behave like a semi-infinite space, the scale for $\bar{z}^*$ is clearly independent of $L$. To determine the scale for $\bar{z}^*$ for such a substrate, we examine the mechanical equilibrium equation, equation \eqref{eq:ap_disp}. We observe that if we consider $\delta\gg k^{\frac{1}{2}}\epsilon^{\frac{1}{2}}$, equation \eqref{eq:ap_disp} simplifies to,
\begin{subequations}
\begin{equation}
\frac{\partial^2 u_{r}}{\partial r^{2}} + \frac{1}{r} \frac{\partial u_{r}}{\partial r} - \frac{u_{r}}{r^2} = 0 \implies \frac{1}{r^2}\frac{\partial}{\partial r}\left(r^3\frac{\partial }{\partial r}\left(\frac{u_r}{r}\right)\right) = 0 \implies u_r = C_1(z)r+\frac{C_2(z)}{r}
\label{eq:ap_rdisp_anomalous}
\end{equation}
\begin{equation}
\frac{\partial^2 u_{\bar{z}}}{\partial r^{2}} + \frac{1}{r} \frac{\partial u_{\bar{z}}}{\partial r} = 0 \implies \frac{1}{r}\frac{\partial}{\partial r}\left(r\frac{\partial u_{\bar{z}}}{\partial r}\right) = 0 \implies u_{\bar{z}} = C_3(z)\ln(r)+C_4(z)
\label{eq:ap_ydisp_anomalous}
\end{equation}
\label{eq:ap_disp_anomalous}
\end{subequations}
Applying the boundary conditions \eqref{eq:ap_deform_farend} and  \eqref{eq:ap_deform_centerline}, we get the solution for $u_r$ and $u_{\bar{z}}$ as,
\begin{subequations}
\begin{equation}
u_r = C_1(z)r
\label{eq:usolnr_anomalous}
\end{equation}
\begin{equation}
u_{\bar{z}} = C_4(z)
\label{eq:usolny_anomalous}
\end{equation}
\label{eq:usoln_anomalous}
\end{subequations}
The solution for $u_{\bar{z}}$ is independent of $r$, which is unrealistic. Hence, we deduce that $\delta\gg k^{\frac{1}{2}}\epsilon^{\frac{1}{2}}$ is not a realistic scaling for $\bar{z}^*$. For such a sufficiently thick substrate, $\delta = k^{\frac{1}{2}}\epsilon^{\frac{1}{2}}$. Collating these scaling arguments, we prescribe $\delta = \min(\beta,k^{\frac{1}{2}}\epsilon^{\frac{1}{2}})$. We now turn our attention to $\theta$. Since the substrate deforms in response to the pressure at the fluid-substrate interface, and given that the deformed shape of the fluid-substrate interface is sufficiently close to horizontal, the scale for $\theta$ is obtained from the $\bar{z}$-component of the traction balance condition, equation \eqref{eq:ap_ytrac}. Scaling the leading-order terms of its LHS and RHS, which stand for the force at the fluid-substrate interface from the substrate domain and from the fluid domain respectively, we get $\displaystyle \theta = \frac{\delta\Pi}{(\lambda+2G)}$. Examining the expression, we observe that for an incompressible substrate material, $\nu = 0.5$ gives $\lambda \rightarrow \infty$ and therefore, $\displaystyle \theta = \frac{\delta\Pi}{(\lambda+2G)}$ becomes zero. However, for substrates that are not significantly thin, the deformation for an incompressible substrate is not necessarily zero. To remedy this issue, for incompressible substrates, we take the expression for $\theta$ as $\displaystyle \frac{\delta\Pi}{(\lambda_{\text{C}}+2G)}$, where $\lambda_{\text{C}}$ is calculated from $G$ as per the expression $\displaystyle \frac{2G\nu_{\text{C}}}{1-2\nu_{\text{C}}}$ (or from $E_{\text{Y}}$ as per the expression $\displaystyle \frac{E_{\text{Y}}\nu_{\text{C}}}{(1+\nu_{\text{C}})(1-2\nu_{\text{C}})}$) with $\nu_{\text{C}} = 0.49$. The consequence of this work-around is that the non-dimensional fluid-substrate deflection $l$ gets diminished (clarified in section \ref{subsubsec:simp}) - this diminution is small for value of $\nu$ close to 0.5, i.e 0.49. However, the diminution of $l$ does not creep into its dimensional expression $l^*$, because $\theta$ (which gets multiplied to $l$ in order to obtain $l^*$) is magnified to the same extent to which $l$ is diminished. Lastly, we emphasize that the current formulation, particularly the substrate deformation constitutive model, is applicable for small strains only. This implies the restriction,
\begin{equation}
\label{eq:smalldefl}
\theta \ll \delta \implies \Pi \ll \lambda+2G
\end{equation}
It should be noted that for some of the cases studied, $\theta$ assumes high enough values for equation \eqref{eq:smalldefl} to not hold true, but the deformation still remains in the infinitesimal-strain limit. This happens when the interplay of the different pressure components and the substrate deflection lead to the actual pressure staying significantly smaller than $\Pi$. For such cases, we \textit{a posteriori} check that the obtained deflection should be sufficiently smaller than the $\bar{z}^*$-scale, and admit the solution if this check is fulfilled. \\
\subsubsection{Pressure Components and Total Pressure}\label{subsec:nonhd}
The pressure appearing in the traction balance condition, equation \eqref{eq:ap_trac}, is the total pressure, $p$, which is the sum of the hydrodynamic pressure (obtained from the solution of the flow dynamics) and two non-hydrodynamic pressure components emerging from DLVO forces,
\begin{equation}
\label{eq:pdim}
p^*=p_{\text{hd}}^*+p_{\text{EDL}}^*+p_{\text{vdW}}^*
\end{equation}
These non-hydrodynamic pressure components are EDL disjoining and van der Waals pressure. EDL disjoining pressure is the osmotic pressure arising out of the non-uniform distribution of electrolytic species between two surface when the intervening fluid is an electrolytic solution. The non-uniform distribution occurs because of the interplay between electrostatic and entropic effects. This osmotic pressure becomes significant as the EDLs get closer when the surfaces approach each other. \\
When studying electrokinetic effects along with flow dynamics, the contribution of electrokinetics to the flow field is quantified by a body force term in the fluid momentum conservation equation, called the Maxwell stress (the product of free charge density and electric field) . Including the Maxwell stress term components and under lubrication approximation, equations \eqref{eq:ap_rmom} and \eqref{eq:ap_zmom}, convert to the components of the Stokes equation as,
\begin{subequations}
\label{eq:StokesMaxwell}
\begin{equation}
0 = -\frac{\partial p}{\partial r} + \frac{\partial^2 v_{r}}{\partial z^{2}} + \frac{\varepsilon\varepsilon_0\psi_S^2}{\mu\omega\alpha\epsilon R^2}\frac{\partial \psi}{\partial r}\frac{\partial^2 \psi}{\partial z^2},
\label{eq:ap_rmom_simp_withMx}
\end{equation}
\begin{equation}
0 = -\frac{\partial p}{\partial z} + \frac{\varepsilon\varepsilon_0\psi_S^2}{\mu\omega\alpha\epsilon R^2}\frac{\partial \psi}{\partial z}\frac{\partial^2 \psi}{\partial z^2}.
\label{eq:ap_zmom_simp_withMx}
\end{equation}
\end{subequations}
In these equations, $\psi$ is the screening potential developing in the fluid, which is non-dimensionalized with $\psi_S$, the surface (zeta) potential; $\varepsilon$ and $\varepsilon_0$ are the relative permittivity of the electrolytic solution and permittivity of free space and $q$ is elementary charge. Note that the axial scale for the screening potential is Debye length for non-overlapping EDL and flow-domain $z^*$-scale for overlapping EDL, either of which is significantly smaller than the $r^*$-scale. Hence, lubrication approximation applies to the EDL and the resultant screening potential. We have therefore dropped the $\displaystyle \frac{1}{r}\left(r\frac{\partial \psi}{\partial r}\right)$ term in equation \eqref{eq:StokesMaxwell}.\\
These equations are coupled with the Poisson's equation for electrical potential and the electrolytic species conservation equation (or the Nernst-Planck equation) for electrolytic species transport. These equations, subjected to the simplifications of lubrication approximation and negligible transient and inertial effects, are,
\begin{subequations}
\label{eq:EDLgdes}
\begin{equation}
\frac{\partial^2 \psi}{\partial z^2} = -\frac{\epsilon^2 R^2 q}{\varepsilon\varepsilon_0 \psi_{S}}(c_{+}-c_{-}),
\label{eq:ap_Poissons_simp}
\end{equation}
\begin{equation}
0 = \frac{\partial}{\partial z}\left[\frac{\partial c_{i}}{\partial z}+\frac{q\psi_Sc_{i}}{k_{\text{B}}T}\frac{\partial \psi}{\partial z}\right],~~~i=\pm,
\label{eq:ap_NernstPlanck_simp}
\end{equation}
\end{subequations}
where $k_{\text{B}}$ is Boltzmann constant and $T$ is the fluid temperature. \\
We consider an aqueous 1:1 electrolytic solution as the intervening fluid between sphere and substrate. Since equations \eqref{eq:ap_Poissons_simp} and \eqref{eq:ap_NernstPlanck_simp} are independent of the flow-field, the EDL is in quasi-equilibrium, i.e. behaves as if in a static fluid. The solution for $n_\pm$ from equation \eqref{eq:EDLgdes} is,
\begin{equation}
c_\pm = c_0\exp\left(\mp\frac{q\psi_S\psi}{k_{\text{B}}T}\right),
\label{eq:ap_n}
\end{equation}
where $c_0$ is the electroneutral number density of the dissolved salt. This is the midplane concentration for non-overlapping EDL and the reservoir concentration for overlapping EDL \cite{Das2011}. Substituting equation \eqref{eq:ap_n} in equation \eqref{eq:StokesMaxwell} and with some algebra, we get the equations,
\begin{subequations}
\label{eq:StokesMaxwell1}
\begin{equation}
0 = -\frac{\partial }{\partial r}\left[p-\frac{2k^2\epsilon c_0 k_{\text{B}}T}{\mu\omega\alpha\psi_S}\left(\cosh\left(\frac{q\psi_S\psi}{k_{\text{B}}T}\right)-1\right)\right] + \frac{\partial^2 v_{r}}{\partial z^{2}},
\label{eq:ap_rmom_simp_withMx_1}
\end{equation}
\begin{equation}
0 = -\frac{\partial }{\partial z}\left[p-\frac{2k^2\epsilon c_0 k_{\text{B}}T}{\mu\omega\alpha\psi_S}\left(\cosh\left(\frac{q\psi_S\psi}{k_{\text{B}}T}\right)-1\right)\right].
\label{eq:ap_zmom_simp_withMx_1}
\end{equation}
\end{subequations}
Replacing $\displaystyle p-\frac{2k^2\epsilon c_0 k_{\text{B}}T}{\mu\omega\alpha\psi_S}\left(\cosh\left(\frac{q\psi_S\psi}{k_{\text{B}}T}\right)-1\right)$ with $p_{\text{hd}}$, we have,
\begin{subequations}
\label{eq:StokesMaxwell2}
\begin{equation}
0 = - \frac{\partial p_{\text{hd}}}{\partial r} + \frac{\partial^2 v_{r}}{\partial z^{2}},
\label{eq:ap_rmom_simp_withMx_11}
\end{equation}
\begin{equation}
0 = - \frac{\partial p_{\text{hd}}}{\partial z}.
\label{eq:ap_zmom_simp_withMx_11}
\end{equation}
\end{subequations}
Equation \eqref{eq:StokesMaxwell2} is identical to the Stokes equation, that is obtained by simplifying equation \eqref{eq:ap_mom} under considerations of small Wommersley number and lubrication approximation, except the pressure term has an expression added to it. This implies that electrokinetic effects can be accounted for, when considering the flow domain's interaction with any adjoining domain, by adding an osmotic pressure term $\displaystyle \frac{2k^2\epsilon n_0 k_{\text{B}}T}{\mu\omega\alpha\psi_S}\left(\cosh\left(\frac{q\psi_S\psi}{k_{\text{B}}T}\right)-1\right)$ to the $p_{\text{hd}}$ obtained from solving a purely hydrodynamic system, i.e. equation \eqref{eq:ap_mom}. This osmotic pressure is commonly called EDL disjoining pressure \cite{Trefalt2016,Urzay2010,Israelachvili2011}.  \\
We re-emphasize that such a simplification of the governing equations has become possible only because of smallness of non-steady and advection terms as well as lubrication approximation being applicable to both the fluid momentum conservation equations and electrolytic species conservation equation. In physical terms, the nature of the flow dynamics allow the EDL and the associated electrokinetic effects to remain practically equilibriated, thereby contributing to the force interactions in the system in the form of an osmotic pressure, the EDL disjoining pressure. The expression for EDL disjoining pressure for a 1:1 electrolyte at separations larger than Debye length is \cite{Lyklema2005,Urzay2010,Israelachvili2011},
\begin{equation}
p_{\text{EDL}}^* = 64c_{0}k_{\text{B}}T\Psi^{2}\exp(-\kappa (H^*+l^*)),
\label{eq:ap_piDL_a}
\end{equation}
where $\displaystyle \Psi = \tanh\left(\frac{q\psi_S}{4 k_{\text{B}}T}\right)$, and $\kappa$ is the inverse of Debye length, given as $\displaystyle \kappa = \sqrt\frac{2c_{0}q^2}{\varepsilon\varepsilon_{0} k_{\text{B}}T}$.\\
A small-seperation limit also exists and is given as \cite{Israelachvili2011,Philipse2013,Ghosal2016},
\begin{equation}
p_{\text{EDL}}^*|_{\kappa (H^*+l^*)<<1} = \frac{2\sigma_S k_{\text{B}}T}{q(H^*+l^*)},
\label{eq:ap_piDL_c}
\end{equation}
where, $\sigma_S$ is the surface charge magnitude. However, we employ equation \eqref{eq:ap_piDL_a} for our analysis for all separations. Furthermore, we emphasize that owing to the analysis presented above, although there isn't an analytically derived expression for EDL disjoining pressure for all separations, a closed-form expression is still obtainable by curve-fitting the numerical result for the quasi-equilibriated EDL. \\
On the other hand, van der Waals force is the force between two surfaces that occurs as the cumulative consequence of electrostatic interactions between induced dipoles on the surfaces, the intervening fluid and the bulks behind the surfaces. This force per unit area, which can also be called as van der Waals pressure, between two sufficiently large surfaces (i.e. surface size being substantially larger than their separation) varies with the inverse cube of their separation. It approaches significant values at small separations (of the order of a few nanometers). The expression for van der Waals pressure is,
\begin{equation}
p_{\text{vdW}}^* = -\frac{A_{\text{sfw}}}{6\pi} \frac{1}{(H^*+l^*)^3}, \label{eq:pivdW}
\end{equation}
where, $\displaystyle A_{\text{sfw}}$ is the Hamaker's constant, which is typically of the order of $10^{-21}$ Joules \cite{Israelachvili2011}.\\
Note that similar to EDL disjoining pressure (equation \eqref{eq:ap_piDL_a}), the system's van der Waals force interactions are quantified using van der Waals pressure (equations \eqref{eq:pivdW}.\\
We non-dimensionalize $p_{\text{DL}}^*$ and $p_{\text{vdW}}^*$ with $2n_0k_{\text{B}}T$ and $\displaystyle \frac{A_{\text{sfw}}}{6\pi k^3\epsilon^3R^3}$ respectively, giving us,
\begin{equation}
\label{eq:p_nhd}
\begin{split}
& p_{\text{EDL}} = 32{\Psi_{\text{S}}}\exp\left(-k\epsilon KR(H+\Gamma l)\right), \\
& p_{\text{vdW}} = -\frac{1}{(H+\Gamma l)^3}. \\
\end{split}
\end{equation}
Non-dimensionalizing equation \eqref{eq:pdim}, we get,
\begin{equation}
\label{eq:p}
p = \frac{\mu\omega\alpha}{k^2\epsilon\Pi}p^{\star}+\frac{2n_0k_{\text{B}}T}{\Pi}p_{\text{DL}}+\frac{A_{\text{sfw}}}{6k^3\pi\epsilon^3R^3\Pi}p_{\text{vdW}}.
\end{equation}
We now obtain the scale of $p^*$, i.e. $\Pi$. Since total pressure is the sum of the hydrodynamic and non-hydrodynamic pressure components, its scale depends on the dominant pressure component(s). For certain imposed dynamic parameters, the pressure components might swap dominance over the sphere's range of motion. Therefore, we prescribe $\Pi$ such that the maximum magnitude $p$ assumes during the complete range of motion of the sphere is unity-scaled. For any particular case, we take $\Pi$ to be close to maximum pressure for its rigid-substrate counterpart, which occurs when the sphere is at $z=D-h_0$. We thus have $\Pi$ as,
\begin{equation}
\label{eq:Pi}
\Pi = \frac{\mu\omega\alpha}{\epsilon(1-\alpha)^2}+2n_0 k_\text{B} T + \frac{1}{(1-\alpha)^3}\frac{A_{\text{sfw}}}{6\pi\epsilon^3R^3}.
\end{equation}
\subsubsection{Reynolds Equation and Hankel-space Pressure-Deflection Relation}\label{subsubsec:simp}
Looking at equation \eqref{eq:H}, it is clear that the range of motion for the sphere implies that a maximum time, $\displaystyle t_{\text{max}}^*=\frac{2}{\omega}$ or $t_{\text{max}}=2$, applies to approach and recession loading. No such time-limit applies for oscillatory loading. \\
The non-dimensionalized flow domain equations and boundary conditions are solved as per the traditional lubrication approach. Looking at equation \eqref{eq:k}, and given $\alpha\not \gg 1$ and the limitation of maximum time of $t_{\max} = 2$ for approach and recession loading, it is evident that $k$ as well as $\displaystyle \frac{dk}{dt}$ is either much smaller than or of the same order as 1. Hence, examining equation \eqref{eq:ap_mom}, the lubrication approximation applies straighaway given $\epsilon$ is small and Stokes' flow assumption (i.e. small unsteady and advection terms) applies given the restriction on the system dimensions and fluid properties that Wommersley number is small, i.e.,
\begin{equation}
\label{eq:smallWo}
\frac{\epsilon^2\rho\omega R^2}{\mu} \ll 1 \implies D \ll \sqrt{\frac{\mu}{\rho\omega}}.
\end{equation} 
With these conditions, we follow the standard approach to lubrication problems \cite{Urzay2010} and get the velocity fields,
\begin{equation}
\label{eq:ap_vel}
\begin{split}
& v_r = \frac{1}{2}\frac{\partial p_{\text{hd}}}{\partial r}\left[z^2-(H-\Gamma l)z-\Gamma Hl\right], \\
& -v_z = -V_{\text{S}}+\frac{1}{12r}\frac{\partial }{\partial r}\left[r\frac{\partial p_{\text{hd}}}{\partial r}\left\{2(z^3-H^3)-3(H-\Gamma l)(z^2-H^2)-6\Gamma Hl(z-H) \right\}\right],
\end{split}
\end{equation}
and the Reynolds equation,
\begin{equation}
\label{eq:Re_eq}
\frac{1}{12r}\frac{d}{dr}\left[r\left(H+\Gamma l\right)^3 \frac{dp_{\text{hd}}}{dr}\right] = \frac{\theta}{\alpha\epsilon}\left[\frac{\partial l}{\partial t}-\left(\frac{l}{\theta}\frac{d\theta}{dt}+\frac{r}{2k}\frac{dk}{dt}\frac{\partial l}{\partial r}\right)\right]+V_{\text{S}},
\end{equation}
subjected to the boundary conditions,
\begin{equation}
\label{eq:phd_bcs}
\left.\frac{d p_{\text{hd}}}{d r}\right|_{r=0} = 0,~~~~\left.p_{\text{hd}}\right|_{r\rightarrow \infty} \rightarrow 0.
\end{equation}
In equations \eqref{eq:ap_vel} and \eqref{eq:Re_eq}, $V_{\text{S}}$ is the sphere $z$-velocity, which is $-\sin(t)$ for oscillatory loading, $-1$ for approach loading and $1$ for recession loading. \\
We now derive the Hankel-space deflection-pressure relation. Following \cite{Harding1945}, we consider the Airy stress function, $\Phi$, defined as,
\begin{equation}
\label{eq:ap_airy}
\begin{split}
& u_r = -\frac{\delta}{k^{\frac{1}{2}}\epsilon^{\frac{1}{2}}}\left(1+\frac{\lambda}{G}\right)\frac{\partial^2 \Phi}{\partial r \partial \bar{z}}, \\
& u_{\bar{z}} = \frac{\partial^2 \Phi}{\partial \bar{z}^2}+\frac{\delta^2}{k\epsilon} \frac{1}{r}\frac{\partial }{\partial r}\left(r\frac{\partial \Phi}{\partial r}\right).
\end{split}
\end{equation}
Substituting these expressions in equation \eqref{eq:ap_disp}, the former gets trivially satisfied and the latter converts into the non-dimensional biharmonic equation, 
\begin{equation}
\label{eq:ap_biharmonic}
\text{B}\Phi = 0,
\end{equation}
where, $\text{B}$ is the non-dimensional Biharmonic operator,
\begin{equation}
\label{eq:ap_biharmonic_operator}
\text{B} = \text{L}^2 = \left[\frac{\partial^2}{\partial \bar{z}^2} + \frac{\delta^2}{\epsilon}\left(\frac{\partial^2}{\partial r^2}+\frac{1}{r}\frac{\partial}{\partial r}\right)\right]^2,
\end{equation}
Following an analogous approach to that applied by Harding and Sneddon (1945) \cite{Harding1945} for the dimensional version of equation \eqref{eq:ap_biharmonic_operator}, we obtain the governing equation for the zero-eth order Hankel transform of $\Phi$, $\tilde\Phi$, as,
\begin{equation}
\label{eq:ap_gde_Phi_H}
\left(\frac{\partial^2}{\partial \bar{z}^2}-\chi^2\right)\tilde{\Phi} = 0,
\end{equation}
where $\displaystyle \chi$ is $\displaystyle \frac{\delta x}{k^{\frac{1}{2}}\epsilon^{\frac{1}{2}}}$, $x$ is the Hankel space counterpart of $r$. Equation \eqref{eq:ap_gde_Phi_H} admits the solution of the form,
\begin{equation}
\label{eq:ap_sol_Phi_H}
\tilde{\Phi} = (\mathbb{A}+\mathbb{B}y)\exp(\chi z)+(\mathbb{C}+\mathbb{D}y)\exp(-\chi z).
\end{equation}
The constants of integration, $\mathbb{A}$ to $\mathbb{D}$ are obtained from the boundary conditions. This equation, whose foundation was laid by Harding and Sneddon \cite{Harding1945} and Muki \cite{Muki1960}, has been used to solve multiple problems of axisymmetric as well as general deformation field in setups ranging from coatings to semi-infinite media to stratified soft layers to continuously graded solid domains \cite{Dhaliwal1970,Chan1974,Giannakopoulos1997,Li1997,Selvadurai1998,Yang1998,Ai2002,Yang2003,Gacoin2006,Zhao2009,Selvadurai2009,Leroy2011,Wang2017a}. For the substrate-coating of arbitrary thickness undergoing axisymmetric loading that we are studying, boundary conditions \eqref{eq:ap_deform_centerline} and \eqref{eq:ap_deform_farend} get trivially satisfied in the Hankel space, and boundary conditions \eqref{eq:ap_trac} and \eqref{eq:ap_deform_platform} provide the values of $\mathbb{A}$ to $\mathbb{D}$ . \\
Plugging in the expression for $\theta$ and retaining only the leading order terms, equation \eqref{eq:ap_trac} simplifies to,
\begin{subequations}	
\label{eq:ap_trac_simp}
\begin{equation}
\frac{\partial u_{r}}{\partial y} + \frac{\delta}{k^{\frac{1}{2}}\epsilon^{\frac{1}{2}}}\frac{\partial u_{y}}{\partial r} = 0
\label{eq:ap_rtrac_simp}
\end{equation}
\begin{equation}
\frac{\partial u_{y}}{\partial y} + \frac{\delta}{k^{\frac{1}{2}}\epsilon^{\frac{1}{2}}}\left(\frac{\lambda}{\lambda+2G}\right)\left(\frac{\partial u_{r}}{\partial r}+\frac{u_{r}}{r}\right) = -p
\label{eq:ap_ytrac_simp}
\end{equation}
\end{subequations}
These simplified equations are equivalent to the condition of $\displaystyle \sigma_{(S)ry} = 0, \sigma_{(S)yy} = q$ at the substrate domain boundary with distributed load $q$, that was applied by Harding and Sneddon (1945) and Li and Chou (1997) \cite{Harding1945,Li1997}. Hence following similar approach to theirs and employing equations \eqref{eq:ap_trac_simp} and \eqref{eq:ap_deform_platform}, the expressions for $\mathbb{A}$ to $\mathbb{D}$ are obtained as,
\begin{equation}
\label{eq:ap_coeffs}
\begin{split}
\mathbb{A} = & \frac{b\epsilon^{\frac{3}{2}}\Pi}{k^\frac{3}{2}\delta^2\theta}\frac{\tilde{p}}{2x^3}\exp\left(-\gamma x\right) \cdot \\
& \left[\frac{\left\{(\lambda+G)^3(2\gamma^2+1)+2\lambda(\lambda+G)\gamma x+G(\lambda+3G)\right\}\exp(\gamma x) + \lambda(\lambda+3G)\exp(-\gamma x) }{2(\lambda+G)^3(2\gamma^2+1)+4G(\lambda+G)(\lambda+2G)+(\lambda+G)^2(\lambda+3G)\left\{\exp(2\gamma x)+\exp(-2\gamma x)\right\}}\right], \\
\mathbb{B} = & -\frac{b\epsilon^{\frac{3}{2}}\Pi}{k^\frac{3}{2}\delta^2\theta}\frac{\tilde{p}}{2x^3}\exp\left(-\gamma x\right) \cdot \\
& \left[\frac{(\lambda+G)(2\gamma x+1)\exp(\gamma x)+(\lambda+3G)\exp(-\gamma x)}{2(\lambda+G)^3(2\gamma^2+1)+4G(\lambda+G)(\lambda+2G)+(\lambda+G)^2(\lambda+3G)\left\{\exp(2\gamma x)+\exp(-2\gamma x)\right\}}\right], \\
\mathbb{C} = & -\frac{b\epsilon^{\frac{3}{2}}\Pi}{k^\frac{3}{2}\delta^2\theta}\frac{\tilde{p}}{2x^3}\exp\left(-\gamma x\right) \cdot \\
& \left[\frac{\left\{(\lambda+G)^3(2\gamma^2+1)-2\lambda(\lambda+G)\gamma x+G(\lambda+3G)\right\}\exp(-\gamma x) + \lambda(\lambda+3G)\exp(\gamma x) }{2(\lambda+G)^3(2\gamma^2+1)+4G(\lambda+G)(\lambda+2G)+(\lambda+G)^2(\lambda+3G)\left\{\exp(2\gamma x)+\exp(-2\gamma x)\right\}}\right], \\
\mathbb{D} = & \frac{b\epsilon^{\frac{3}{2}}\Pi}{k^\frac{3}{2}\delta^2\theta}\frac{\tilde{p}}{2x^3}\exp\left(-\gamma x\right) \cdot \\
& \left[\frac{(\lambda+G)(2\gamma x-1)\exp(-\gamma x)-(\lambda+3G)\exp(\gamma x)}{2(\lambda+G)^3(2\gamma^2+1)+4G(\lambda+G)(\lambda+2G)+(\lambda+G)^2(\lambda+3G)\left\{\exp(2\gamma x)+\exp(-2\gamma x)\right\}}\right], \\
\end{split}
\end{equation}
where $\displaystyle \gamma = \frac{\beta}{k^{\frac{1}{2}}\epsilon^{\frac{1}{2}}}$ and $\tilde{p}$ is the zero-eth order Hankel transform of pressure $p$. \\
The first-order Hankel transform of $u_r$, $\tilde{\tilde{u}}_r$ and zeroeth-order Hankel transform of $u_{\bar{z}}$, $\tilde{u}_{\bar{z}}$, in terms of $\tilde{\Phi}$ and its derivatives with $\bar{z}$ are,
\begin{equation}
\label{eq:ap_disp_H}
\begin{split}
& \tilde{\tilde{u}}_r =  -\frac{\delta x}{k^{\frac{1}{2}}\epsilon^{\frac{1}{2}}}\frac{\partial \tilde{\Phi}}{\partial \bar{z}} \\
& \tilde{u}_{\bar{z}} =  \frac{\partial^2 \tilde{\Phi}}{\partial \bar{z}^2}-\left(2+\frac{\lambda}{G}\right)\frac{\delta^2x^2}{k\epsilon}\frac{\partial \tilde{\Phi}}{\partial \bar{z}}
\end{split}
\end{equation}
The expression for zero-eth order Hankel transform of deflection $l$, $\tilde{l}$ is obtained by taking the expression for $\tilde{u}_{\bar{z}}$ at $\bar{z}=0$, giving us,
\begin{equation}
\label{eq:l_H0}
\tilde{l} = X\tilde{p},
\end{equation}
where $\displaystyle X$ is the Hankel-space compliance function, which represents the combined effect of substrate's compliance properties and its thickness. It is given as,
\begin{multline}
\label{eq:X}
X = \frac{\beta}{\delta}\left(\frac{\lambda+2G}{G}\right)\left[\frac{\Lambda_1\left(1-\exp\left(-4\gamma x\right)\right)-\Lambda_2\gamma x\exp\left(-2 \gamma x\right)}{\Lambda_3\gamma x \left(1+\exp\left(-4\gamma x\right)\right)+\left\{\Lambda_4 \gamma x+\Lambda_5 \gamma^3x^3\right\}\exp\left(-2 \gamma x\right)}\right], \\
\text{ with }X = -\frac{\beta}{\delta}\left(\frac{\lambda+2G}{G}\right)\frac{\Lambda_2}{2\Lambda_3+\Lambda_4}\text{ at }x=0,
\end{multline}
with,
\begin{equation}
\label{eq:coeffs}
\begin{split}
\Lambda_1 & = (\lambda+3G)(\lambda+2G), \\
\Lambda_2 & = 4(\lambda+G)(\lambda+2G), \\
\Lambda_3 & = 2(\lambda+3G)(\lambda+G), \\
\Lambda_4 & = 4\left(\lambda^2+4\lambda G+5G^2\right), \\
\Lambda_5 & = 8\left(\lambda+G\right)^2.
\end{split}
\end{equation}\\
Here, we consider the case of an incompressible substrate. As discussed in section \ref{subsubsec:domain}, using the regular expression for $\theta$ (i.e. $\displaystyle \frac{\delta\Pi}{\lambda+2G}$) leads to its value becoming zero (because $\lambda \rightarrow \infty$). On the other hand, examining equation \eqref{eq:X}, we see that $X$ approaches infinity - this occurs because the power of $\lambda$ is higher (three) in the denominator and lower (two) in the numerator of $X$. Hence, the product of $\theta$ and $X$ becomes unobtainable but cannot be ruled to be necessarily zero - this product is of interest here because it occurs in obtaining the dimensional expression of deflection $l^*$. Therefore, using the regular expression for $\theta$ renders the solution unobtainable. To remedy this, we set the expression for $\theta$ to $\displaystyle \frac{\delta\Pi}{\lambda_{\text{C}}+2G}$, as specified in section \ref{subsubsec:domain}. This results in equation \eqref{eq:ap_ytrac_simp} changing to,
\begin{equation}
\frac{\partial u_{y}}{\partial y} + \frac{\delta}{k^{\frac{1}{2}}\epsilon^{\frac{1}{2}}}\left(\frac{\lambda}{\lambda+2G}\right)\left(\frac{\partial u_{r}}{\partial r}+\frac{u_{r}}{r}\right) = -\left(\frac{\lambda_{\text{C}}+2G}{\lambda+2G}\right)p,
\label{eq:ap_ytrac_simp_incompr}
\end{equation}
due to which, the expressions for $\mathbb{A}$ to $\mathbb{D}$ also change, leading the expression for $X$ being,
\begin{multline}
\label{eq:X_incompr}
X = \frac{\beta}{\delta}\left(\frac{\lambda_{\text{C}}+2G}{G}\right)\left[\frac{\Lambda_1\left(1-\exp\left(-4\gamma x\right)\right)-\Lambda_2\gamma x\exp\left(-2 \gamma x\right)}{\Lambda_3\gamma x \left(1+\exp\left(-4\gamma x\right)\right)+\left\{\Lambda_4 \gamma x+\Lambda_5 \gamma^3x^3\right\}\exp\left(-2 \gamma x\right)}\right], \\
\text{ with }X = -\frac{\beta}{\delta}\left(\frac{\lambda+2G}{G}\right)\frac{\Lambda_2}{2\Lambda_3+\Lambda_4}\text{ at }x=0,
\end{multline}
where $\boldmath{\Lambda} = [\Lambda_i]_{i=1...5} = [1, 4, 2, 4, 8]$.  Note that the terms in the square braces of equation \eqref{eq:X_incompr} still has $\lambda$, which is approaching infinity. However, since the power of $\lambda$ is now two in both numerator and denominatior, we divide each by $\lambda^2$ and take the limit $\lambda^{-1} \rightarrow 0$. The expression for $X$ in equation \eqref{eq:X_incompr} compared to that in equation \eqref{eq:X} shows that due to the work-around of using $\lambda_{\text{C}}$, $X$ gets diminished to the same extent as $\theta$ gets magnified. Hence, $l$, obtained as per equation \eqref{eq:l_H0}, also gets diminished to the same extent.\\ 
\section{Solution Methodology}\label{sec:Soln}
Equations \eqref{eq:Re_eq} and \eqref{eq:l_H0}, subject to the boundary conditions \eqref{eq:phd_bcs}, with the expressions in equation \eqref{eq:p_nhd} constitute the set of equations representing any physical system we study. The solution for these equations is obtained by marching on a dicretized $t$-grid. For any particular case (i.e. any particular set of parameters), we perform the following checks. We check that condition in equation \eqref{eq:smallWo} is satisfied before proceeding with obtaining the solutions. We \textit{a posteriori} check that the deflection is significantly smaller than the $\bar{z}^*$-scale before admitting the obtained solution. We fulfil the condition in equation \eqref{eq:avoiddeflscale} by affirming their satisfaction at each time-step during computation before proceeding to compute the next time-step. The solution methodology is presented ahead.
\subsection{EHD Coupling Regimes}\label{subsec:Coupling}
For any particular case being studied, at any particular instance of the oscillation/range-of-motion, the EHD interaction between pressure and deflection can either be `One-Way Coupled' (OWC), i.e. the pressure is independent of the deflection, or `Two-Way Coupled' (TWC), i.e the pressure is dependent on the deflection. Examining equations \eqref{eq:Re_eq} and \eqref{eq:p_nhd}, it is clear that when both $\displaystyle \Gamma$ and $\displaystyle \frac{\theta}{\alpha\epsilon}\cdot\max\left(1,\left|\frac{1}{\theta}\frac{d\theta}{dt}\right|,\left|\frac{1}{k}\frac{dk}{dt}\right|\right)$ are much smaller than unity, the EHD interaction is OWC, else it is TWC. Hence, this is the criterion we apply for switching between the two coupling regimes. \\
For OWC, the hydrodynamic pressure, $p^{\star}$, which is solved straightaway using equation \eqref{eq:Re_eq} as,
\begin{equation}
\label{eq:p_hd_0}
p_{\text{hd}} = \frac{3\sin(t)}{H^2} \implies p_{\text{hd}}^* = \frac{3\mu\omega h_0 R\sin(\omega t^*)}{H^{*2}}.
\end{equation}
Similarly, the non-hydrodynamic pressure components are given as their terms in equation \eqref{eq:p_nhd}, with the $\displaystyle \Gamma l$ term dropped for each, i.e.,
\begin{equation}
\label{eq:p_nhd_0}
\begin{split}
& p_{\text{EDL}} = 32{\Psi_{\text{S}}}\exp\left(-k\epsilon KRH\right), \\
& p_{\text{vdW}} = -\frac{1}{H^3}.
\end{split}
\end{equation}
Once the pressure components and thus the total pressure is obtained, the solution for deflection is obtained using equation \eqref{eq:l_H0} by obtaining the Hankel-space pressure $\tilde{p}$, then multiplying with Hankel-space compliance parameter $X$, and then performing inverse-Hankel transformation.
\subsection{Methodology for Two-Way Coupling}\label{sec:ap_TWC}
For TWC, the solution is obtained by numerically solving the discretized versions of the equations \eqref{eq:Re_eq} and \eqref{eq:l_H0} and boundary conditions \eqref{eq:phd_bcs} and \eqref{eq:l_bcs}, employing the expressions in equation \eqref{eq:p_nhd}. Note that when solving for the deflection in TWC using the computational scheme, we explicitly apply the boundary conditions,
\begin{equation}
\label{eq:l_bcs}
\left.\frac{d l}{d r}\right|_{r=0} = 0,~~~~\left.l\right|_{r\rightarrow \infty} \rightarrow 0,
\end{equation}
which has been obtained from equations \eqref{eq:ap_deform_centerline} and \eqref{eq:ap_deform_farend}. We do this because it helps in improving the convergence of the numerical scheme as well as obtaining smoother solutions when dealing with relatively crude grid.\\
The multi-variable Newton-Raphson scheme is employed to solve the discretized equations, where the Hankel transformation kernel and the radial grid discretization are done in keeping with the discrete Hankel transformation formalism presented by Baddour and Chouinard (2005) \cite{Baddour2015}.\\
For equation \eqref{eq:l_H0}, we use the discrete Hankel transformation formalism presented by Baddour and Chouinard (2005) \cite{Baddour2015}. Presented ahead is a concise version of their study that is pertinent to our problem. Their formalism utilizes the Fourier-Bessel series form of functions, according to which, any function $f(r)$ defined on a finite range $r=0$ to $r=r_{\text{max}}$ can be expanded in terms of a Fourier-Bessel series as,
\begin{equation}
\label{eq:ap_FourierBessel}
f(r) = \sum_{\text{k}=1}^{\infty}f_\text{k} \mathcal{J}_\text{n}\left(\frac{j_{\text{n,k}}r}{r_{\text{max}}}\right).
\end{equation}
In equation, \eqref{eq:ap_FourierBessel}, $\mathcal{J}$ represents Bessel function, $\text{n}$ is the order of the Bessel function which is  arbitrary, $j_{\text{nk}}$ is the $\text{k}$th root of the Bessel function of order $\text{n}$, and $f_\text{k}$ is the value of the function $f(r)$ at the point $r=r_k$ on the discretized grid, $\displaystyle r_{\text{n,k}} = \frac{j_{\text{n,k}}r_{\text{max}}}{j_{\text{n,N}}}$. Subsequent algebra yields the Hankel transformation kernel, $\text{Y}$ as,
\begin{equation}
\label{eq:Hankel_kernel}
\text{Y}_{\text{m,k}}^{\text{n,N}} = \frac{2}{j_{\text{n,N}}\mathcal{J}_{\text{n+1}}^2(j_{\text{n,k}})}\mathcal{J}_{\text{n}}\left(\frac{j_{\text{n,m}}j_{\text{n,m}}}{j_{\text{n,N}}}\right)~~~~~~~~~~~1\le \text{m},~~~~~~\text{k}\le \text{N-1}
\end{equation}  
This transformation kernel is applicable for forward and inverse transformation between discretized functions $f(r)$ and its $\text{n}$th-order Hankel transform $\tilde{f}_{\text{n}}(x)$, defined on the discretized grids, $\displaystyle r_{\text{n,k}} = \frac{j_{\text{n,k}}r_{\text{max}}}{j_{\text{n,N}}}$ and $\displaystyle x_{\text{n,m}} = \frac{j_{\text{n,m}}}{r_{\text{max}}}$ respectively, each grid having $\text{N}$ points. The forward transformation is defines as,
\begin{equation}
\label{eq:Hankel_forward}
\tilde{f}_{\text{n}}(x_{\text{n,m}}) = \frac{r_{\text{max}}^2}{j_{\text{n,N}}}\sum_{\text{k}=1}^{\text{N-1}}\text{Y}_{\text{m,k}}^{\text{n,N}}f(r_{\text{n,k}}),
\end{equation}  
and the inverse transform is defines as,
\begin{equation}
\label{eq:Hankel_inverse}
f(x_{\text{n,k}}) = \frac{j_{\text{n,N}}}{r_{\text{max}}^2}\sum_{\text{m}=1}^{\text{N-1}}\text{Y}_{\text{k,m}}^{\text{n,N}}\tilde{f}_{\text{n}}(r_{\text{n,m}}).
\end{equation}  
Henceforth, we drop the subscripts superscripts $n$ and $N$, with $n$ being zero and $N$ being implicitly considered. \\
Since the grid-sizes of discretized $r$ and $x$ are the same, $\text{N}$, we discretize equation \eqref{eq:Re_eq} in the real-space and equation \eqref{eq:l_H0} in the Hankel space. Therefore, the discretized versions of these equations are,
\begin{equation}
\label{eq:ap_discretized}
\begin{split}
r_{\text{k}}(H_{\text{k}}+\Gamma l_{\text{k}})^3 \left.\frac{d^2p_{\text{hd}}}{dr^2}\right|_{\text{k}} + & \left[(H_{\text{k}}+\Gamma l_{\text{k}})^3+3r(H_{\text{k}}+\Gamma l_{\text{k}})^2\left(r_{\text{k}}+\Gamma\left.\frac{dl}{dr}\right|_{\text{k}}\right)\right]\left.\frac{dp_\text{hd}}{dr}\right|_{\text{k}} \\
& = 12\frac{r_{\text{k}}\theta}{\alpha\epsilon}\left[\left.\frac{dl}{dt}\right|_{\text{k}}-\left(\frac{l_{\text{k}}}{\theta}\frac{d\theta}{dt}+\frac{r_{\text{k}}}{2k}\frac{dk}{dt}\left.\frac{\partial l}{\partial r}\right|_{\text{k}}\right)\right]+12r_{\text{k}}V_{\text{S}} \\
\sum_{\text{k}=1}^{\text{N}-1}Y_{\text{m,k}}\left(l_{\text{k}}-X_{\text{m}}p_{\text{k}}\right) = 0
\end{split}
\end{equation}
The discretized equations \eqref{eq:ap_discretized} apply to the node points $\text{k=2}$ to $\text{k=N-1}$ and  $\text{m=2}$ to $\text{m=N-1}$. At the centerline, i.e. $\text{k=1}$, the symmetric pressure and symmetric deflection boundary conditions apply, i.e.
\begin{equation}
\label{eq:ap_discretized_symbc}
\begin{split}
& \left.\frac{dp_{\text{hd}}}{dr}\right|_{\text{k}} = 0, \\
& \left.\frac{dl}{dr}\right|_{\text{k}} = 0.  \\
\end{split}
\end{equation}
At the far-end of the lubrication zone, i.e. $\text{k=N}$, the zero pressure and zero deflection boundary conditions apply, i.e.
\begin{equation}
\label{eq:ap_discretized_farbc}
\begin{split}
& p_{\text{hd,k}} = 0, \\
& l_{\text{k}} = 0.  \\
\end{split}
\end{equation}
We solve equations \eqref{eq:ap_discretized} to \eqref{eq:ap_discretized_farbc} using the multivariable Newton-Raphson scheme \cite{Press1986}. For this scheme, $p_{\text{hd,k}}$, $\text{k=1}$ to $\text{k=N}$ constitute the first $\text{N}$ unknowns and $l_{\text{k}}$, $\text{k=1}$ to $\text{k=N}$ constitute the next $\text{N}$ unknowns, totalling $\text{2N}$ unknowns. The top equations in equation \eqref{eq:ap_discretized_symbc}, \eqref{eq:ap_discretized} \eqref{eq:ap_discretized_farbc} respectively constitute the first $\text{N}$ required equations, and the bottom equations constitute the next $\text{N}$ required equations, thus closing the system of equations. The expressions for the residuals are,
\begin{equation}
\label{eq:ap_Residuals}
\begin{split}
& \text{R}_{\text{k}} =  \left.\frac{dp_{\text{hd}}}{dr}\right|_{\text{k}} 
~~~~~~~~~~~~~~~~~~~~~~~~~~~~~~~~~~~~~~~~~~~~~~~~~~~~~~~~~~~~~~~~~~~~~~~~~~ \text{k=1}\\
& \text{R}_{\text{k}} =  r_{\text{k}}(H_{\text{k}}+\Gamma l_{\text{k}})^3 \left.\frac{d^2p_{\text{hd}}}{dr^2}\right|_{\text{k}} +  \left[(H_{\text{k}}+\Gamma l_{\text{k}})^3+3r(H_{\text{k}}+\Gamma l_{\text{k}})^2\left(r_{\text{k}}+\Gamma\left.\frac{dl}{dr}\right|_{\text{k}}\right)\right]\left.\frac{dp_{\text{hd}}}{dr}\right|_{\text{k}} \\
& - 12\frac{r_{\text{k}}\theta}{\alpha\epsilon}\left[\left.\frac{dl}{dt}\right|_{\text{k}}-\left(\frac{l_{\text{k}}}{\theta}\frac{d\theta}{dt}+\frac{r_{\text{k}}}{2k}\frac{dk}{dt}\left.\frac{\partial l}{\partial r}\right|_{\text{k}}\right)\right]-12r_{\text{k}}V_{\text{S}} 
~~~~~~~~~~~~~~~~~~~~~~~~ \text{k=2 to k=N-1}\\
\end{split}
\end{equation}
\begin{equation}
\label{eq:ap_Residuals_nextpart}
\begin{split}
& \text{R}_{\text{k}} =  p_{\text{hd,k}} 
~~~~~~~~~~~~~~~~~~~~~~~~~~~~~~~~~~~~~~~~~~~~~~~~~~~~~~~~~~~~~~~~~~~~~~~~~~~~~~~~	 \text{k=N}\\
& \text{R}_{\text{n+k}} =  \left.\frac{dl}{dr}\right|_{\text{k}} 
~~~~~~~~~~~~~~~~~~~~~~~~~~~~~~~~~~~~~~~~~~~~~~~~~~~~~~~~~~~~~~~~~~~~~~~~~~ \text{k=1}\\
& \sum_{\text{k}=1}^{\text{N}-1}Y_{\text{m,k}}\left(l_{\text{k}}-X_{\text{m}}p_{\text{k}}\right) 
~~~~~~~~~~~~~~~~~~~~~~~~~~~~~~~~~~~~~~~~~~~~~~~~~~~~~~~~~~~~~~~~ \text{k=2 to k=N-1}\\
& \text{R}_{\text{k}} =  l_{\text{k}} 
~~~~~~~~~~~~~~~~~~~~~~~~~~~~~~~~~~~~~~~~~~~~~~~~~~~~~~~~~~~~~~~~~~~~~~~~~~~~~~~~~~	 \text{k=N}\\
\end{split}
\end{equation}
and the expressions for the Jacobians are,
\begin{equation}
\label{eq:ap_Jacobians}
\begin{split}
& \text{J}_{\text{k,k}+\kappa} =  \text{a}_{\text{k},\kappa}^{\text{FD}(r)}
~~~~~~~~~~~~~~~~~~~~~~~~~~~~~~~~~~~~~~~~~~~~~~~~~~~~~~~~~~~~~~~~~~~~~~~ \text{k=1}\\
& \text{J}_{\text{k,k}+\kappa} =  r_{\text{k}}(H_{\text{k}}+\Gamma l_{\text{k}})^3 \text{a}_{\text{k},\kappa}^{\text{CD}(rr)} +  \left[(H_{\text{k}}+\Gamma l_{\text{k}})^3+3r(H_{\text{k}}+\Gamma l_{\text{k}})^2\left(r_{\text{k}}+\Gamma\left.\frac{dl}{dr}\right|_{\text{k}}\right)\right]\text{a}_{\text{k},\kappa}^{\text{CD}(r)} \\
& 
~~~~~~~~~~~~~~~~~~~~~~~~~~~~~~~~~~~~~~ 
~~~~~~~~~~~~~~~~~~~~~~~~~~~~~~~~~~~~~~~~~~~~~~~~~~~ \text{k=2 to k=N-1}\\
& \text{J}_{\text{k,N+k}+\kappa} =  3r_{\text{k}}(H_{\text{k}}+\Gamma l_{\text{k}})^2 \left.\frac{d^2p_{\text{hd}}}{dr^2}\right|_{\text{k}}\text{a}_{\text{k},\kappa}^{\text{CD}(0)} + \\ & \left[3(H_{\text{k}}+\Gamma l_{\text{k}})^2 \text{a}_{\text{k},\kappa}^{\text{CD(0)}} + \Gamma\left(r_{\text{k}}+\Gamma\left.\frac{dl}{dr}\right|_{\text{k}}\right)\text{a}_{\text{k},\kappa}^{\text{CD}(0)} + 3r(H_{\text{k}} + \Gamma l_{\text{k}})\text{a}_{\text{k},\kappa}^{\text{CD}(r)}\right]\left.\frac{dp_{\text{hd}}}{dr}\right|_{\text{k}} \\
& - 12\frac{r_{\text{k}}\theta}{\alpha\epsilon}\left[\text{a}_{\text{it},0}^{\text{BD}(t)}-\left(\frac{\text{a}_{\text{k},\kappa}^{\text{CD}(0)}}{\theta}\frac{d\theta}{dt}+\frac{\text{a}_{\text{k},\kappa}^{\text{CD}(r)}r_{\text{k}}}{2k}\frac{dk}{dt}\right)\right]
~~~~~~~~~~~~~~~~~~~~~~~~~~~~ \text{k=2 to k=N-1}\\
& \text{J}_{\text{k,k}} =  1 
~~~~~~~~~~~~~~~~~~~~~~~~~~~~~~~~~~~~~~~~~~~~~~~~~~~~~~~~~~~~~~~~~~~~~~~~~~~~~~~~	 \text{k=N}\\
& \text{J}_{\text{n+k,n+k}+\kappa} =  \text{a}_{\text{k},\kappa}^{\text{FD}(r)} 
~~~~~~~~~~~~~~~~~~~~~~~~~~~~~~~~~~~~~~~~~~~~~~~~~~~~~~~~~~~~~~~~~~ \text{k=1}\\
& \text{J}_{\text{n+m,k}} = -Y_{\text{m,k}}X_{\text{m}}
~~~~~~~~~~~~~~~~~~~~~~~~~~~~~~~~~~~~~~~~~~~~~~~~~~~~~~~~~~~~~~~~~~ \text{m=2 to m=N-1}\\
\end{split}
\end{equation}
\begin{equation}
\label{eq:ap_Jacobians_nextpart}
\begin{split}
& \text{J}_{\text{n+m,n+k}} = Y_{\text{m,k}}\left(1-X_{\text{m}}\left.\frac{\mathcal{D}p_{\text{nhd}}}{\mathcal{D}l}\right|_{\text{k}}\right)
~~~~~~~~~~~~~~~~~~~~~~~~~~~~~~~~~~~~~~~~~~~~~~~~ \text{m=2 to m=N-1}\\
& \text{J}_{\text{N+k,N+k}} =  1 
~~~~~~~~~~~~~~~~~~~~~~~~~~~~~~~~~~~~~~~~~~~~~~~~~~~~~~~~~~~~~~~~~~~~~~~~~~	 \text{k=N}\\
\end{split}
\end{equation}
In the expressions for $\text{J}$, terms like $\text{a}_{\text{k},\kappa}^{\text{CD(r)}}$ stands for the co-efficients in the finite-difference approximations applied to different derivatives. For example, the $r$-derivative of an arbitrary dependent variable $\mathbb{Y}$ is,
\begin{equation}
\label{eq:ap_deriv}
\left.\frac{\partial \mathbb{Y}}{\partial r}\right|_{\text{k}} = \sum_{\kappa=-2}^{\kappa=2}\text{a}_{\text{k},\kappa}^{\text{CD}(r)}\mathbb{Y}_{\text{k,k}+\kappa}.
\end{equation} 
The superscript `BD',`CD' and `FD' imply backward, central and forward differentiation. The term in bracket in the superscript implies the derivation, i.e. `$r$' implies first derivative with $r$, `$rr$' implies second derivative with $r$, and $0$ implies no derivative, i.e. $\text{a}_{\text{k},\kappa}^{\text{CD(0)}}$ is 1 for $\kappa = 0$ and $0$ otherwise. Lastly, in the second last expression presented in equation \eqref{eq:ap_Jacobians}, $\mathcal{D}$ denotes variational derivative, and $\displaystyle\frac{\mathcal{D}p_{\text{nhd}}}{\mathcal{D}l}$ denotes the variational derivative of total non-hydrodynamic pressure with $l$, i.e. the sum of the variational derivatives of each non-hydrodynamic pressure component with $l$,
\begin{equation}
\label{eq:ap_dpdl}
\begin{split}
& \frac{\mathcal{D} p_{\text{DL}}}{\mathcal{D} l} = -k\Gamma\epsilon KR p_{\text{DL}}\\
& \frac{\mathcal{D} p_{\text{vdW}}}{\mathcal{D} l} = -\frac{3\Gamma p_{\text{vdW}}}{(H+\Gamma l)} \\
& \frac{\mathcal{D} p_{\text{S}}}{\mathcal{D} l} = -\frac{k\Gamma\epsilon R p_{\text{S}}}{\sigma}\left[1+2\pi\tan\left(\frac{2\pi k\epsilon R \left(H+\Gamma l\right)}{\sigma}+\phi\right)\right] \\
& \frac{\mathcal{D} p}{\mathcal{D} l} = \frac{2n_0k_{\text{B}}T}{\Pi}\frac{\mathcal{D} p_{\text{DL}}}{\mathcal{D} l}+\frac{A_{\text{sfw}}}{6\pi k^3\epsilon^3R^3\Pi}\frac{\mathcal{D} p_{\text{vdW}}}{\mathcal{D} l}+\frac{A}{\Pi}\frac{\mathcal{D} p_{\text{S}}}{\mathcal{D} l}
\end{split}
\end{equation}
The solution for $p_{\text{hd}}$ and $l$ is obtained by interating over the equations,
\begin{equation}
\label{eq:ap_NRiter}
\begin{split}
& p_{\text{hd,k}} = p_{\text{hd,k}}-\left(\text{J}^{-1}\text{R}\right)_{\text{k}}, \\
& l_{\text{k}} = l_{\text{k}}-\left(\text{J}^{-1}\text{R}\right)_{\text{N+k}},
\end{split}
\end{equation}
till $\displaystyle ||\text{R}|| = \min\left[|\text{R}_{\text{k}}|\right]_{\text{k}=1...\text{2N}}$ reduces below a threshold. 
\subsection{Thickness and Compressibility Limits}\label{subsec:limits}
The expression for $X$ as given in equation \eqref{eq:X} is a generic expression applicable for any substrate thickness and Poisson's ratio. However, its approximate expressions for various limiting cases are insightful. \\
For the limiting case of thin substrate i.e. $\delta\ll k^{\frac{1}{2}}\epsilon^{\frac{1}{2}}$ and therefore $\delta=\beta$ and $\gamma\ll 1$. This implies,
\begin{equation}
\label{eq:thin_approx}
\begin{split}
& \exp(-4\gamma x)\approx 1-4\gamma x, \\
& \exp(-2\gamma x)\approx 1-2\gamma x.
\end{split}
\end{equation}
Substituting these approximations in equation \eqref{eq:X} and keeping only the respectively highest ordered terms in the numerator and the denominator, the approximate expression for $X$ is,
\begin{equation}
\label{eq:X_thin}
X = 
\left\{ \begin{array}{ll}
\displaystyle 1 & 0 \leq \nu < 0.5 \\
\displaystyle \frac{\lambda_{\text{C}}+2G}{\lambda+2G} = 0, &  \nu = 0.5 \\
\end{array} \right.
\end{equation}
Substituting this expression in equation \eqref{eq:l_H0} gives,
\begin{equation}
\label{eq:l_thin}
\tilde{l} = 
\left\{ \begin{array}{ll}
\displaystyle \tilde{p} &  0 \leq \nu < 0.5 \\
\displaystyle 0, &  \nu = 0.5 \\
\end{array} \right.
\end{equation}
This equation can be straightaway inverted to give,
\begin{equation}
\label{eq:l_thin_real}
l = 
\left\{ \begin{array}{ll}
\displaystyle p &  0 \leq \nu < 0.5 \\
\displaystyle 0, &  \nu = 0.5 \\
\end{array} \right. \implies
l^* = 
\left\{ \begin{array}{ll}
\displaystyle \frac{Lp^*}{\lambda+2G} &  0 \leq \nu < 0.5 \\
\displaystyle 0, &  \nu = 0.5 \\
\end{array} \right.
\end{equation}
The deflection-pressure relation in equation \eqref{eq:l_thin_real} matches those obtained in other EHD and soft-lubrication studies involving thin soft coatings \cite{Naik2017,Mahadevan2004,Karan2019}.\\
On the other hand, for the limiting case of a semi-infinite substrate i.e. $\delta = k^{\frac{1}{2}}\epsilon^{\frac{1}{2}}$ and $\gamma\gg 1$, implying,
\begin{equation}
\label{eq:semiinfinite_approx}
\gamma x \gg \gamma^3 x^3\exp\left(-2\gamma x\right) \gg \gamma x\exp\left(-2\gamma x\right) \gg \gamma x \exp\left(-4\gamma x\right) \gg \exp\left(-4\gamma x\right).
\end{equation}
Substituting these approximations in \eqref{eq:X} and keeping only the respectively highest ordered terms in the numerator and the denominator, the approximate expression for $X$ is,
\begin{equation}
\label{eq:X_semiinfinite}
X = 
\left\{ \begin{array}{ll}
\displaystyle \frac{(\lambda+2G)^2}{2G(\lambda+G)\gamma x} &  0 \leq \nu < 0.5 \\
\displaystyle \frac{(\lambda+2G)(\lambda_{\text{C}}+2G)}{2G(\lambda+G)\gamma x} = \frac{(\lambda_{\text{C}}+2G)}{2G\gamma x}, &  \nu = 0.5 \\
\end{array} \right.
\end{equation}
Substituting this expression in equation \eqref{eq:l_H0} gives,
\begin{equation}
\label{eq:l_semiinfinite}
\tilde{l} = 
\left\{ \begin{array}{ll}
\displaystyle \frac{(\lambda+2G)^2}{2G(\lambda+G)}\frac{\tilde{p}}{x} &  0 \leq \nu < 0.5 \\
\displaystyle \frac{(\lambda_{\text{C}}+2G)}{2G}\frac{\tilde{p}}{x}, &  \nu = 0.5 \\
\end{array} \right.
\end{equation}
Evidently, equation \eqref{eq:l_semiinfinite} cannot be inverted straightaway to get a real-space relation between pressure and deflection. The presence of $x$ in the RHS indicates that for a semi-infinite substrate, the deflection at a particular point (on the $r$-axis) is dependent on the pressure at that point as well as its neighbourhood, and theoretically the entire radial span. This dependence persists for intermediate thicknesses a well, down until the substrate thickness reaches the thin-coating limit (equations \eqref{eq:l_thin} and \eqref{eq:l_thin_real}), when the deflection at a radial point becomes solely dependent on the pressure at that radial point. \\
We now turn attention to substrate compressibility. The compressibility of any material is characterized by Poisson's ratio. For most linear-elastic materials, Poisson's ratio has a value between 0 and 0.5, with Poisson's ratio of 0.5 for incompressible materials. Materials with Poisson's ratio closer to 0.5 exhibit lower compressibility. Hence, we consider two limits of material compressibility, a `perfectly incompressible limit (PIL)' having $\nu = 0.5$ and a corresponding opposite limit that we henceforth refer to as `perfectly compressible limit (PCL)', for which $\nu=0$. These limits correspond to $\lambda\rightarrow\infty$ and $\lambda=0$ respectively, given that its expression is $\displaystyle \lambda = \frac{\nu E_{\text{Y}}}{(1+\nu)(1-2\nu)} = \frac{2\nu G}{1-2\nu}$. On the other hand, $G$, whose expression in terms of $E_{\text{Y}}$ and $\nu$ is $\displaystyle \frac{E_{\text{Y}}}{2(1+\nu)}$, varies merely from $\displaystyle \frac{E_{\text{Y}}}{3}$ to $\displaystyle \frac{E_{\text{Y}}}{2}$ as $\nu$ varies from 0 to 0.5. Therefore, PCL corresponds to $\displaystyle \frac{\lambda}{G}=0$ and PIL corresponds to $\displaystyle \frac{G}{\lambda} = 0$. The PIL expressions of $X$ are presented in \eqref{eq:X_incompr}, and bottom expression of equations \eqref{eq:X_thin} and \eqref{eq:X_semiinfinite} for general thickness, and thin and semi-infinite limits respectively. The PCL can be obtained by simply putting $\lambda = 0$ in equations \eqref{eq:X}, and top expression of equations \eqref{eq:X_thin} and \eqref{eq:X_semiinfinite} for general thickness, and thin and semi-infinite limits respectively. All the limiting expressions for $X$ are summarized in table \ref{tab:limits}. The limiting expressions for $\theta$ are also presented for reference. Examining the expressions for $X$ and $\theta$ for thin and semi-infinite cases of PCL and PIL, we infer while incompressibility can be seen to impart rigidity to a thin substrate (since $X = 0\implies l=0 \implies l^* = 0$), its effect on a semi-infinite substrate is much more ameliorated as the dimensional deflection $l^*$ gets reduced to half for the same $G$ (since the product of $X$ and $\delta$ has a ratio of 1/2 for same $G$ between PCL and PIL for semi-infinite thickness), or to three-fourth for the same $E_{\text{Y}}$ (since changing $\nu$ from 0 to 0.5 for the same $E_{\text{Y}}$ leads to decrease in $G$ by a factor of 2/3, and so, 1/2 divided by 2/3 gives 3/4). This factor is exactly recovered only when the pressure-deflectio is OWC for the complete range of $\nu$, which leads to the pressure being same for the two limiting cases being considered. Otherwise, the deflection alters the pressure and therefore, a comparison doesn't remain as straightforward. \\
\begin{table}[H]
\def\arraystretch{2}
\small
\centering
\caption{Expressions for $X$ and $\theta$ in thickness and compressibility limits}
\label{tab:limits}
\begin{tabular*}{1.00\textwidth}{@{\extracolsep{\fill}}||l||c|c|c||c||}
\hline
\hline
~  																																							& 	
\multicolumn{3}{c||}{$X$}																																	&
~																																							\\
\hline
~ 																																							& 	
\textbf{Thin }																																				&		
\textbf{Thick }																																				&		
\textbf{Semi-Infinite }																																		&	~																																							\\
~ 																																							& 	
\textbf{($\gamma\ll 1$)}																																	&		
\textbf{($\gamma\sim 1$)}																																	&		
\textbf{($\gamma\gg 1$)}																																	&	$\theta$																																					\\
\hline
\hline
\textbf{PCL} $\displaystyle \left(\nu=0\right)$ 																						&	
1																																							&	
\begin{tabular}{@{}c@{}} equation \eqref{eq:X} with \\ $\Lambda = [\Lambda_i] = [3, 4, 3, 10, 4]$ \end{tabular}												&	 
$\displaystyle \left.\frac{2}{x}\right|_{x\ne0}, \left.-\frac{1}{4}\right|_{x=0}$	&	$\displaystyle \frac{\delta \Pi}{2G}$								\\
\hline
\textbf{Normal}																																		&	
1																					 																		&	 
\begin{tabular}{@{}c@{}} equation \eqref{eq:X} with \\ $\Lambda = [\Lambda_i]$ as per  \eqref{eq:coeffs} \end{tabular}										&	 
$\displaystyle \left.\frac{(\lambda+2G)^2}{2G(\lambda+G)x}\right|_{x\ne0},\left.-\frac{\lambda^2+3\lambda G+2G^2}{2\lambda^2+9\lambda G+11G^2}\right|_{x=0}$&	
$\displaystyle \frac{\delta \Pi}{\lambda+2G}$						\\
\hline
\textbf{PIL} $\displaystyle \left(\nu=0.5\right)$ 																			&	
$\displaystyle 0$																																			&	
\begin{tabular}{@{}c@{}} equation \eqref{eq:X_incompr} with \\ $\Lambda = [\Lambda_i] = [1, 4, 2, 4, 8]$ \end{tabular}										&	 
$\displaystyle \left.\frac{(\lambda_{\text{c}}+2G)}{2Gx}\right|_{x\ne0}, \left.-\frac{1}{2}\right|_{x=0}$													&	
$\displaystyle \frac{\delta \Pi}{\lambda_{\text{c}}+2G}$																									\\
\hline
\hline
\end{tabular*}
\end{table}
\section{Results and Discussion}\label{sec:Results}
The system variables of interest for the physical setup we are studying here are force between sphere and substrate and the latter's deflection. In some scenarios, it is illumiating to asssess the total pressure and its individual components as well. While it is common practice to straight-away present the non-dimensional variables in the mathematical formulation, the peculiarity of the non-dimensionalization approach used in this study (particularly the transience of many system variable scales) renders assessment of system behaviour and trends in terms of these non-dimensional variables unnecessarily contorted. Therefore, we introduce `normalized variables', which are simply the dimensional variables non-dimensionalized with terms that do not vary with time, and are therefore more amenable to assessment. These normalized variables are annotated with $\hat{}$ , and are defined as - $\displaystyle \hat{l}=\frac{l^*}{(1-\alpha)\epsilon R}=\frac{\hat{\alpha}l^*}{\epsilon R}$, $\displaystyle \hat{p}=\frac{p^*}{\frac{\mu\omega\alpha}{(1-\alpha)^2\epsilon }}=\frac{p^*}{\frac{2\pi\mu\hat{\omega}\alpha\hat{\alpha}^2}{\epsilon}}$, $\displaystyle \hat{p}_{\text{hd}}=\frac{p_{\text{hd}}^*}{\frac{\mu\omega\alpha}{(1-\alpha)^2\epsilon }}=\frac{p_{\text{hd}}^*}{\frac{2\pi\mu\hat{\omega}\alpha\hat{\alpha}^2}{\epsilon}}$, $\displaystyle \hat{p}_{\text{EDL}}=\frac{p_{\text{EDL}}^*}{\frac{\mu\omega\alpha}{(1-\alpha)^2\epsilon }}=\frac{p_{\text{EDL}}^*}{\frac{2\pi\mu\hat{\omega}\alpha\hat{\alpha}^2}{\epsilon}}$, $\displaystyle \hat{p}_{\text{vdW}}=\frac{p_{\text{vdW}}^*}{\frac{\mu\omega\alpha}{(1-\alpha)^2\epsilon }}=\frac{p_{\text{vdW}}^*}{\frac{2\pi\mu\hat{\omega}\alpha\hat{\alpha}^2}{\epsilon}}$, $\displaystyle \hat{F}=\frac{F^*}{\frac{\mu\omega\alpha R^2}{(1-\alpha)}}=\frac{F^*}{2\pi\mu\hat{\omega}\alpha\hat{\alpha} R^2}$, and $\hat{t} = \omega t^* = 2\pi\hat{\omega}t^* = t$, where, $\displaystyle \hat{\omega} = \frac{\omega}{2\pi}$ and $\displaystyle \hat{\alpha} = \frac{1}{1-\alpha}$. \\
A few comments regarding presentation of the results ahead in this section are in order. First, we present the variables using one of the two forms, i.e. dimensional or normalized. In some scenarios, we have presented the product of normalized variables with some other parameters rather than as is - the reason for the same is ease of assessment in such scenarios, elaborated whereever such plots are presented. Second, normalized time $\hat{t}$ is the same as non-dimensionalized time $t$. This is because time has been non-dimensionalized with the constant $\omega$, hence continuing with the same expression does not cause any problems. Third, the pressure (total and individual components) and deflection presented henceforth are their values at the origin (i.e. at $r=0$). Fourth, in a number of parametric plots pertaining to oscillatory loading, we present the magnitudes of maximum attractive (or negative), maximum repulsive (or positive) and mean values of the presented variables over the complete range of motion - the complete range of motion for low-frequency oscillations is essentially one complete oscillation, but for high-frequency oscillations includes an initial quasi-transience as well, elaborated in section \ref{subsubsec:results_freq}. And lastly,  the parameter values considered - intervening fluid and substrate material properties, system geometry, imposed dynamics, and DLVO force parameters - are in ranges that occur often in many industrial, engineering and biological scenarios, including scanning probe microscopy (SPM) setups, and are suitable for delineating different aspects of the physical setup being studied.\\
The results presented ahead are arranged as follows. In section \ref{subsubsec:results_limits}, we present the deflection characteristics occuring in OWC for the limiting systems discussed in section \ref{subsec:limits}. We restrict the solution to OWC in order to recover the different ratios of deflection as expected for the limiting cases. This restriction is enforced by taking sufficiently high value of the substrate material Young's modulus $E_{\text{Y}}$. Next, in section \ref{subsubsec:results_rep}, we assess the temporal evolution of total pressure, its components, and deflection, for some representative cases for the systems that are used in the parametric analyses of upcoming subsections. Next, in section \ref{subsec:results_apprec}, we study approach and recession loading for different combinations of substrate thickness, substrate compressibility (quantified by Poisson's ratio $\nu$), DLVO pressure component parameters, and approach/recession speed. Subsequently, in sections \ref{subsubsec:results_L} to \ref{subsubsec:results_DLVO}, we assess effects of substrate thickness, substrate material Poisson's ratio (which effectively determines the material compressibility), and variation of DLVO force parameters in the context of low frequency oscillatory loading. Lastly, in section \ref{subsubsec:results_freq}, we study the effects of variation in oscillation frequency, in conjugation with oscillation amplitude and DLVO forces. \\
\subsection{Demonstrative Case}\label{subsec:results_demonstrative}
\subsubsection{Limiting Cases}\label{subsubsec:results_limits}
\begin{figure}[!htb]
\centering
\begin{subfigure}[b]{0.495\textwidth}
\centering
\includegraphics[width=\textwidth]{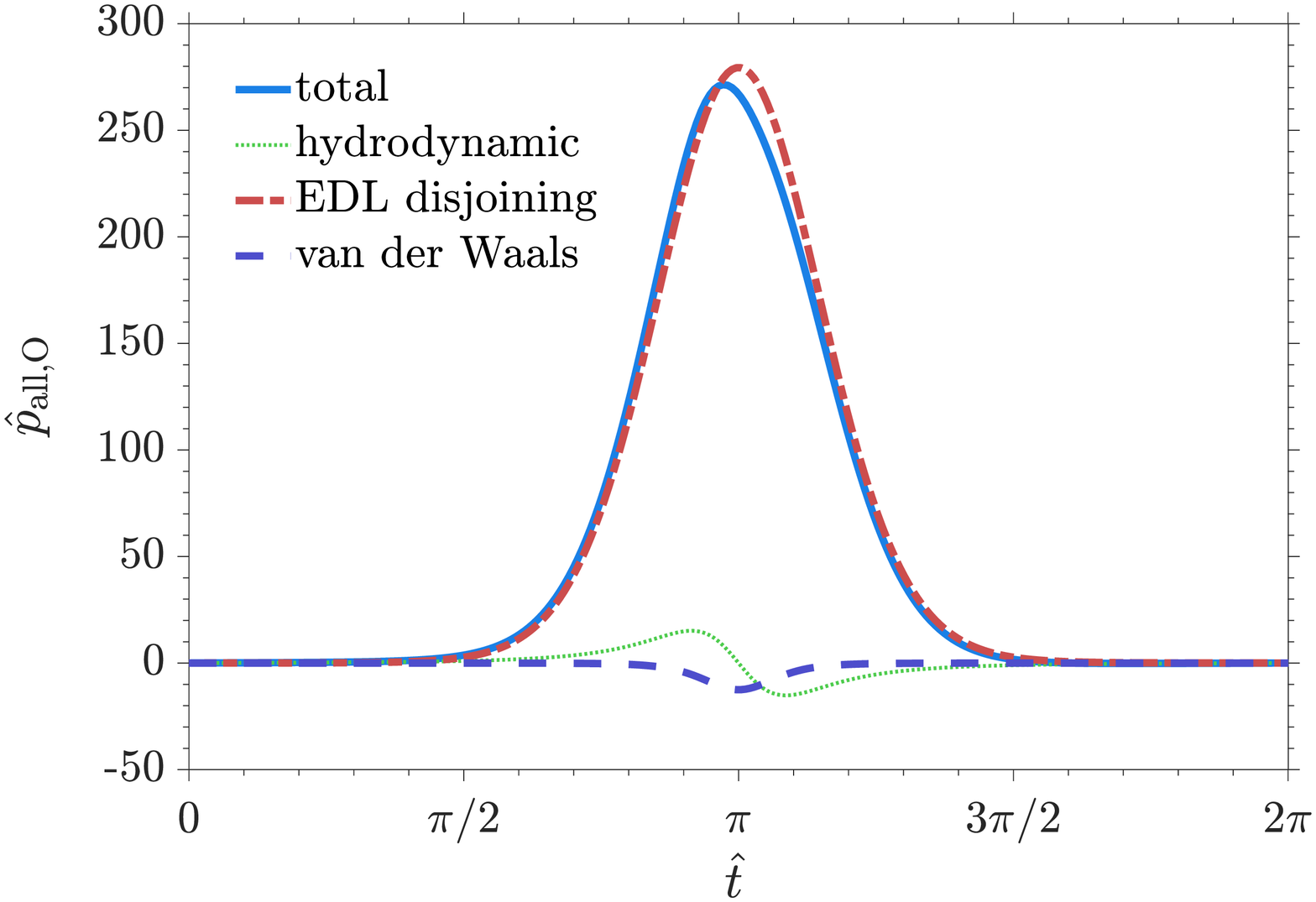}
\caption{}
\label{subfig:pall_vs_t_lim}
\end{subfigure}
\begin{subfigure}[b]{0.495\textwidth}
\centering
\includegraphics[width=\textwidth]{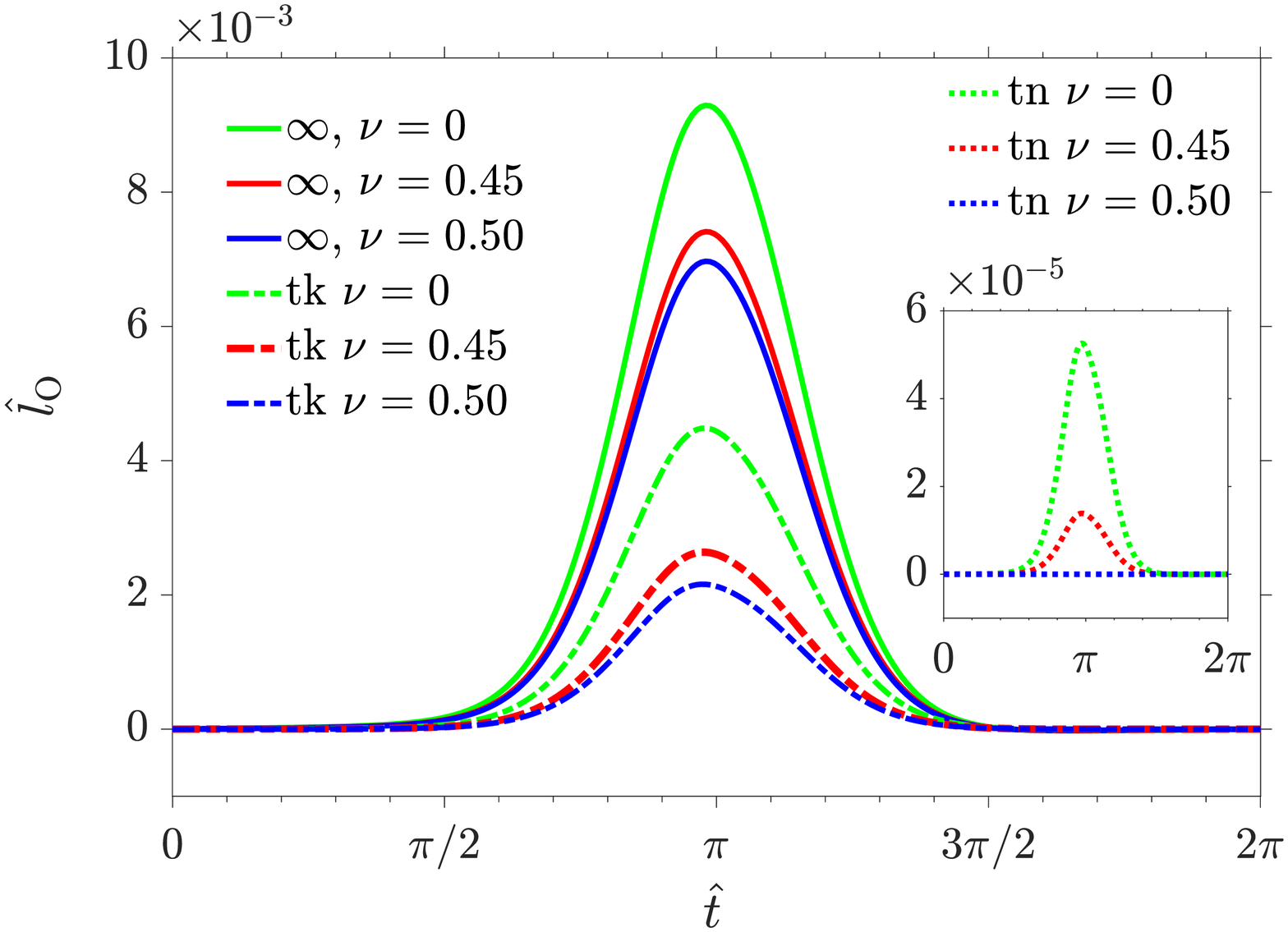}
\caption{}
\label{subfig:l_vs_t_lim}
\end{subfigure}
\caption{Variation of (a) normalized pressure components $\hat{p}_{\text{hd}}$, $\hat{p}_{\text{EDL}}$ and $\hat{p}_{\text{vdW}}$ and normalized total pressure $\hat{p}$ (labelled `tot'), and (b) normalized deflection $\hat{l}$ at origin, with normalized time $\hat{t}$ for oscillatory loading for demonstration of limiting cases discussed in subsection \ref{subsec:limits} and presented in table \ref{tab:limits}; other system parameters are: $R$ = 1 mm, $D$ = 50 nm, $h_0$ = 45 nm, $L$ = [0.5 mm (semi-infinite, labelled `$\infty$'), 5 $\mu$m (thick, labelled `tk'), 50 nm (thin, labelled `tn')],  $\omega$ = 2$\pi$ rad/s, $E_{\text{Y}}$ = 17.5 GPa, $\nu$ = [0.00 (PCL), 0.45 (normal), 0.50 (PIL)], $\mu$ = 1 mPa-s, $A_{\text{sfw}}$ = $10^{-20}$ J, $\psi_S$ = 2500 mV}
\label{fig:limit}
\end{figure}
We first present solutions corresponding to the limiting cases discussed in section \ref{subsec:limits} (each of the limiting cases presented table \ref{tab:limits}). For this, we study the case of oscillatory loading and choose a set of parameter values such that the FSI coupling remains OWC for the entire oscillation. Since the coupling is one-way, the total pressure and individual pressure components exhibit time evolution (presented in figure \ref{subfig:pall_vs_t_lim} in normalized form) which are identical for all the limiting cases. The  time evolution of normalized deflection for the different limiting cases is presented in figure \ref{subfig:l_vs_t_lim}. We emphasize that for each limiting case, the deflection plot obtained using the corresponding limiting expressions for $X$ (as given in table \ref{tab:limits}) and using the complete expression for $X$ (i.e. equation \eqref{eq:X} with $\Lambda = [\Lambda_i]$ as per \eqref{eq:coeffs} and equation \eqref{eq:X_incompr} with $\Lambda = [\Lambda_i] = [1,4,2,4,8]$ for the incompressible case) coincide. The trends are in keeping with expectations we obtain from table \ref{tab:limits}. The deflection for thin substrate in PIL is zero. The ratio of deflection for thin substrate in PCL and normal zone is expected to be $\displaystyle \frac{X_{\text{thin,PCL}}\cdot\theta_{\text{PCL}}}{X_{\text{thin,normal}}\cdot\theta_{\text{normal}}} = \frac{1\cdot\delta\Pi\cdot\frac{1}{2G_{\text{PCL}}}}{1\cdot\delta\Pi\cdot\frac{1}{(\lambda+2G)_{\text{normal}}}} = \frac{\frac{1}{E_{\text{Y}}}\cdot(1+\nu_{\text{PCL}})}{\frac{1}{E_{\text{Y}}}\cdot\frac{(1+\nu_{\text{normal}})(1-2\nu_{\text{normal}})}{(1-\nu_{\text{normal}})}} = \frac{(1+0.0)}{\frac{(1+0.45)(1-2\cdot0.45)}{(1-0.45)}} = 3.79$, which matches the ratio $\displaystyle \frac{\hat{l}_{\max,\text{PCL}}}{\hat{l}_{\max,\text{normal}}} = \frac{5.26}{1.38} = 3.79$. The ratio of deflection for semi-infinite substrate in PCL to normal zone to PIL is expected to be 
$\displaystyle \left[xX\right]_{\text{semi-infinite,PCL}}\cdot\theta_{\text{PCL}} : \left[xX\right]_{\text{semi-infinite,normal}}\cdot\theta_{\text{normal}} : \left[xX\right]_{\text{semi-infinite,PIL}}\cdot\theta_{\text{PIL}} = 2\cdot\left[\frac{1}{2G}\right]_{\text{PCL}}\cdot\delta\Pi : \left[\frac{(\lambda+2G)^2}{2G(\lambda+G)}\cdot\frac{1}{(\lambda+2G)}\right]_{\text{normal}}\cdot\delta\Pi : \left[\frac{(\lambda_{\text{C}}+2G)}{2G}\cdot\frac{1}{(\lambda_{\text{C}}+2G)}\right]_{\text{PCL}}\cdot\delta\Pi = \left[\frac{1}{G}\right]_{\text{PCL}} : \left[\frac{(\lambda+2G)}{2G(\lambda+G)}\right]_{\text{normal}} : \left[\frac{1}{2G}\right]_{\text{PCL}} = \frac{2(1+\nu_{\text{PCL}})}{E_{\text{Y}}} : \frac{2(1-\nu_{\text{normal}})(1+\nu_{\text{normal}})}{E_{\text{Y}}} : \frac{(1+\nu_{\text{PIL}})}{E_{\text{Y}}} = 2(1+\nu_{\text{PCL}}) : 2(1-\nu_{\text{normal}})(1+\nu_{\text{normal}}) : (1+\nu_{\text{PIL}}) = 2(1+0.0) : 2(1-0.45)(1+0.45) : (1+0.5) = 2:1.6:1.5$ which matches the ratio $\displaystyle \hat{l}_{\max,\text{PCL}} : \hat{l}_{\max,\text{normal}} : \hat{l}_{\max,\text{PCL}} = 9.29 : 7.41 : 6.97 = 2:1.6:1.5$. Lastly, we emphasize that such a comparison of ratios has been possible only because of OWC between deflection and pressure. In case of TWC, while the expectations of deflection being higher as the substrate thickness approaches semi-infinite and as the substrate material increases in compressibility are qualitatively met, quantitative comparisons are not possible. Therefore, for the rest of the results, which delves primarily into TWC regime, we refrain from making a quantiative assessment of these limiting ratios. 
\subsubsection{Illustrative Cases}\label{subsubsec:results_rep}
\begin{figure}[!htb]
\centering
\begin{subfigure}[b]{0.495\textwidth}
\centering
\includegraphics[width=\textwidth]{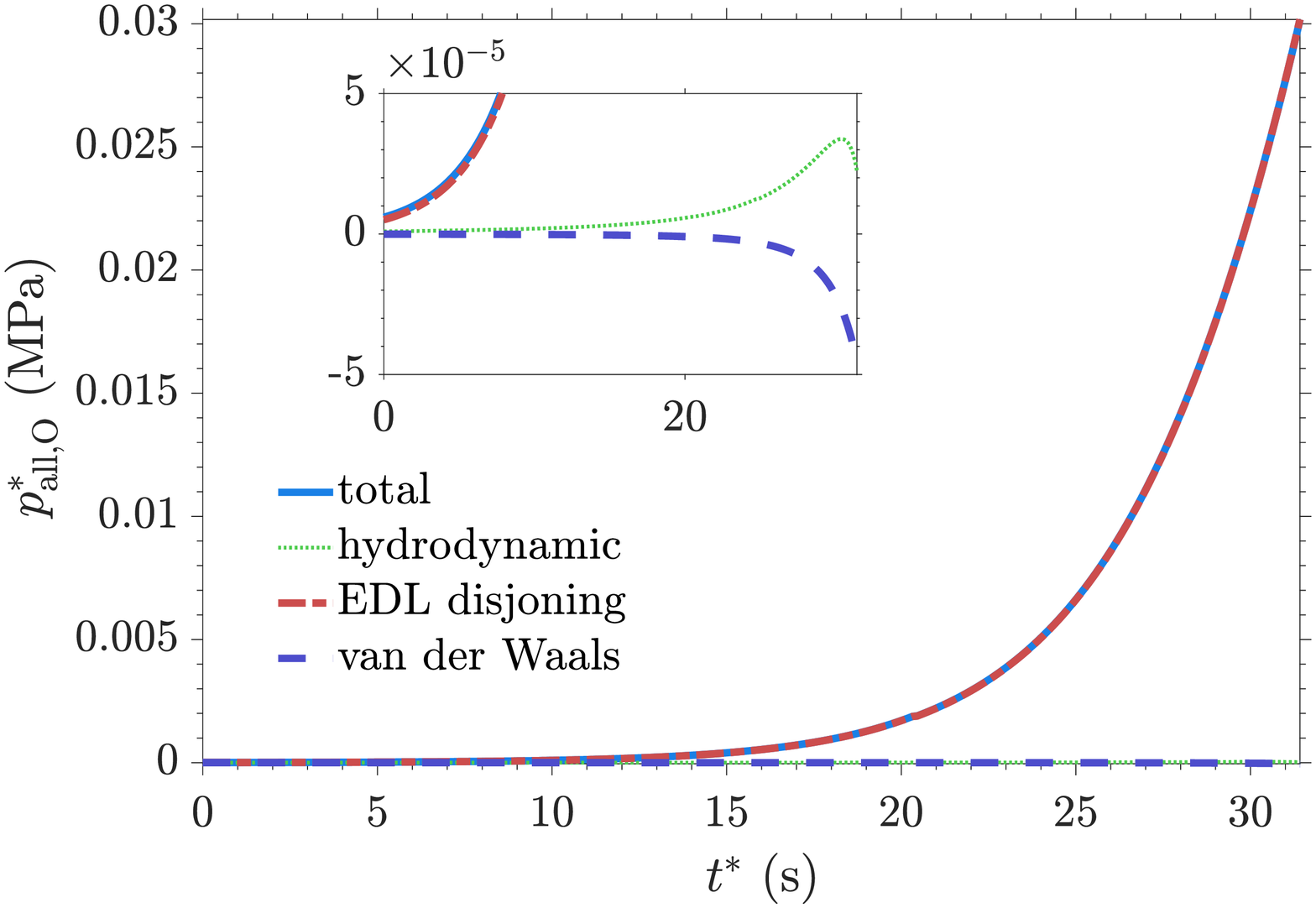}
\caption{}
\label{subfig:pall_vs_t_app_dlvo_mid}
\end{subfigure}
\begin{subfigure}[b]{0.495\textwidth}
\centering
\includegraphics[width=\textwidth]{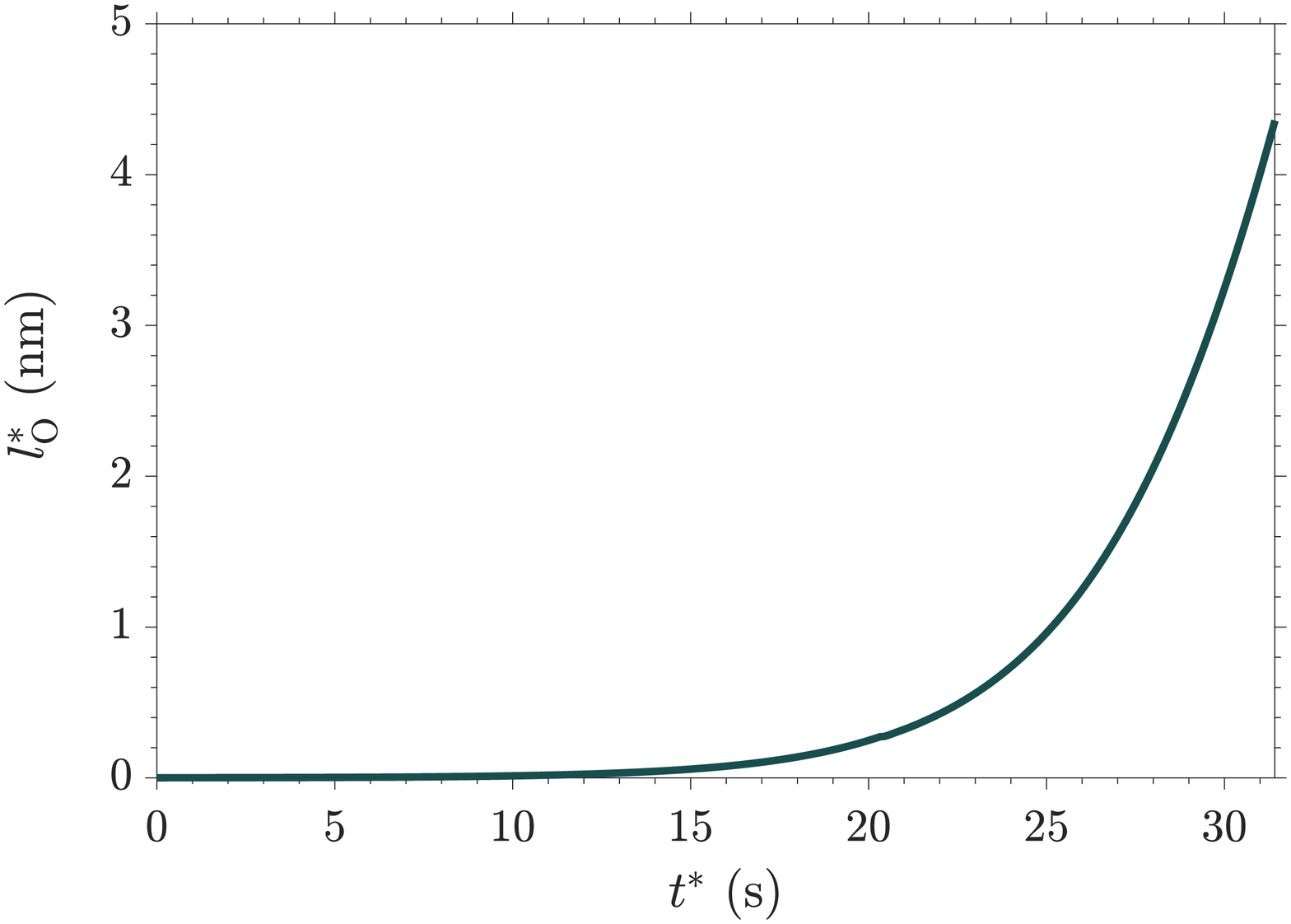}
\caption{}
\label{subfig:l_vs_t_app_dlvo_mid}
\end{subfigure}
\caption{Variation of (a) pressure components $p_{\text{hd}}^*$, $p_{\text{EDL}}^*$ and $p_{\text{vdW}}^*$ and total pressure $p^*$, and (b) deflection $l^*$ at origin, with time $t^*$ for representative case of approach loading; other system parameters are: $R$ = 1 mm, $D$ = 50 nm, $h_0$ = 45 nm, $L$ = 5 $\mu$m,  $\omega$ = 2$\pi\times 10^{-2}$ rad/s, $E_{\text{Y}}$ = 17.5 MPa, $\nu$ = 0.45, $\mu$ = 1 mPa-s, $A_{\text{sfw}}$ = $10^{-21}$ J, $\psi_S$ = 100 mV}
\label{fig:app}
\end{figure}
\begin{figure}[!htb]
\centering
\begin{subfigure}[b]{0.495\textwidth}
\centering
\includegraphics[width=\textwidth]{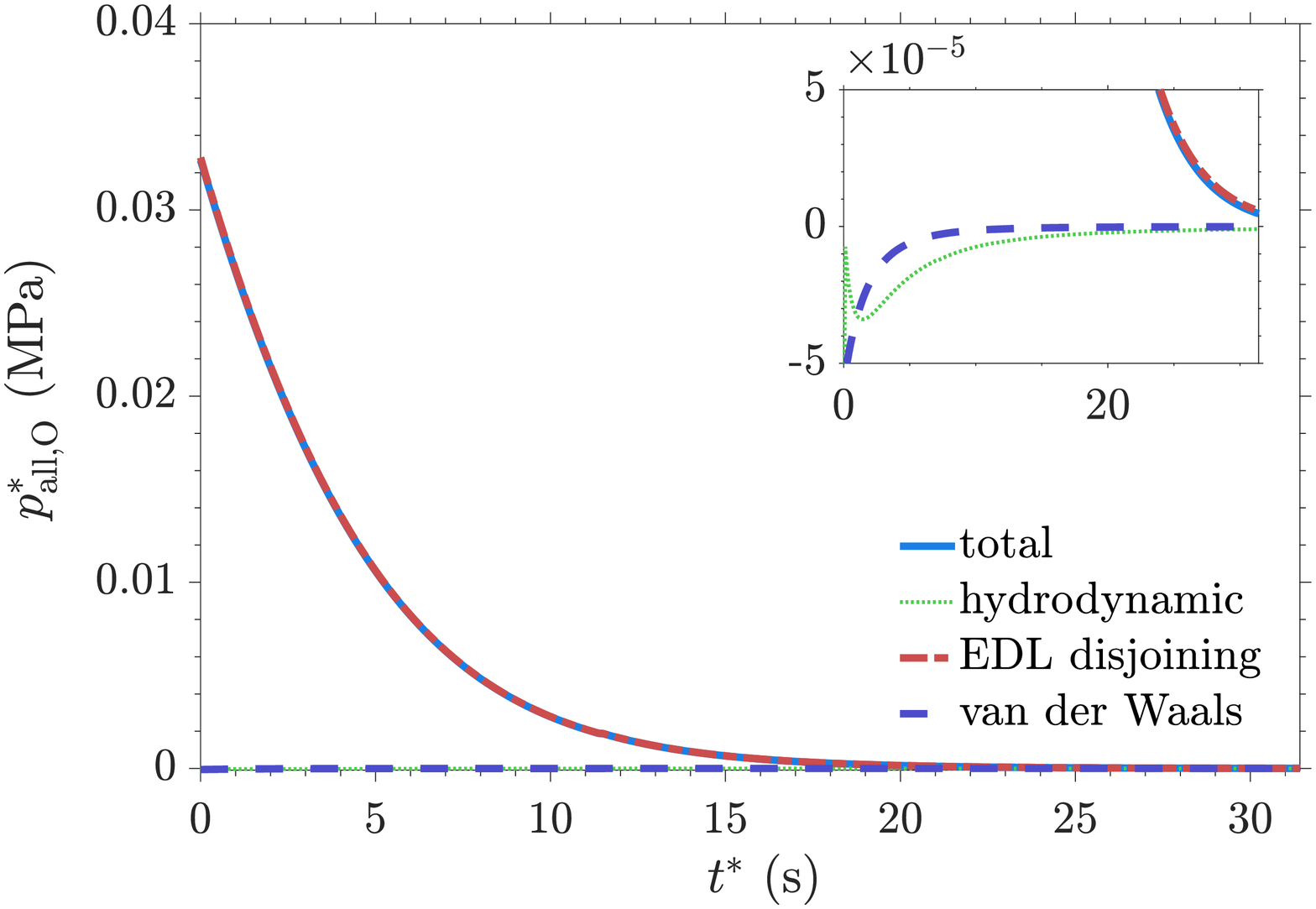}
\caption{\centering}
\label{subfig:pall_vs_t_rec_dlvo_mid}
\end{subfigure}
\begin{subfigure}[b]{0.495\textwidth}
\centering
\includegraphics[width=\textwidth]{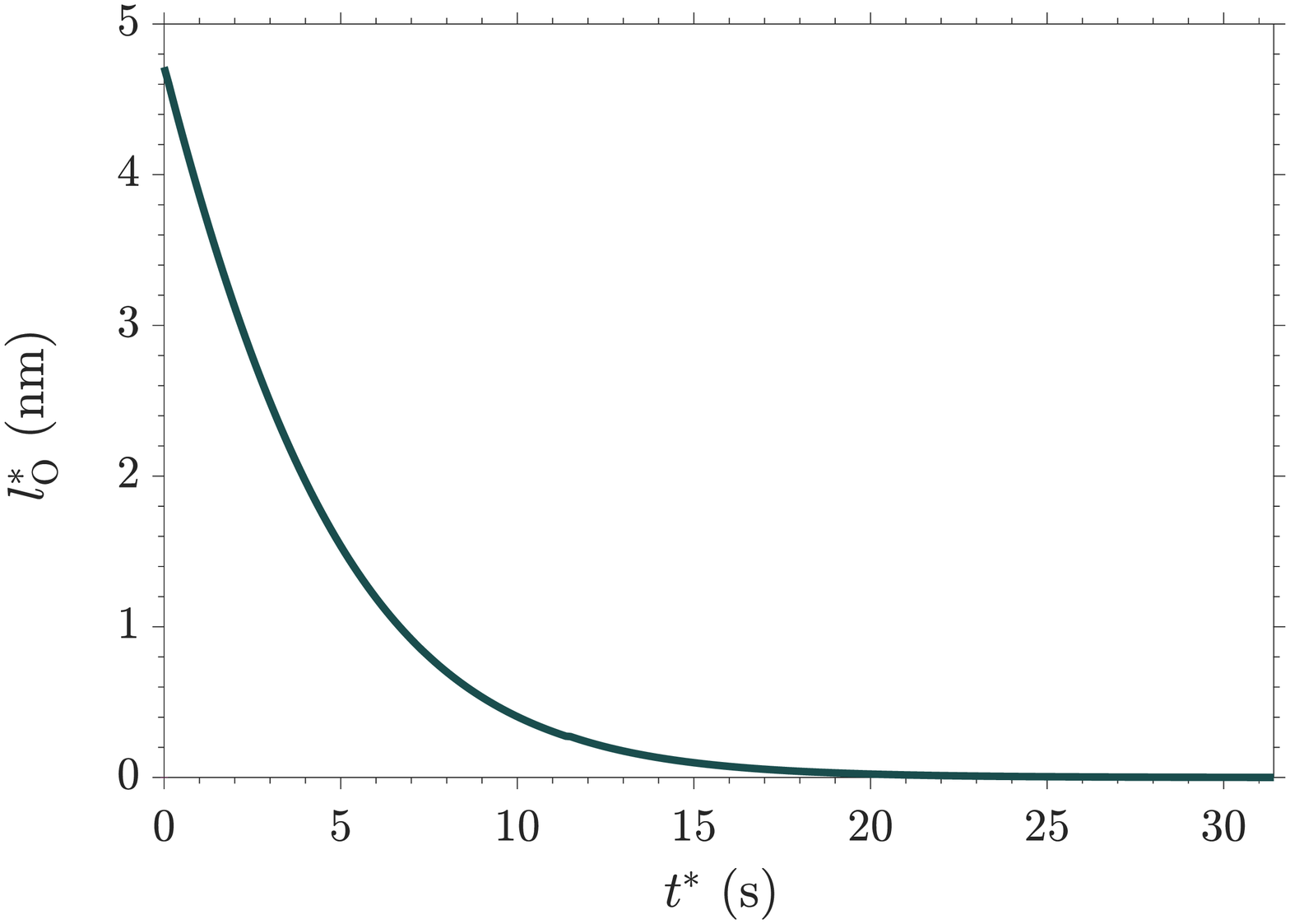}
\caption{\centering}
\label{subfig:l_vs_t_rec_dlvo_mid}
\end{subfigure}
\caption{Variation of (a) pressure components $p_{\text{hd}}^*$, $p_{\text{EDL}}^*$ and $p_{\text{vdW}}^*$ and total pressure $p^*$, and (b) deflection $l^*$ at origin, with time $t^*$ for representative case of recession loading; other system parameters are: $R$ = 1 mm, $D$ = 50 nm, $h_0$ = 45 nm, $L$ = 5 $\mu$m,  $\omega$ = 2$\pi\times 10^{-2}$ rad/s, $E_{\text{Y}}$ = 17.5 MPa, $\nu$ = 0.45, $\mu$ = 1 mPa-s, $A_{\text{sfw}}$ = $10^{-21}$ J, $\psi_S$ = 100 mV}
\label{fig:rec}
\end{figure}
\begin{figure}[!htb]
\centering
\begin{subfigure}[b]{0.495\textwidth}
\centering
\includegraphics[width=\textwidth]{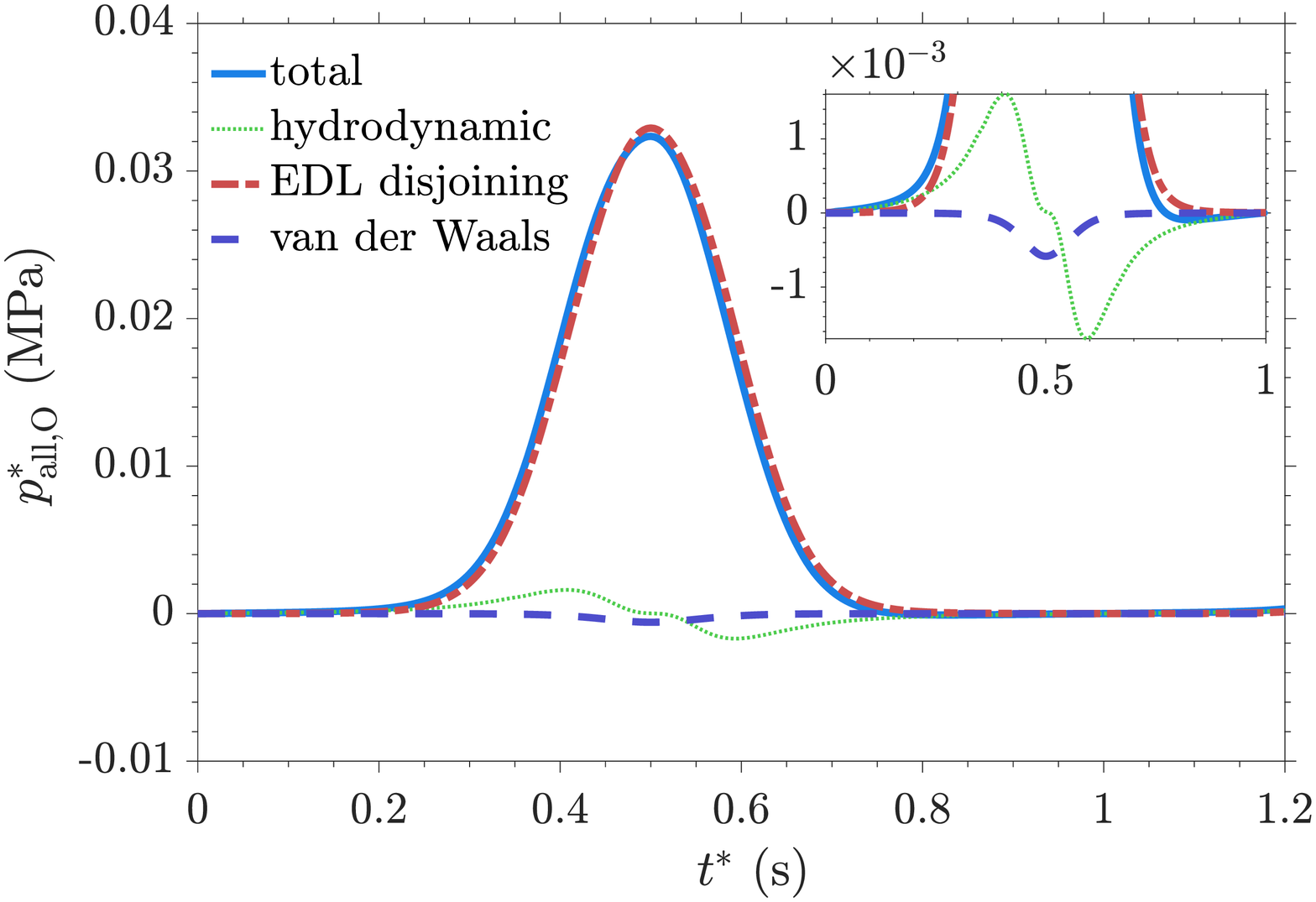}
\caption{\centering}
\label{subfig:pall_vs_t_osc_dlvo_mid}
\end{subfigure}
\begin{subfigure}[b]{0.495\textwidth}
\centering
\includegraphics[width=\textwidth]{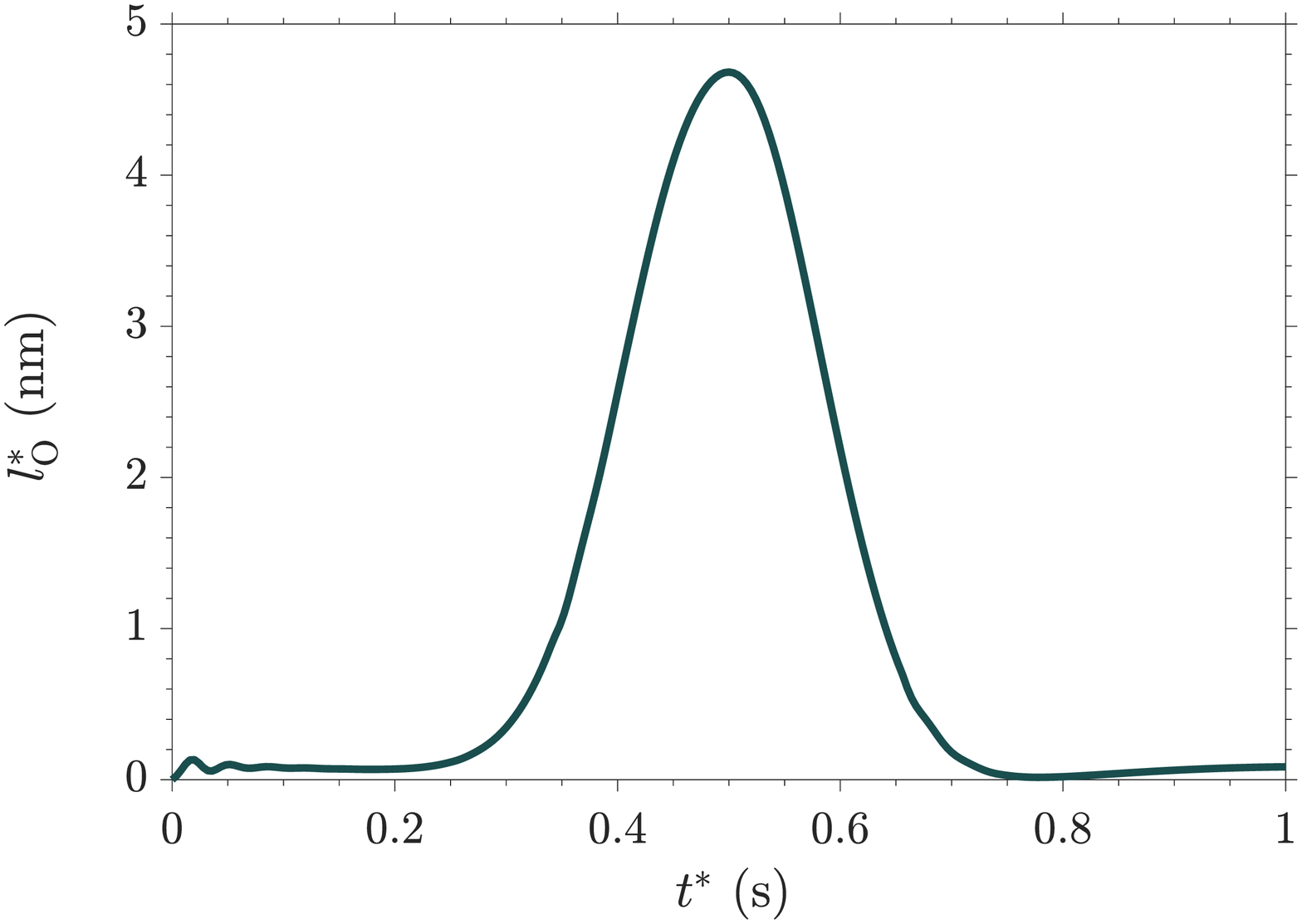}
\caption{\centering}
\label{subfig:l_vs_t_osc_dlvo_mid}
\end{subfigure}
\caption{Variation of (a) pressure components $p_{\text{hd}}^*$, $p_{\text{EDL}}^*$ and $p_{\text{vdW}}^*$ and total pressure $p^*$, and (b) deflection $l^*$ at origin, with time $t^*$ for representative case of oscillatory loading; other system parameters are: $R$ = 1 mm, $D$ = 50 nm, $h_0$ = 45 nm, $L$ = 5 $\mu$m,  $\omega$ = 2$\pi$ rad/s, $E_{\text{Y}}$ = 17.5 MPa, $\nu$ = 0.45, $\mu$ = 1 mPa-s, $A_{\text{sfw}}$ = $10^{-20}$ J, $\psi_S$ = 100 mV}
\label{fig:osc}
\end{figure}
In this subsection, we present the time-evolutions of pressure and deflection for one representative case each for the three loading types. These cases correspond to the approximate median parametric values that are used to compute the parametric variations that are presented in the upcoming sections. The parametric values for each representative case are presented in its figure caption. \\
In figures \ref{fig:app} and \ref{fig:rec}, the representative case for approach and recession loading each are presented. The maximum deflection for these representative cases occurs when the sphere is at smallest separation from sphere, which occurs at end time for approach loading and at start time for recession loading. This because the DLVO pressure components, that dominate the total pressure and hence deflection, are strongest at least separation. The maximum deflection can be seen to be a little lower than $\sim$ 5 nm, i.e. they approach values close to $D-h_0$. Hence, we deduce that these representative cases are exhibiting strong TWC. While the deflection and DLVO pressure components for approach and recession loading are approximately mirror images of each other about the mid-time, same is not true for hydrodynamic pressure (compare the dashed green lines in insets of figures \ref{subfig:pall_vs_t_app_dlvo_mid} and \ref{subfig:pall_vs_t_rec_dlvo_mid}). Hydrodynamic pressure in both cases decreases near the point of least separation of sphere from origin. However, it exhibits a local maxima a bit before reaching the point of least separation of sphere from origin for approach loading. This is feature is absent in recession loading. As for oscillatory loading (figure \ref{fig:osc}), we observe the maximum pressure and deflection are the same in magnitude to those for approach and recession loading (figures \ref{fig:app} and \ref{fig:rec}) - however, the maxima occurs at mid-oscillation. Lastly, we note that although maximum values are similar for oscillatory, approach and recession loading, the profiles (of pressure components, total pressure, and deflection) for oscillatory loading are not a simple superimposition of the profiles for approach and recession loading. This is an outcome of the harmonic speed of sphere for oscillatory loading as opposed to constant speed for approach and recession loading.\\ 
\subsection{Approach and Recession Loading}\label{subsec:results_apprec}
\begin{figure}[!htb]
\centering
\begin{subfigure}[b]{0.495\textwidth}
\centering
\includegraphics[width=\textwidth]{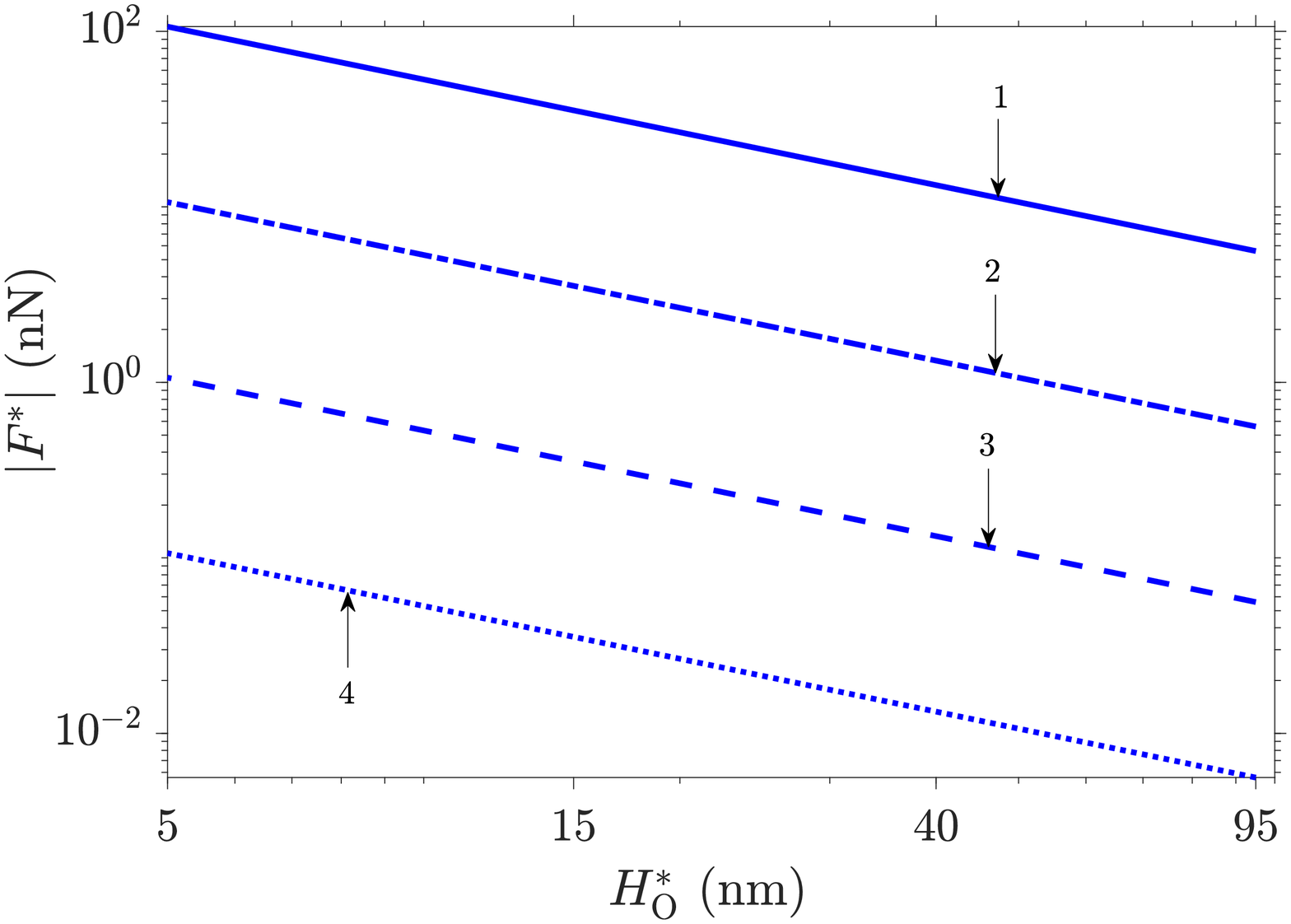}
\caption{\centering}
\label{subfig:F_vs_H0_hd_diff_speed}
\end{subfigure}
\begin{subfigure}[b]{0.495\textwidth}
\centering
\includegraphics[width=\textwidth]{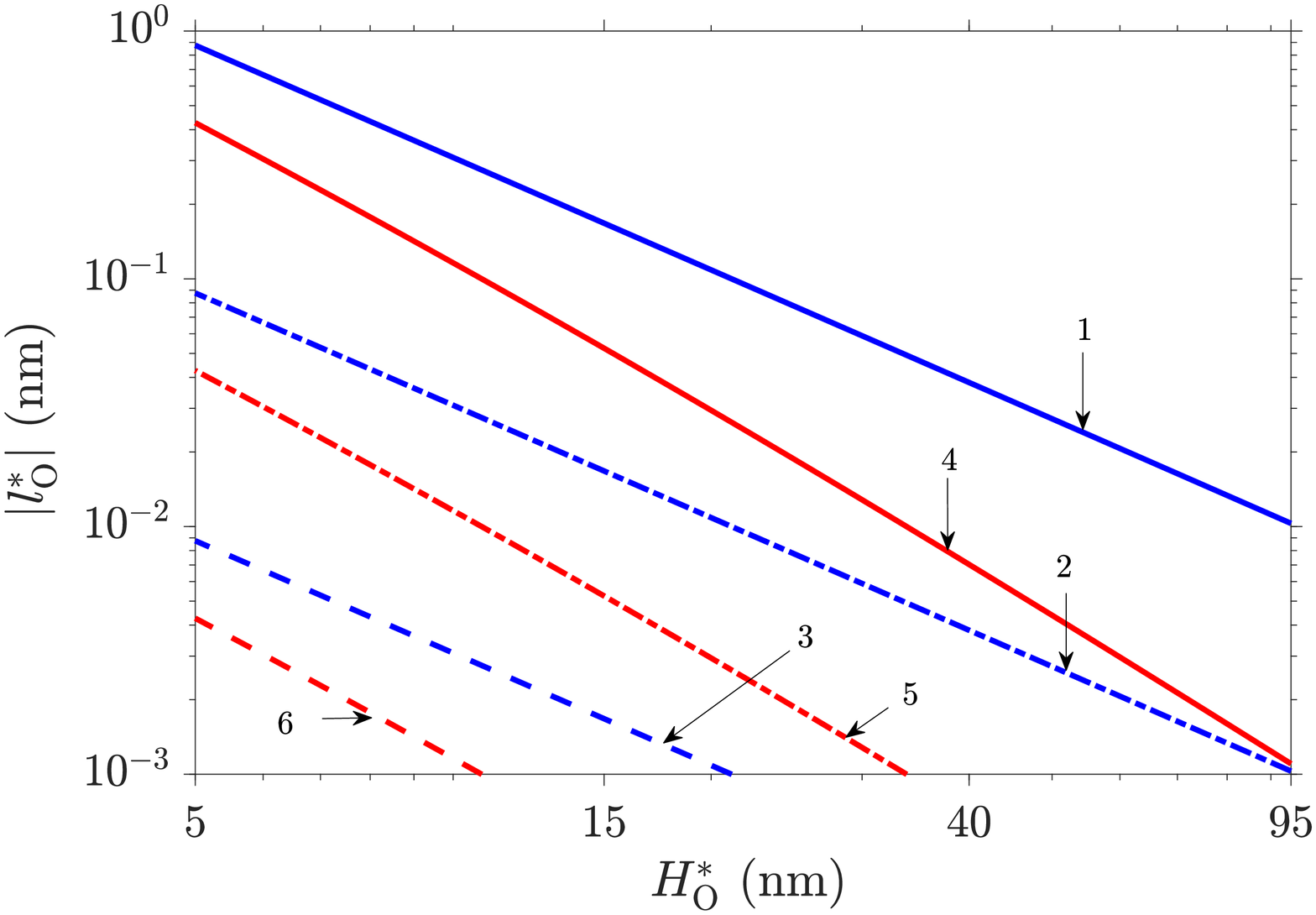}
\caption{\centering}
\label{subfig:l_vs_H0_hd_diff_speed}
\end{subfigure}
\caption{Variation of (a) magnitude of force $|F^*|$, and (b) magnitude of deflection at origin $|l_{\text{O}}^*|$ with separation of sphere from origin $H_{\text{O}}^*$ for approach as well as recession loading, sphere speed is $\omega h_0 = 2\pi \hat{\omega} h_0$; DLVO forces are not considered; force as well as deflection are positive for approach loading and negative for recession loading; other system parameters are: $R$ = 1 mm, $D$ = 50 nm, $h_0$ = 45 nm, $L$ = [0.5 mm (semi-infinite), 5 $\mu$m (thick), 50 nm (thin)], $\hat{\omega}$ = $[10^{-4}, 10^{-3}, 10^{-2}, 10^{-1}]$ Hz, $E_{\text{Y}}$ = 17.5 MPa, $\nu$ = 0.45, $\mu$ = 1 mPa-s; substrate thickness does not affect force; deflection is negligible (smaller than $10^{-3}$ nm) for all $\hat{\omega}$ for thin substrate and for $\hat{\omega} = 10^{-4}$ for thick and semi-infinite substrates; the plot-line labels for subfigure a mean: 1 - $\hat{\omega}$ = $10^{-1}$ Hz, 2 - $\hat{\omega}$ = $10^{-2}$ Hz, 3 - $\hat{\omega}$ = $10^{-3}$ Hz, 4 - $\hat{\omega}$ = $10^{-4}$ Hz; the plot-line labels for subfigure b mean: 1 - $\hat{\omega}$ = $10^{-1}$ Hz for semi-infinite substrate, 2 - $\hat{\omega}$ = $10^{-2}$ Hz for semi-infinite substrate, 3 - $\hat{\omega}$ = $10^{-3}$ Hz for semi-infinite substrate, 4 - $\hat{\omega}$ = $10^{-1}$ Hz for thick substrate, 5 - $\hat{\omega}$ = $10^{-2}$ Hz for thick substrate, 6 - $\hat{\omega}$ = $10^{-3}$ Hz for thick substrate; both the panels are log-scaled on both horizontal and vertical axes}
\label{fig:V_apprec_hd}
\end{figure}
\begin{figure}[!htb]
\centering
\begin{subfigure}[b]{0.495\textwidth}
\centering
\includegraphics[width=\textwidth]{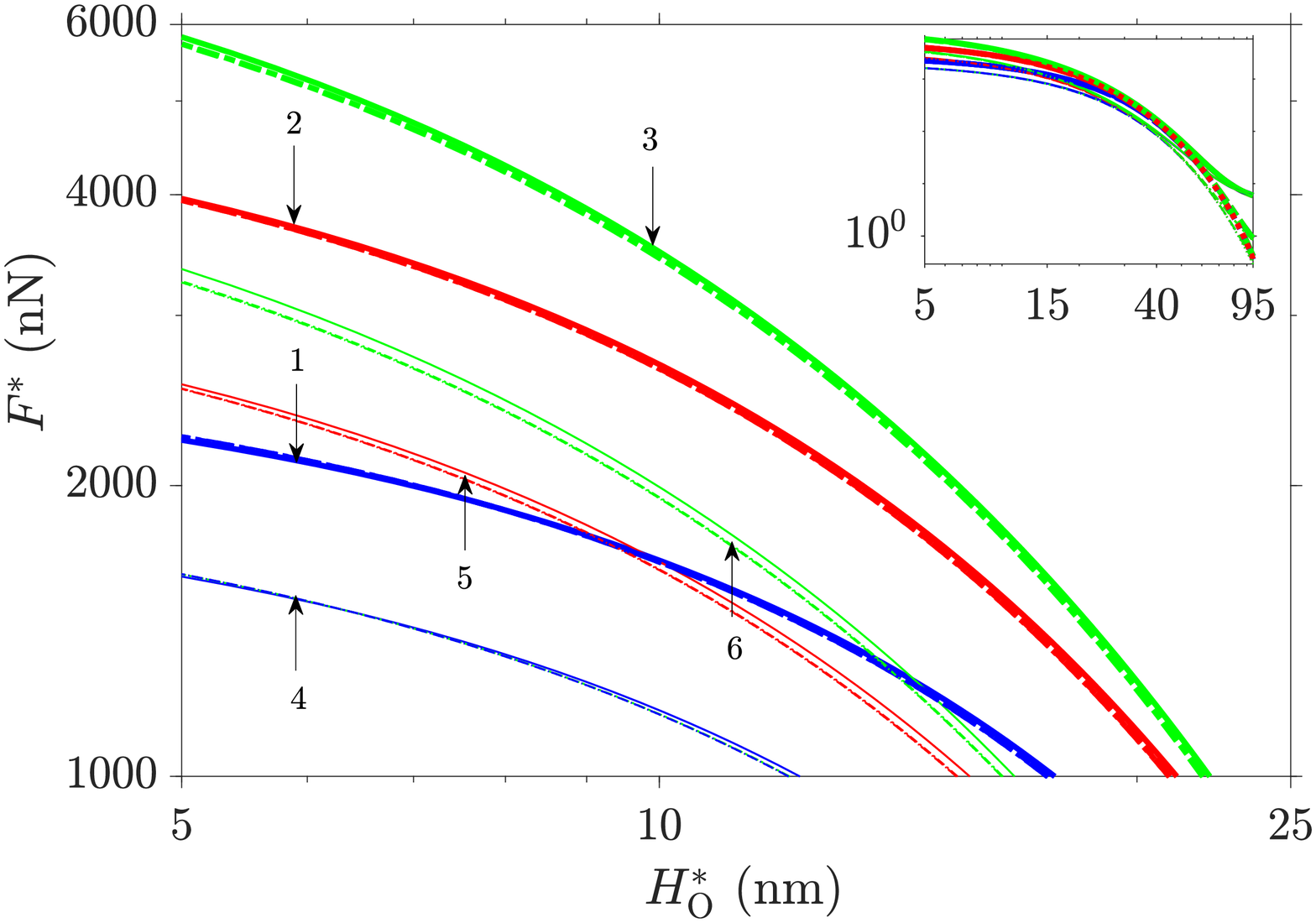}
\caption{\centering}
\label{subfig:F_vs_H0_dlvo_diff_speed}
\end{subfigure}
\begin{subfigure}[b]{0.495\textwidth}
\centering
\includegraphics[width=\textwidth]{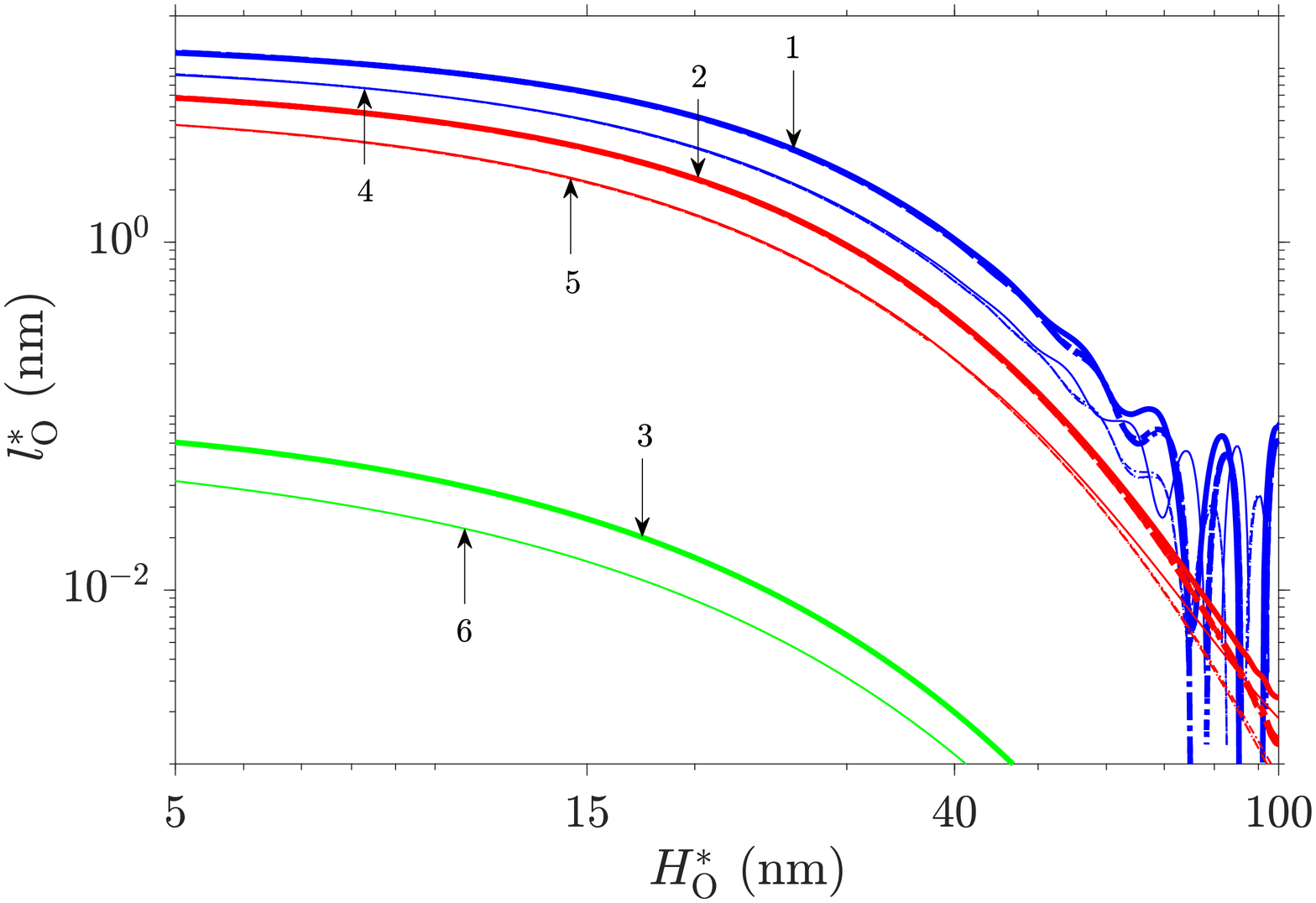}
\caption{\centering}
\label{subfig:l_vs_H0_dlvo_diff_speed}
\end{subfigure}
\caption{Variation of (a) force $F^*$, and (b) deflection at origin $l_{\text{O}}^*$ with separation of sphere from origin $H_{\text{O}}^*$ for approach as well as recession loading, sphere speed is $\omega h_0 = 2\pi\hat{\omega} h_0$; DLVO forces are considered with two sets of parameter values - $A_{\text{sfw}} = 10^{-21}$ J, $\psi_S = 100$ mV for `moderate DLVO', and, $A_{\text{sfw}} = 10^{-20}$ J, $\psi_S = 2500$ mV for `strong DLVO'; other system parameters are: $R$ = 1 mm, $D$ = 50 nm, $h_0$ = 45 nm, $L$ = [0.5 mm (semi-infinite), 5 $\mu$m (thick), 50 nm (thin)], $\hat{\omega}$ = $[10^{-4}, 10^{-3}, 10^{-2}, 10^{-1}]$ Hz, $E_{\text{Y}}$ = 17.5 MPa, $\nu$ = 0.45, $\mu$ = 1 mPa-s; effect of sphere speed on both force and deflection is negligible and is perceptible only for $H_{\text{O}}^*>$ 70 nm; the plot-line labels for each subfigure mean: 1 - semi-infinite substrate with strong DLVO, 2 - thick substrate with strong DLVO, 3 - thin substrate with strong DLVO, 4 - semi-infinite substrate with moderate DLVO, 5 - thick substrate with moderate DLVO, 6 - thin substrate with moderate DLVO; both the panels are log-scaled on both horizontal and vertical axes}
\label{fig:V_apprec_dlvo}
\end{figure}
In this subsection, we present some crucial cases of approach and recession loading. Approach as well as recession loading for three cases of thickness - `thin' i.e. thickness much smaller compared to lubrication zone radial length scale, `thick' i.e. thickness comparable to lubrication zone radial length scale and `semi-infinite' i.e. thickness much larger compared to lubrication zone radial length scale, are studied. For each thickness, we study four speeds of approach/recession, given as $\omega h_0$ = $2\pi\hat{\omega}h_0$ where $\hat{\omega} = [10^{-4},10^{-3},10^{-2},10^{-1}]$. For each combination of approach or recession, substrate thickness and speed, three categories are studied - when DLVO forces are not considered, when DLVO forces are considered with DLVO force parameters amounting to moderately larger DLVO pressure components compared to hydrodynamic pressure (labelled `moderate DLVO' for brevity), when DLVO forces are considered with DLVO force parameters amounting to dominating DLVO pressure components compared to hydrodynamic pressure (labelled `strong DLVO' for brevity). The rest of the system parameters are identical for all the cases studied, and are presented in the captions of figure \ref{fig:V_apprec_hd} as well as \ref{fig:V_apprec_dlvo}. \\ 
We first consider the case where DLVO forces are not present, i.e. the system is purely hydrodynamic (presented in figure \ref{fig:V_apprec_hd}). Variation of magnitude of force and magnitude of deflection for the different cases in this category are presented in figures \ref{subfig:F_vs_H0_hd_diff_speed} and \ref{subfig:l_vs_H0_hd_diff_speed} respectively. The same plots apply for both approach and recession loading - approach loading leads to positive values and recession loading leads to negative values, magnitude is same. Looking at figure \ref{subfig:F_vs_H0_hd_diff_speed}, the force is dependent only on approach/recession speed, and is expectedly higher in magnitude for higher approach/recession speed. The pressure and deflection coupling is predominantly OWC. Therefore, the magnitude of force follows the expression $\displaystyle |F^*| = \frac{6\pi\mu\omega h_0 R^2}{DH_{\text{O}}^*}$, which leads to linear variation of the logarithms of $\log(|F^*|)$ with $\log(H^*)$ with slope of $-1$ with different intercepts for different speeds (because of different $\omega$), as can be seen in figure \ref{subfig:F_vs_H0_hd_diff_speed}. Looking at figure \ref{subfig:l_vs_H0_hd_diff_speed}, we expectedly find that higher approach/recession speed leads to higher deflection. An interesting feature is that magnitude of deflection for thick substrate drops faster than semi-infinite substrate (lines 4, 5, and 6).\\
We next consider the case where DLVO forces are present, with both moderate and strong DLVO pressure components (presented in figure \ref{fig:V_apprec_dlvo}). Variation of magnitude of force and magnitude of deflection for the different cases are presented in figures \ref{subfig:F_vs_H0_dlvo_diff_speed} and \ref{subfig:l_vs_H0_dlvo_diff_speed} respectively. The same plots apply for both approach and recession loading - both magnitude and sign are same. This occurs because DLVO pressure components, which are not dependent on direction of motion but only on separation, are the main contributor to total pressure, and thus force and deflection. Looking at both figures \ref{subfig:F_vs_H0_dlvo_diff_speed} and \ref{subfig:l_vs_H0_dlvo_diff_speed}, the force as well as deflection is dependent only on DLVO pressure components and substrate thickness. The thick lines (lines 1, 2, 3), which correspond to strong DLVO, are expectadly higher than the thin lines (lines 3, 4, 5), which correspond to moderate DLVO. The force increases as one moves from semi-infinite substrate to thin substrate (line 1 then line 2 then line 3 for strong DLVO and line 4 then line 5 then line 6 for moderate DLVO). This order is however reversed for deflection. This contrast between the ordering of force and deflection with thickness is an outcome of the `self-diminishing' nature of repulsive pressure. This nature is explained as follows. The nature of repulsive pressure components being studied here, i.e. EDL disjoining pressure for all situations and hydrodynamic pressure in certain situations, is of increase with decrease in separation. However, the deflection caused by a repulsive pressure is positive, leading to increase in separation. Therefore, any feature of the substrate domain that leads to effectively softer substrate (i.e. allows for higher deflection for same pressure), larger thickness in this case, leads to larger deflection and thus smaller magnitude of the repulsive pressure when the interaction of pressure and deflection is TWC. \\
\subsection{Low Frequency Oscillatory Loading}\label{subsec:results_lowfreq}
In this subsection, we focus on delineating the effects of substrate thickness, substrate material compressibility and presence of DLVO forces. Therefore, we study low frequency oscillations having $\omega = 2\pi$ rad/s as these effects do not appear clearly discernible for higher frequencies. The other system parameter values are presented in captions of the respective figure of the different sets of results presented.\\
\subsubsection{Effect of Substrate Thickness}\label{subsubsec:results_L}
\begin{figure}[!htb]
\centering
\begin{subfigure}[b]{0.495\textwidth}
\centering
\includegraphics[width=\textwidth]{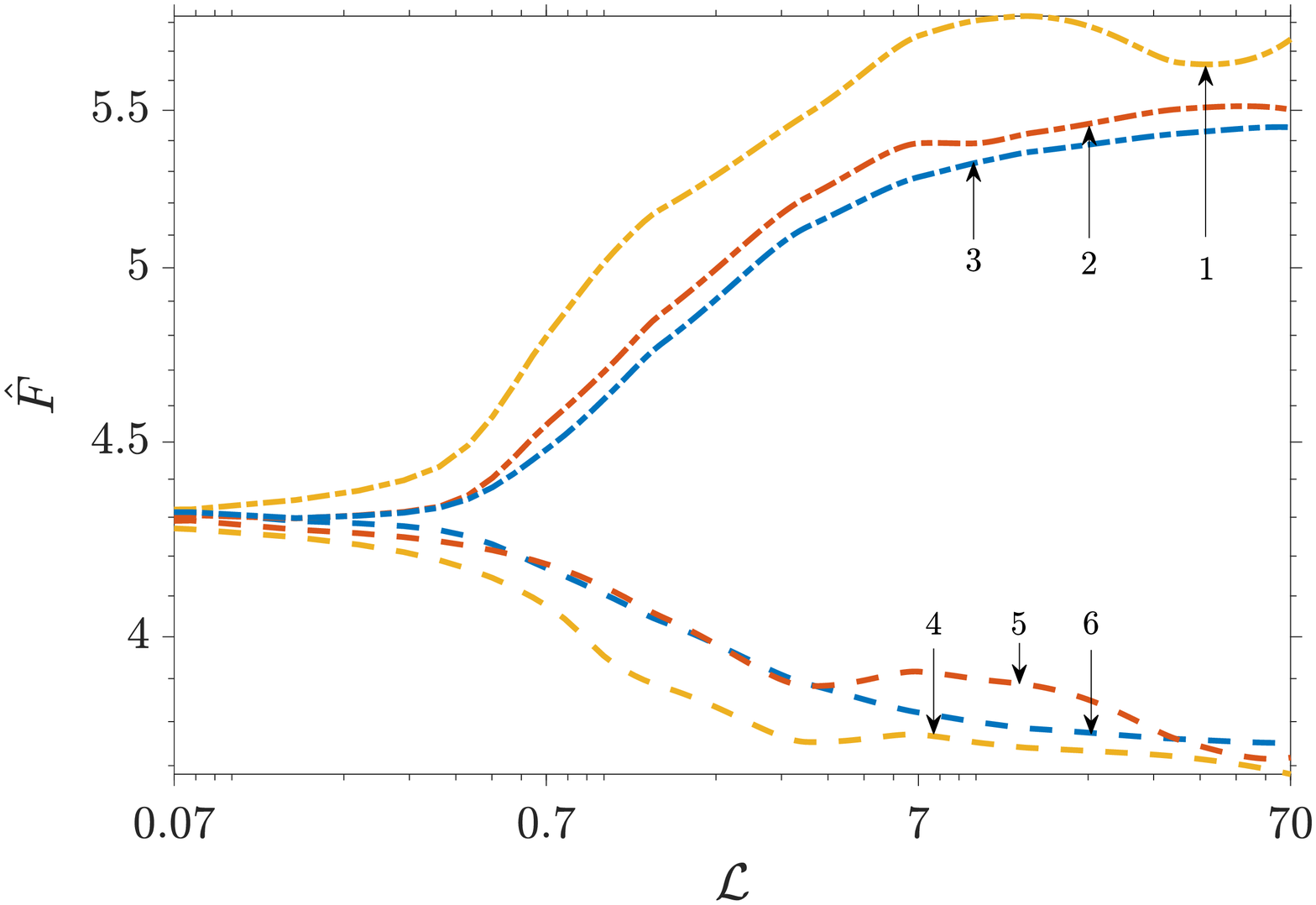}
\caption{}
\label{subfig:F_L_hd}
\end{subfigure}
\begin{subfigure}[b]{0.495\textwidth}
\centering
\includegraphics[width=\textwidth]{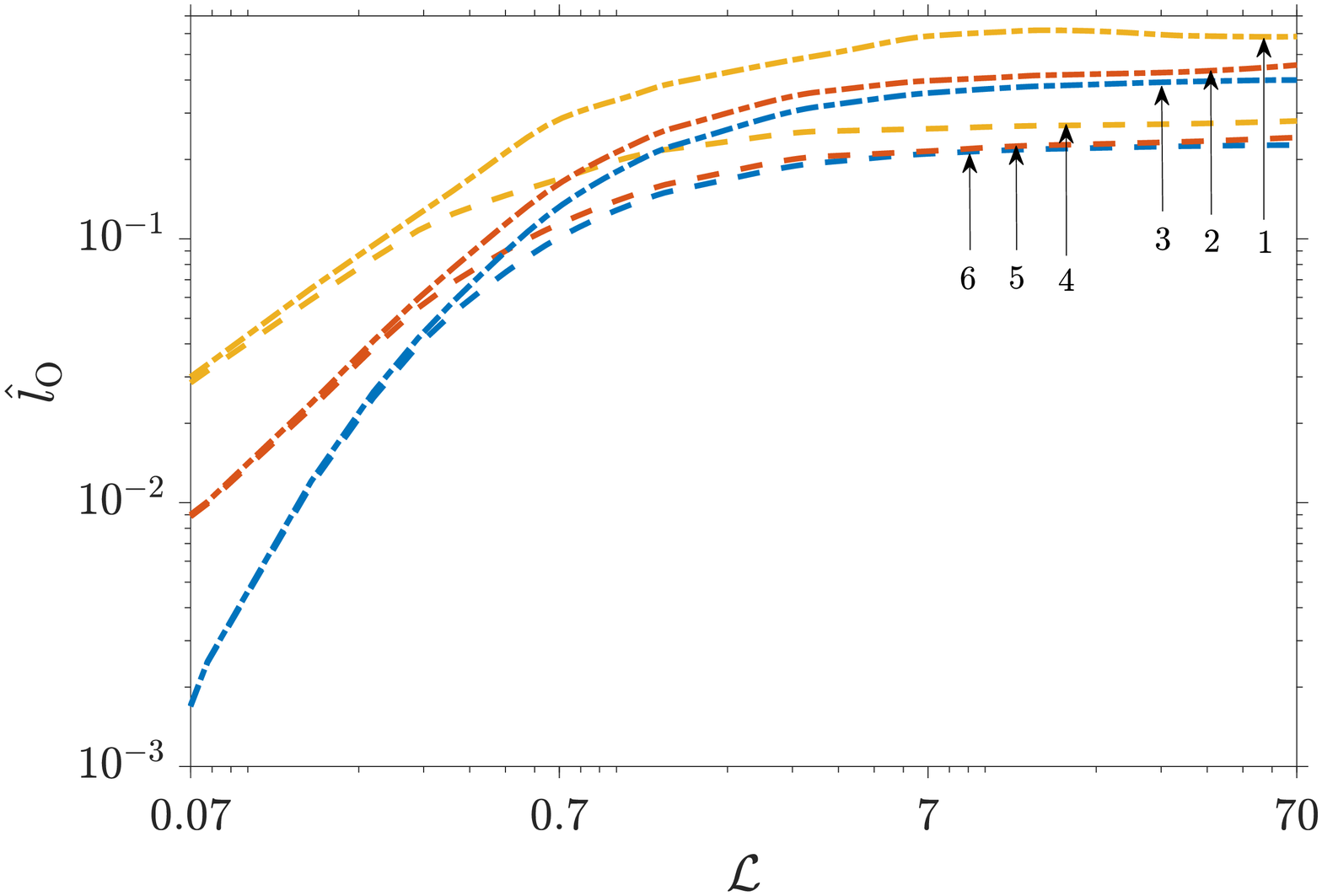}
\caption{}
\label{subfig:l_L_hd}
\end{subfigure}
\begin{subfigure}[b]{0.495\textwidth}
\centering
\includegraphics[width=\textwidth]{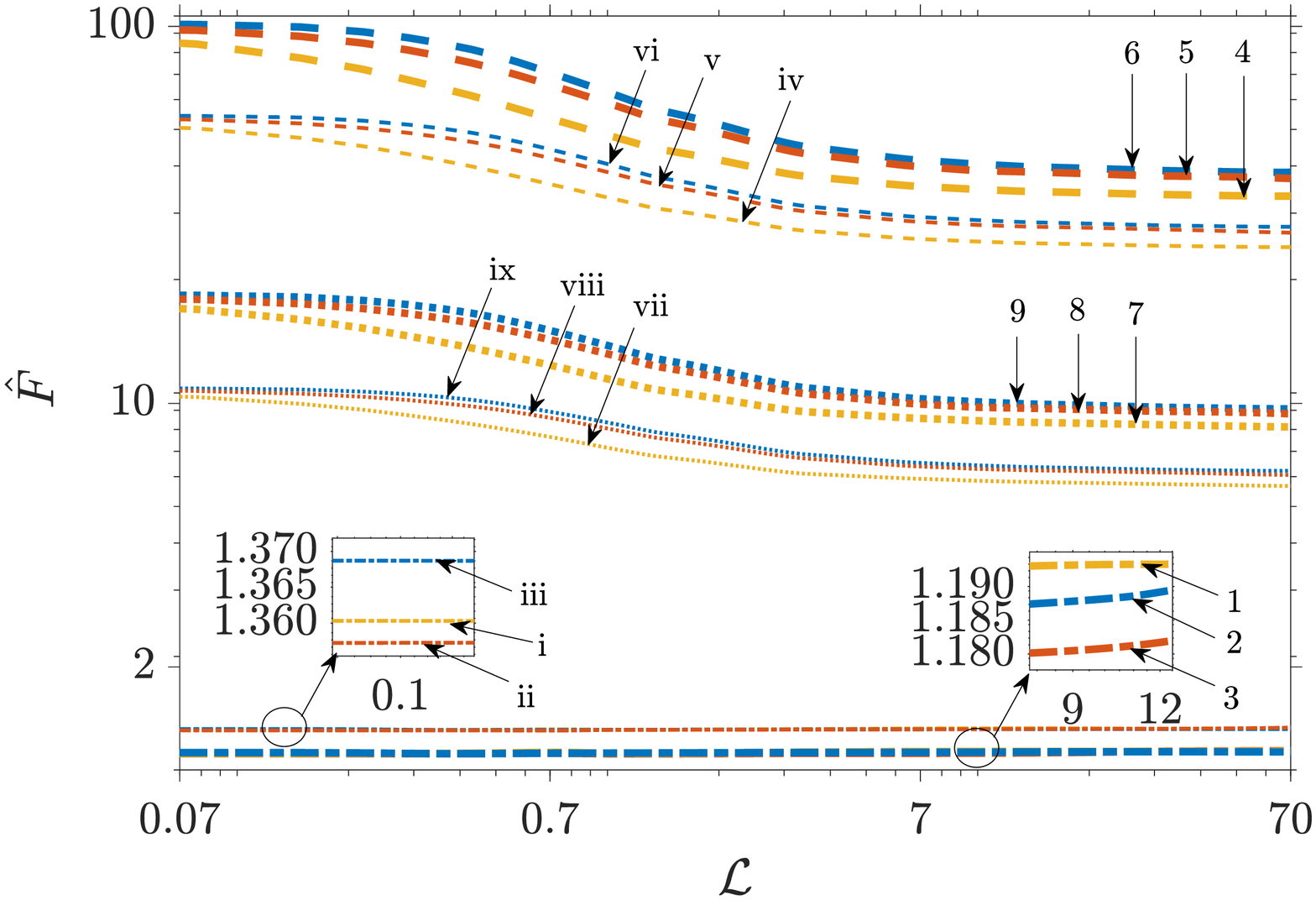}
\caption{}
\label{subfig:F_L_dlvo}
\end{subfigure}
\begin{subfigure}[b]{0.495\textwidth}
\centering
\includegraphics[width=\textwidth]{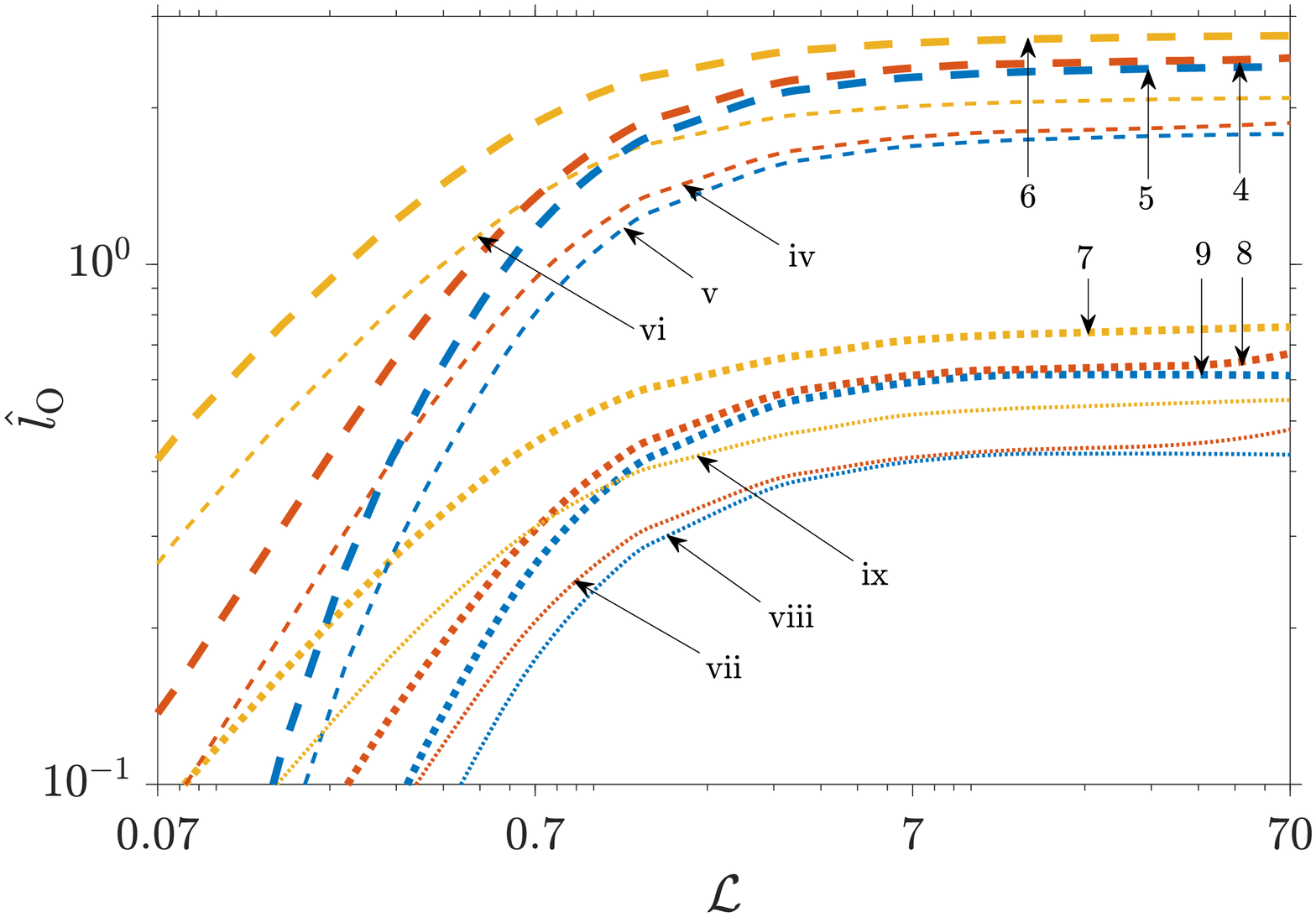}
\caption{}
\label{subfig:l_L_dlvo}
\end{subfigure}
\caption{Variation of magnitude of maximum attractive (negative), maximum repulsive (positive) and mean (a,c) normalized Force $\hat{F}$ and (b,d) normalized deflection at origin $\hat{l}_{\text{O}}$, over one complete oscillation, with $\mathcal{L}$ for the case with (a,b) only hydrodynamic pressure, and, (c,d) hydrodynamic and DLVO pressure components; other system parameters are: $R$ = 1 mm, $D$ = 50 nm, $h_0$ = 45 nm, $\omega$ = 2$\pi$ rad/s, $E_{\text{Y}}$ = 17.5 MPa, $\nu$ = [0.00 (compressible), 0.45 (normal), 0.50 (incompressible)], $\mu$ = 1 mPa-s; the plot-line labels for each subfigure mean: 1 - maximum attractive characteristics for compressible substrate material, 2 - maximum attractive characteristics for normal substrate material, 3 - maximum attractive characteristics for incompressible substrate material, 4 - maximum repulsive characteristics for compressible substrate material, 5 - maximum repulsive characteristics for normal substrate material, 6 - maximum repulsive characteristics for incompressible substrate material, 7 - mean characteristics for compressible substrate material, 8 - mean characteristics for normal substrate material, 9 - mean characteristics for incompressible substrate material; in the subfigures on bottom (i.e. c and d), the labels 1 to 9 correspond to `strong DLVO' ($A_{\text{sfw}}$ = $10^{-20}$ J i.e. $\mathcal{H} = 0.7$, $\psi_S$ = 2500 mV i.e. $\mathcal{R} = 700$) while the labels i to ix correspond to their respective arabic numbers' in description but for `moderate DLVO' ($A_{\text{sfw}}$ = $10^{-21}$ J i.e. $\mathcal{H} = 0.07$, $\psi_S$ = 100 mV i.e. $\mathcal{R} = 28$); all the panels are log-scaled on both horizontal and vertical axes}
\label{fig:F_L}
\end{figure}
As discussed in the scaling analysis in section \ref{subsubsec:domain}, the behaviour of substrate thickness is dependent on the radial length scale of lubrication region - significantly thicker substrates are practically semi-infinite and significantly thinner substrates behave like Winkler foundation (i.e. deflection at a particular radial point is linearly related to the pressure at that radial point). Therefore, to assess the effects of substrate thickness, we utilize the non-dimensional parameter $\displaystyle \mathcal{L} = \frac{L}{\sqrt{RD}}$, which compares the substrate thickness to the lubrication zone radial length scale at mid-oscillation. We take the range of $\mathcal{L}$ as 0.07 to 70, which amounts to varying the substrate thickness from thin (followed by thick) to semi-infinite. The system parameter values are presented in caption of figure \ref{fig:F_L}.\\
We assess the parametric variation with $\mathcal{H}$ of magnitudes of maximum attractive, maximum repulsive and mean, over one complete oscillation, of force and deflection, presented in figure \ref{fig:F_L}. We consider three scenarios - when there is only hydrodynamic pressure, when hydrodynamic as well as DLVO pressure components are present and the latter are moderate ($A_{\text{sfw}} = 10^{-21}$ J i.e. $\mathcal{H} = 0.07$, $\zeta$ = 100 mV i.e. $\mathcal{R} = 28$), labelled `DLVO moderate', and, when hydrodynamic as well as DLVO pressure components are present and the latter are strong ($A_{\text{sfw}} = 10^{-20}$ J i.e. $\mathcal{H} = 0.7$, $\zeta$ = 2500 mV i.e. $\mathcal{R} = 700$), labelled `DLVO strong'. \\
We first consider the case when DLVO forces are absent ($A_{\text{sfw}} = 0$ i.e. $\mathcal{H} = 0$, $\psi_S = 0$ i.e. $\mathcal{R} = 0$), presented in figures \ref{subfig:F_L_hd} and \ref{subfig:l_L_hd}. \\
\begin{figure}[!htb]
\centering
\begin{subfigure}[b]{0.495\textwidth}
\centering
\includegraphics[width=\textwidth]{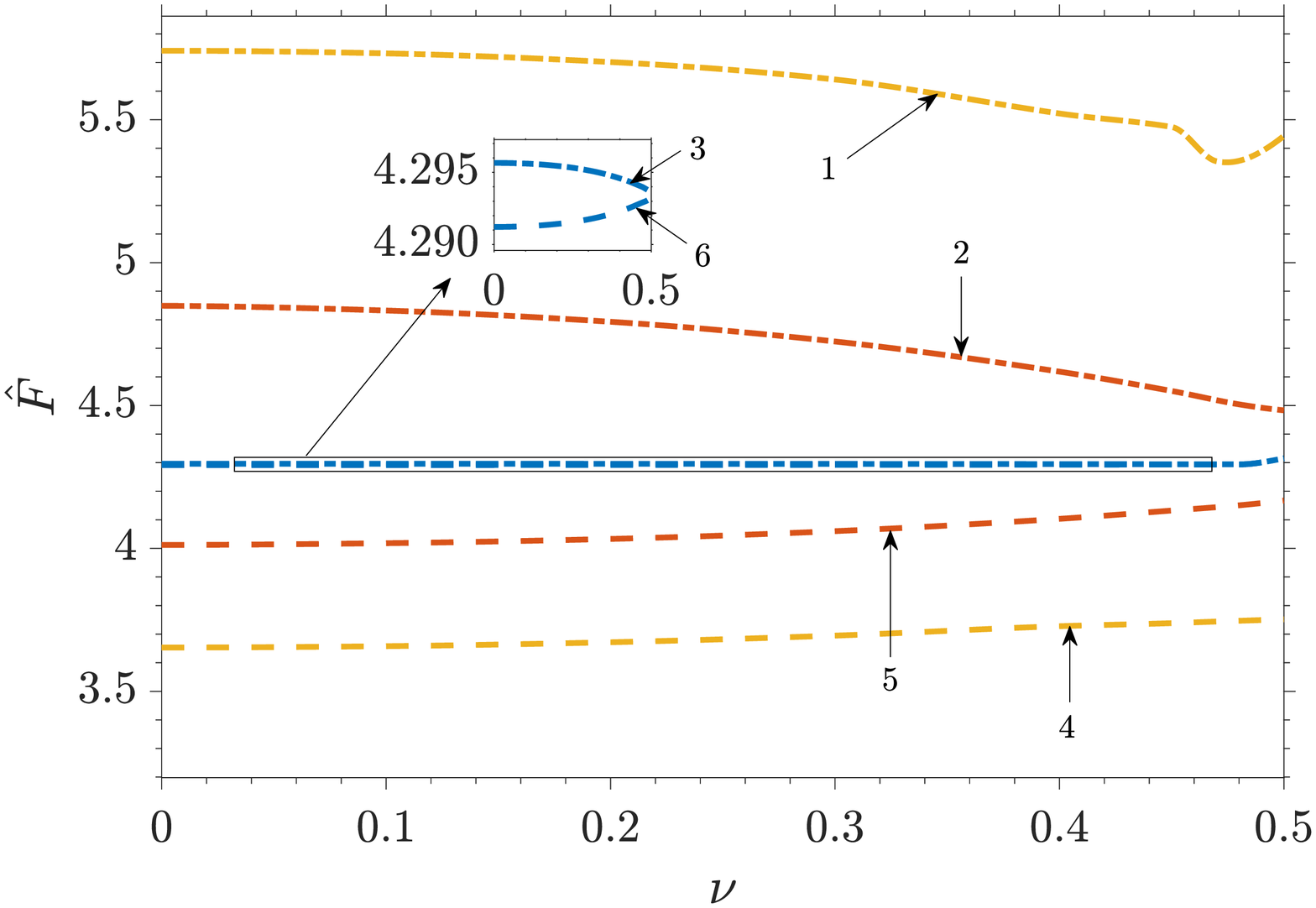}
\caption{}
\label{subfig:F_nu_hd}
\end{subfigure}
\begin{subfigure}[b]{0.495\textwidth}
\centering
\includegraphics[width=\textwidth]{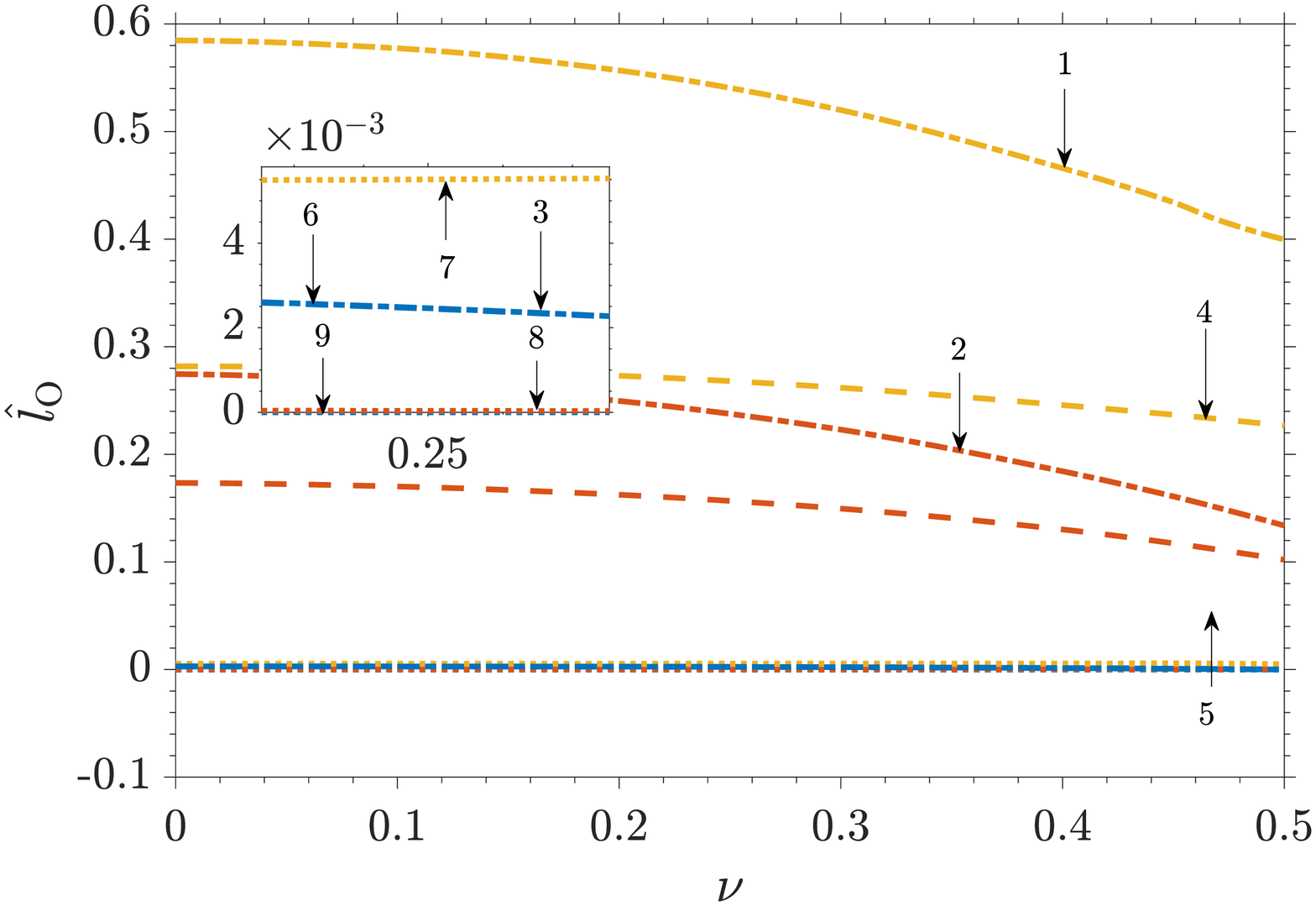}
\caption{}
\label{subfig:l_nu_hd}
\end{subfigure}
\begin{subfigure}[b]{0.495\textwidth}
\centering
\includegraphics[width=\textwidth]{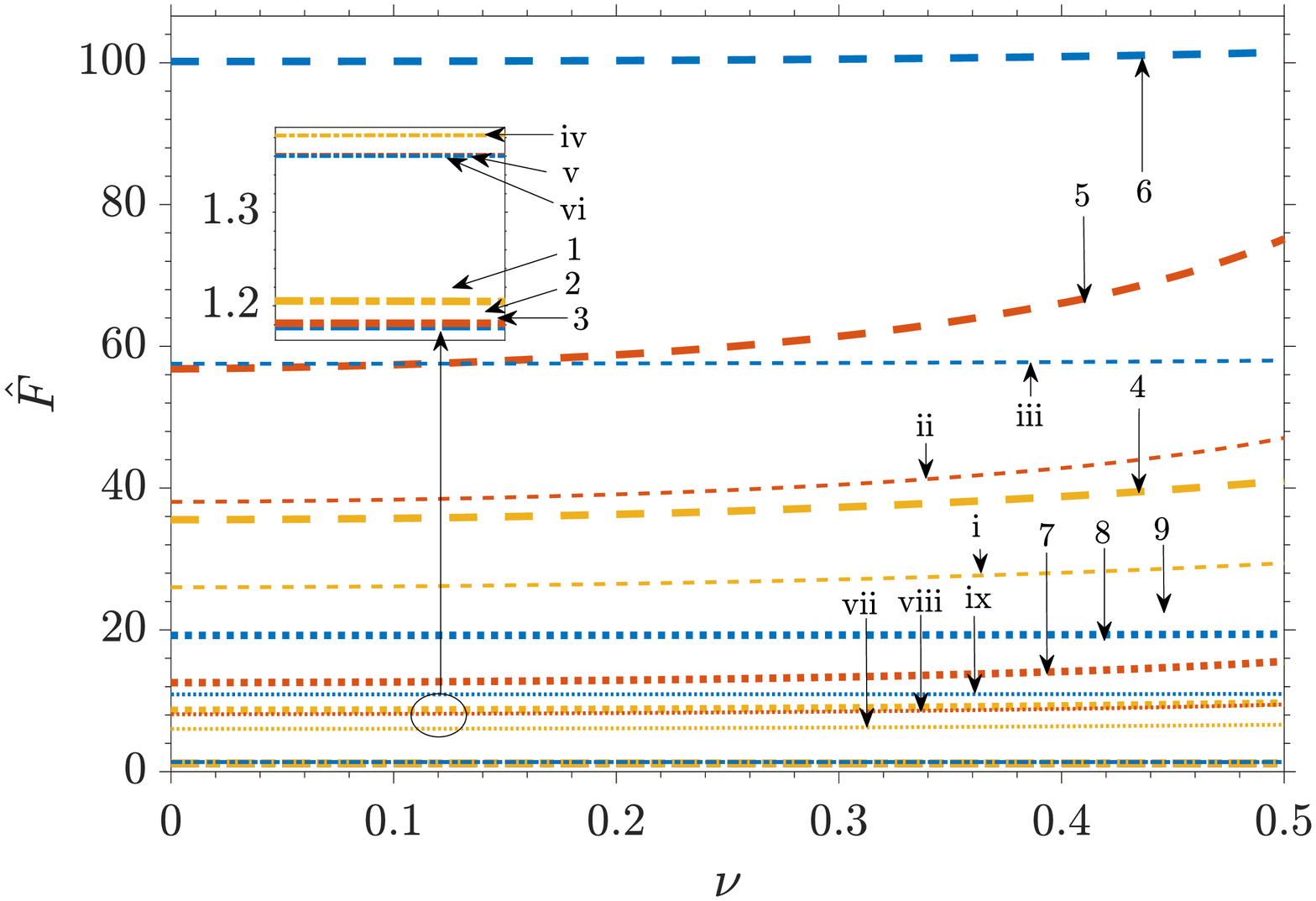}
\caption{}
\label{subfig:F_nu_dlvo}
\end{subfigure}
\begin{subfigure}[b]{0.495\textwidth}
\centering
\includegraphics[width=\textwidth]{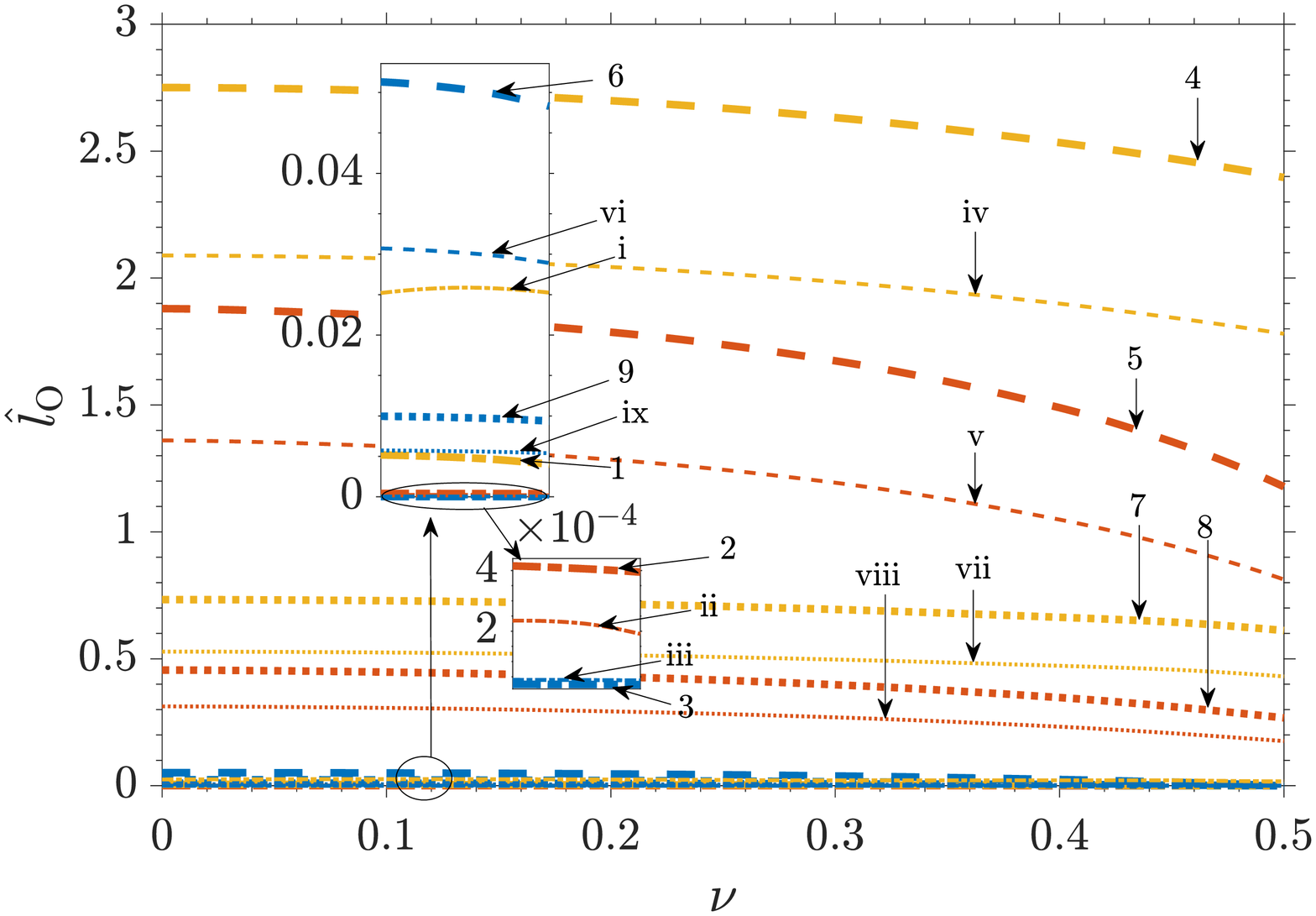}
\caption{}
\label{subfig:l_nu_dlvo}
\end{subfigure}
\caption{Variation of magnitude of maximum attractive (negative), maximum repulsive (positive) and mean (a,c) normalized Force $\hat{F}$ and (b,d) normalized deflection at origin $\hat{l}_{\text{O}}$, over one complete oscillation, with the Poisson's ratio $\nu$ for the case with (a,b) only hydrodynamic pressure, and, (c,d) hydrodynamic and DLVO pressure components; other system parameters are: $R$ = 1 mm, $D$ = 50 nm, $h_0$ = 45 nm, $L$ = [0.5 mm (semi-infinite), 5 $\mu$m (thick), 50 nm (thin)], $\omega$ = 2$\pi$ rad/s, $E_{\text{Y}}$ = 17.5 MPa, $\mu$ = 1 mPa-s; the plot-line labels for each subfigure mean: 1 - maximum attractive characteristics for semi-infinite substrate, 2 - maximum attractive characteristics for thick substrate, 3 - maximum attractive characteristics for thin substrate, 4 - maximum repulsive characteristics for semi-infinite substrate, 5 - maximum repulsive characteristics for thick substrate, 6 - maximum repulsive characteristics for thin substrate, 7 - mean characteristics for semi-infinite substrate, 8 - mean characteristics for normal thick substrate, 9 - mean characteristics for thin substrate; in the subfigures on bottom (i.e. c and d), the labels 1 to 9 correspond to `strong DLVO' ($A_{\text{sfw}}$ = $10^{-20}$ J i.e. $\mathcal{H} = 0.7$, $\psi_S$ = 2500 mV i.e. $\mathcal{R} = 700$) while the labels i to ix correspond to their respective arabic numbers' in description but for `moderate DLVO' ($A_{\text{sfw}}$ = $10^{-21}$ J i.e. $\mathcal{H} = 0.07$, $\psi_S$ = 100 mV i.e. $\mathcal{R} = 28$)}
\label{fig:F_nu}
\end{figure}
We assess the force characteristics first, presented in figure \ref{subfig:F_L_hd}. The mean force is significantly smaller than the maximum attractive force and maximum repulsive force, because of the anti-symmetric variation of hydrodynamic pressure with time. The mean force characteristics are not presented. Looking at the contrast between maximum attractive force characteristics and maximum repulsive force characteristics, we can see that with increasing substrate thickness (which allows for higher deflection, i.e. acts to effectively increase substrate softness), maximum attractive force increases whereas maximum repulsive force decreases. This is an outcome of `self-magnifying' nature of attractive pressure (higher the deflection caused by attractive pressure, more it contributes to the attractive pressure) which is the converse of the `self-diminishing' nature of repulsive pressure - attractive presure also increases in magnitude with decreasing seperation of sphere from fluid-substrate interface, however, its effect is to decrease this seperation and hence attractive pressure is self-magnifying. This effect is also observed in the contrast between maximum attractive force characteristics for incompressible (line 3), normal (line 2) and compressible (line 1) substrate material, since the substrate allows more deflection as we move from former to latter. However, similar effect is not recovered for maximum repulsive characteristics where we observe a crossover between the lines for incompressible (line 6) and normal (line 5) substrate materials, a peculiar behaviour.  Although we obtain the discussed contrasts with increasing substrate thickness, the maximum attractive and maximum repulsive force characteristics remain within a factor of 2, indicating that the deflection does not lead to significant deviation in the force characteristics. We also observe that the plot-lines exhibit `wiggles', and even deviate from monotonic variation for maximum attractive characteristics for semi-infinite substrate and maximum repulsive characteristics for thick substrate. This is another peculiar behaviour. Both the peculiarities in the system behaviour occur because of the inherent complexities of the coupling between hydrodynamic pressure and deflection as opposed to DLVO pressure components and deflection. In TWC, while DLVO pressure components depend on only the deflection, hydrodynamic pressure is coupled with the radial and temporal gradients of deflection as well (see equation \eqref{eq:Re_eq}). Lastly, we observe that the force characteristics approach saturation as the substrate thickness approaches semi-infinite.\\
We assess the deflection characteristics next, presented in figure \ref{subfig:l_L_hd}. The deflection characteristics are simpler in comprison to force characteristics. The trend of increase in deflection with thickness and the contrast between compressible, normal and incompressible, both follow expected trends - any feature that effectively increases the softness of substrate (thickness and compressibility) leads to higher deflection. The deflection characteristics can be seen to saturate as the substrate approaches semi-infinite thickness - essentially, the characteristics lose sensitivity to substrate thickness when it is large enough to effectively be semi-infinite.\\ 
We next consider the cases where DLVO forces are present, presented in figures \ref{subfig:F_L_dlvo} and \ref{subfig:l_L_dlvo}. The two cases of `moderate DLVO'  ($A_{\text{sfw}} = 10^{-21}$ J i.e. $\mathcal{H} = 0.07$, $\zeta$ = 100 mV i.e. $\mathcal{R} = 28$) and `strong DLVO' ($A_{\text{sfw}} = 10^{-20}$ J i.e. $\mathcal{H} = 0.7$, $\zeta$ = 2500 mV i.e. $\mathcal{R} = 700$) are distinguised by using thin lines and roman number labels for the former and thick lines and arabic number labels for the latter. \\
We assess the force characteristics first, presented in figure \ref{subfig:F_L_dlvo}. With increasing substrate thickness and moving from compressible to incompressible substrate material, both of which lead to higher deformation, the maximum repulsive force characteristics decrease. Similar trends are observed for the mean force characterstics. It can be seen that although present, the maximum attractive force characterstics are much smaller than the maximum repulsive force characterstics and the mean force characteristics. Furthermore, the mean force characteristics exhibit significant magnitude and are positive. This indicates that EDL disjoining pressure strongly dominates both hydrodynamic pressure when its negative and van der Waals pressure. Lastly, we can see that the trends saturate as the substrate thickness approaches semi-infinite. \\
We assess the deflection characteristics next, presented in figure \ref{subfig:l_L_dlvo}. As expected, the deflection is higher with increasing thickness as well as increasing compressibility. The trends saturate as substrate thickness approaches semi-infinite.\\
\subsubsection{Effect of Substrate Compressibility}\label{subsubsec:results_nu}
As discussed in section \ref{subsec:limits}, the substrate material varies from the hypothetical `perfectly-compressible' to incompressible behaviour as we vary $\nu$ from 0 to 0.5. In this subsection, we see the variation of system response with $\nu$, taking its range from 0 to 0.5. The system parameter values are presented in caption of figure \ref{fig:F_nu}.\\
We assess the parametric variation with $\nu$ of magnitudes of maximum attractive, maximum repulsive and mean, over one complete oscillation, of force and deflection, presented in figure \ref{fig:F_nu}. We consider the same three scenarios as done in section \ref{subsubsec:results_L}, i.e. purely hydrodynamic, moderate DLVO and strong DLVO. \\
We first consider the case where DLVO forces are absent ($A_{\text{sfw}} = 0$ i.e. $\mathcal{H} = 0$, $\psi_S = 0$ i.e. $\mathcal{R} = 0$), presented in figures \ref{subfig:F_nu_hd} and \ref{subfig:l_nu_hd}. The force characteristics (presented in figure \ref{subfig:F_nu_hd}) for this case are fairly straightforward. The maximum attractive characteristics (lines 1, 2, 3) are higher for thicker substrates and for lower values of $\nu$, i.e. they are higher for substrates that are effectively softer. In contrast, the maximum repulsive characteristics (lines 4, 5, 6) are lower for thicker substrates and lower values of $\nu$, i.e. they are lower for substrates that are effectively softer. Note that the contrast is negligible for thin substrates (lines 3 and 6). The mean force is expectedly negligible, and so its characteristics are not presented. We observe a peculiarity in the maximum repulsive force characteristics of semi-infinite substrate close to $\nu$ = 0.5 - a local minima is observable. This is again attributed to the complex interaction of hydrodynamic pressure with deflection, particularly its direct interaction with the temporal and radial gradients of deflection. The deflection characteristics (presented in figure \ref{subfig:l_nu_hd}) are also straightforward to interpret. With increasing $\nu$, the deflection decreases. The maximum attractive deflection characteristics (lines 1, 2, and 3) are higher than maximum repulsive deflection characteristics (lines 4, 5, and 6). Except maximum repulsive and maximum attractive deflection characteristics for thick and semi-infinite substrates, the other deflection characteristics are very small.\\
We next consider the cases where DLVO forces are present, presented in figures \ref{subfig:F_nu_dlvo} and \ref{subfig:l_nu_dlvo}. The two cases of moderate DLVO ($A_{\text{sfw}} = 10^{-21}$ J i.e. $\mathcal{H} = 0.07$, $\zeta$ = 100 mV i.e. $\mathcal{R} = 28$) and strong DLVO ($A_{\text{sfw}} = 10^{-20}$ J i.e. $\mathcal{H} = 0.7$, $\zeta$ = 2500 mV i.e. $\mathcal{R} = 700$) are distinguised by using thin lines and roman number labels for the former and thick lines and arabic number labels for the latter. We look at force charactersitics first (figure \ref{subfig:F_nu_dlvo}). The maximum repulsive force characteristics (lines 4, 5, and 6) and the mean force characteristics (lines 7, 8, and 9) are significantly higher in magnitude than maximum attractive force characteristics (lines 1, 2, and 3), an outcome of the dominance of EDL disjoining pressure over hydrodynamic pressure and van der Waals pressure. Also, maximum repulsive force gets higher with increasing $\nu$, i.e. as the substrate gets more rigid - this is the outcome of the self-diminishing nature of repulsive pressure, acting the opposite manner i.e. more rigid substrate is allowing lower deflection and thus higher repulsive pressure. The deflection characteristics (presented in figure \ref{subfig:l_nu_dlvo}) are analogous to the force characteristics, except that all deflection characteristics decrease with increasing $\nu$.\\
\subsubsection{Effect of DLVO Forces}\label{subsubsec:results_DLVO}
\begin{figure}[!htb]
\centering
\begin{subfigure}[b]{0.495\textwidth}
\centering
\includegraphics[width=\textwidth]{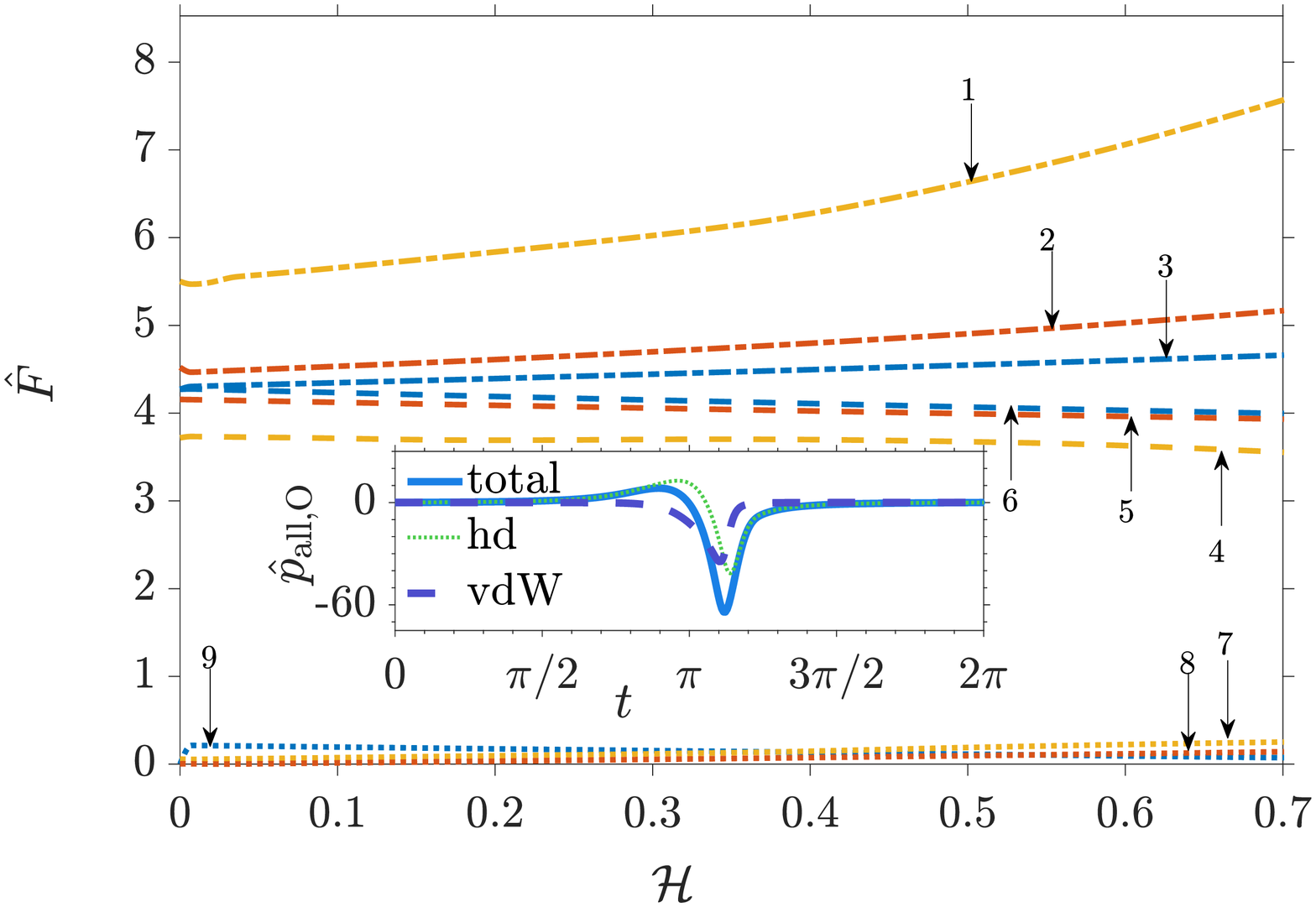}
\caption{}
\label{subfig:F_Asfw_no_DL}
\end{subfigure}
\begin{subfigure}[b]{0.495\textwidth}
\centering
\includegraphics[width=\textwidth]{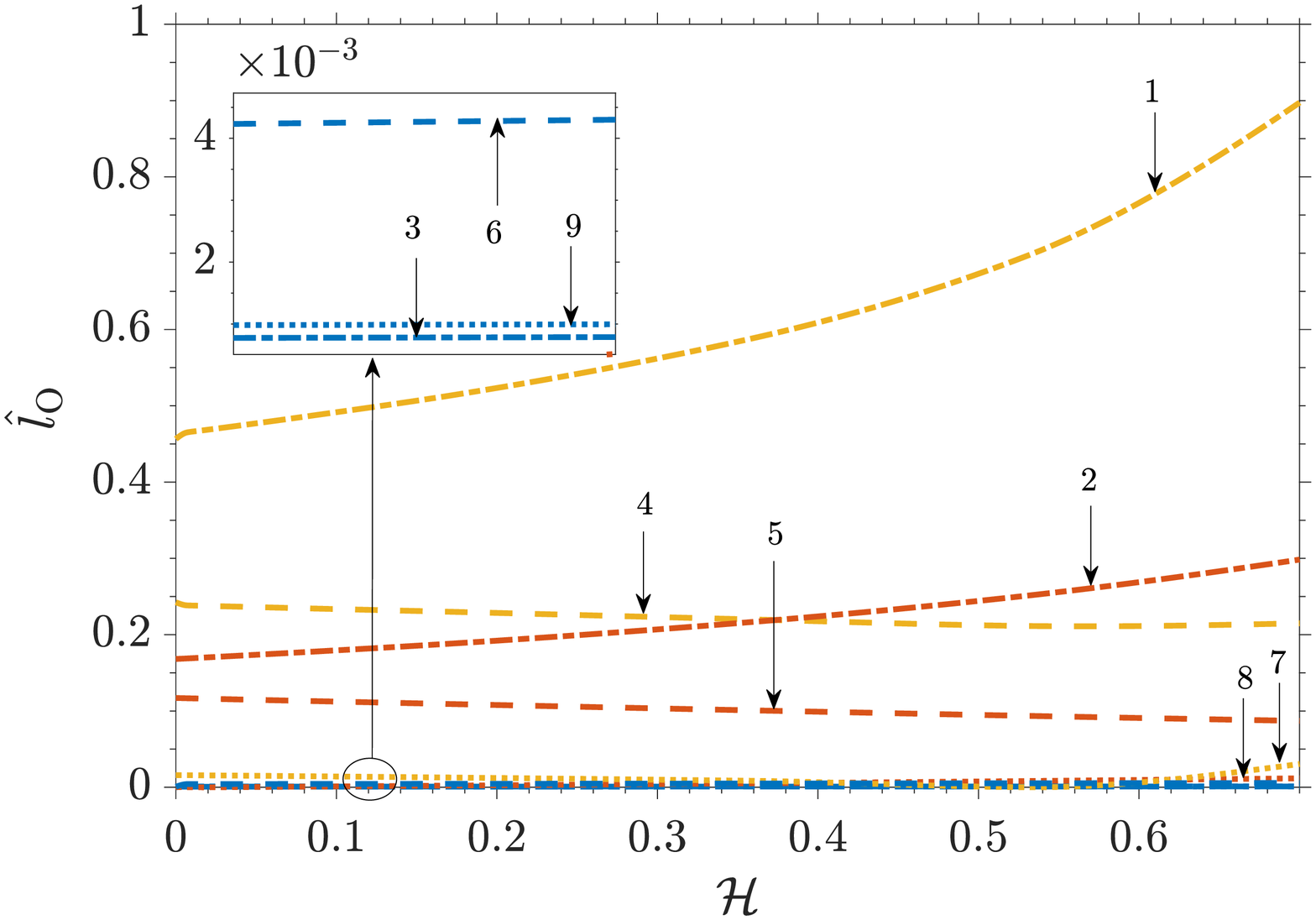}
\caption{}
\label{subfig:l_Asfw_no_DL}
\end{subfigure}
\begin{subfigure}[b]{0.495\textwidth}
\centering
\includegraphics[width=\textwidth]{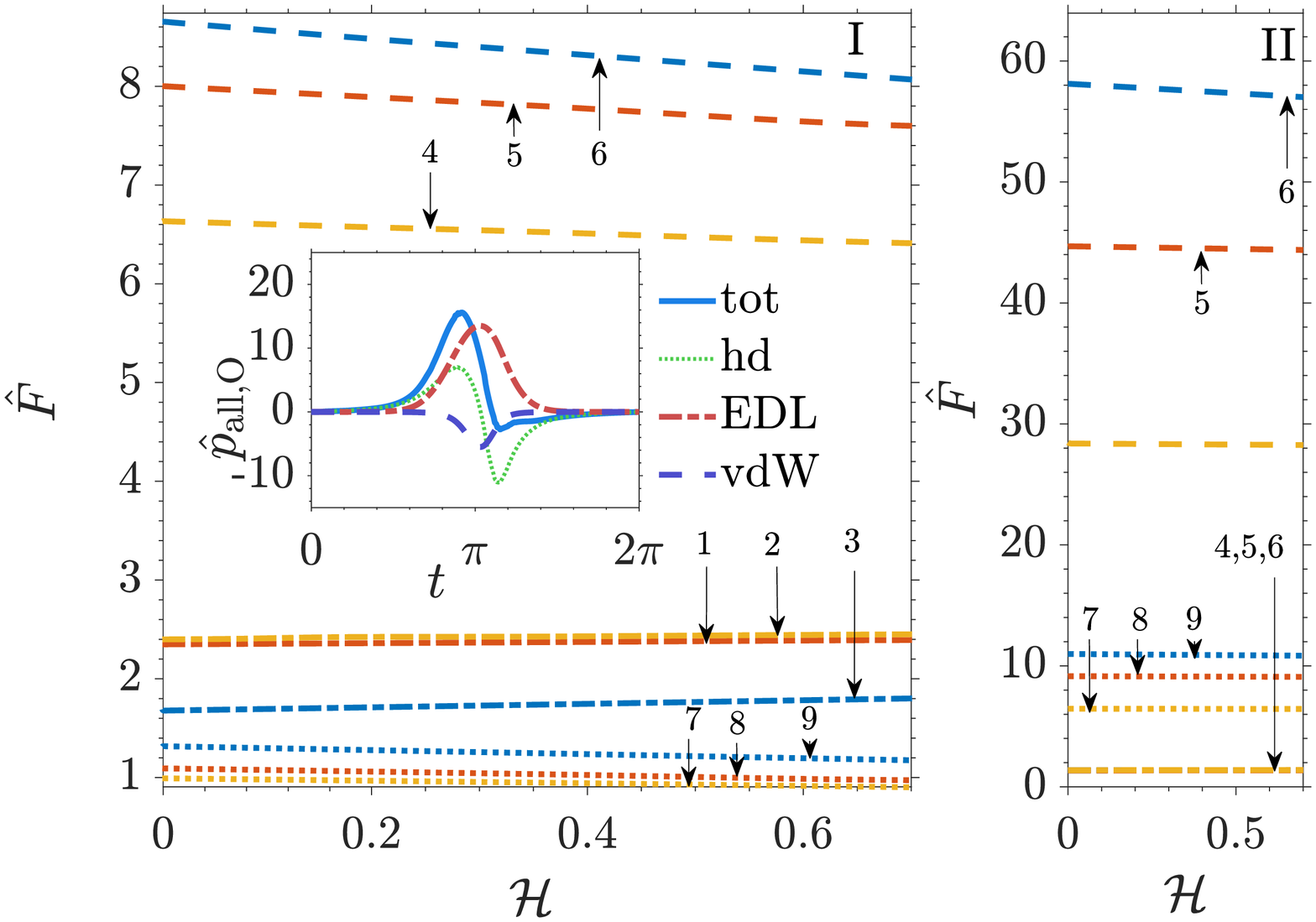}
\caption{}
\label{subfig:F_Asfw_25mV_DL}
\end{subfigure}
\begin{subfigure}[b]{0.495\textwidth}
\centering
\includegraphics[width=\textwidth]{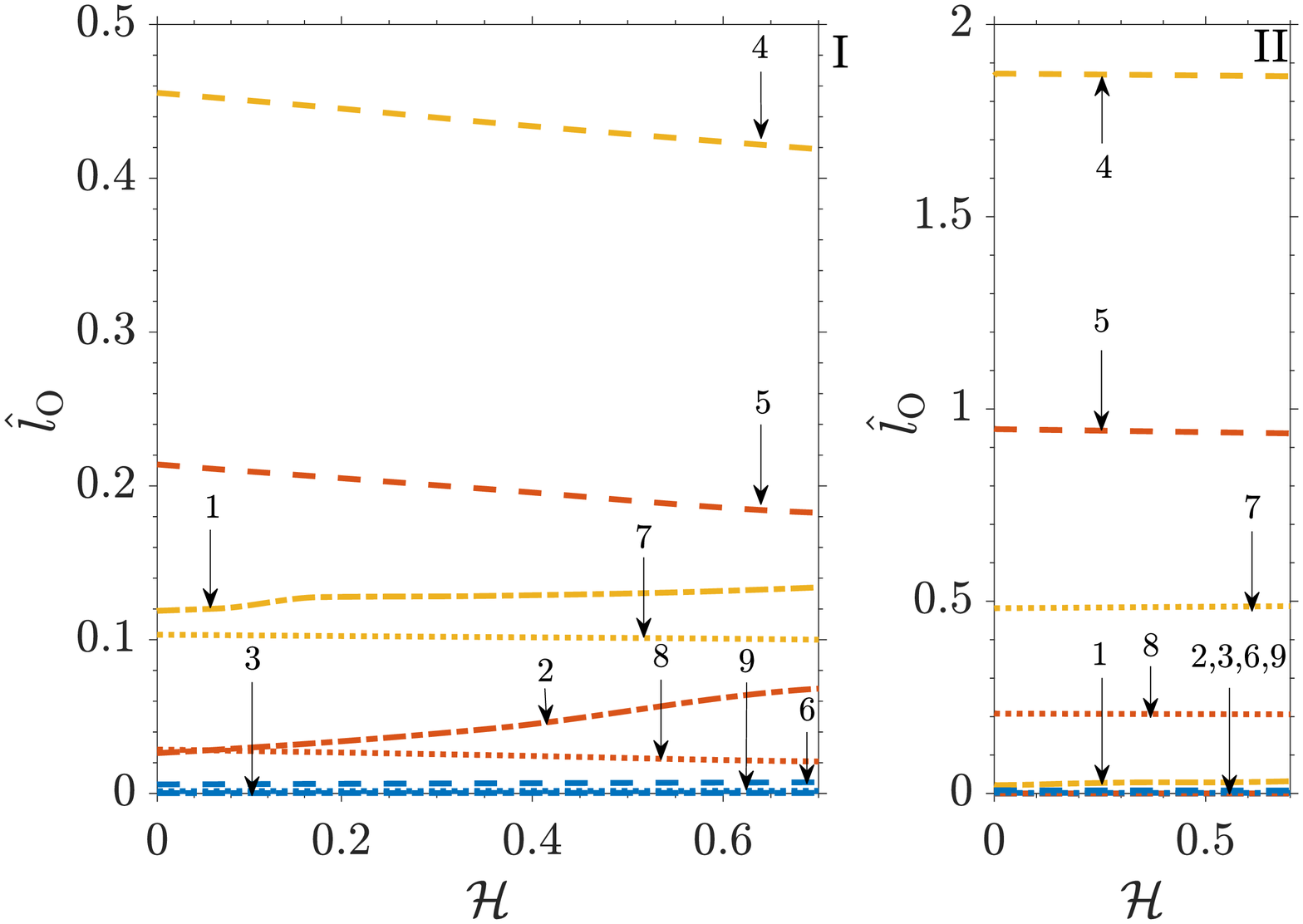}
\caption{}
\label{subfig:l_Asfw_25mV_DL}
\end{subfigure}
\caption{Variation of magnitude of maximum attractive (negative), maximum repulsive (positive) and mean (a,c) normalized Force $\hat{F}$ and (b,d) normalized deflection at origin $\hat{l}_{\text{O}}$, over one complete oscillation, with $\mathcal{H}$ for (a,b) $\mathcal{R}$ = 0 ($\psi_S = 0$), and, (c,d) $\mathcal{R}$ = 7 ($\psi_{S}$ = 25 mV) for the larger panels on the left labelled I and $\mathcal{R}$ = 700 ($\psi_{S}$ = 2.5 V) for the smaller panels on the right labelled II; other system parameters are: $R$ = 1 mm, $D$ = 50 nm, $h_0$ = 45 nm, $L$ = [0.5 mm (semi-infinite), 5 $\mu$m (thick), 50 nm (thin)], $\omega$ = 2$\pi$ rad/s, $E_{\text{Y}}$ = 17.5 MPa, $\nu$ = 0.45, $\mu$ = 1 mPa-s; in the insets of subfigure a and c-I, the time evolution of normalized pressure components and total pressure at origin for $\mathcal{H} = 0.7$ (i.e. $A_{\text{sfw}}$ = $10^{-20}$ J) for semi-infinite substrate is presented for reference; the plot-line labels for each subfigure mean: 1 - maximum attractive characteristics for semi-infinite substrate, 2 - maximum attractive characteristics for thick substrate, 3 - maximum attractive characteristics for thin substrate, 4 - maximum repulsive characteristics for semi-infinite substrate, 5 - maximum repulsive characteristics for thick substrate, 6 - maximum repulsive characteristics for thin substrate, 7 - mean characteristics for semi-infinite substrate, 8 - mean characteristics for thick substrate, 9 - mean characteristics for thin substrate}
\label{fig:F_Asfw}
\end{figure}
\begin{figure}[!htb]
\centering
\begin{subfigure}[b]{0.495\textwidth}
\centering
\includegraphics[width=\textwidth]{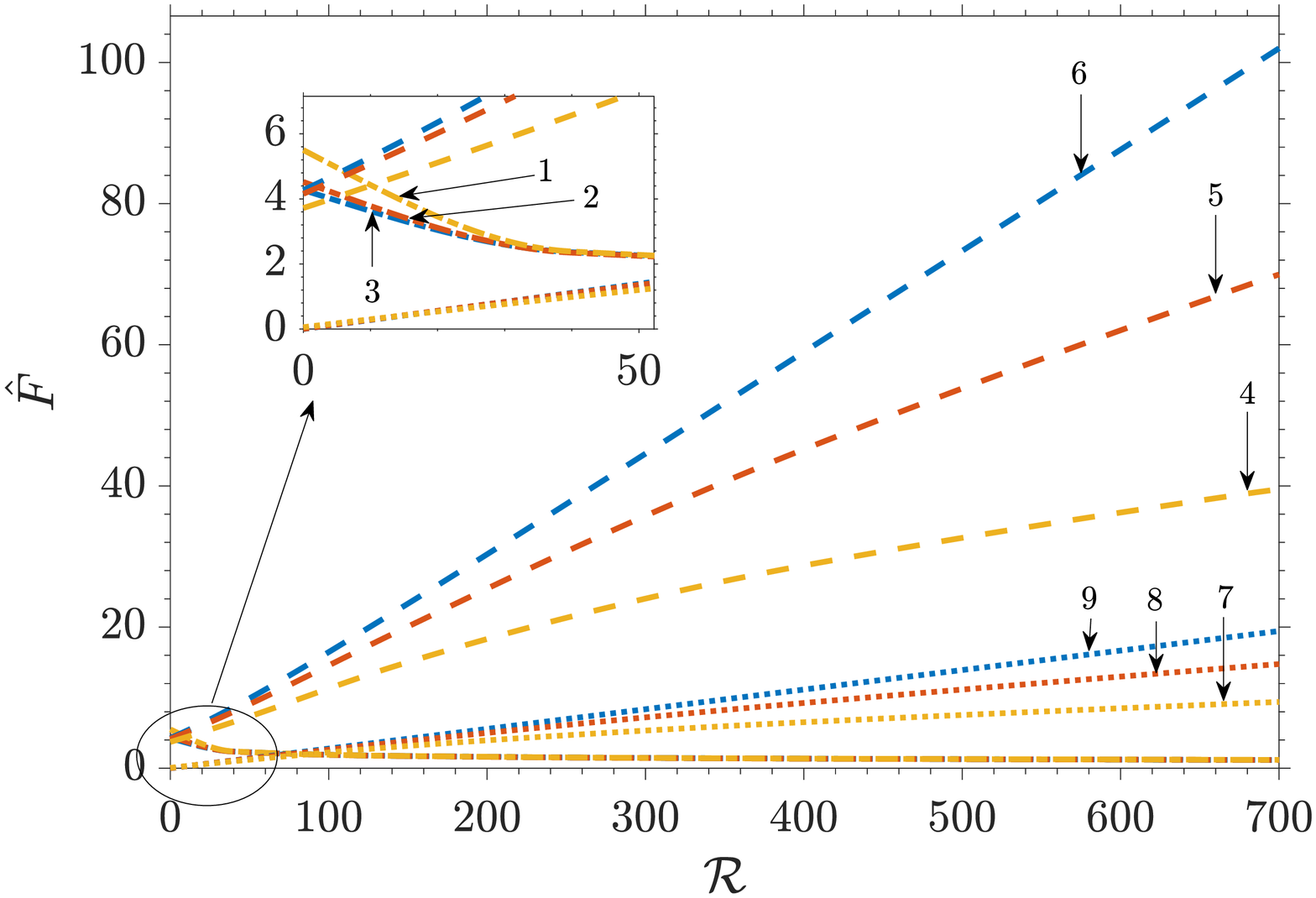}
\caption{}
\label{subfig:F_zeta_no_vdW}
\end{subfigure}
\begin{subfigure}[b]{0.495\textwidth}
\centering
\includegraphics[width=\textwidth]{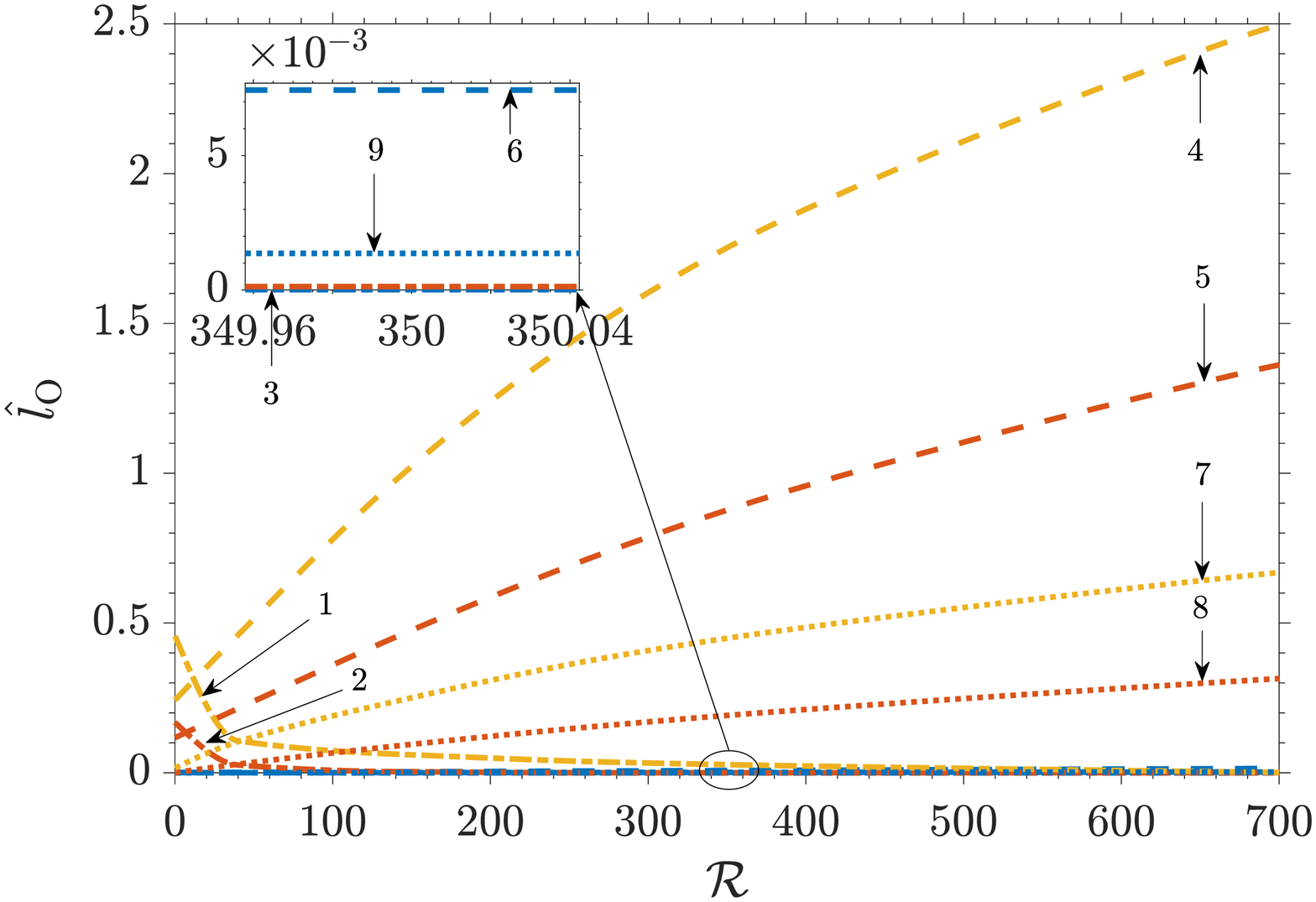}
\caption{}
\label{subfig:l_zeta_no_vdW}
\end{subfigure}
\begin{subfigure}[b]{0.495\textwidth}
\centering
\includegraphics[width=\textwidth]{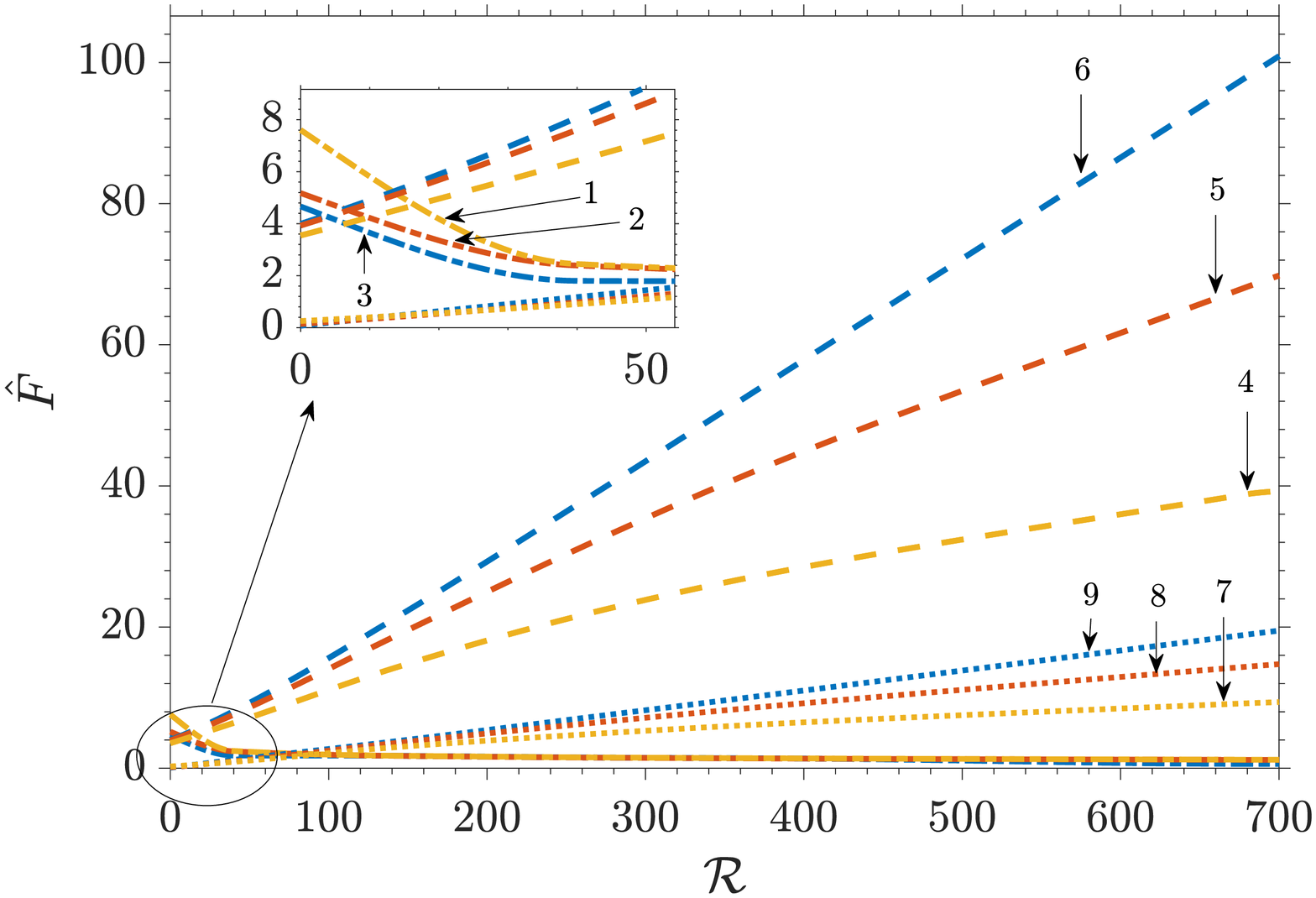}
\caption{}
\label{subfig:F_zeta_1em20J_vdW}
\end{subfigure}
\begin{subfigure}[b]{0.495\textwidth}
\centering
\includegraphics[width=\textwidth]{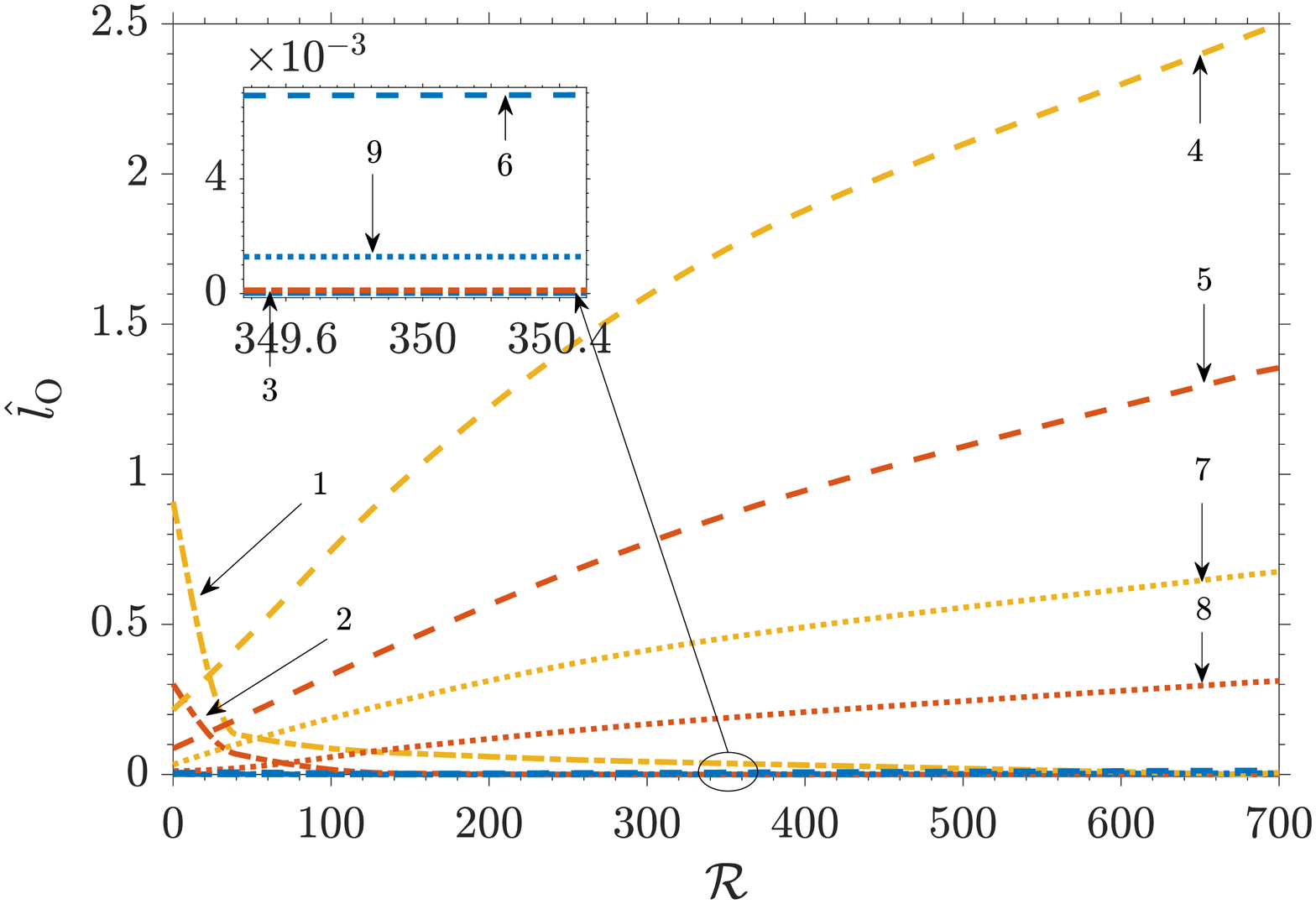}
\caption{}
\label{subfig:l_zeta_1em20J_vdW}
\end{subfigure}
\caption{Variation of magnitude of maximum attractive (negative), maximum repulsive (positive) and mean (a,c) normalized Force $\hat{F}$ and (b,d) normalized deflection at origin $\hat{l}_{\text{O}}$, over one complete oscillation, with $\mathcal{R}$ for (a,b) $\mathcal{H}$ = 0 ($A_{\text{sfw}}$ = 0), and, (c,d) $\mathcal{H}$ = 0.7 ($A_{\text{sfw}}$ = $10^{-20}$ J); other system parameters are: $R$ = 1 mm, $D$ = 50 nm, $h_0$ = 45 nm, $L$ = [0.5 mm (semi-infinite), 5 $\mu$m (thick), 50 nm (thin)], $\omega$ = 2$\pi$ rad/s, $E_{\text{Y}}$ = 17.5 MPa, $\nu$ = 0.45, $\mu$ = 1 mPa-s; the plot-line labels for each subfigure mean: 1 - maximum attractive characteristics for semi-infinite substrate, 2 - maximum attractive characteristics for thick substrate, 3 - maximum attractive characteristics for thin substrate, 4 - maximum repulsive characteristics for semi-infinite substrate, 5 - maximum repulsive characteristics for thick substrate, 6 - maximum repulsive characteristics for thin substrate, 7 - mean characteristics for semi-infinite substrate, 8 - mean characteristics for thick substrate, 9 - mean characteristics for thin substrate}
\label{fig:F_zeta}
\end{figure}
DLVO force consists of the two pressure components - EDL disjoining pressure and van der Waals pressure. We assess the effect of DLVO forces utilizing two non-dimensional parameters. The first is $\displaystyle \mathcal{H} = \frac{A_{\text{sfw}}}{\epsilon^2R^3\mu\omega\alpha}$, a parameter that compares van der Waals pressure to hydrodynamic pressure (quantified primarily by Hamaker's constant, hence denoted by $\mathcal{H}$); the second is $\displaystyle \mathcal{R} = \frac{32\epsilon c_{0} k_{\text{B}}T\Psi_{S}^2}{\mu\omega\alpha}$, a parameter that compares EDL disjoining pressure to van hydrodynamic pressure (represents the repulsive component of DLVO force, hence denoted by $\mathcal{R}$). This scheme of non-dimensionalization is similar to the one used by Davis and Seryssol \cite{Davis1986}. For the parametric variations presented in this subsection, we take the range of $A_{\text{sfw}}$ as $10^{-22}$ J to $10^{-20}$ J and zero, and the range of $\psi_S$ as 25 mV to 2500 mV and zero. These values of Hamaker's constant and surface potential represent the range of typical values that are observed in many common engineering abd biological materials \cite{Hunter1986,Israelachvili2011}. The pertinent system parameter values are presented in caption of figure \ref{fig:F_Asfw} and figure figure \ref{fig:F_zeta}.\\
We assess the effect of van der Waals pressure first, presented in figure \ref{fig:F_Asfw}. The parametric variation with $\mathcal{H}$ of magnitudes of maximum attractive, maximum repulsive and mean, over one complete oscillation, of force and deflection are presented. We consider three scenarios - when there is no EDL disjoining pressure, when there is comparable EDL disjoining pressure and when there is dominant EDL disjoining pressure. Before delving into each of these scenarios, we make an observation about van der Waals pressure. The range of values of $A_{\text{sfw}}$ considered here amount to $\mathcal{H}$ varying from 0 to $\sim$0.7. This indicates that the van der Waals pressure does not strongly dominate hydrodynamic pressure for the entire range of $A_{\text{sfw}}$ considered. \\
We first consider the case where EDL disjoining pressure is absent ($\psi_S = 0$ i.e. $\mathcal{R} = 0$). \\
We assess the force characteristics first, presented in figure \ref{subfig:F_Asfw_no_DL}. We make some key observations.
\begin{enumerate}
\item the maximum attractive force dominates the maximum repulsive force and mean force, for all three thicknesses considered
\item the magnitude of maximum attractive force is highest for semi-infinite substrate (line 1), lower for thick substrate (line 2) and lowest for thin substrate (line 3)
\item with increasing $\mathcal{H}$, there is increase in not only the magnitude but also the rate of increase of the magnitude, of maximum attractive force
\item 
\begin{enumerate}
\item the maximum repulsive force (lines 4,5,6) is of the same order of magnitude as maximum attractive force
\item however, they remain of approximately the same magnitude (decrease slightly) with increasing $\mathcal{H}$
\end{enumerate}
\item the contrast between the semi-infinite, thick and thin substrates is opposite for maximum repulsive force compared to maximum attractive force
\item the mean force (lines 7,8,9) is significantly smaller than the maximum attractive and maximum repulsive force
\item the maximum repulsive force for semi-infinite substrate (line 4), thick substrate (line 5), and thin substrate (line 6), are significantly close together
\end{enumerate}
Now, we explore the underlying interations between the pressure components and the deflection that lead to these observations. Obervations 2, 3 and 5 are simply the outcome of the self-diminishing nature of repulsive pressure (here being applied to total pressure i.e. sum of hydrodynamic pressure and DLVO pressure components) and self-magnifying nature of attractive pressure (here being applied to total pressure i.e. sum of hydrodynamic pressure and DLVO pressure components). Note that this effect is what contributes to deformation-induced adhesion behaviour in setups involving soft bodies, a theoretical study on the same was conducted by Urzay \cite{Urzay2010}. Now we take a look at the interplay of the two pressure components - van der Waals pressure and hydrodynamic pressure. The van der Waals pressure is always attractive, and acts only for a small duration near the mid-oscillation. On the other hand, hydrodynamic pressure is positive for the first-half and negative for the second half. For reference, the time evolution of normalized pressure components and total pressure at origin for $\mathcal{H}=0.7$ (i.e. $A_{\text{sfw}} = 10^{-20}$ J) for semi-infinite substrate is presented in the inset of figure \ref{subfig:F_Asfw_no_DL}. As can be seen, the van der Waals pressure and hydrodynamic pressure act together and add up at a time-instant a little after mid-oscillation, which is where we get the maximum attractive force and deflection. On the other hand, they act against each other before mid-oscillation (and subtract from each other), which is where we get maximum repulsive force (as the positive hydrodynamic pressure overpowers total pressure for a small duration). This interplay between the two pressure components explains observation 1. Furthermore, the maximum positive and maximum negative peaks of the total pressure can be seen to be of the same order of magntiude, although the former is much smaller than the latter. However, van der Waals pressure dies out quicker with increasing separation than hydrodynamic pressure - resultantly, the maximum repulsive pressure, although smaller in magnitude than maximum attractive pressure, extends to a larger radial span. Therefore, the dominance of attractive over repulsive is not as strong for force as for pressure at origin because force includes integration over the radial span. This explains observation 4(a). As higher value of $\mathcal{H}$ implies more cancelling out of the positive hydrodynamic pressure, increase values of $\mathcal{H}$ lead to the slightly decreasing values of maximum repulsive force, explaining observation 4(b). Lastly, we note that the substrate deflection is significant only when the force is attractive (the maximum repulsive deflection for semi-infinite substrate, line 4 of figure \ref{subfig:l_Asfw_no_DL}, does not grow higher than 0.25 in normalized form). Resultantly, deflection (and therefore substrate thickness) doesn't strongly affect maximum repulsive force characteristics. This explains observation 6. \\
We assess the deflection characteristics next, presented in figure \ref{subfig:l_Asfw_no_DL}. The deflection characteristics exhibit trends similar to force characteristics, albiet a bit more pronounced. This occurs because the deflection characteristics presented correspond to deflection at the origin, where van der Waals pressure is the strongest. In contrast, force depends on the force characteristics along the increasing radial distances as well, where van der Waals pressure drops faster than hydrodynamic pressure. \\ 
We next consider the case where EDL disjoining pressure is comparable to van der Waals pressure, for which $\psi_S = 25$ mV ($\mathcal{R} = 7$), i.e. they are of the same order of magnitude. \\
We assess the force characteristics first, presented in figure \ref{subfig:F_Asfw_25mV_DL}-I. In contrast to the case without EDL disjoining pressure, i.e. figure \ref{subfig:F_Asfw_no_DL}, the maximum repulsive and maximum attractive forces have interchanged in terms of dominance. This has occured because of the introduction of EDL disjoining pressure - EDL disjoining pressure varies exponentially with separation and its exponential decay length is such that it dies out slower over radial span than van der Waals pressure and in some cases even hydrodynamic pressure. As a result, if EDL disjoining pressure is present, it tends to dominate the system behaviour, even if the value of $\psi_S$ is small (25 mV in this case). The comparison between semi-infinite, thick and thin substrates is similar to that seen in figure \ref{subfig:F_Asfw_no_DL}, and hence the same explanation holds. For reference, the time evolution of normalized pressure components and total pressure at origin for $\mathcal{H}=0.7$ (i.e. $A_{\text{sfw}} = 10^{-20}$ J) for semi-infinite substrate is presented in the inset of figure \ref{subfig:F_Asfw_25mV_DL}-I. The deflection behaviour, presented in figure \ref{subfig:l_Asfw_25mV_DL}-II, is similar to the force behaviour. Also, similar explanation for the contrast of deflection characteristics with force characteristics applies here that applied for the case of zero EDL disjoining pressure. \\
Lastly, the force and deflection characteristics for the case of $\psi_S$ = 100 mV (i.e. $\mathcal{R}$ = 28) is presented in figures \ref{subfig:F_Asfw_25mV_DL}-II and \ref{subfig:l_Asfw_25mV_DL}-II. As can be seen, the characteristics exhibit negligible dependence on $\mathcal{H}$. This indicates that for values of $\psi_S$ as 100 mV and higher, EDL disjoining pressure is strong enough to dictate the force and deflection behaviour of the system, with the effect of van der Waals pressure and hydrodynamic pressure being much smaller in comparison. \\
We assess the effect of EDL disjoining pressure next, presented in figure \ref{fig:F_zeta}. The parametric variation with $\mathcal{R}$ of magnitudes of maximum attractive, maximum repulsive and mean, over one complete oscillation, of force and deflection are presented. We consider two scenarios - when there is no van der Waals pressure and when $A_{\text{sfw}} = 10^{-20}$ J ($\mathcal{H} = 0.7$). Before discussing these scenarios, we make an observation about EDL disjoining pressure. The range of values of $\psi_{S}$ considered here amount to $\mathcal{R}$ varying from 0 to $\sim$700. This indicates that except for sufficiently low values of $\psi_S$ ($\approx$ 25 mV and lower), EDL disjoining pressure strongly dominates van der Waals pressure and hydrodynamic pressure, and is the prime and virtually sole contributor to the deflection and pressure characteristics. This observation is in keeping with the expectations arrived at when assessing the effects of van der Waals pressure.\\
We now take a look at the force characteristics of the two scenarios. These are presented in figures \ref{subfig:F_zeta_no_vdW} and \ref{subfig:F_zeta_1em20J_vdW}, for the scenario of no van der Waals pressure and $A_{\text{sfw}} = 10^{-20}$ J ($\mathcal{H} = 0.7$) respectively. The contrast between the two is restricted to maximum attractive characteristics and occus appreciably for values of $\mathcal{R}$ smaller than $\sim$ 25 (i.e. $\psi_S$ = 400 mV), where the maximum attractive force is higher for the latter scenario compared to the former. As for the maximum repulsive force characteristics, they increase with increasing $\mathcal{R}$, increase as substrate thickness moves from semi-infinite to thin and are close to identical for the two scenarios (of $A_{\text{sfw}} = 0$ and $A_{\text{sfw}} = 10^{-20}$ J). These are expected behaviour, the second one being attributed to the self-diminishing nature of repulsive pressure. A crucial distinction between maximum attractive force in figure \ref{subfig:F_Asfw_no_DL} and maximum repulsive force in \ref{subfig:F_zeta_no_vdW} (or equivalently \ref{subfig:F_zeta_1em20J_vdW}) is that the former grows inside the same order of magnitude with increasing $\mathcal{H}$ while the latter grows between one to two orders of magnitude with increasing $\mathcal{R}$. This is a consequence of the strong growth of $\mathcal{R}$ compared to $\mathcal{H}$ - even though both $A_{\text{sfw}}$ and $\psi_S$ grow by two orders of magnitude (in the respective assessments of van der Waals pressure and EDL disjoining pressure), $\mathcal{H}$ only grows from 0 to 0.7, but $\mathcal{R}$ grows from 0 to 700. Looking at the deflection characteristics, presented in figures \ref{subfig:l_zeta_no_vdW} and \ref{subfig:l_zeta_1em20J_vdW}, we observe that they are analogous to the corresponding force characteristics. \\
Lastly, going through the force characteristics in subfigure a for each of figures \ref{fig:F_L} to \ref{fig:F_zeta}, we observe that the presence of DLVO forces leads to magnification in force characteristics of upto two orders of magnitude and in deflection characteristics of upto an order of magnitude. 
\subsection{High Frequency Oscillatory Loading}\label{subsec:results_freq}
\begin{figure}[!htb]
\centering
\includegraphics[width=\textwidth]{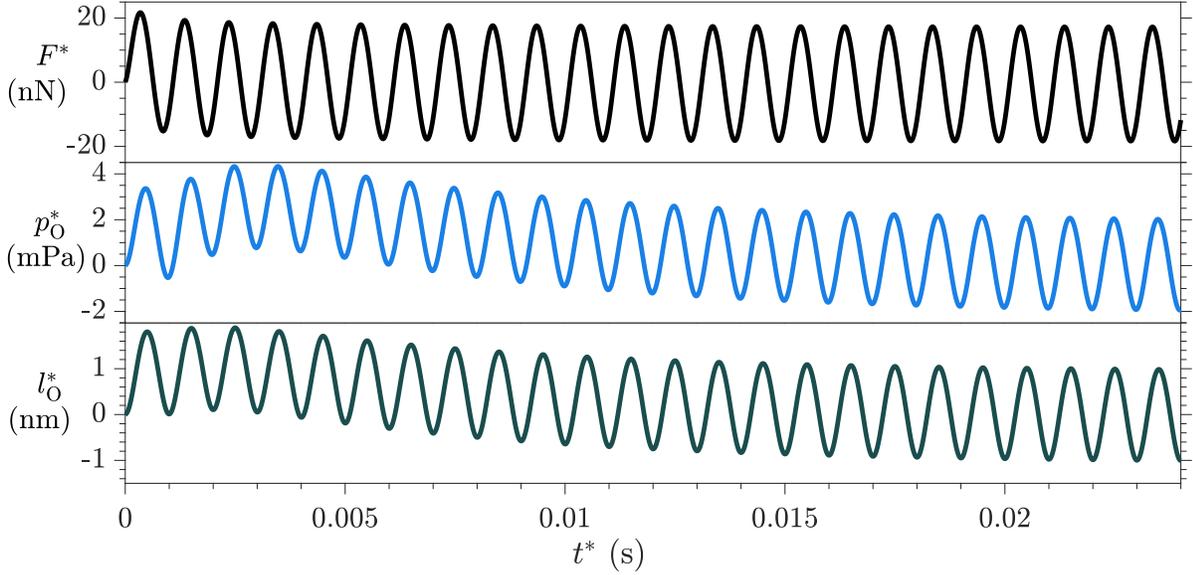}
\caption{Variation of (top to bottom) force $F^*$, pressure at origin $p_{\text{O}}^*$ and deflection at origin $l_{\text{O}}^*$ with time $t^*$ for high-frequency low-amplitude oscillatory loading; DLVO forces are not considered; other system parameters are: $R$ = 10 $\mu$m, $D$ = 10 nm, $\hat{\omega} = 10^3$ Hz, $h_0$ = 1 nm, $L$ = 1 $\mu$m, $E_{\text{Y}}$ = 1 MPa, $\nu$ = 0.45, $\mu$ = 100 mPa-s}
\label{fig:1e3Hz_D_10nm_h0_1nm}
\end{figure}
\begin{figure}[!htb]
\centering
\begin{subfigure}[b]{0.495\textwidth}
\centering
\includegraphics[width=\textwidth]{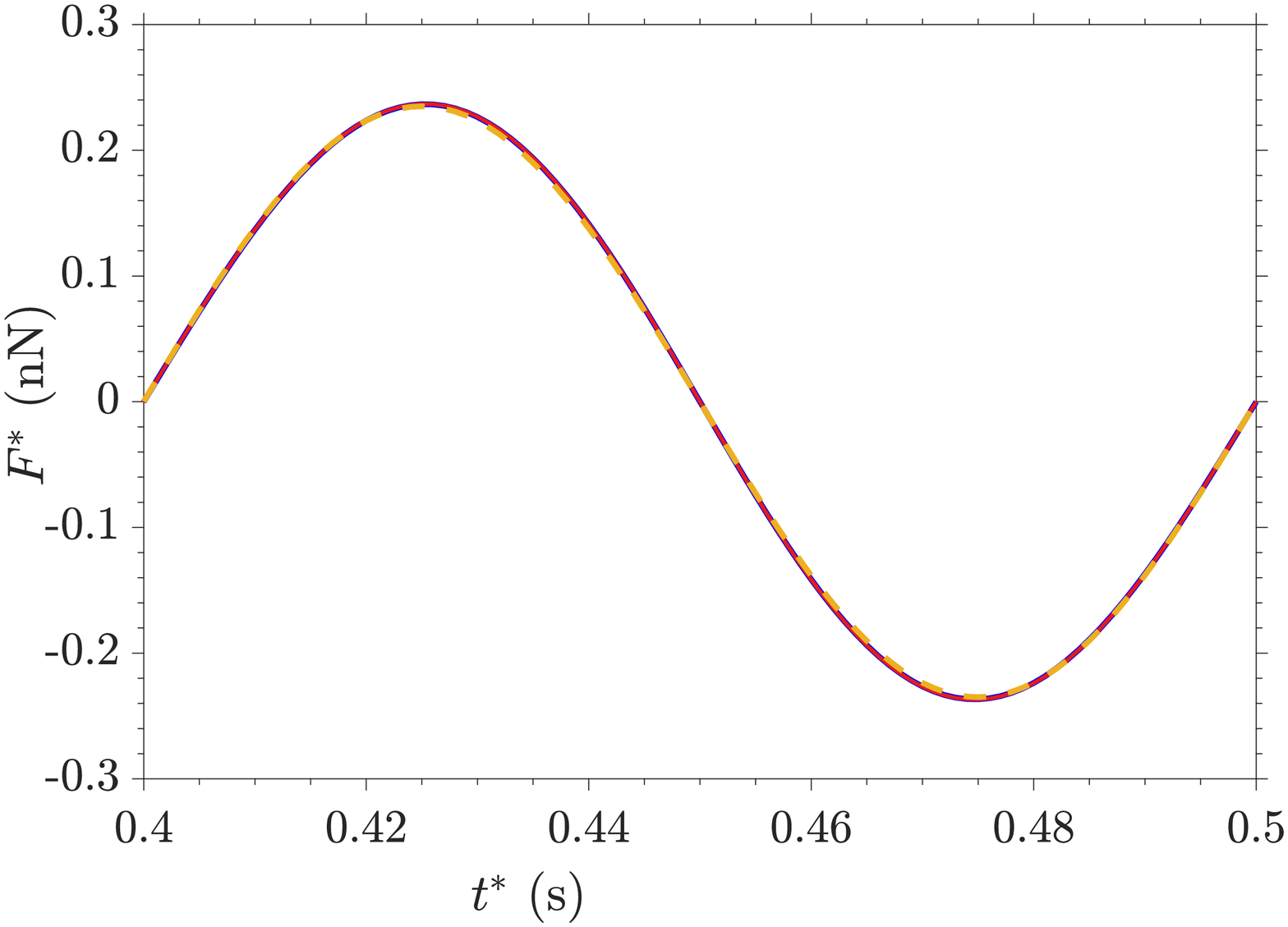}
\caption{}
\label{subfig:F_vs_t_low_omega_low_h0}
\end{subfigure}
\begin{subfigure}[b]{0.495\textwidth}
\centering
\includegraphics[width=\textwidth]{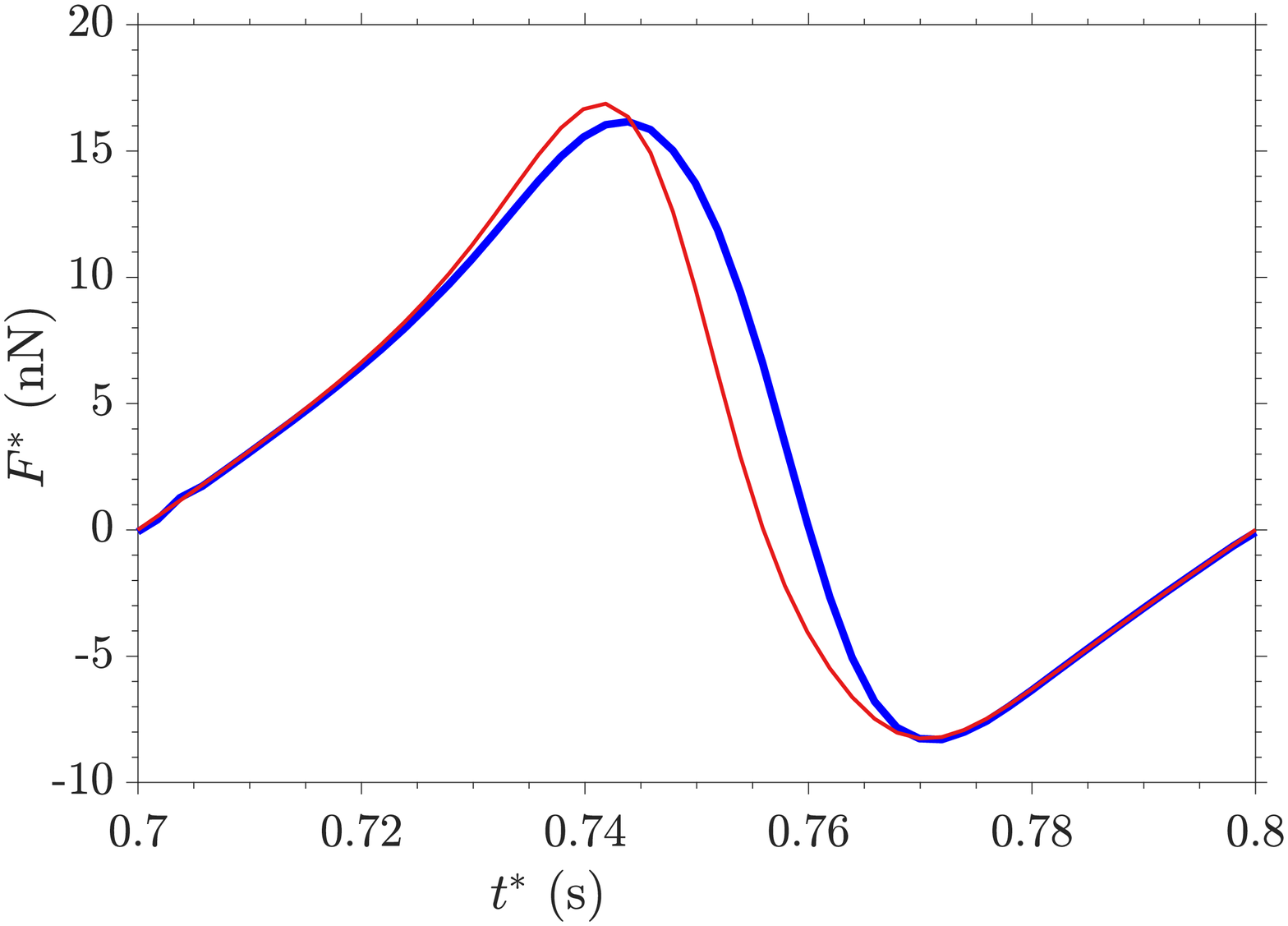}
\caption{}
\label{subfig:F_vs_t_low_omega_high_h0_with_DLVO}
\end{subfigure}
\begin{subfigure}[b]{0.495\textwidth}
\centering
\includegraphics[width=\textwidth]{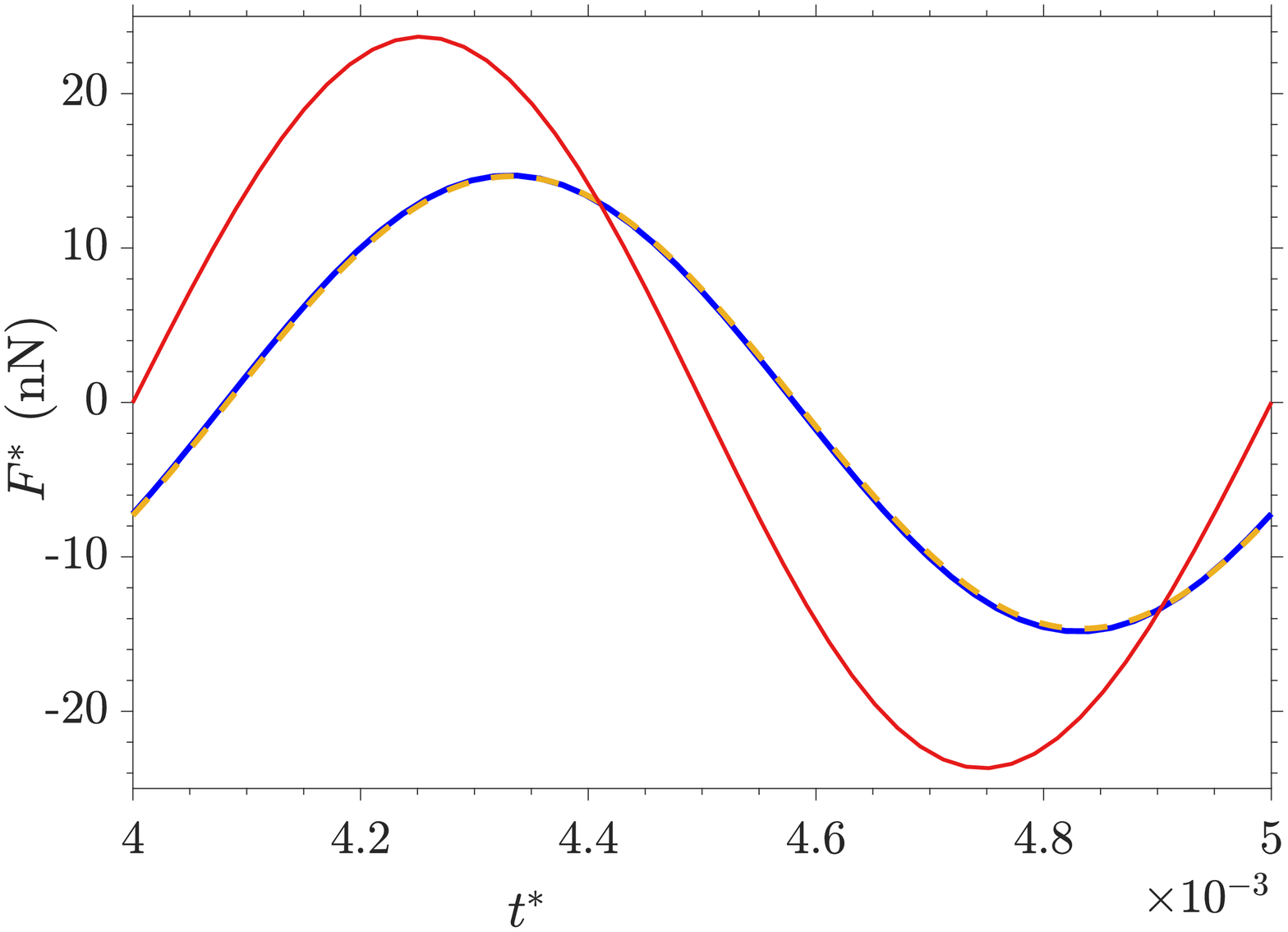}
\caption{}
\label{subfig:F_vs_t_high_omega_low_h0}
\end{subfigure}
\begin{subfigure}[b]{0.495\textwidth}
\centering
\includegraphics[width=\textwidth]{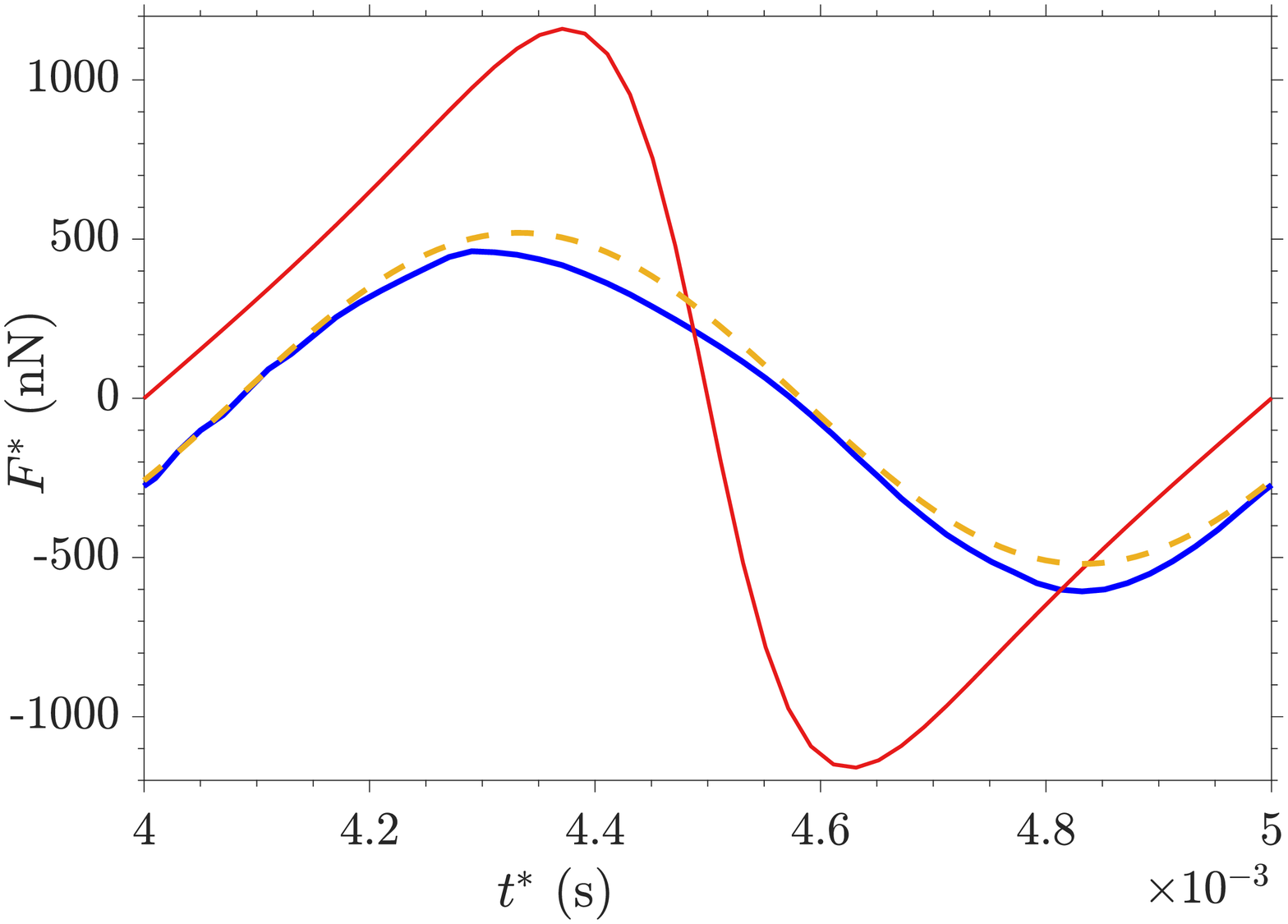}
\caption{}
\label{subfig:F_vs_t_high_omega_high_h0}
\end{subfigure}
\caption{Variation of force $F^*$, with time $t^*$ after the system has reached quasi-steady state, for (a) low $\hat{\omega}$, low $h_0$ ($\hat{\omega} = 10^{1}$ Hz, $h_0 = 1$ nm) with or without DLVO forces, (b) low $\hat{\omega}$, high $h_0$ ($\hat{\omega} = 10^{1}$ Hz, $h_0 = 35$ nm) with DLVO forces (c) high $\hat{\omega}$, low $h_0$ ($\hat{\omega} = 10^{3}$ Hz, $h_0 = 1$ nm) with or without DLVO forces, (d) high, $\hat{\omega}$ high $h_0$ ($\hat{\omega} = 10^{3}$ Hz, $h_0 = 35$ nm) with or without DLVO forces; the solid blue lines are the computed solution, the dashed yellow lines are the best sinusiondal fit to the computed solution, the thin solid red lines are the expected trend if the deflection-pressure interaction is hypothetically constrained to be OWC; other system parameters are: $R$ = 10 $\mu$m, $D$ = 50 nm, $L$ = 1 $\mu$m, $E_{\text{Y}}$ = 1 MPa, $\nu$ = 0.45, $\mu$ = 100 mPa-s}
\label{fig:F_vs_t}
\end{figure}
\begin{figure}[!htb]
\centering
\begin{subfigure}[b]{0.245\textwidth}
\centering
\includegraphics[width=\textwidth]{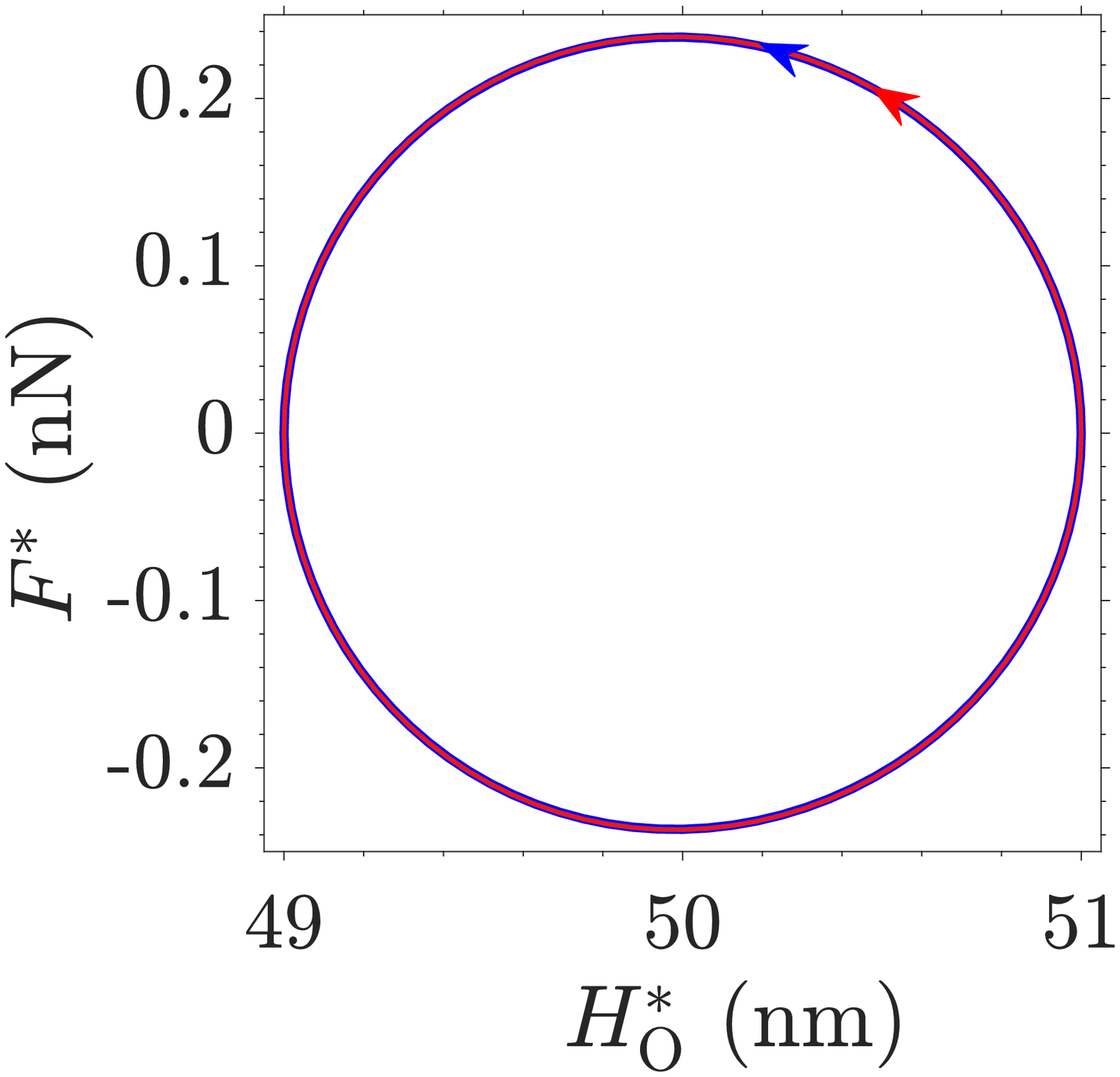}
\caption{}
\label{subfig:F_vs_H0_low_omega_low_h0}
\end{subfigure}
\begin{subfigure}[b]{0.245\textwidth}
\centering
\includegraphics[width=\textwidth]{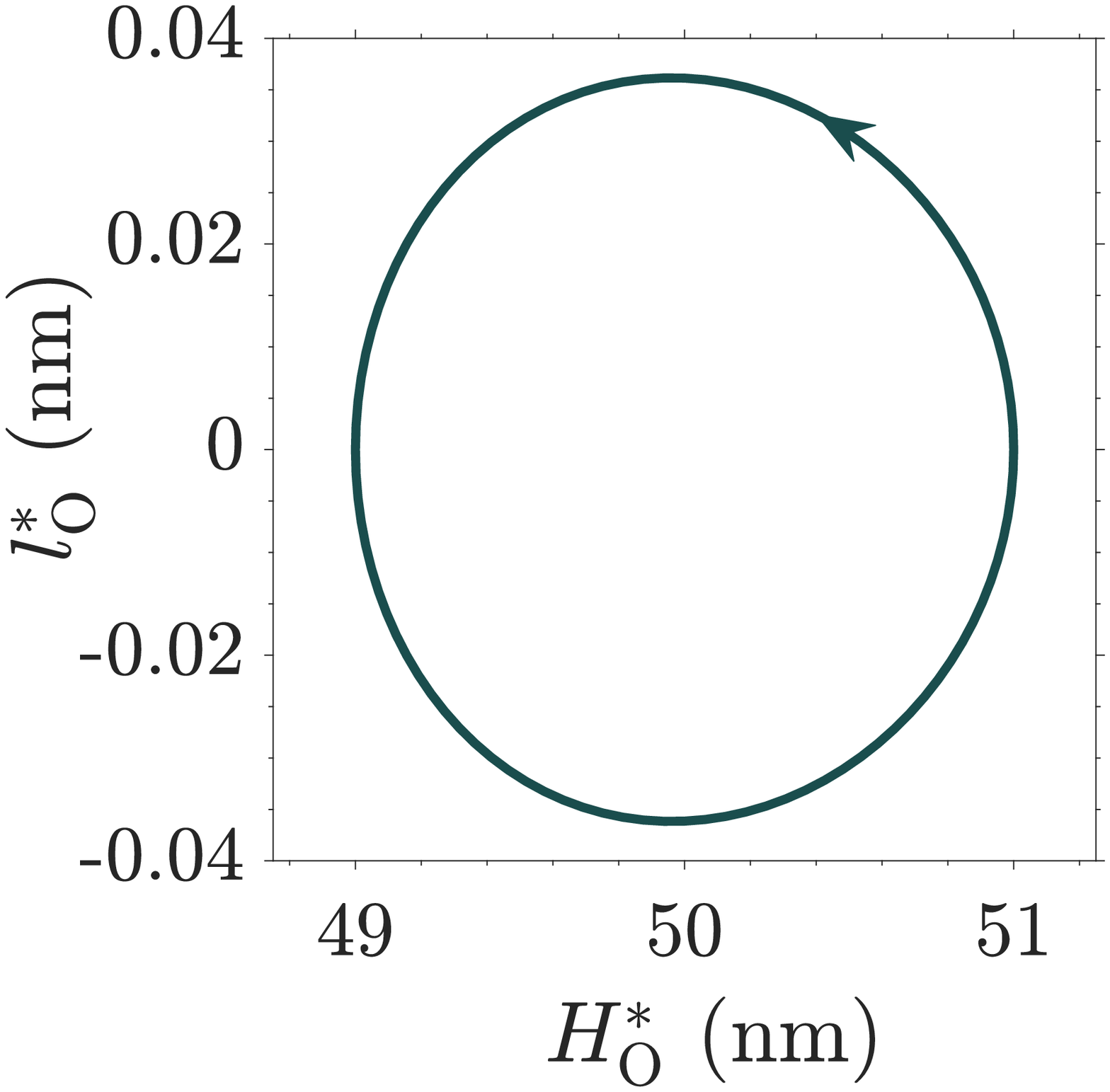}
\caption{}
\label{subfig:l_vs_H0_low_omega_low_h0}
\end{subfigure}
\begin{subfigure}[b]{0.245\textwidth}
\centering
\includegraphics[width=\textwidth]{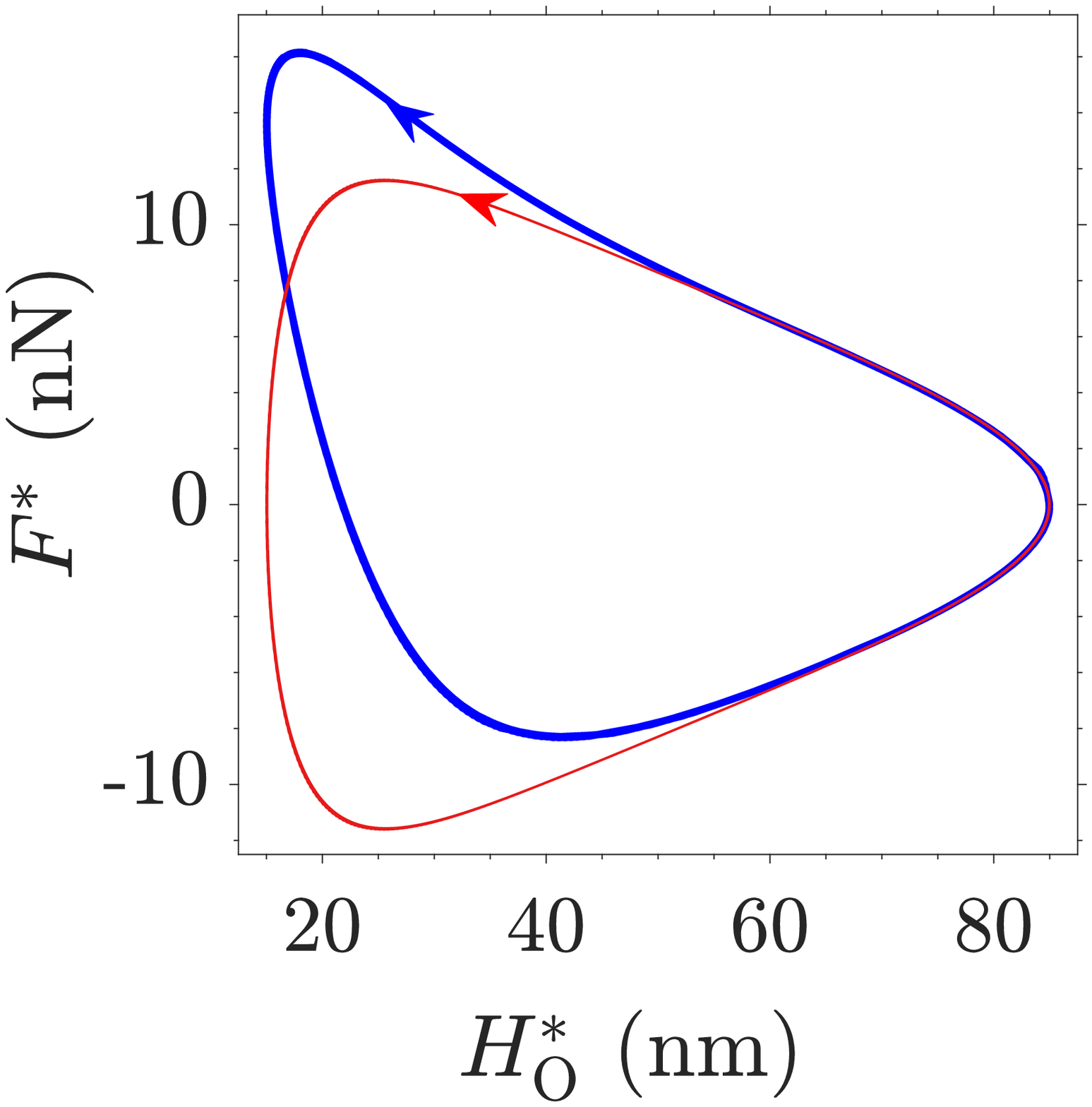}
\caption{}
\label{subfig:F_vs_H0_low_omega_high_h0_with_DLVO}
\end{subfigure}
\begin{subfigure}[b]{0.245\textwidth}
\centering
\includegraphics[width=\textwidth]{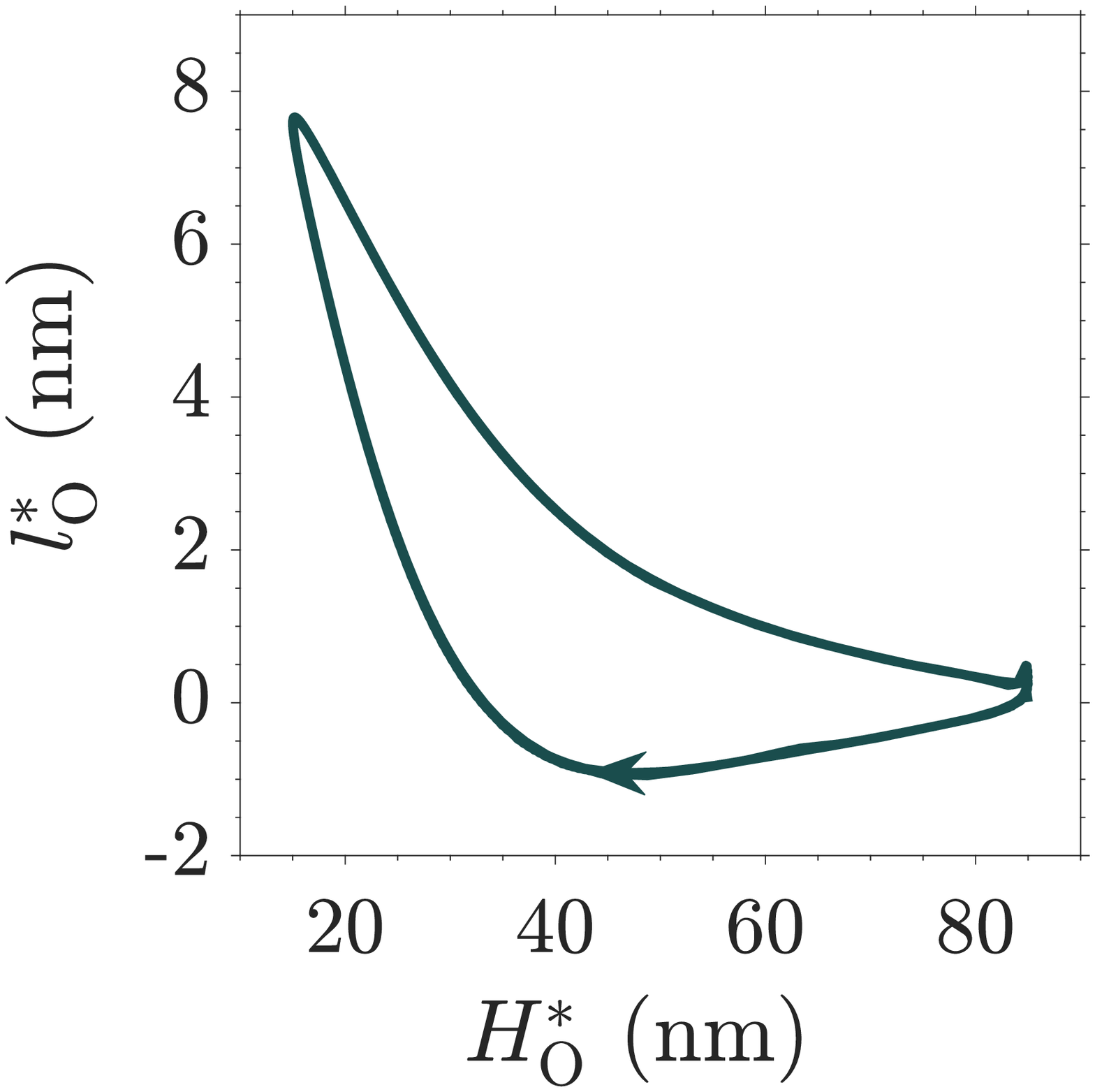}
\caption{}
\label{subfig:l_vs_H0_low_omega_high_h0_with_DLVO}
\end{subfigure}
\begin{subfigure}[b]{0.245\textwidth}
\centering
\includegraphics[width=\textwidth]{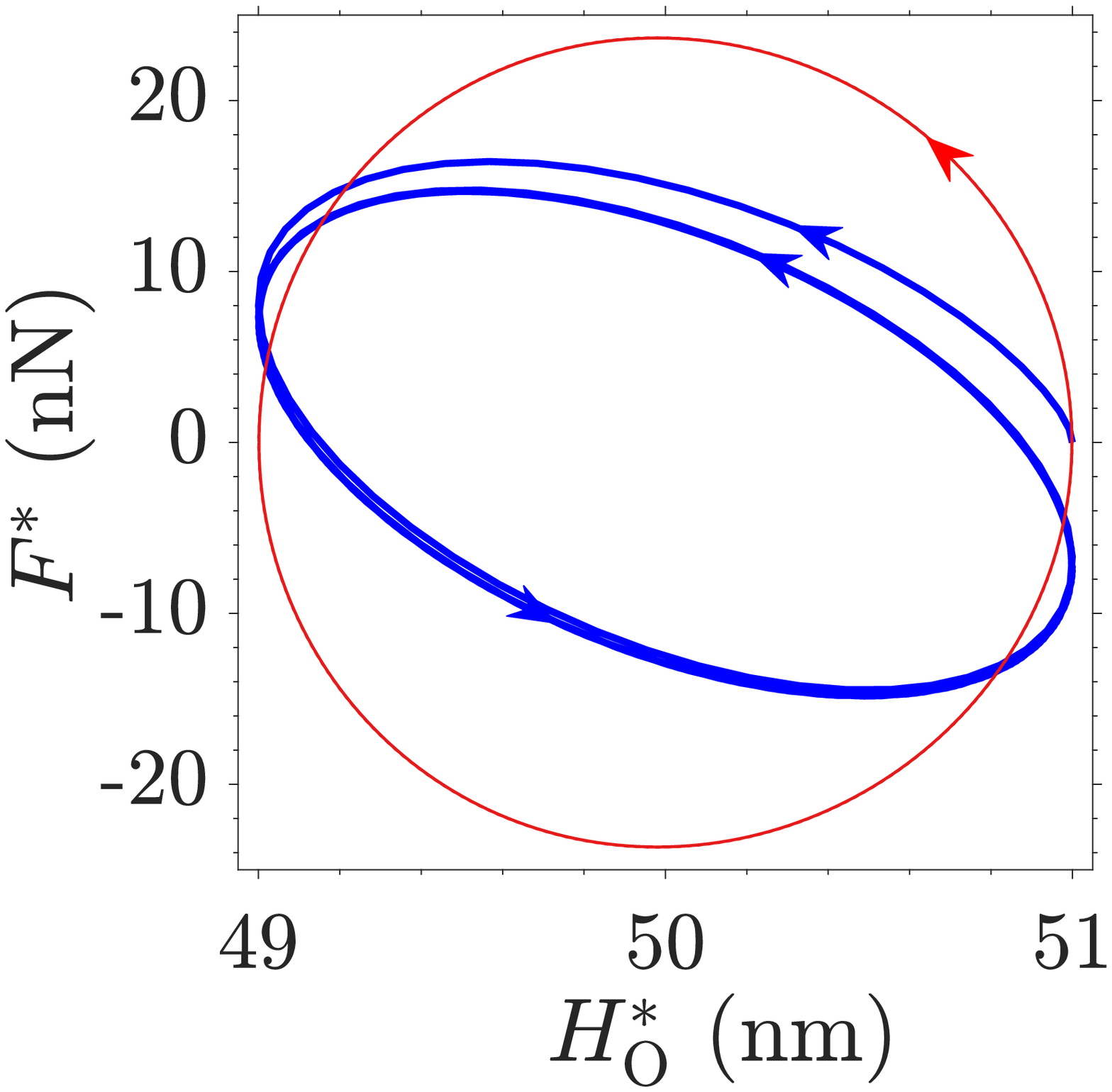}
\caption{}
\label{subfig:F_vs_H0_high_omega_low_h0}
\end{subfigure}
\begin{subfigure}[b]{0.245\textwidth}
\centering
\includegraphics[width=\textwidth]{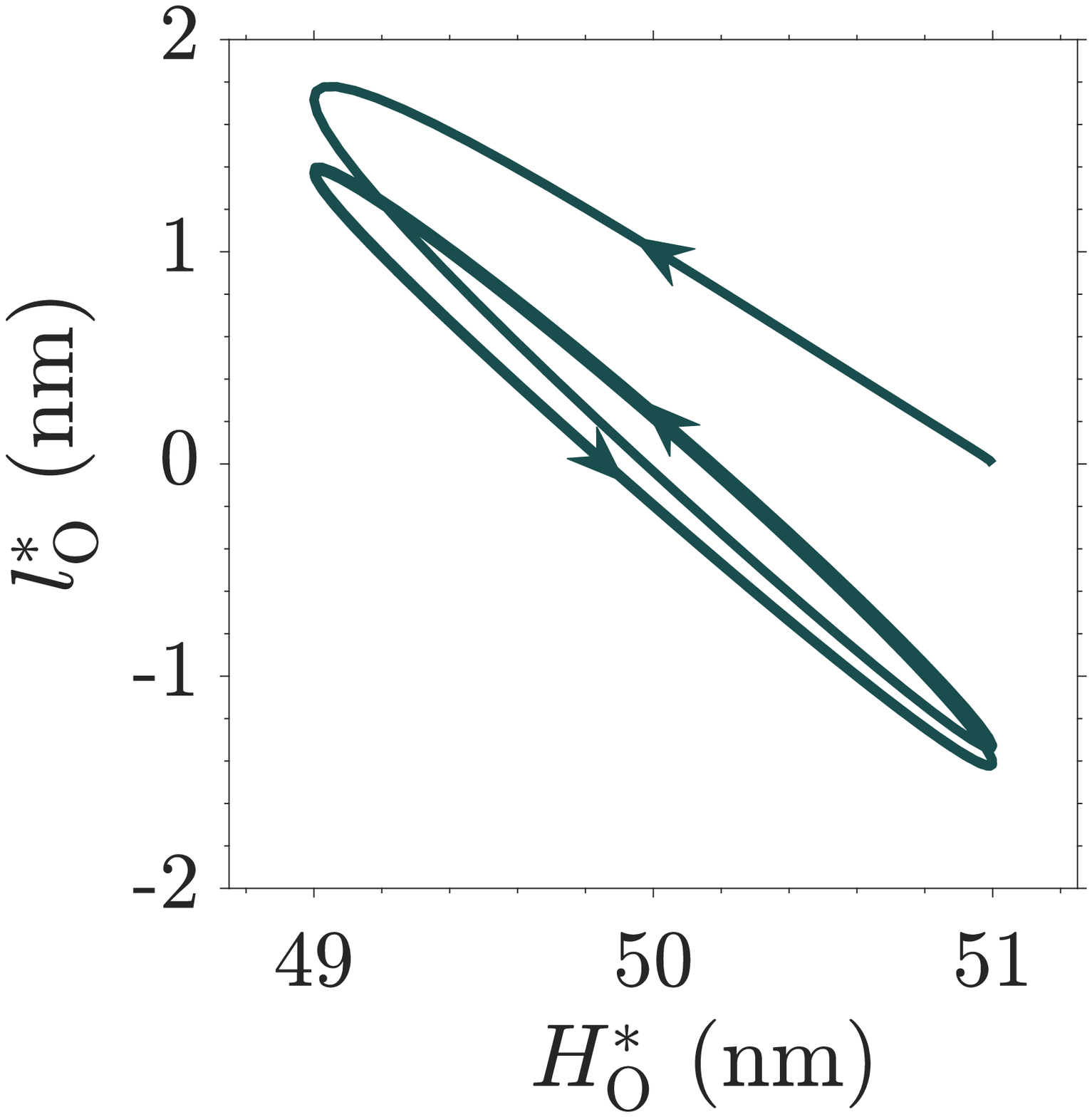}
\caption{}
\label{subfig:l_vs_H0_high_omega_low_h0}
\end{subfigure}
\begin{subfigure}[b]{0.245\textwidth}
\centering
\includegraphics[width=\textwidth]{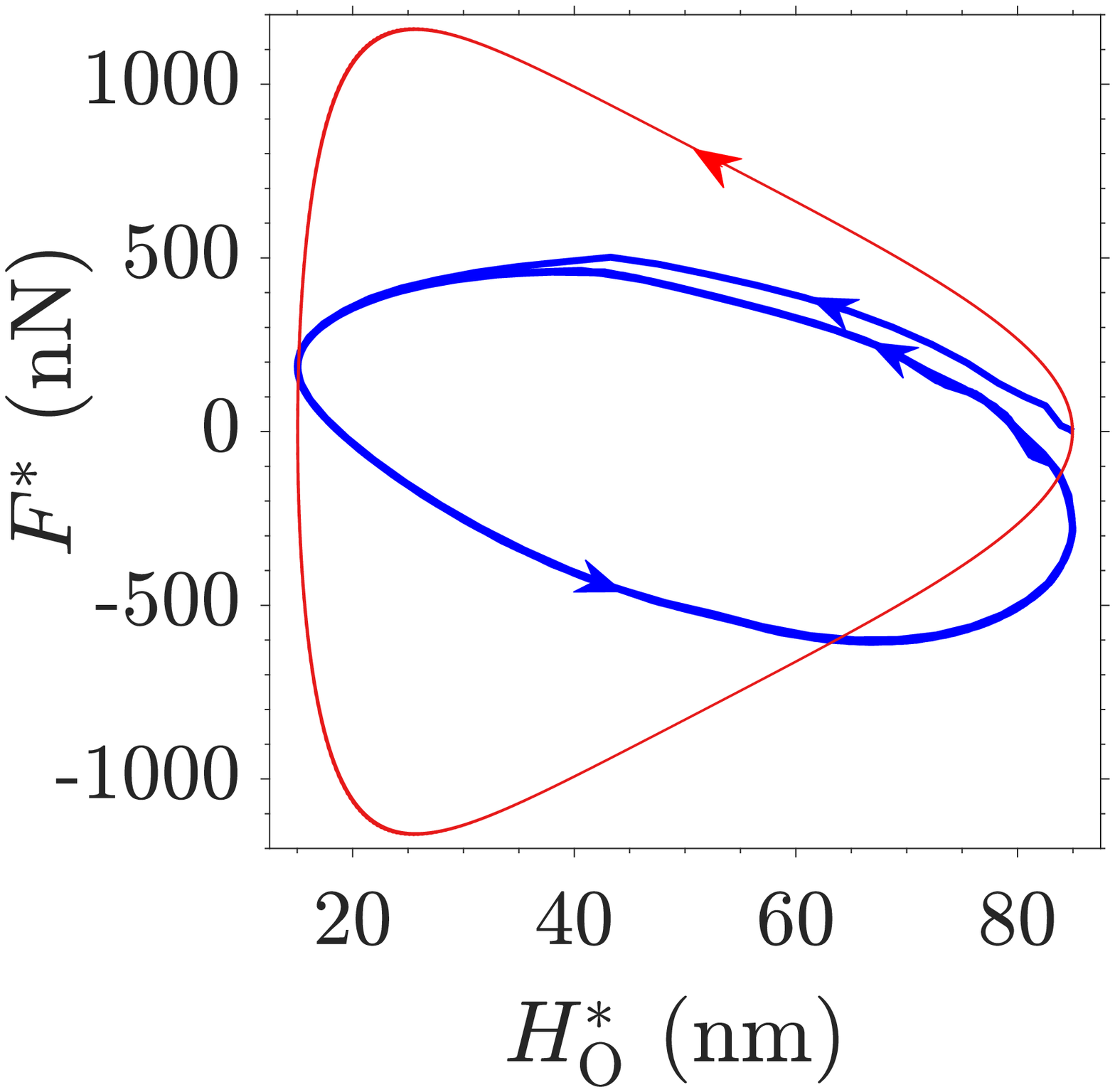}
\caption{}
\label{subfig:F_vs_H0_high_omega_high_h0}
\end{subfigure}
\begin{subfigure}[b]{0.245\textwidth}
\centering
\includegraphics[width=\textwidth]{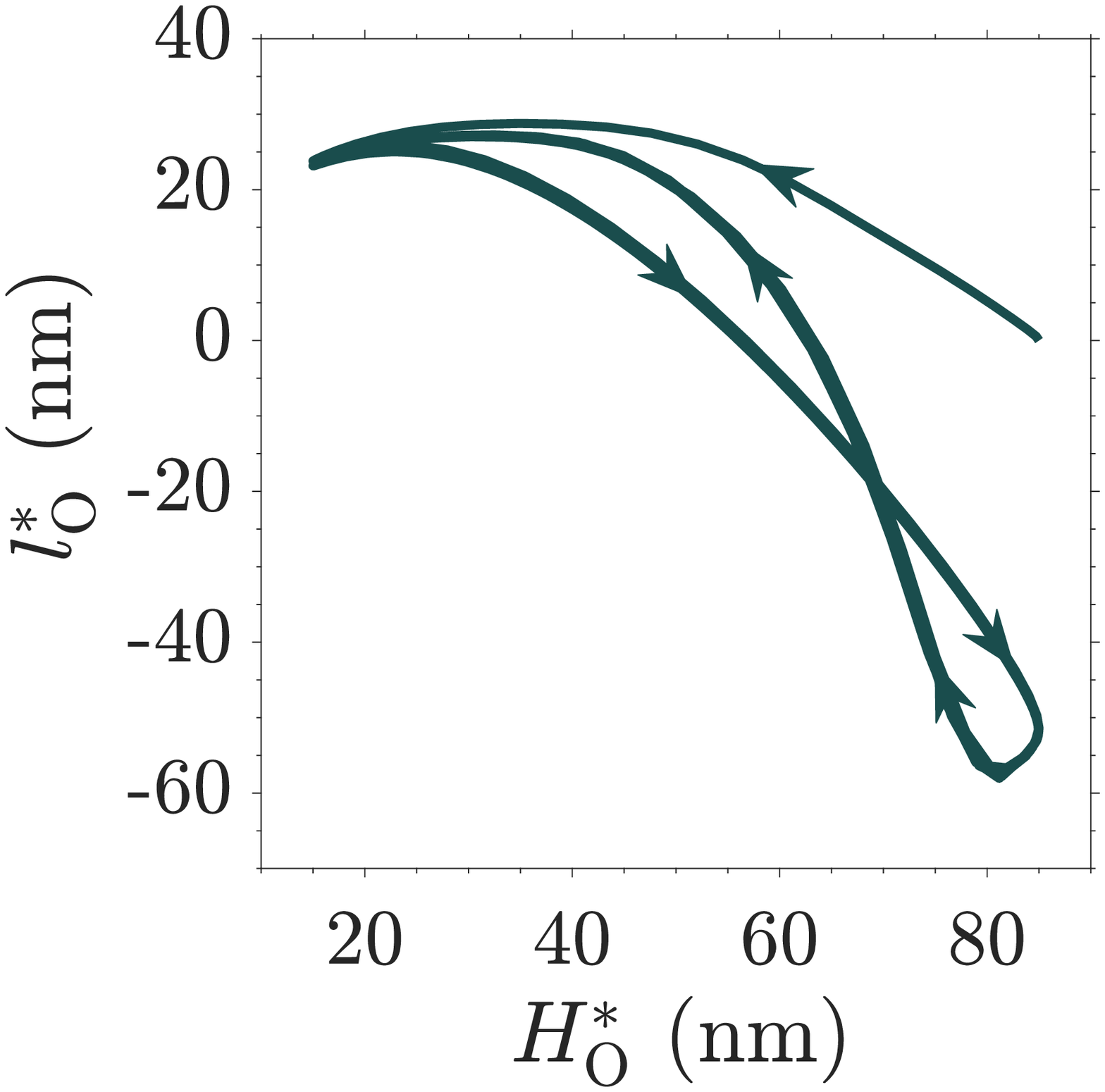}
\caption{}
\label{subfig:l_vs_H0_high_omega_high_h0}
\end{subfigure}
\caption{(a,b) low $\hat{\omega}$, low $h_0$ ($\hat{\omega} = 10^{1}$ Hz, $h_0 = 1$ nm) with or without DLVO forces, (c,d) low $\hat{\omega}$, high $h_0$ ($\hat{\omega} = 10^{1}$ Hz, $h_0 = 35$ nm) with DLVO forces (e,f) high $\hat{\omega}$, low $h_0$ ($\hat{\omega} = 10^{3}$ Hz, $h_0 = 1$ nm) with or without DLVO forces, (g,h) high, $\hat{\omega}$ high $h_0$ ($\hat{\omega} = 10^{3}$ Hz, $h_0 = 35$ nm) with or without DLVO forces; in subfigures (a,c,e,g) the solid blue lines are the computed solution, the thin solid red lines are the expected trend if the deflection-pressure interaction is hypothetically constrained to be OWC; other system parameters are: $R$ = 10 $\mu$m, $D$ = 50 nm, $L$ = 1 $\mu$m, $E_{\text{Y}}$ = 1 MPa, $\nu$ = 0.45, $\mu$ = 100 mPa-s}
\label{fig:F_vs_H0}
\end{figure}
\begin{figure}[!htb]
\centering
\begin{subfigure}[b]{0.495\textwidth}
\centering
\includegraphics[width=\textwidth]{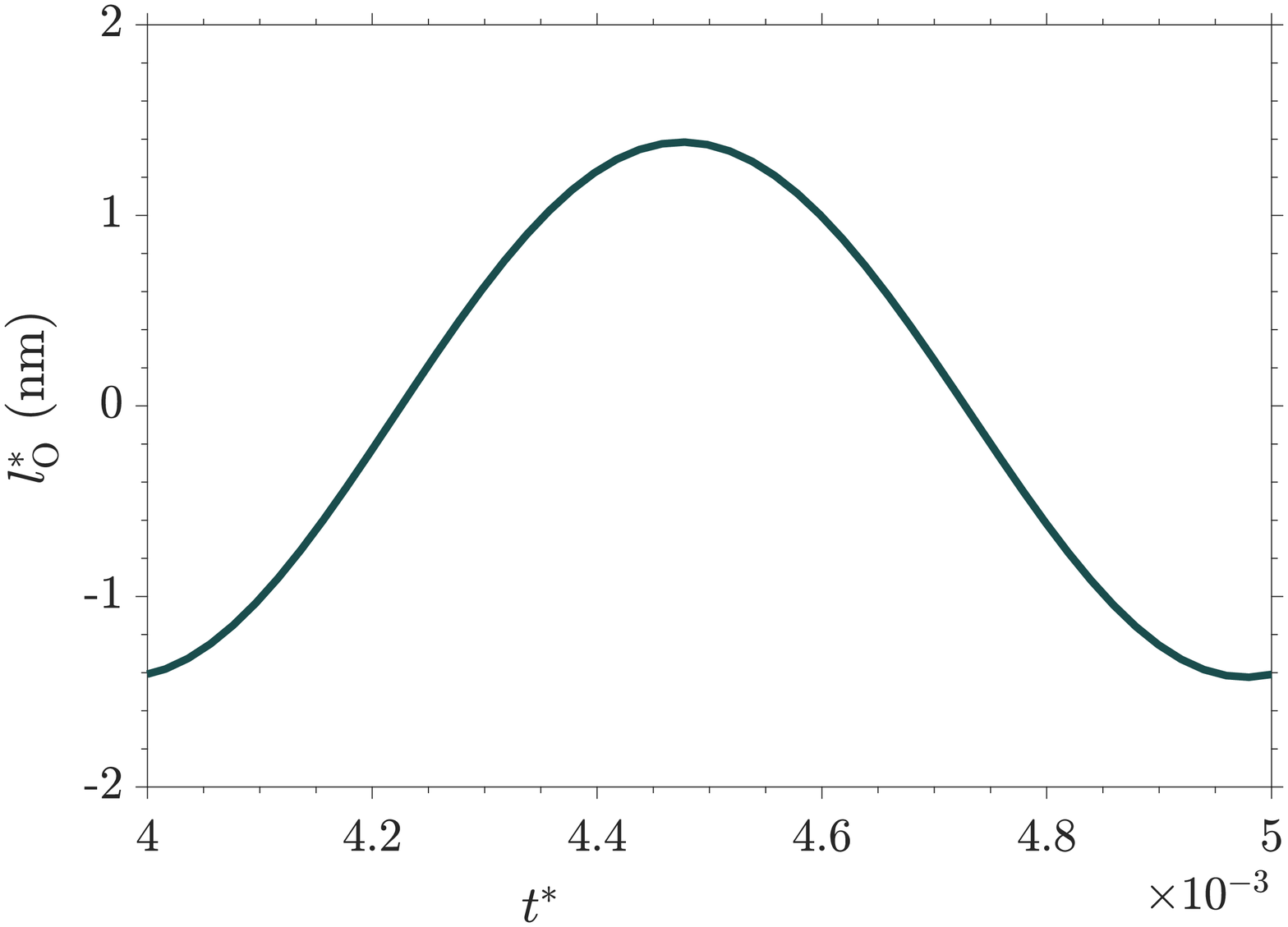}
\caption{}
\label{subfig:l_vs_t_high_omega_low_h0}
\end{subfigure}
\begin{subfigure}[b]{0.495\textwidth}
\centering
\includegraphics[width=\textwidth]{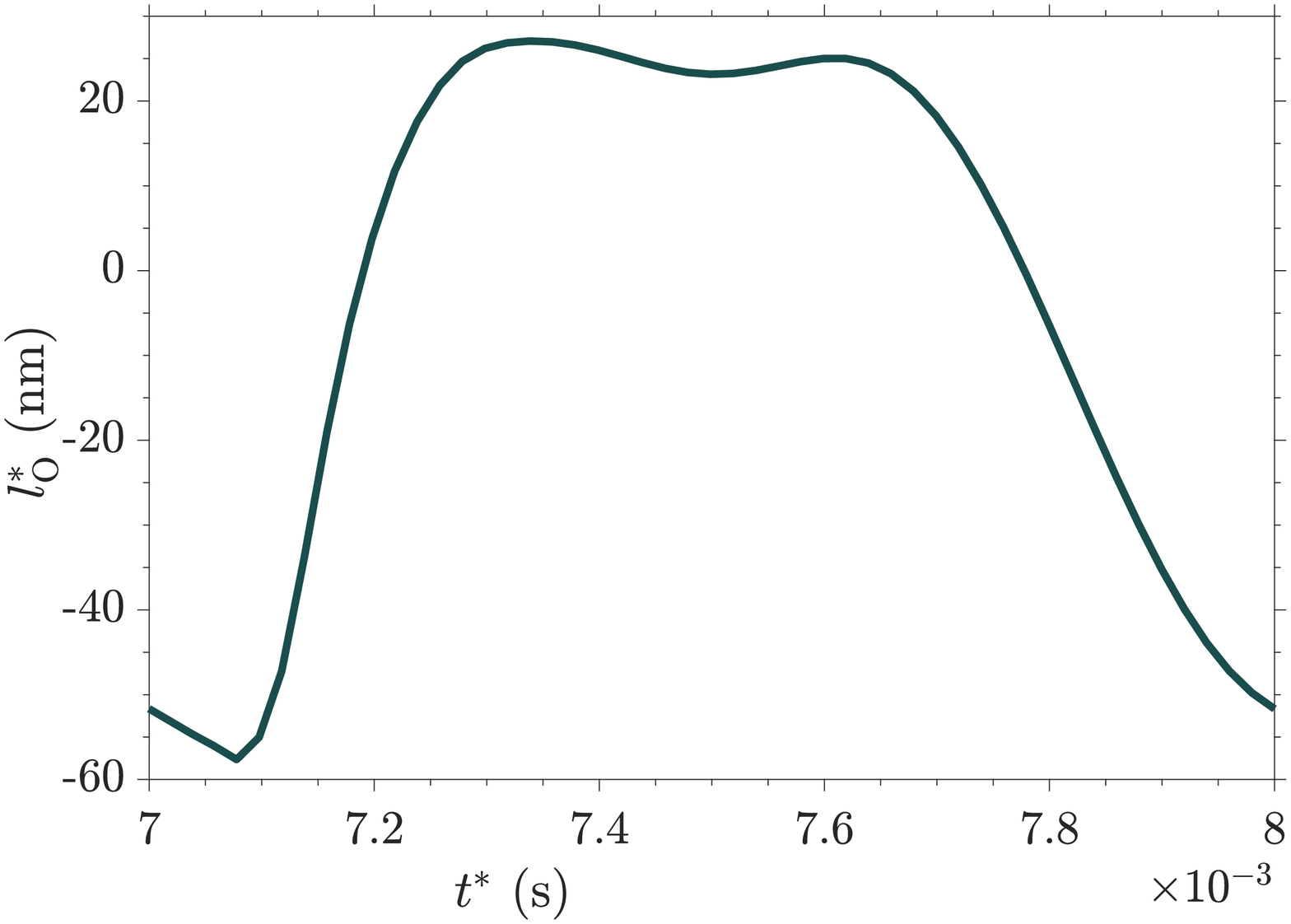}
\caption{}
\label{subfig:l_vs_t_high_omega_high_h0}
\end{subfigure}
\caption{Variation of deflection at origin $l_{\text{O}}^*$, with time $t^*$ after the system has reached quasi-steady state, for (a) high $\hat{\omega}$ low $h_0$ ($\hat{\omega} = 10^{3}$ Hz, $h_0 = 1$ nm) with or without DLVO forces, and, (b) high $\hat{\omega}$ high $h_0$ ($\hat{\omega} = 10^{3}$ Hz, $h_0 = 35$ nm) with or without DLVO forces; other system parameters are: $R$ = 10 $\mu$m, $D$ = 50 nm, $L$ = 1 $\mu$m, $E_{\text{Y}}$ = 1 MPa, $\nu$ = 0.45, $\mu$ = 100 mPa-s}
\label{fig:l_vs_t_freq}
\end{figure}
\begin{figure}[!htb]
\centering
\begin{subfigure}[b]{0.495\textwidth}
\centering
\includegraphics[width=\textwidth]{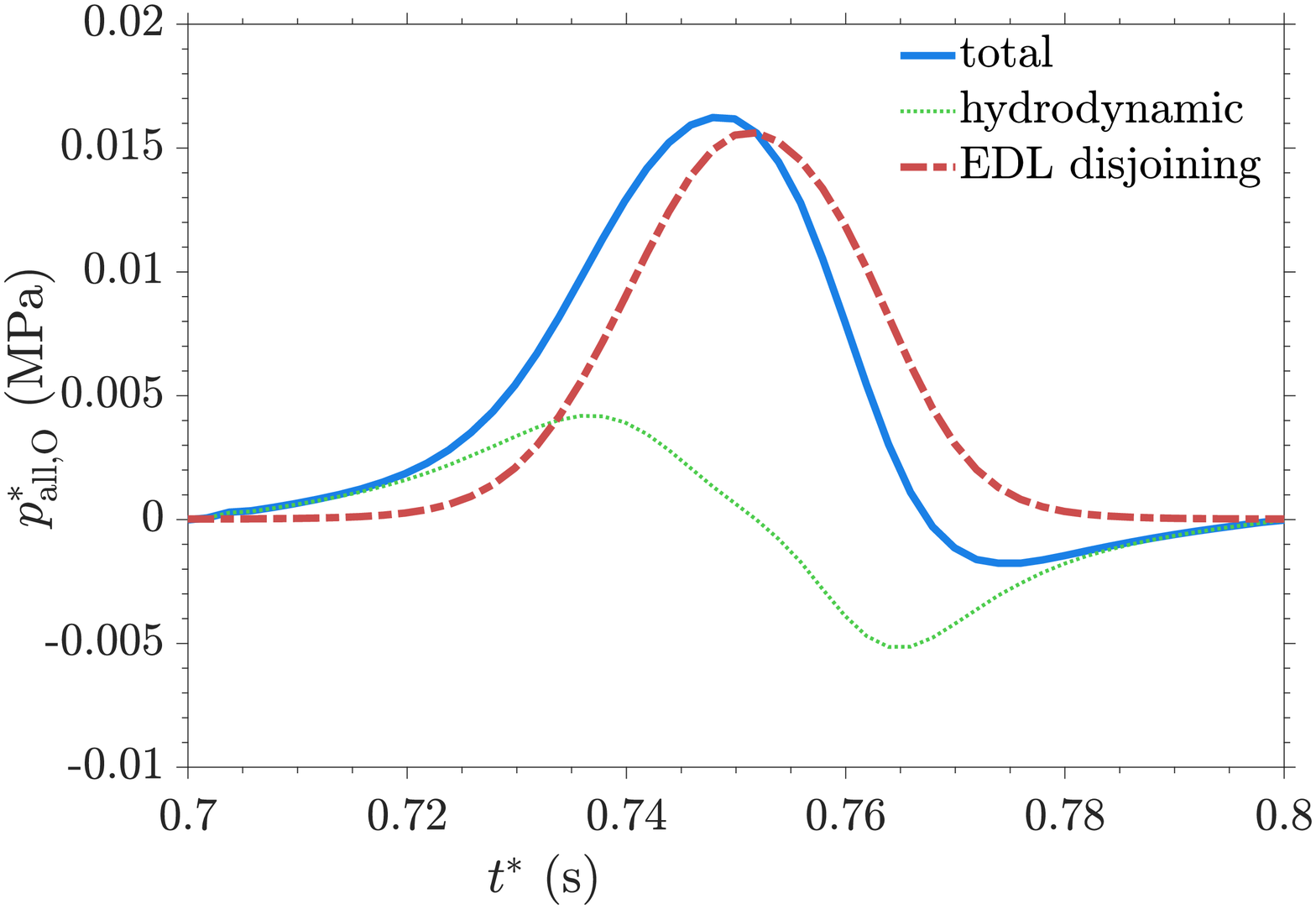}
\caption{}
\label{subfig:pall_vs_t_low_omega_high_h0_with_DLVO}
\end{subfigure}
\begin{subfigure}[b]{0.495\textwidth}
\centering
\includegraphics[width=\textwidth]{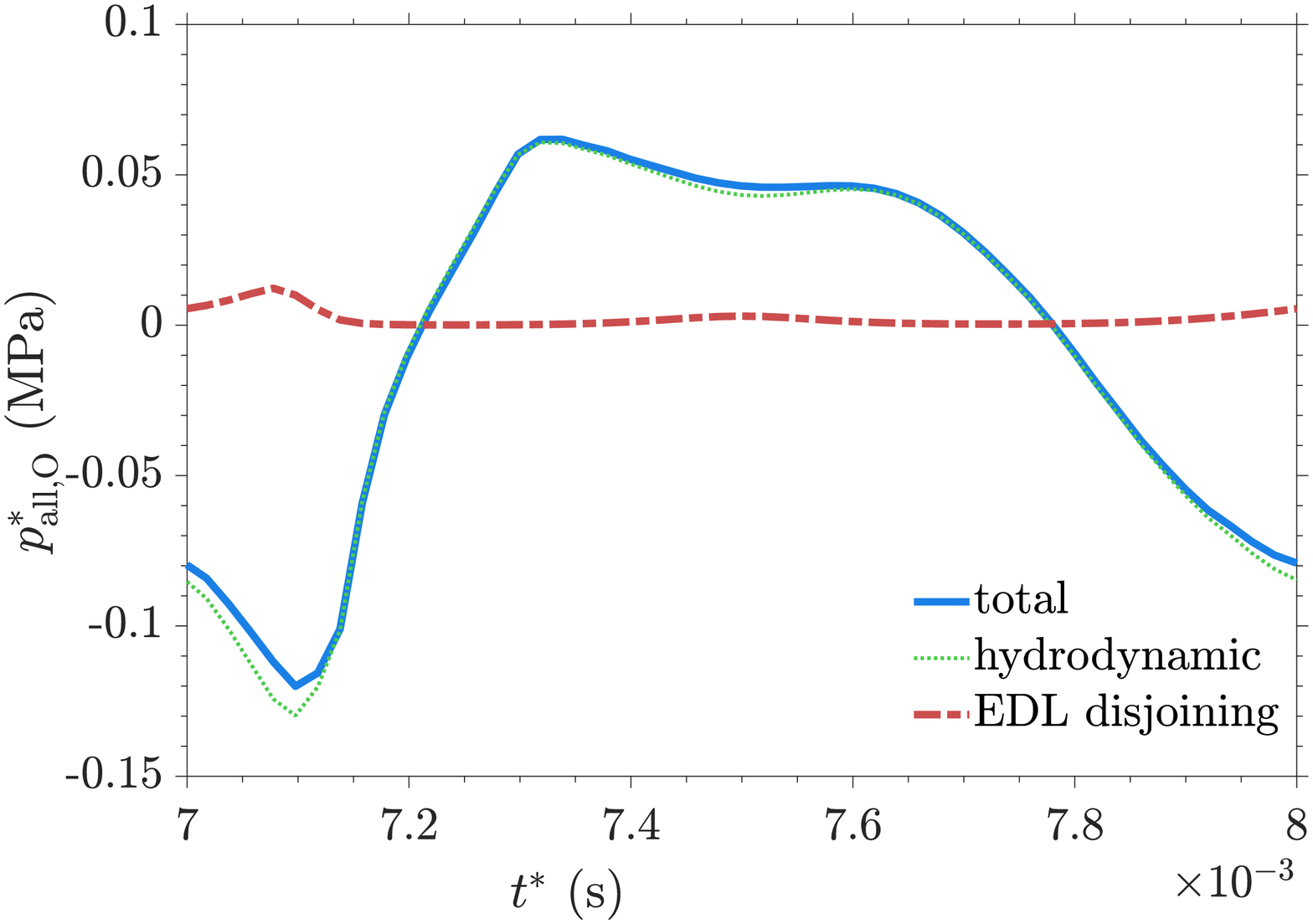}
\caption{}
\label{subfig:pall_vs_t_high_omega_high_h0_with_DLVO}
\end{subfigure}
\caption{Variation of pressure components at origin $p_{\text{hd(O)}}^*$, $p_{\text{EDL(O)}}^*$, $p_{\text{vdW(O)}}^*$, and total pressure at origin $p_{\text{O}}^*$, with time $t^*$ after the system has reached quasi-steady state, for (a) low $\hat{\omega}$ high $h_0$ ($\hat{\omega} = 10^{1}$ Hz, $h_0 = 35$ nm) with DLVO forces, and, (b) high $\hat{\omega}$ high $h_0$ ($\hat{\omega} = 10^{3}$ Hz, $h_0 = 35$ nm) with DLVO forces; other system parameters are: $R$ = 10 $\mu$m, $D$ = 50 nm, $L$ = 1 $\mu$m, $E_{\text{Y}}$ = 1 MPa, $\nu$ = 0.45, $\mu$ = 100 mPa-s}
\label{fig:pall_vs_t}
\end{figure}
\begin{figure}[!htb]
\centering
\begin{subfigure}[b]{0.495\textwidth}
\centering
\includegraphics[width=\textwidth]{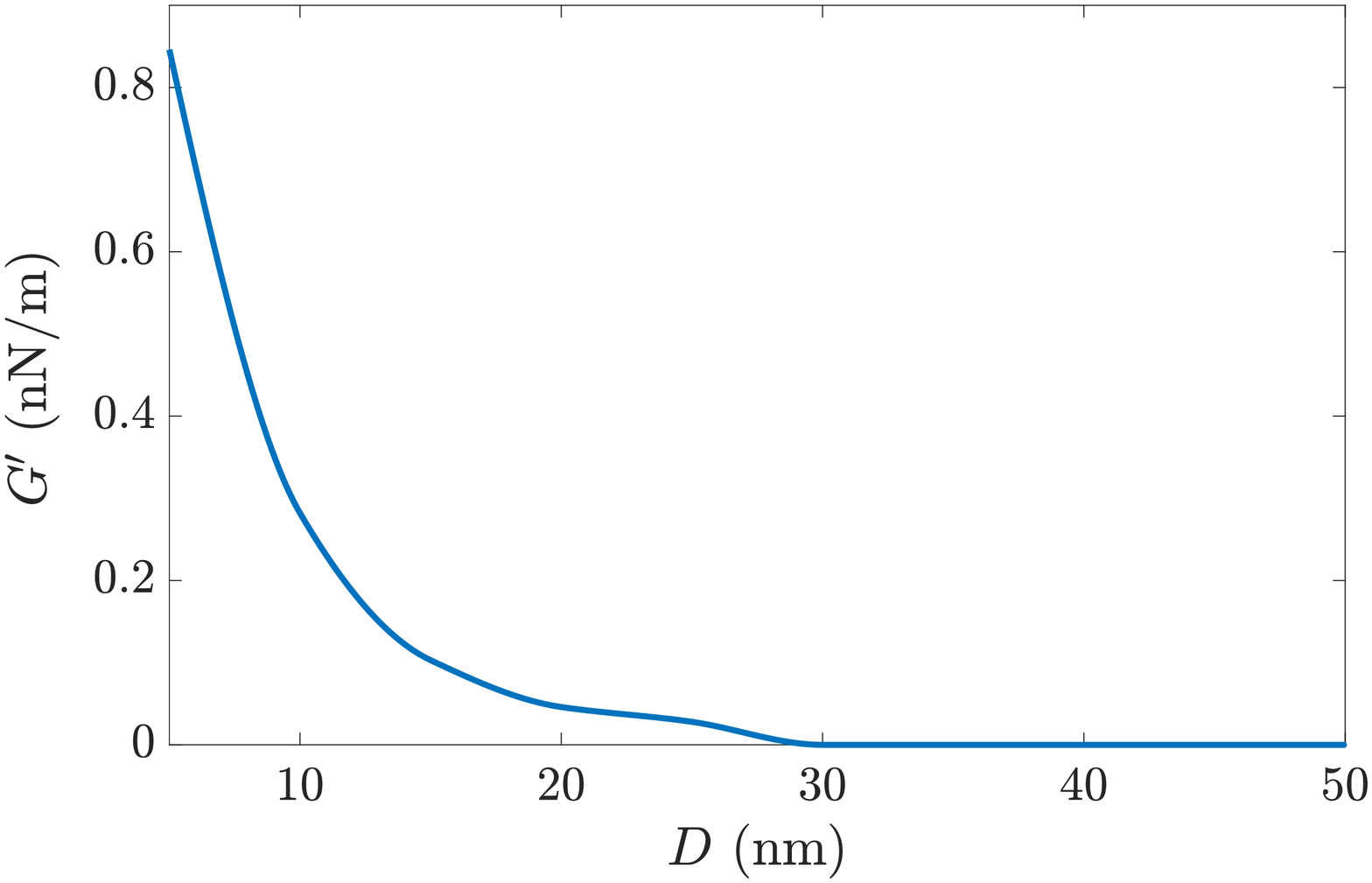}
\caption{}
\label{subfig:Gpr_1e1Hz}
\end{subfigure}
\begin{subfigure}[b]{0.495\textwidth}
\centering
\includegraphics[width=\textwidth]{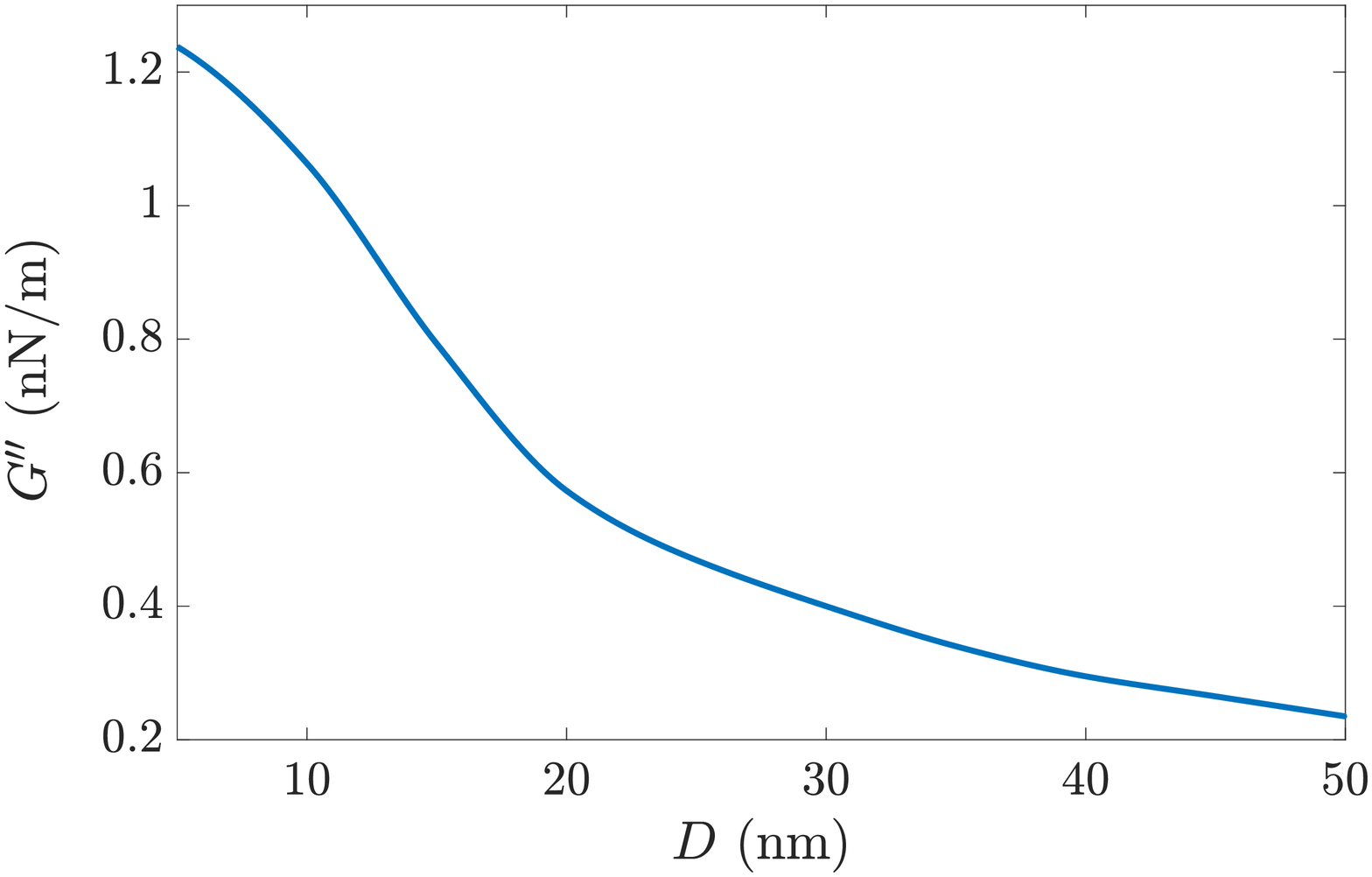}
\caption{}
\label{subfig:Gprpr_1e1Hz}
\end{subfigure}
\begin{subfigure}[b]{0.495\textwidth}
\centering
\includegraphics[width=\textwidth]{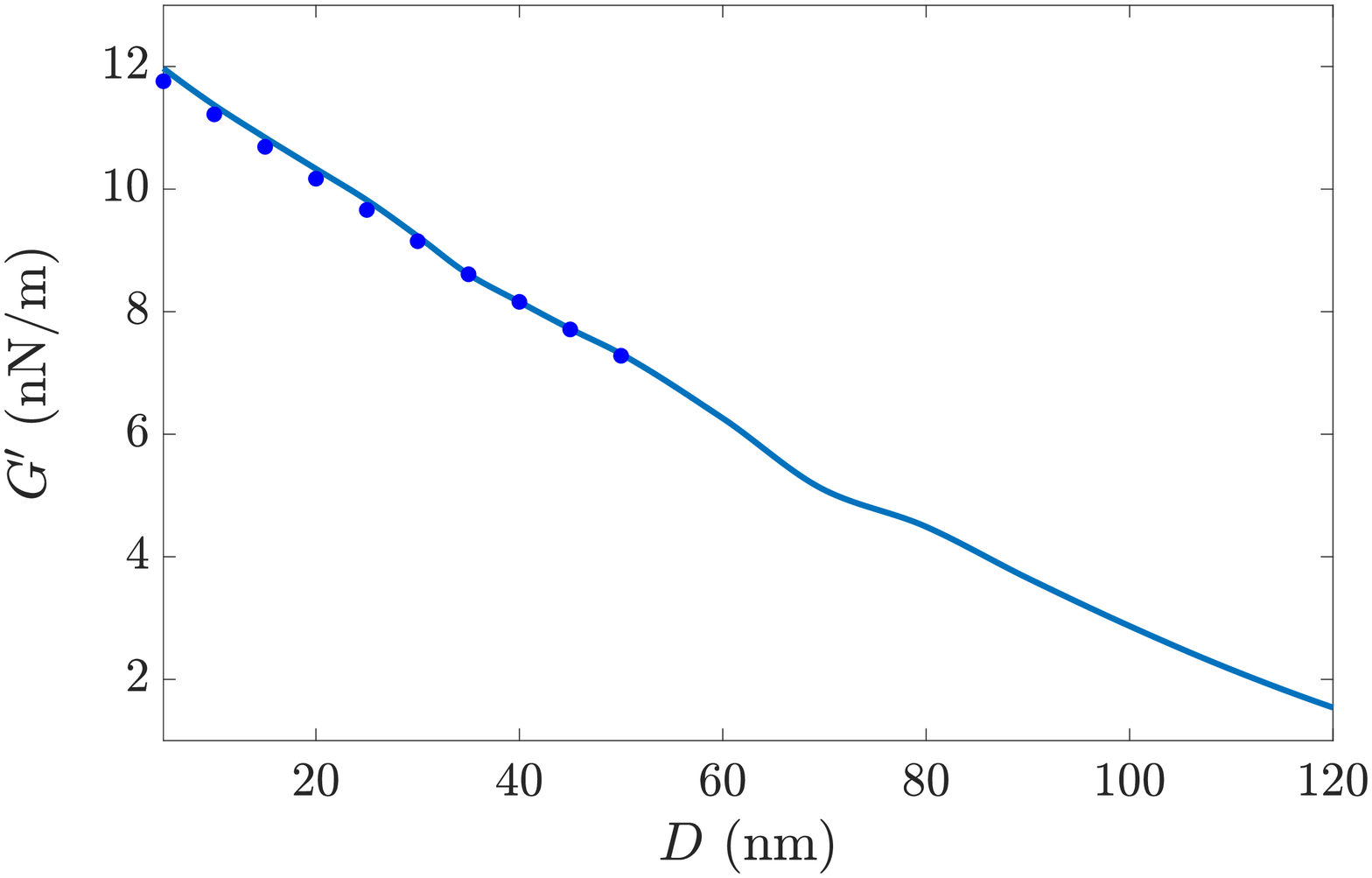}
\caption{}
\label{subfig:Gpr_1e3Hz}
\end{subfigure}
\begin{subfigure}[b]{0.495\textwidth}
\centering
\includegraphics[width=\textwidth]{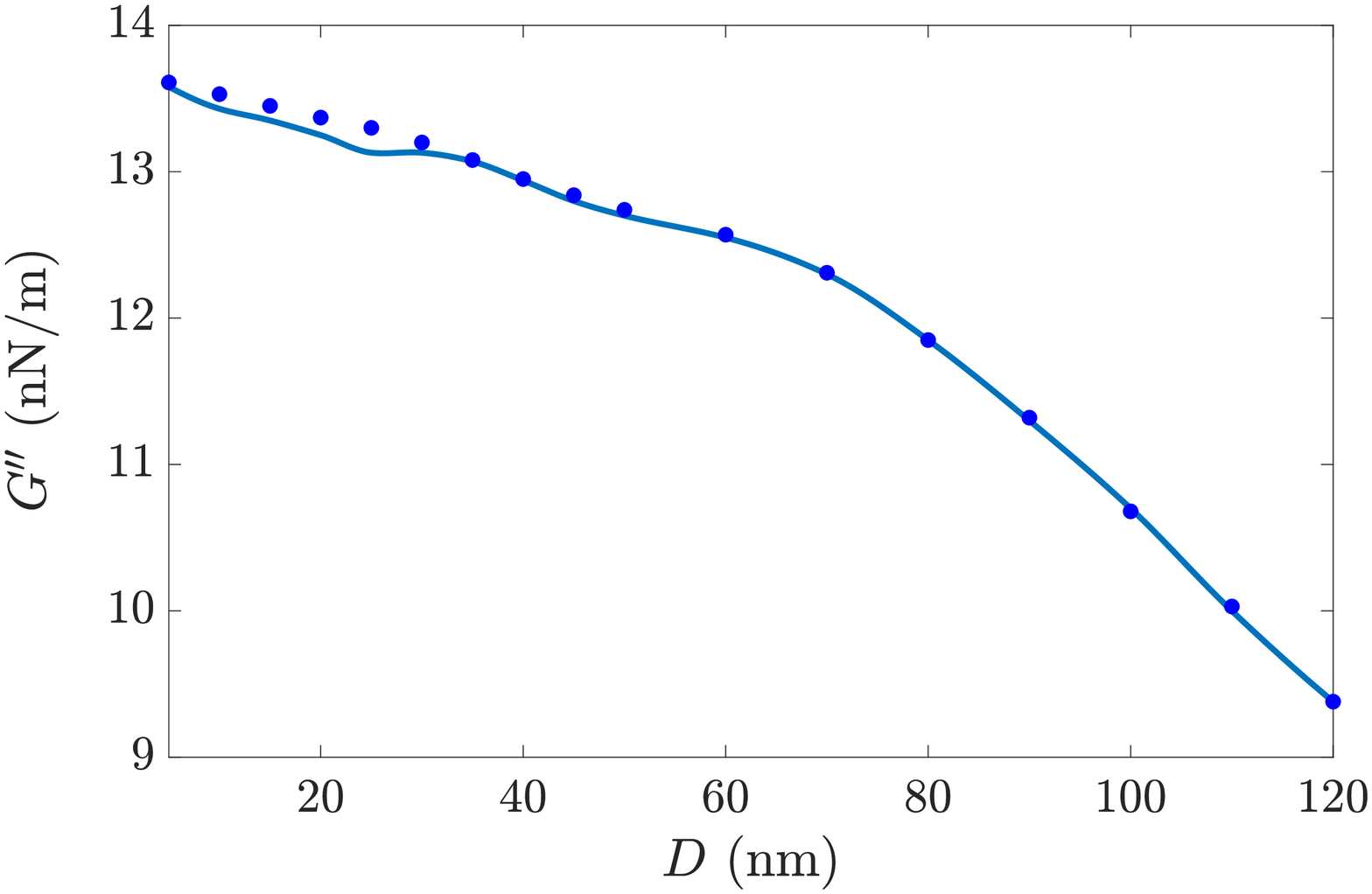}
\caption{}
\label{subfig:Gprpr_1e3Hz}
\end{subfigure}
\begin{subfigure}[b]{0.495\textwidth}
\centering
\includegraphics[width=\textwidth]{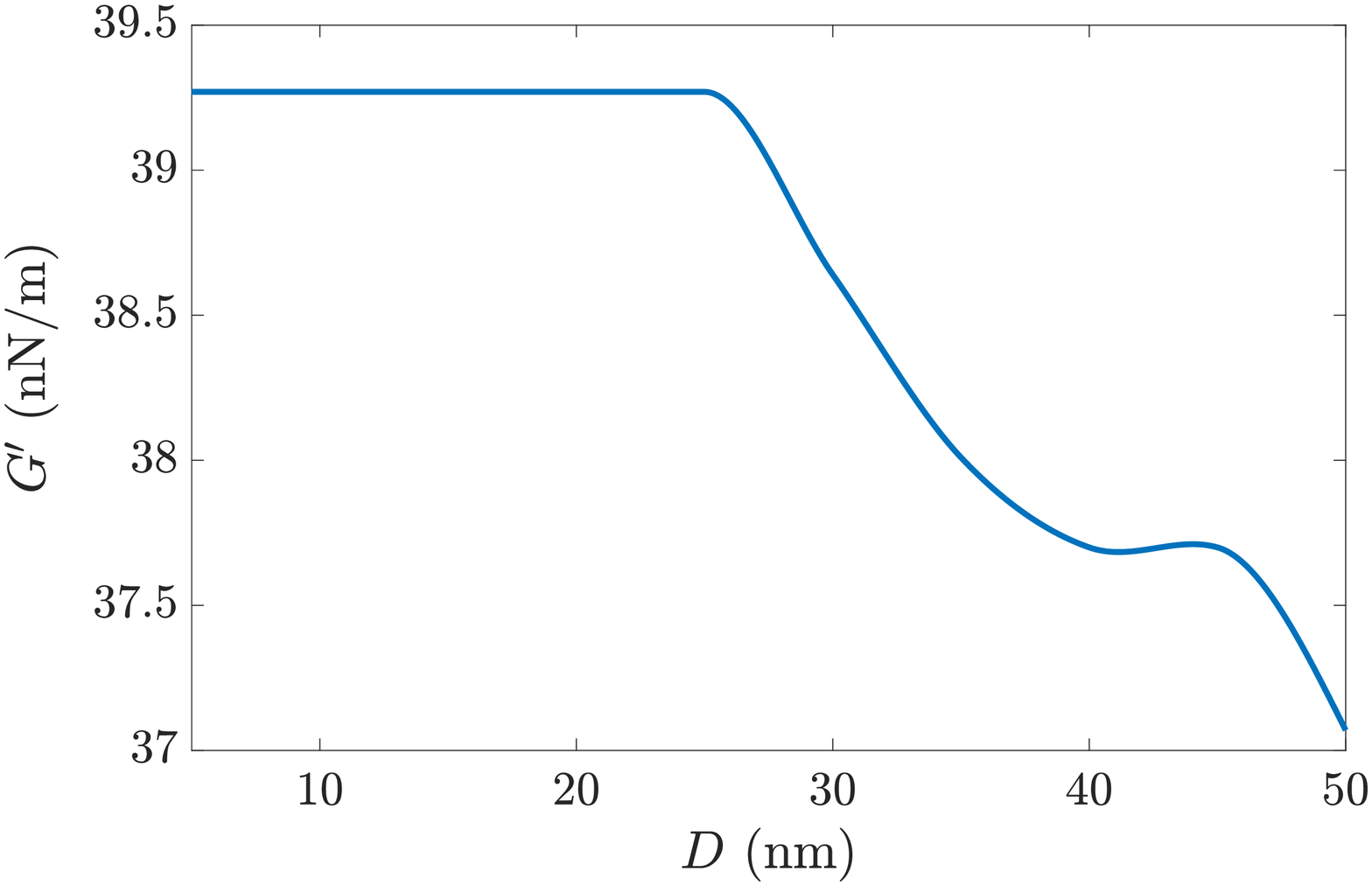}
\caption{}
\label{subfig:Gpr_1e4Hz}
\end{subfigure}
\begin{subfigure}[b]{0.495\textwidth}
\centering
\includegraphics[width=\textwidth]{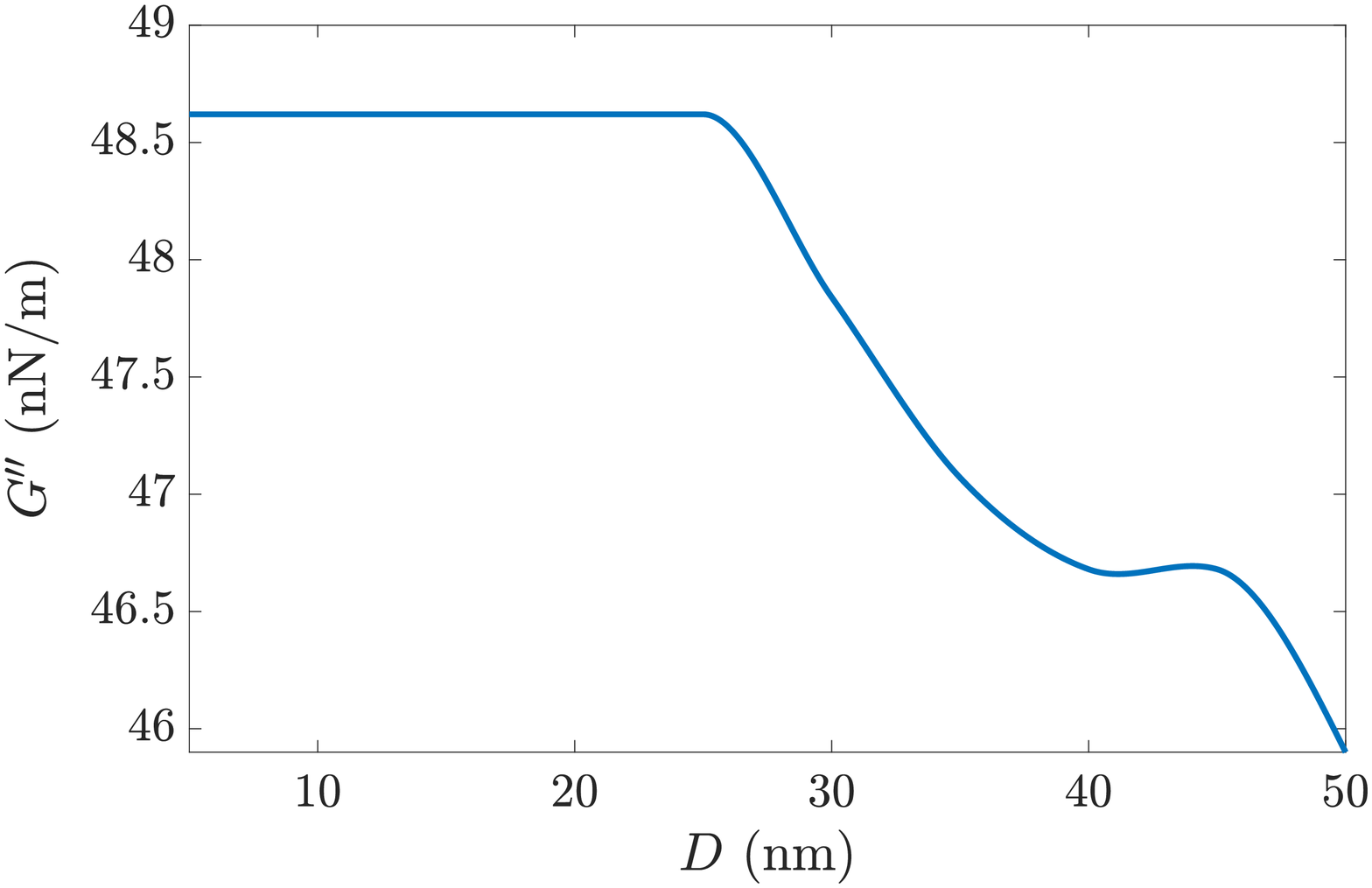}
\caption{}
\label{subfig:Gprpr_1e4Hz}
\end{subfigure}
\caption{Variation of (a,c,e) $G^{\prime}$, and (b,d,f) $G^{\prime\prime}$ with $D$, for $\hat{\omega}$ = (a,b) $10^1$ Hz, (c,d)  $10^3$ Hz, and (e,f)  $10^4$ Hz; DLVO forces are not considered; other system parameters are: $R$ = 10 $\mu$m, $h_0$ = 1 nm, $L$ =1 $\mu$m, $E_{\text{Y}}$ = 1 MPa, $\nu$ = 0.45, $\mu$ = 100 mPa-s; the circle markers in subfigures c and d are the values obtained by Leroy and Charlaix \cite{Leroy2011}}
\label{fig:G}
\end{figure}
\begin{figure}[!htb]
\centering
\includegraphics[width=0.35\textwidth]{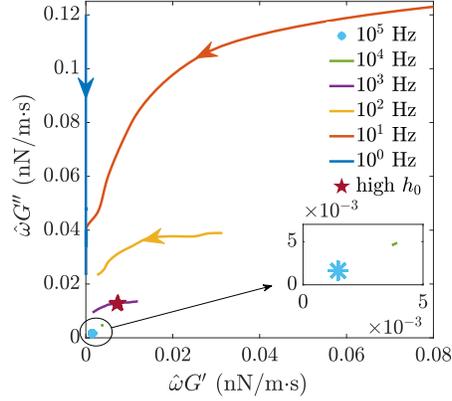}
\caption{Phase plot of $\hat{\omega} \bar{G} = \hat{\omega}G^{\prime}+\iota \hat{\omega}G^{\prime\prime}$ in the complex plane for $\hat{\omega}$ = [$10^0$, $10^1$, $10^2$, $10^3$, $10^4$, $10^5$] Hz; DLVO forces are not considered for $\omega = 2\pi$ rad/s $\times$ [$10^0$, $10^1$] and are inconsequential for the rest of the frequencies; direction of the arrows indicated increasing value of $D$; other system parameters are: $R$ = 10 $\mu$m, $h_0$ = 1 nm, $L$ =1 $\mu$m, $E_{\text{Y}}$ = 1 MPa, $\nu$ = 0.45, $\mu$ = 100 mPa-s; the magenta star marker is the solution for varying $h_0$ from 1 nm to 35 nm for $\hat{\omega}$ = $10^3$ Hz and D = 50 nm}
\label{fig:G_zplane}
\end{figure}
\begin{figure}[!htb]
\centering
\begin{subfigure}[b]{0.495\textwidth}
\centering
\includegraphics[width=\textwidth]{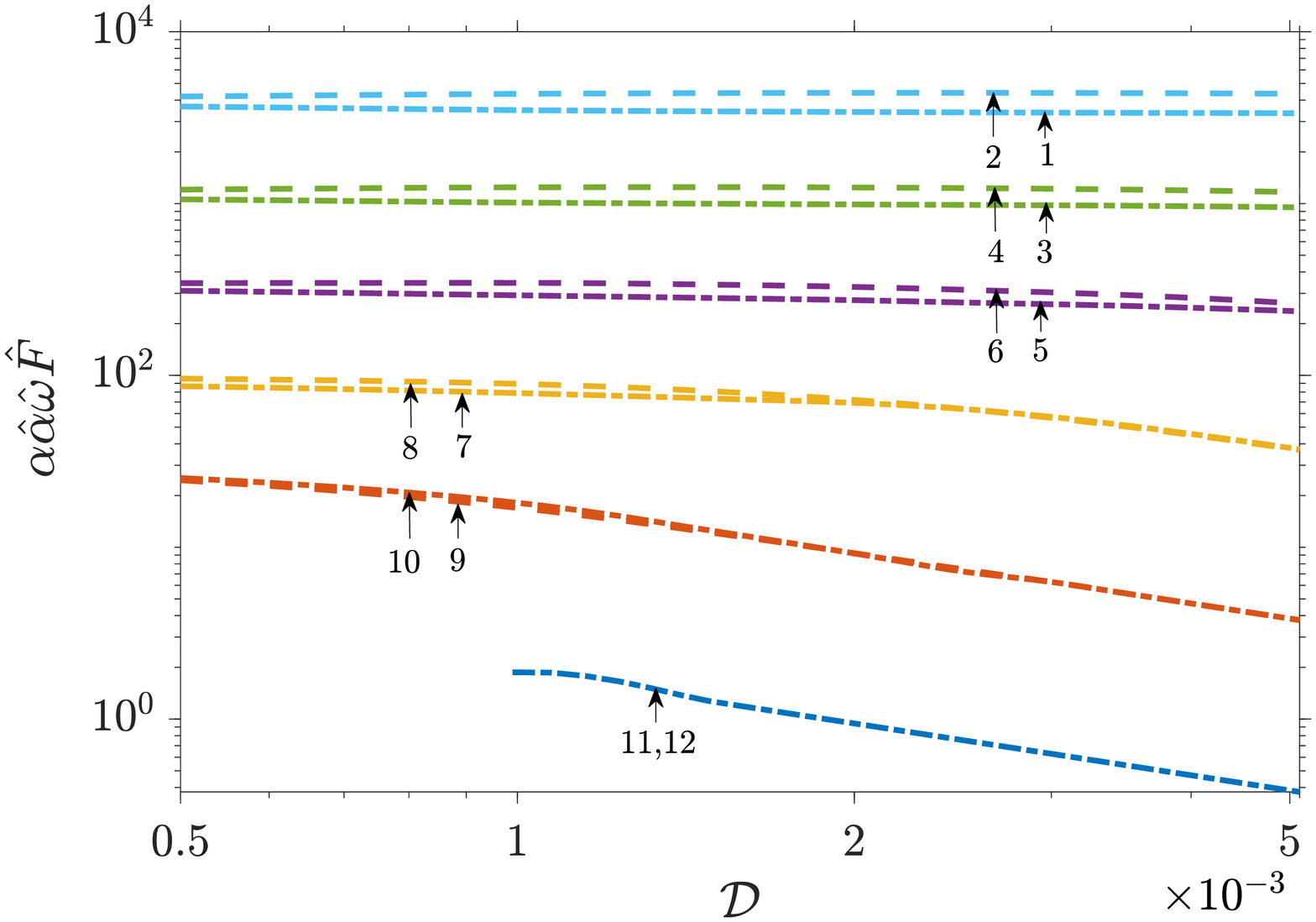}
\caption{}
\label{subfig:F_vs_D_Freq}
\end{subfigure}
\begin{subfigure}[b]{0.495\textwidth}
\centering
\includegraphics[width=\textwidth]{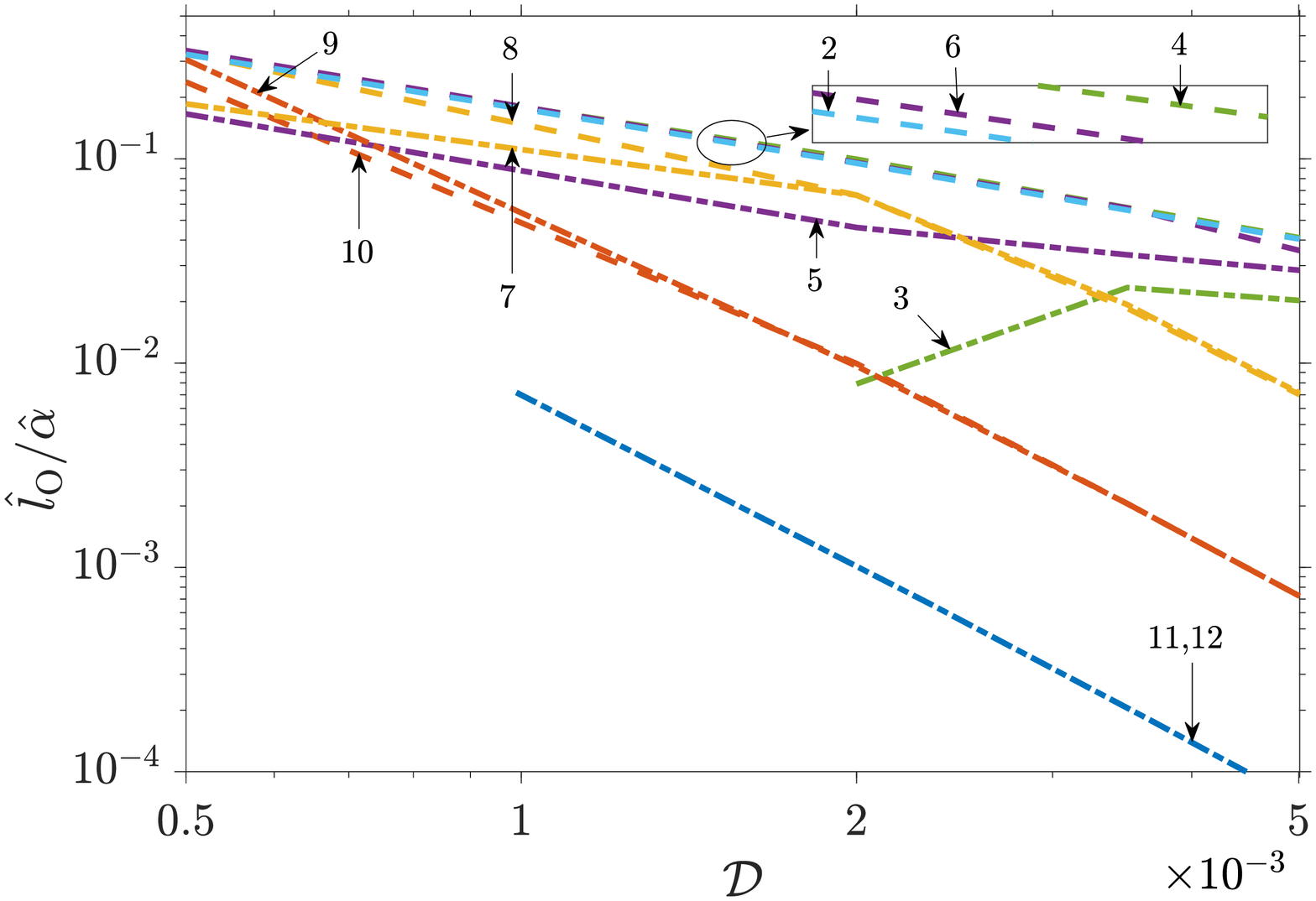}
\caption{}
\label{subfig:l_vs_D_Freq}
\end{subfigure}
\caption{Variation of magnitude of maximum attractive (negative), and, maximum repulsive (positive) (a) re-normalized Force $\alpha\hat{\alpha}\hat{\omega}\hat{F}$ and (b) re-normalized deflection at origin $\hat{l}/\hat{\alpha}$ over the computed range of motion (i.e. the initial quasi-transience and the latter quasi steady state), with $\mathcal{D}$; DLVO forces are not considered; other system parameters are: $R$ = 10 $\mu$m, $h_0$ = 1 nm, $L$ = 1 $\mu$m, $E_{\text{Y}}$ = 1.0 MPa, $\nu$ = 0.45, $\mu$ = 1 mPa-s; the plot-line labels for each subfigure mean: 1 - maximum attractive for $\hat{\omega} = 10^{5}$ Hz, 2 - maximum repulsive for $\hat{\omega} = 10^{5}$ Hz, 1 - maximum attractive for $\hat{\omega} = 10^{5}$ Hz, 2 - maximum repulsive for $\hat{\omega} = 10^{5}$ Hz, 3 - maximum attractive for $\hat{\omega} = 10^{4}$ Hz, 4 - maximum repulsive for $\hat{\omega} = 10^{4}$ Hz, 5 - maximum attractive for $\hat{\omega} = 10^{3}$ Hz, 6 - maximum repulsive for $\hat{\omega} = 10^{3}$ Hz, 7 - maximum attractive for $\hat{\omega} = 10^{2}$ Hz, 8 - maximum repulsive for $\hat{\omega} = 10^{2}$ Hz, 9 - maximum attractive for $\hat{\omega} = 10^{1}$ Hz, 10 - maximum repulsive for $\hat{\omega} = 10^{1}$ Hz, 11 - maximum attractive for $\hat{\omega} = 10^{0}$ Hz, 12 - maximum repulsive for $\hat{\omega} = 10^{0}$ Hz; both the panels are log-scaled on both horizontal and vertical axes}
\label{fig:D}
\end{figure}
\begin{figure}[!htb]
\centering
\begin{subfigure}[b]{0.495\textwidth}
\centering
\includegraphics[width=\textwidth]{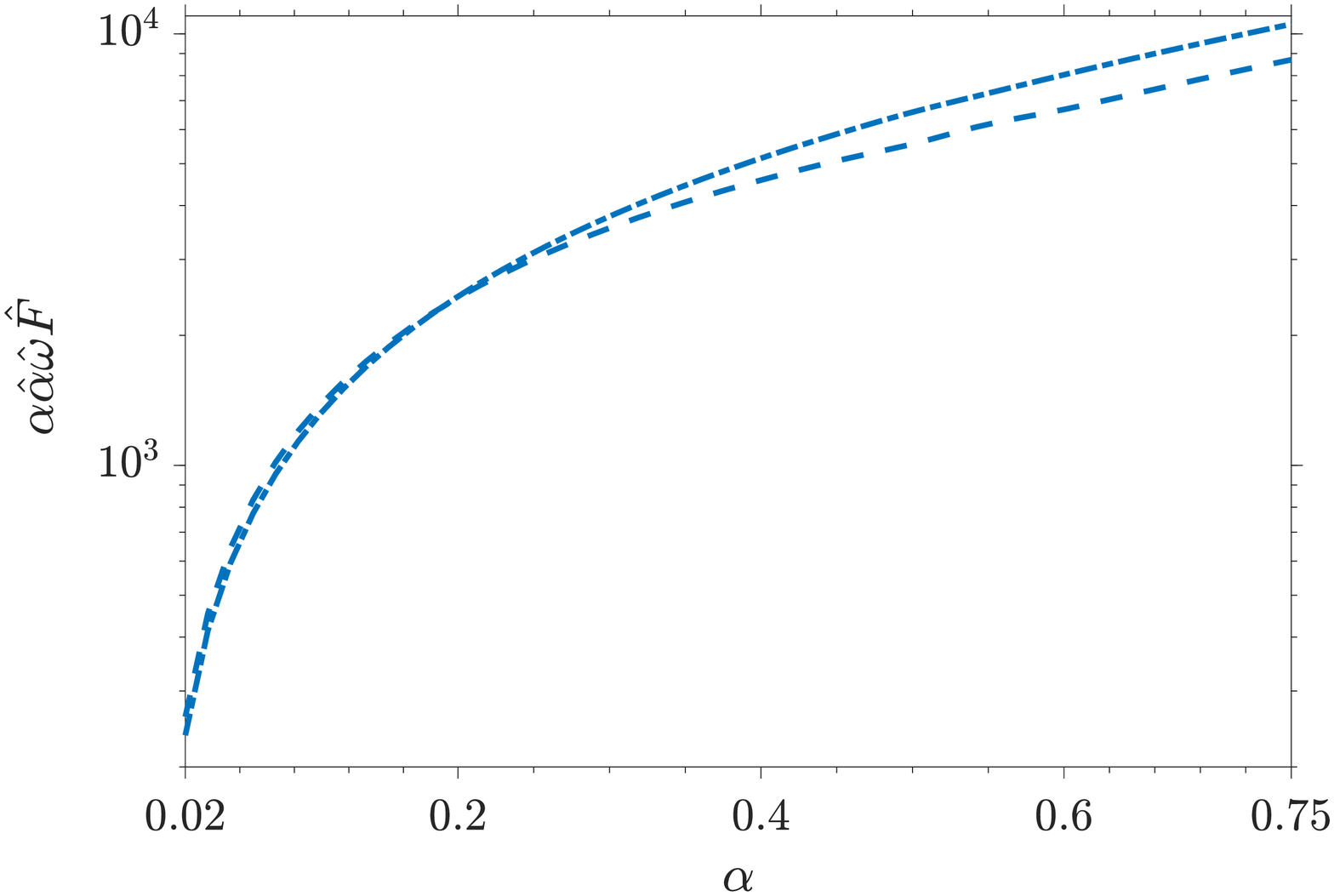}
\caption{}
\label{subfig:F_vs_alpha_HighAmp}
\end{subfigure}
\begin{subfigure}[b]{0.495\textwidth}
\centering
\includegraphics[width=\textwidth]{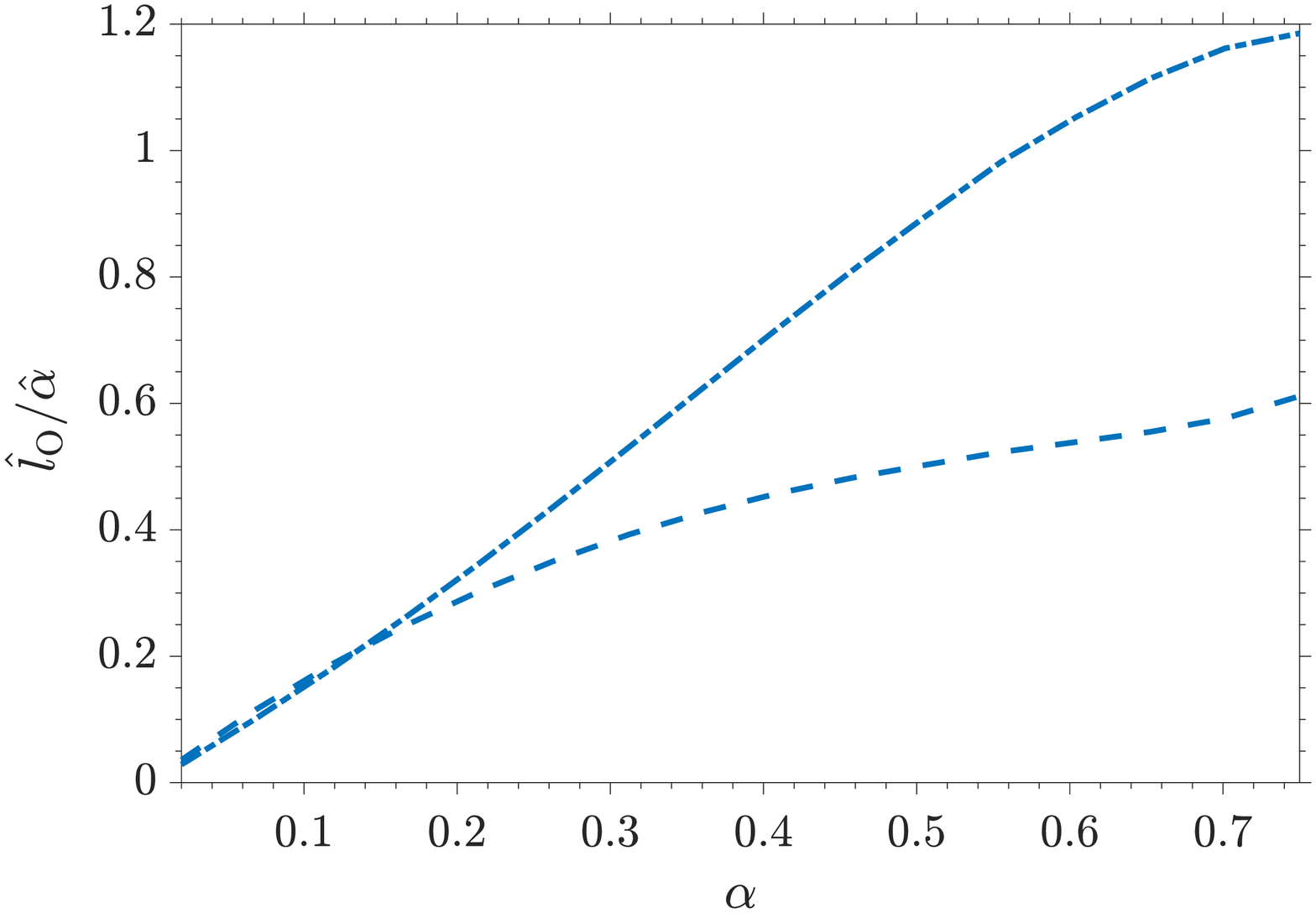}
\caption{}
\label{subfig:l_vs_alpha_HighAmp}
\end{subfigure}
\caption{Variation of magnitude of maximum attractive (negative, denoted by dashed-dot line), and, maximum repulsive (positive, denoted by dashed line) (a) re-normalized Force $\alpha\hat{\alpha}\hat{\omega}\hat{F}$ and (b) re-normalized deflection at origin $\hat{l}/\hat{\alpha}$ over the computed range of motion (i.e. the initial quasi-transience and the latter quasi steady state), with $\alpha$; DLVO forces are not considered; $h_0$ is varied from 1 nm to 35 nm; other system parameters are: $R$ = 10 $\mu$m, $D$ = 50 nm, $L$ = 1 $\mu$m, $\hat{\omega} = 10^{3}$ Hz, $E_{\text{Y}}$ = 1.0 MPa, $\nu$ = 0.45, $\mu$ = 1 mPa-s}
\label{fig:F_alpha_highamp}
\end{figure}
\begin{figure}[!htb]
\centering
\begin{subfigure}[b]{0.495\textwidth}
\centering
\includegraphics[width=\textwidth]{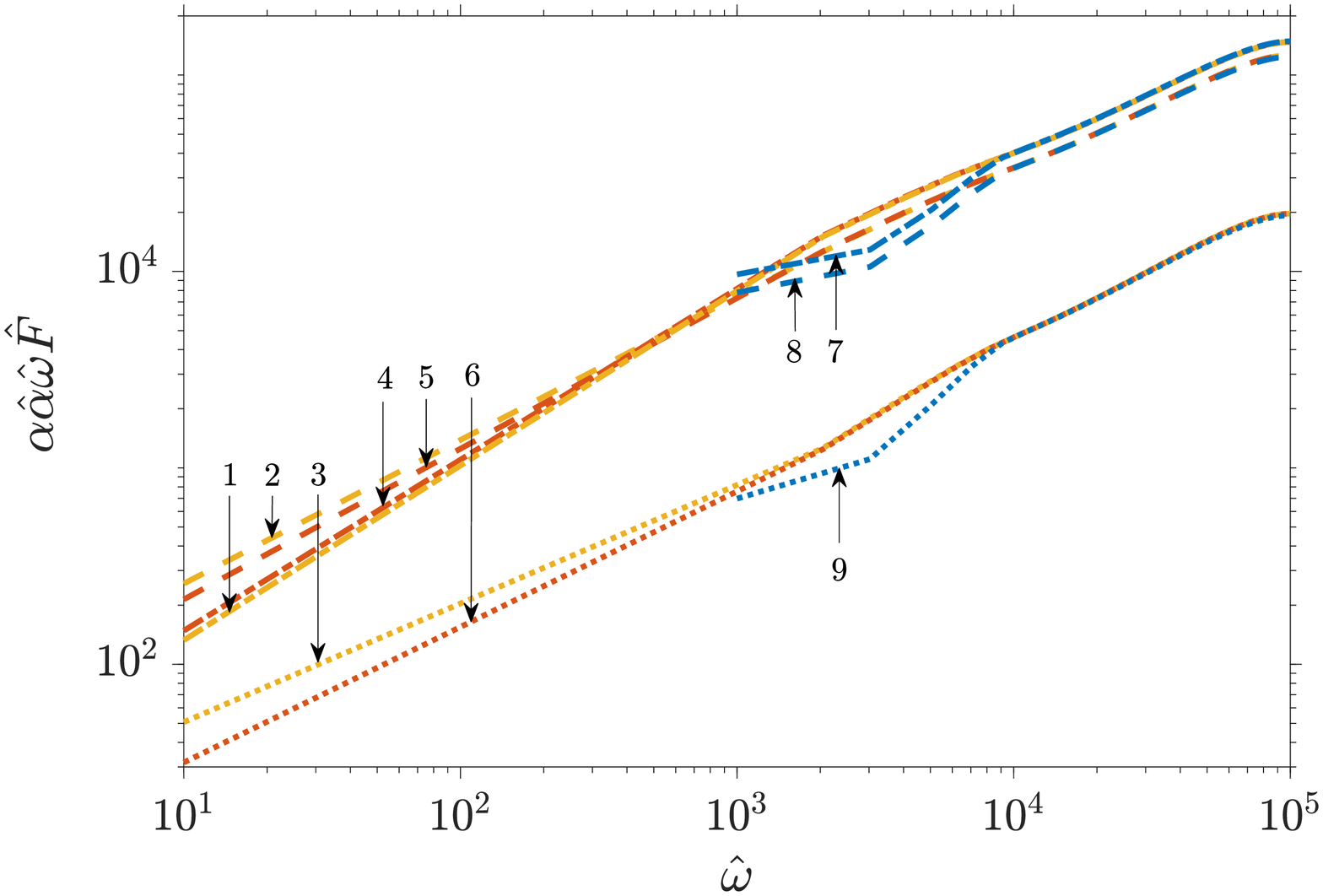}
\caption{}
\label{subfig:F_vs_omga_FreqDLVO}
\end{subfigure}
\begin{subfigure}[b]{0.495\textwidth}
\centering
\includegraphics[width=\textwidth]{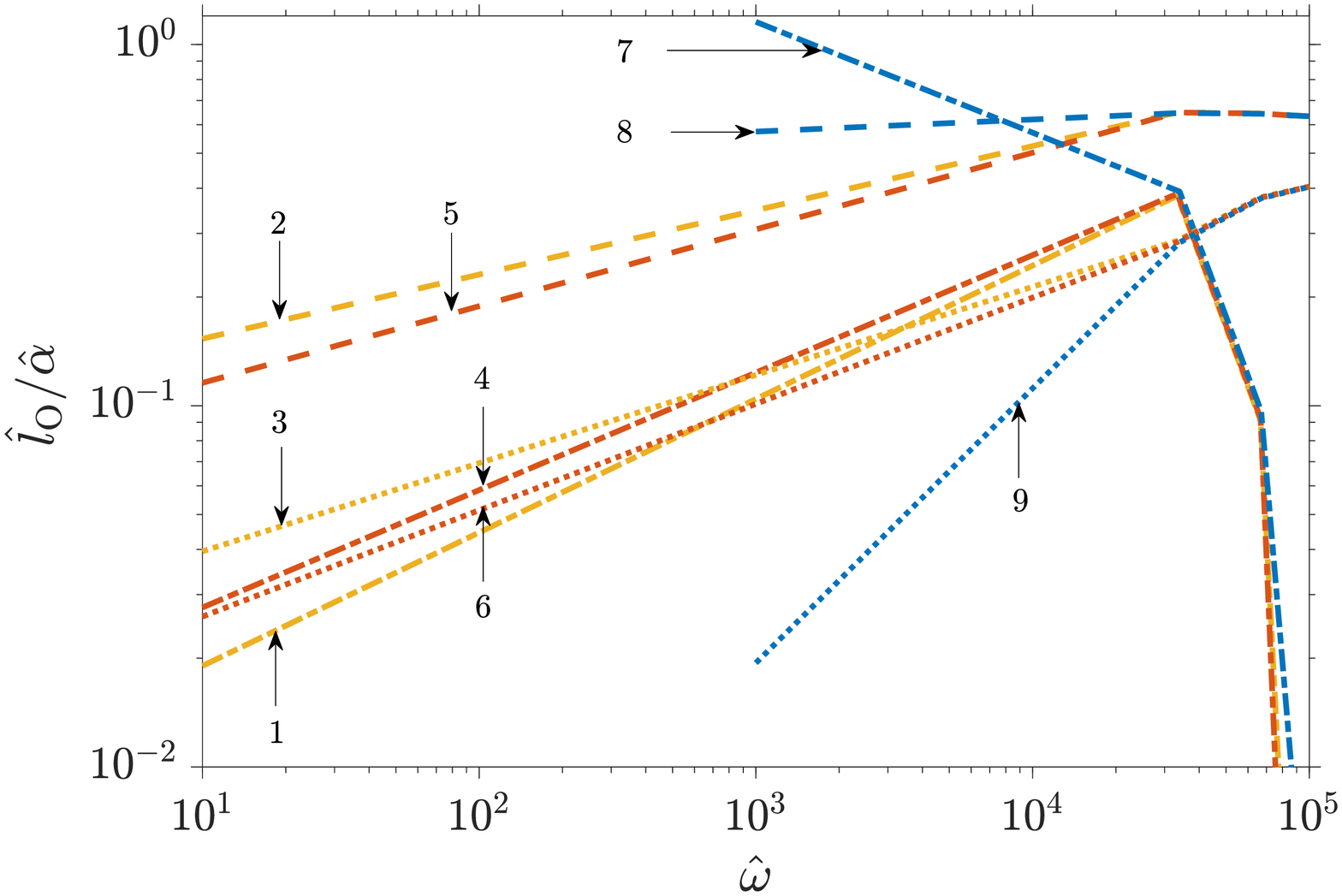}
\caption{}
\label{subfig:l_vs_omga_FreqDLVO}
\end{subfigure}
\caption{Variation of magnitude of maximum attractive (negative), and, maximum repulsive (positive), of (a) re-normalized Force $\alpha\hat{\alpha}\hat{\omega}\hat{F}$ and (b) re-normalized deflection at origin $\hat{l}/\hat{\alpha}$ over the computed range of motion (i.e. the initial quasi-transience and the latter quasi steady state), with $\hat{\omega}$; other system parameters are: $R$ = 10 $\mu$m, $D$ = 50 nm, $h_0$ = 35 nm, $L$ = 1 $\mu$m, $\hat{\omega} = 10^{3}$ Hz, $E_{\text{Y}}$ = 1.0 MPa, $\nu$ = 0.45, $\mu$ = 1 mPa-s; characteristics for system without DLVO force are not obtainable below $\hat{\omega}$ smaller than $10^3$ Hz because of adhesion-like behaviour; the plot-line labels for each subfigure mean: 1 - maximum attractive for strong DLVO, 2 - maximum repulsive for strong DLVO, 3 - mean for strong DLVO, 4 - maximum attractive for moderate DLVO, 5 - maximum repulsive for moderate DLVO, 6 - mean for moderate DLVO, 7 - maximum attractive without DLVO forces, 8 - maximum repulsive without DLVO forces, 9 - mean without DLVO forces; both the panels are log-scaled on both horizontal and vertical axes}
\label{fig:omga_DLVO}
\end{figure}
The results presented in section \ref{subsec:results_lowfreq} are for $\hat{\omega} = 1$ Hz, i.e. for low oscillation frequency. However, SPMs are often actuated with high-frequency oscillations for a variety of applications. The characterization of viscoelastic behaviour of many materials is done using high-frequency oscillations \cite{Phan2017}. The associated theoretical principle of this approach was utilized by Leroy and Charlaix (2011) \cite{Leroy2011} in developing a methodology for fluid-mediated non-contact elasticity characterisation of delicate substances - together, the fluid and the deformable substance act like a viscoelastic material. The standard approach to viscoelasticity characterization using this methodology is to recover the complex modulus of the material, where the real part is called the storage modulus and represents the elastic component (energy gets `stored' upon deformation and is recovered when the material returns to undeformed configuration) and the imaginary part is called the loss modulus and represents the viscous component (energy gets `lost' due to viscous dissipation). The key to this methodology is the lag between the force response and the oscillatory loading, both of which are periodic and the oscillatory loading is sinusoidal. The oscillation amplitude is typically restricted to a small value (in comparison to the mean separation of sphere from origin) so that the force response stays appreciably close to sinusoidal and hence, expression of the force response in terms of a complex modulus becomes possible \cite{Leroy2011}. However, in presenting our exhaustive analytical framework, we assess the effects of presence of DLVO forces and high oscillation amplitude in the context of high frequency oscillations as well.\\
Before delving into the individual aspects as mentioned above, we bring out a feature of the solution of the individual cases of high frequency oscillations. Since the simulation of a particular case starts from rest, we observe that the time derivative of deflection term on RHS of equation \eqref{eq:Re_eq} imparts a transience to the solution. This transience is strongly dependent on oscillation frequency and on ratio of oscillation amplitude to mean separation of sphere from origin, i.e. $\hat{\omega}$ and $\alpha$ - the transience persists for a larger number of oscillations for higher values of $\hat{\omega}$ and $\alpha$. We demonstrate this feature in figure \ref{fig:1e3Hz_D_10nm_h0_1nm}, where we show the time-evolution of force (top panel), pressure at origin (middle panel) and deflection at origin (bottom panel) for the case of $\hat{\omega} = 10^{3}$ Hz with $h_0$ and $D$ as 1 nm and 10 nm, i.e. $\alpha$ = 0.1 (other system parameters are presented in the caption of figure \ref{fig:1e3Hz_D_10nm_h0_1nm}). We can see that both pressure $p^*$ and deflection $l^*$ take about 15 oscillations to reach a quasi-steady state. In contrast, force $F^*$ reaches a quasi-steady state in only two oscillations. We have observed that for other cases as well, the force reaches a quasi-steady state faster than pressure and deflection. Lastly, we emphasize that in the assessments presented in the rest of this section, we have considered the variations of different system variables only after the system has reached the quasi-steady state in terms of all three variables - force, pressure and deflection. \\  
\subsubsection{Representative Cases}\label{subsubsec:results_representative}
We first discuss some representative cases that assist us in studying the effects of high oscillation frequency and its conjugation with high oscillation amplitude and presence of DLVO forces. The time evolution of force $F^*$ (after system has reached quasi-steady state) for these representative cases is presented in figure \ref{fig:F_vs_t}. Alongwith this, the phase plot of force $F^*$ and deflection at origin $l_{\text{O}}^*$ with separation of sphere from origin $H_{\text{O}}^*$ as the system evolves is presented in figure \ref{fig:F_vs_H0}. The first representative case is of low frequency oscillations with low amplitude ($\hat{\omega} = 10^{1}$ Hz and $h_0$ = 1 nm), presented in figures \ref{subfig:F_vs_t_low_omega_low_h0}, \ref{subfig:F_vs_H0_low_omega_low_h0} and \ref{subfig:l_vs_H0_low_omega_low_h0}. The coupling of pressure and deflection is OWC as the computed force (solid blue line) co-incides with the hypothetical-OWC solution (thin solid red line). Furthermore, it is perfectly sinusoidal and exhibits a lag of exactly $\pi/2$ to the oscillatory loading (figure \ref{subfig:F_vs_t_low_omega_low_h0})- this manifests as the phase plots of force and deflection (figures \ref{subfig:F_vs_H0_low_omega_low_h0} and \ref{subfig:l_vs_H0_low_omega_low_h0}) being perfect circles. The presence or absence of DLVO forces for this case is inconsequential because the separation is always large enough for them to be negligible. The second representative case is of low frequency oscillations with high amplitude ($\hat{\omega} = 10^{1}$ Hz and $h_0$ = 35 nm), presented in figures \ref{subfig:F_vs_t_low_omega_high_h0_with_DLVO}, \ref{subfig:F_vs_H0_low_omega_high_h0_with_DLVO}, \ref{subfig:l_vs_H0_low_omega_high_h0_with_DLVO} and \ref{subfig:pall_vs_t_low_omega_high_h0_with_DLVO}. For this combination of frequency and amplitude, the system in absence of DLVO forces exhibits adhesive behaviour, and is therefore outside the scope of the framework. On the other hand, the solution with the presence of DLVO forces is expectedly different from a sinusoidal one. Amongst the DLVO pressure components, only EDL disjoining pressure is appreciable in magnitude. However, it is sufficiently strong to dominate hydrodynamic pressure and therefore primarily dictate total pressure and resultantly force (see figure \ref{subfig:pall_vs_t_low_omega_high_h0_with_DLVO}). We note that the computed force (solid blue line) is considerably close to the hypothetical OWC force (thin solid red line). Comparing figures \ref{subfig:F_vs_H0_low_omega_high_h0_with_DLVO} and \ref{subfig:l_vs_H0_low_omega_high_h0_with_DLVO}, we can see that the deflection phase-plot is significantly more `bent' than the force phase-plot. The third representative case is of high frequency oscillations with low amplitude ($\hat{\omega} = 10^{3}$ Hz and $h_0$ = 1 nm), presented in figures \ref{subfig:F_vs_t_high_omega_low_h0}, \ref{subfig:F_vs_H0_high_omega_low_h0}, \ref{subfig:l_vs_H0_high_omega_low_h0} and \ref{subfig:l_vs_t_high_omega_low_h0}. The force can be seen to vary sinusoidally and with a lag (which is not equal to $\pi/2$) with the oscillatory loading (compare phases of the solid blue line and the thin solid red line in figures  \ref{subfig:F_vs_t_high_omega_low_h0} and  \ref{subfig:F_vs_H0_high_omega_low_h0}). Also, analysing figures  \ref{subfig:F_vs_H0_high_omega_low_h0} and  \ref{subfig:l_vs_H0_high_omega_low_h0}, we can see that the quasi-steady state variations of both force and deflection  with $H_{\text{O}}^*$ are ellipses. This implies that both force and deflection vary sinusoidally with time with a (non-$\pi/2$) lag with oscillatory loading. Furthermore, the ellipse for deflection is significantly more eccentric than the one for force. This occurs because the lag for deflection is much smaller than the lag for force, observable by comparing figures \ref{subfig:F_vs_t_high_omega_low_h0} and \ref{subfig:l_vs_t_high_omega_low_h0}. The presence or absence of DLVO forces for this case is inconsequential because the separation is always large enough for them to be negligible.  The fourth representative case is of high frequency oscillations with high amplitude ($\hat{\omega} = 10^{3}$ Hz and $h_0$ = 35 nm), presented in figures \ref{subfig:F_vs_t_high_omega_high_h0}, \ref{subfig:F_vs_H0_high_omega_high_h0}, \ref{subfig:l_vs_H0_high_omega_high_h0}, \ref{subfig:l_vs_t_high_omega_high_h0} and \ref{subfig:pall_vs_t_high_omega_high_h0_with_DLVO}. The force, while not perfectly sinusoidal, is close enough to sinusoidal for a sine-curve to exhibit appreciable similarity (compare the solid blue line and dashed yellow line in figure \ref{subfig:F_vs_t_high_omega_high_h0}). Analysing figures  \ref{subfig:F_vs_H0_high_omega_high_h0} and  \ref{subfig:l_vs_H0_high_omega_high_h0}, we can see that the quasi-steady state variations of force with $H_{\text{O}}^*$ is close to an ellipse (although not a perfect ellipse). However, the quasi-steady state variations of deflection with $H_{\text{O}}^*$ is not an ellipse anymore and assumes a complicated shape - an outcome of the high amplitude. Similarly, the shape of deflection evolution with time is also not sinusoidal but a more complicated shape - see figure \ref{subfig:l_vs_t_high_omega_high_h0}. Lastly, the presence or absence of DLVO pressure components for this case turns out to be inconsequential. The quasi-steady state evolution of the two non-zero pressure components (hydrodynamic and EDL disjoining) and the total pressure is presented in figure \ref{subfig:pall_vs_t_high_omega_high_h0_with_DLVO} - clearly, the total pressure is strongly dominated by hydrodynamic pressure. This dominance occurs because of the high-frequency. \\
\subsubsection{Effect of Varying Frequency}\label{subsubsec:results_freq}
We now assess the effects of oscillation frequency on the behaviour of low amplitude oscillations in the absence of DLVO forces.  We study a range of osciilation frequency from $\hat{\omega} = 10^{5}$ Hz to $\hat{\omega} = 10^{5}$ Hz. The oscillation amplitude ($h_0$) is maintained at 1 nm. For each of the oscillation frequencies, we consider a range of the mean distance of sphere from origin ($D$) - 5 nm to 50 nm (5 nm to 120 nm for $\hat{\omega} = 10^{3}$ Hz). The other system parameters are presented in the caption of figure \ref{fig:G}. The oscillatory loading of the sphere is given as,
\begin{equation}
\label{eq:H_highfreq}
H^* = 1+\frac{r^{*2}}{2R}+h_0\cos(\omega t^*),
\end{equation}
i.e. the transient part of the sphere's oscillation is,
\begin{equation}
\label{eq:h_highfreq}
h^* = h_0\cos(\omega t^*).
\end{equation}
For the case of low amplitude oscillation, i.e. when $h_0 \ll D \implies \alpha \ll 1$, we observe that the force between sphere and substrate remains sinusoidal, and can therefore be expressed as,
\begin{equation}
\label{eq:F_highfreq_lowamp}
F^* = F_0\cos(\omega t^*+\phi).
\end{equation}
Note the presence of $\phi$, which is the lag of force $F^*$ in comparison to the oscillatory loading $h^*$. This lag is $\pi/2$ for the case of sufficiently low-oscillation frequency, where deformation of the substrate is low enough for it to be practically rigid and so force response staying purely viscous. Equations \eqref{eq:h_highfreq} and \eqref{eq:F_highfreq_lowamp} can be expressed in complex notation as,
\begin{equation}
\label{eq:h_complex}
h^* = h_0\exp(\iota \omega t^*),
\end{equation}
\begin{equation}
\label{eq:F_complex}
F^* = \bar{F}_0\exp(\iota \omega t^*),
\end{equation}
where $\bar{F}_0$ is complex and captures the lag. The expressions for the storage modulus $G^{\prime}$ and loss modulus $G^{\prime\prime}$ are given as \cite{Leroy2011},
\begin{equation}
\label{eq:G}
\bar{G} = G^{\prime}+\iota G^{\prime\prime} = \frac{\tilde{F}_0}{h_0}.
\end{equation}
The variation of $G^{\prime}$ and $G^{\prime\prime}$ with $D$ for frequencies $\hat{\omega} = [10^{1}, 10^{3}, 10^{4}]$ Hz are presented in figure \ref{fig:G}. Looking at figures \ref{subfig:Gpr_1e1Hz} and \ref{subfig:Gprpr_1e1Hz}, we observe that for $\hat{\omega} = 10^1$ Hz, the storage modulus $G^{\prime}$ is insensitive to $D$ for values of $D$ higher than $\approx 30$ nm. Similarly, looking at figures \ref{subfig:Gpr_1e4Hz} and \ref{subfig:Gprpr_1e4Hz}, we observe that for $\hat{\omega} = 10^4$ Hz, both storage modulus $G^{\prime}$ and loss modulus $G^{\prime\prime}$ become insensitive to $D$ for values of $D$ lower than $D\approx 25$ nm. These insensitivities are exagerrated as one moves to more extreme frequencies in either direction - for $\hat{\omega} = 10^{1}$ Hz, the storage modulus $G^{\prime}$ is insensitive to $D$ for its considered range, and, for $\hat{\omega} = 10^{5}$ Hz, both the storage modulus $G^{\prime}$ and loss modulus $G^{\prime\prime}$ are insensitive to $D$ for its considered range. Hence, we deduce that $\hat{\omega} = 10^{3}$ Hz is the suitable frequency to recover appreciable sensitivity of $G^{\prime}$ and $G^{\prime\prime}$ to $D$ for the considered range of $D$ and given values of other parameters. The plots for $\hat{\omega} = 10^{3}$ Hz are presented in figures \ref{subfig:Gpr_1e3Hz} and \ref{subfig:Gprpr_1e3Hz}. The variations are more amenable to assessment than other frequencies and exhibit appreciable match with the values obtained by Leroy and Charlaix \cite{Leroy2011}.\\
\subsubsection{Effect of Oscillation Amplitude}\label{subsubsec:results_amp}
Next, we assess the effect of increasing the oscillation amplitude. For this, we consider oscillation frequency of $\hat{\omega} = 10^{3}$ Hz and $D$ = 50 nm. The amplitude $h_0$ is varied from 1 nm to 35 nm. The rest of the system parameter values are the same as those for section \ref{subsubsec:results_freq}. We observe that the effect on force of varying oscillation amplitude $h_0$ is much weaker compared to the effect of varying mean separation of sphere from origin $D$. We do observe some degeneration in the sinusoidality of time-evolution of force, but it remains appreciably close to the force for low amplitude of $h_0$ = 1nm. As a result, the values of $G^{\prime}$ and $G^{\prime\prime}$ remain the same so long as $D$ remains 50 nm, even as we vary $h_0$ from 1 nm to 35 nm. Similarly, considering DLVO forces (as per weak as well as strong DLVO pressure parameters, i.e. $A_{\text{sfw}}$ = $10^{-21}$ J and $\psi_S$ = 100 mV as well as $A_{\text{sfw}}$ = $10^{-20}$ J and $\psi_S$ = 2500 mV) does not affect the force and deflection response appreciably, and hence is inconsequential in determining $G^{\prime}$ and $G^{\prime\prime}$.\\
The obtained results for $\hat{\omega} G^{\prime}$ and $\hat{\omega} G^{\prime\prime}$, for the different oscillation frequencies, and with increasing oscillation amplitude for $\hat{\omega} = 10^{3}$ Hz, are collectively presented in the complex plane in figure \ref{fig:G_zplane}. Note that $\hat{\omega}$ is multiplied as the pressure, and therefore, force response is proportional to $\hat{\omega}$. hence, when comparing the results for different $\hat{\omega}$, we have to multiply with $\hat{\omega}$ so that all the plots are visible simultaneously. For $\hat{\omega}= 10^{0}$ Hz, $G^{\prime}$ is zero and insensitive to $D$ for its considered range. Therefore, its phase plot (blue line) is co-incident with the imaginary axis. Similarly, the phase plot for $\hat{\omega} = 10^{1}$ Hz (orange line) conicides with the imaginary axis for part of its length because $G^{\prime}$ for this frequency is zero and insensitive to $D$ for $D$ higher than $\approx 30$ nm. On the other end, for $\hat{\omega} = 10^{4}$ Hz, both $G^{\prime}$ and $G^{\prime\prime}$ are insensitive to $D$ (but not zero) for values of $D$ lower than $\approx$ 25 nm. Hence its phase-plot (green line) hovers at the starting point with increasing $D$ till $D$ reaches 25 nm, after which it starts moving to exhibit a curve. Since both $G^{\prime}$ and $G^{\prime\prime}$ are insensitive to $D$ (but not zero) for the entirety of its considered range for $\hat{\omega} = 10^{5}$ Hz, its phase-plot (cyan point) is reduced to a point on the imaginary plane. Lastly, varying $h_0$ from 1 nm to 35 nm for $\hat{\omega} = 10^{3}$ Hz gives rise to a small variation in comparison to the phase-plot for $\hat{\omega} = 10^{3}$ Hz with varying $D$ (purple line). Hence, it appears as the magenta star marker at the point on the purple line that corresponds to $D$ = 50 nm. \\
\subsubsection{Parametric Analysis}\label{subsubsec:results_parametric}
Lastly, we assess the parametric variation of maximum attractive, maximum repulsive and mean force and deflection at origin, over the entire range of motion (i.e. the initial transience and the long-time quasi-steady state), , with $\displaystyle \mathcal{D} = \frac{D}{R}$ in figure \ref{fig:D}, $\displaystyle \alpha$ in figure \ref{fig:F_alpha_highamp} and with $\displaystyle \hat{\omega}=\frac{\omega}{2\pi}$ in figure \ref{fig:omga_DLVO}. The first two figures present results restricted to purely hydrodynamic systems (i.e. DLVO forces are absent). The third figure has plots for three cases - DLVO pressure components are absent, DLVO pressure components are present and are of comparable magnitude to hydrodynamic pressure at lower frequencies (labelled `moderate DLVO'), and DLVO pressure components are present and are dominating over hydrodynamic pressure at lower frequencies (labelled `strong DLVO'). For ease of comparison between different plot-lines, we plot $\displaystyle \alpha\hat{\alpha}\hat{\omega}\hat{F} = \frac{F^*}{\mu\omega R^2}$ and $\displaystyle \frac{\hat{l}}{\hat{\alpha}} = \frac{l^*}{D}$, where $\displaystyle \hat{\alpha} = \frac{1}{1-\alpha}$, since in this section $\omega$ and $D$ or $h_0$, i.e. effectively $\alpha$, vary across different cases.   \\
Looking at figure \ref{subfig:F_vs_D_Freq}, we observe that the force increases between two to three order of magnitude as frequency grows from $\hat{\omega} = 10^{1}$ Hz to $\hat{\omega} = 10^{5}$ Hz (red line to cyan line). However, it is significantly smaller for $\hat{\omega} = 10^0$. The maximum repulsive and maximum attractive force characteristics are close to one anther for each frequency, getting closer for lower frquencies and co-inciding perfectly for $\hat{\omega} = 10^0$ Hz. Lastly, with increasing frequency, the lines become closer to horizontal, and they are virtually horizontal for part of $\hat{\omega} = 10^{4}$ Hz and $\hat{\omega} = 10^{5}$ Hz 
\begin{comment}
(the plot being log-scaled and the lines being separated by orders of magnitude has led to the limits of the vertical axis being so large that this effect is not strongly perceptible)
\end{comment}
. This observation is in line with the observation that $G^{\prime}$ and $G^{\prime\prime}$ become insensitive to $D$ for part of range of $D$ for $\hat{\omega} = 10^{4}$ Hz and for complete range of $D$ for $\hat{\omega} = 10^{5}$ Hz. In contrast, the insensitivity of $G^{\prime}$ to $D$ for smaller frequencies (i.e. for part of range of $D$ for $\hat{\omega} = 10^{1}$ Hz and for complete range of $D$ for $\hat{\omega} = 10^{0}$ Hz) is not clearly deductible from their trends (lines 9, 10, 11, 12) because $G^{\prime\prime}$ is still sensitive to $D$ for these frequencies. Looking at figure \ref{subfig:F_vs_D_Freq}, we can see that the deflection characteristics are significantly complicated and clear inferences about the system behaviour cannot be derived. However, we do observe that even with this complicated contrast in deflection characteristics for different frequencies, the contrast in force characteristics for different frequencies is simpler.\\
Looking at figure \ref{subfig:F_vs_alpha_HighAmp}, we observe that the maximum repulsive and maximum attractive force characteristics remain close and increase with increasing $\alpha$ (i.e. increasing oscillation amplitude). Looking at figure \ref{subfig:l_vs_alpha_HighAmp}, we observe that with increasing $\alpha$, the maximum repulsive as well as maximum attractive deflection increases, similar to force charactistics. However, unlike force characteristics, they deviate from each other significantly in magnitude. \\
Looking at figure \ref{subfig:F_vs_omga_FreqDLVO}, we observe force characteristics for the three categories - strong DLVO (lines 1, 2, 3), moderate DLVO (lines 4, 5, 6) and without DLVO forces (lines 7, 8, 9) - are substantially close. All three categories merge together as frequency increases. This is expected as the effect of DLVO pressure components is small for moderately high frequencies (i.e. $\hat{\omega} = 10^3$) and further becomes negligible for higher frequency (i.e. $\hat{\omega} = 10^4, 10^5$). The maximum repulsive force and maximum attractive force characteristics are close in magnitude and about an order higher than mean force characterstics. Looking at figure \ref{subfig:l_vs_omga_FreqDLVO}, we observe that the deflection characteristics for strong DLVO and moderate DLVO are substantially close, similar to what was seen for force characteristics. Similarly, the characteristics strong or moderate DLVO and without DLVO forces merge with increasing $\hat{\omega}$. However, we observe that as $\hat{\omega}$ increases, mean and maximum repulsive deflection characteristics increase whereas the maximum attractive deflection characteristics decreases sharply. Similar to the observation for figure \ref{fig:D}, here also, even with the complicated contrast in deflection characteristics with and without DLVO forces, the contrast in force characteristics is small.\\
\section{Conclusion}\label{sec:Conclusion}
In this study, we have presented a semi-analytical framework for studying the elastohydrodynamics of fluid-mediated axisymmetric loading of a rigid sphere over soft substrate layer of arbitrary thickness coated on a rigid platform. The intervening (i.e. mediating) fluid material is an incompressible homogeneous isotropic Newtonian fluid. The soft substrate material is a homogeneous isotropic linear elastic solid, which can be compressible as well as incompressible. Loading of the rigid sphere can be in three modes - approach with constant speed, recession with constant speed, and harmonic oscillations with one sinusoidal term. The framework is capable of simultaneously incorporating sfor high frequency oscillations (or high speed approach/recession), two-way elastohydrodynamic coupling, and DLVO forces in the form of closed-form expressions of van der Waals pressure and electrical double layer (EDL) disjoining pressure. \\
Using this fremework, we analyse the system behaviour of a wide range of parameters to gain insights about the effect of different aspects of the setup - substrate thickness, substrate material compressibility (quantified by Poisson's ratio), setup geometry and imposed dynamics, and DLVO forces. The fluid and solid material properties, system geometry, and imposed dynamic parameters are taken as those for a representative system occuring in a scanning probe microscopy (SPM) setup. The range of DLVO pressure component parameters (Hamaker's constant varying from zero then $10^{-22}$ J to $10^{-20}$ J and surface potential varying from zero then $100$ mV to $2500$ mV) is taken as representative values that occur for different materials encountered in common biological and engineering situations. We first present solution for the limiting cases presented in section \ref{tab:limits} with one-way coupling between pressure and fluid-substrate interface deflection (termed just `deflection' for brevity), enforced by taking a high value of substrate material Young's modulus, in order to compare the magnitude of deflection for these limiting cases. We next present the time evolution for some representative cases that are presented in the subsequent subsections of section \ref{sec:Results}. We next study approach and recession loading with different loading speeds and with and without DLVO forces. The force and deflection are strongly dependent on loading speed when DLVO forces are absent, but become insensitive to loading speed (except for small part of the complete range of motion of the sphere, corresponding to the large values of separation of sphere from fluid-substrate interface). We next study oscillatory loading with the low oscillation frequency of $2\pi$ rad/s. We assess the effects of substrate thickness, Poisson's ratio (which quatifies the substrate material compressibility), DLVO forces (quantified by the DLVO pressure component parameters, Hamaker's constant for van der Waals pressure and surface potential for EDL disjoining pressure). From the results, we make some key inferences, elaborated ahead. The effect of higher substrate thickness as well as smaller Poisson's ratio is the increase in the effective softness of the substrate, i.e. the substrate allows higher deformation. The effect of substrate thickness saturates in the semi-infinite limit, i.e. when the substrate thickness becomes much larger than the lubrication zone characteristic radial length. Focussing on DLVO forces, we find that van der Waals pressure is comparable to hydrodynamic pressure while EDL disjoining pressure is much stronger. Pertaining to the total pressure, the nature of its constituent components is such that - if total pressure is attractive, it is self-magnifying, and, if total pressure is repulsive, it is self-diminishing. This means that higher the fluid-substrate interface deflection caused by the attractive pressure, higher the attractive pressure and higher the fluid-substrate interface deflection caused by repulsive pressure, lower the repulsive pressure. This occurs because both attractive pressure and repulsive pressure grow with decreasing separation between the sphere and the fluid-substrate interface. However, deflection caused by attractive pressure decreases this separation whereas that caused by repulsive pressure increases it. We emphasize that the effects of substrate thickness and DLVO force are obtained in the aforementioned approach/recession loading with different loading speeds also. We lastly study oscillatory loading with the low to high oscillation frequencies ranging from $2\pi\times 10^{0}$ rad/s to $2\pi\times 10^{5}$ rad/s. By varying the frequency of low-amplitude oscillations for a setup without DLVO forces, we observe that a frequency in the range of  $2\pi\times 10^{2}$ to $2\pi\times 10^{3}$ rad/s is suitable to obtaining better interpretable variation of the storage and loss moduli ($G^{\prime}$ and $G^{\prime\prime}$ respectively) with mean separation of spherical probe from the pre-deformation fluid-substrate interface. We also observe that for the frequency of $2\pi\times 10^{3}$, there is some degeneration in the obtainability of storage and loss moduli when the oscillation amplitude is high, but they remain obtainable. Lastly, it is seen that the presence of DLVO forces does not appreciably effect the force and deflection behaviour for high-amplitude oscillations when frequency is high.\\
Having discussed the framework and the results, we list out some aspects that have not been explored in the implementation of the framework. Apart from DLVO forces, the non-DLVO molecular forces (like hydration force, steric hindrance forces and solvation force) are also often expressed as closed form expressions of pressure and hence amenable to trivial incorporation in the framework. However, we have not studied these forces here. Similarly, wall slip is another phenomenon that is not incorporated here and can exhibit consequential interplay with other effects like electrokinetics and deformability \cite{Karan2018a,Vinogradova2000}. The shape of the probe is considered as only spherical in this study.  However, it could be of interest to consider more general axisymmetric shapes as well \cite{Zhao2020}. We also list out some of the limitations of the framework itself. The separation of sphere from the substrate is restricted to maintain the lubrication approximation valid. However, this could cease to be the case for different sedimentation phenomena and even in some scanning probe microscopy situations. Similarly, axisymmetry is another restriction which should be relaxed in a more general framework. The framework is also not conducive to tackle problems where substrate deformation could cause adhesion-like behaviour. Lastly, apart from EDL disjoining pressure, electroviscous effect is another effect of electrokinetic phenomena in the intervening fluid (as discussed in section \ref{sec:Introduction}). The current framework does not incorporate electroviscous effects. It is expected that incorporating any of these effects will require significant and non-trivial modifications to the framework. The listed out unexplored aspects of the implementation and limitations of the framework serve as future directions for enhancement of this study. \\
In closing, we expect that our study functions as a launchpad for developing more versatile analytical and computational frameworks aimed as soft-lubrication studies of a wide range of objects near a wall setups, finding major applications in areas ranging from scanning probe microscopy to biomimetics and targetted drug delivery.\\
\bibliographystyle{siamplain}
\bibliography{refs_short}
\end{document}